\documentclass[12pt,twoside]{report}
\usepackage{epsfig}
\usepackage{amssymb}
\usepackage{slashed}
\usepackage{chapterbib}
\sectionbib{\section*}{subsubsection}
\newtheorem{exercise}{Exercise}[section]

\begin{document}
\begin{titlepage}
\title{A Chiral Perturbation Theory Primer}
\author{Stefan Scherer\thanks{scherer@kph.uni-mainz.de,
http://www.kph.uni-mainz.de/T/} and Matthias R.~Schindler\thanks{
schindle@kph.uni-mainz.de, http://www.kph.uni-mainz.de/T/}
\\Institut f\"ur Kernphysik\\
Johannes Gutenberg-Universit\"at Mainz\\
J.~J.~Becher Weg 45\\
D-55099 Mainz\\
Germany}
\date{May 2005}
\end{titlepage}
\maketitle
\newpage
\thispagestyle{empty}
\noindent{}
\chapter*{Preface}
\thispagestyle{empty}

  The present text is based on lectures given in the context of
the ECT$^\ast$ Doctoral Training Programme 2005 (Marie Curie Training Site) {\em
Hadronic Physics} at the European Centre for Theoretical Studies in Nuclear
Physics and Related Areas (ECT$^\ast$) in Trento, Italy.

    The course was addressed to PhD students with both rather different
interests and background in experimental and theoretical nuclear
and particle physics.
   The students were assumed to be familiar with elementary concepts
of field theory and relativistic quantum mechanics.
   The goal of the course was to provide a {\em pedagogical introduction} to
the basic concepts of chiral perturbation theory (ChPT) in the
mesonic and baryonic sectors.
   We have tried to also work out those pieces which by the ``experts'' are
considered as well known.
   In particular, we have often included intermediate steps in derivations
in order to facilitate the understanding of the origin of the
final results.
   We have tried to keep a reasonable balance between mathematical rigor
and illustrations by means of simple examples.
   Some of the topics not directly related to ChPT were covered
in extra lectures in the afternoon.
   By preparing numerous
exercises, covering a wide range of difficulty---from very easy to
quite difficult---, we hoped to take the different individual
levels of experience into account.
  Ideally, at the end of the course, a participant (or a reader of
these notes) should be able to perform simple calculations in the
framework of ChPT and to read the current literature.

   These lecture notes include the following topics.
   Chapter 1 deals with QCD and its global symmetries in the chiral
limit, explicit symmetry breaking in terms of the quark masses, and the concept
of Green functions and Ward identities reflecting the underlying chiral symmetry.
   In Chapter 2 the idea of a spontaneous breakdown of a global symmetry
is discussed and its consequences in terms of the Goldstone
theorem are demonstrated.
   Chapter 3 deals with mesonic chiral perturbation theory and
the principles entering the construction of the chiral Lagrangian
are outlined.
   In Chapter 4 the methods are extended to include the interaction
between Goldstone bosons and baryons in the single-baryon sector.
   Sections marked with an asterisk may be omitted in a first reading.

\vspace{2em}

\noindent Mainz and Trento, May 2005
\begin{flushright} Stefan Scherer and Matthias R.~Schindler\end{flushright}

\newpage
\thispagestyle{empty}

   Readers interested in the present status of applications are
referred to lecture notes and review articles
\cite{Leutwyler:pf:1991mz,Bijnens:pf:xi,Meissner:pf:1993ah,Leutwyler:pf:1994fi,%
Bernard:pf:1995dp,deRafael:pf:1995zv,Pich:pf:1995bw,Ecker:pf:1995gg,Manohar:pf:1996cq,%
Pich:pf:1998xt,Burgess:pf:1998ku,Scherer:pf:2002tk} as well as conference
proceedings
\cite{Bernstein:pf:zq,Bernstein:pf:pm,Bernstein:pf:2002,Meissner:pf:2003hr}.

\thispagestyle{empty}

\noindent {\bf Acknowledgements}

   The authors would like to thank the co-ordinator Matthias F.~M.~Lutz and
the co-organizers Michael Birse and W.~Vogelsang for the invitation to
participate in the program.
   Moreover, we would like to thank Prof.\ \mbox{J.~-~P.~Blaizot}, Prof.~G.~Ripka,
Stefania Campregher, and Donatella Rosetti for the hospitality and the pleasant
working conditions at ECT$^\ast$.
   M.~R.~S.~acknowledges the support through a Marie Curie fellowship.
   S.~S.~would like to thank the participants for their enthusiasm, staying
power, and patience to survive 24 lectures in one week.

\newpage
\thispagestyle{empty}
\noindent{}
\newpage
\setcounter{page}{1}
\pagenumbering{roman}
\tableofcontents
\newpage
\pagenumbering{arabic}
\chapter{QCD and Chiral Symmetry}
\section{Some Remarks on SU(3)}
\label{sec_srsu3}

The group SU(3) plays an important role in the context of the
strong interactions, because
\begin{enumerate}
\item it is the gauge group of quantum chromodynamics (QCD);
\item flavor SU(3) is approximately realized as a global symmetry
of the hadron spectrum, so that the observed (low-mass) hadrons
can be organized in approximately degenerate multiplets fitting
the dimensionalities of irreducible representations of SU(3);
\item the direct product $\mbox{SU(3)}_L\times\mbox{SU(3)}_R$ is
the chiral-symmetry group of QCD for vanishing $u$-, $d$-, and
$s$-quark masses.
\end{enumerate}
Thus, it is appropriate to first recall a few basic properties of
SU(3) and its Lie algebra su(3).

   The group SU(3) is defined as the set of all
unitary, unimodular, $3 \times 3$ matrices $U$, i.e.\
$U^\dagger U=1$,\footnote{Throughout these lectures we often
adopt the convention
that 1 stands for the unit matrix in $n$ dimensions. It should be clear
from the respective context which dimensionality actually applies.}
and $\mbox{det}(U)=1$.
   In mathematical terms, SU(3) is an eight-parameter,
simply connected, compact Lie group.
   This implies that any group element can be parameterized by a set of
eight independent real parameters $\Theta=(\Theta_1,\cdots, \Theta_8)$
varying over a continuous range.
   The Lie-group property refers to the fact that the group
multiplication of two elements $U(\Theta)$ and $U(\Psi)$
is expressed in terms of eight {\em analytic} functions
$\Phi_i(\Theta;\Psi)$, i.e.\ $U(\Theta)U(\Psi)=U(\Phi)$, where
$\Phi=\Phi(\Theta;\Psi)$.
   It is simply connected because every element can be connected to the
identity by a continuous path in the parameter space and compactness requires
the parameters to be confined in a finite volume.
   Finally, for compact Lie groups, every finite-dimensional representation
is equivalent to a unitary one and can be decomposed into a direct
sum of irreducible representations (Clebsch-Gordan series).

   Elements of SU(3) are conveniently written in terms
of the exponential representation\footnote{In our notation, the
indices denoting group parameters and generators will appear as subscripts or
superscripts depending on what is notationally convenient.
   We do not distinguish between upper and lower indices, i.e., we
abandon the methods of tensor analysis.}
\begin{equation}
\label{2:1:uexp}
U(\Theta)=\exp\left(-i\sum_{a=1}^8 \Theta_a \frac{\lambda_a}{2}\right),
\end{equation}
with $\Theta_a$ real numbers, and where the eight linearly independent
matrices $\lambda_a$ are the
so-called Gell-Mann matrices, satisfying
\begin{eqnarray}
\label{2:1:gmme1}
\frac{\lambda_a}{2}&=&i\frac{\partial U}{\partial \Theta_a}(0,\cdots,0),\\
\label{2:1:gmme2}
\lambda_a&=&\lambda_a^\dagger,\\
\label{2:1:gmme3}
\mbox{Tr}(\lambda_a \lambda_b)&=&2\delta_{ab},\\
\label{2:1:gmme4}
\mbox{Tr}(\lambda_a)&=&0.
\end{eqnarray}
    The Hermiticity of Eq.\ (\ref{2:1:gmme2}) is responsible for
$U^\dagger=U^{-1}$. On the other hand, since
$\mbox{det}[\exp(C)]=\exp[\mbox{Tr}(C)]$, Eq.\ (\ref{2:1:gmme4})
results in $\mbox{det}(U)=1$.
   An explicit representation of the Gell-Mann matrices is given by
\begin{eqnarray}
\label{2:1:gmm}
&&\lambda_1=\left(\begin{array}{rrr}
0&1&0\\1&0&0\\0&0&0
\end{array}
\right),\quad
\lambda_2=\left(\begin{array}{rrr}
0&-i&0\\i&0&0\\0&0&0
\end{array}
\right),\quad
\lambda_3=\left(\begin{array}{rrr}
1&0&0\\0&-1&0\\0&0&0
\end{array}
\right),\nonumber\\
&&\lambda_4=\left(\begin{array}{rrr}
0&0&1\\0&0&0\\1&0&0
\end{array}
\right),\quad
\lambda_5=\left(\begin{array}{rrr}
0&0&-i\\0&0&0\\i&0&0
\end{array}
\right),\quad
\lambda_6=\left(\begin{array}{rrr}
0&0&0\\0&0&1\\0&1&0
\end{array}
\right),\nonumber\\
&&
\lambda_7=\left(\begin{array}{rrr}
0&0&0\\0&0&-i\\0&i&0
\end{array}
\right),\quad
\lambda_8=\sqrt{\frac{1}{3}}\left(\begin{array}{rrr}
1&0&0\\0&1&0\\0&0&-2
\end{array}
\right).
\end{eqnarray}
   The set $\{i\lambda_a\}$ constitutes a basis of the
Lie algebra su(3) of SU(3), i.e., the set of all complex,
traceless, skew-Hermitian, $3\times 3$ matrices.
   The Lie product is then defined in terms of ordinary matrix multiplication
as the commutator of two elements of su(3).
   Such a definition naturally satisfies the Lie properties of
anti-commutativity
\begin{equation}
\label{2:1:anticom}
[A,B]=-[B,A]
\end{equation}
as well as the Jacobi identity
\begin{equation}
\label{2:1:jacobi}
[A,[B,C]]+[B,[C,A]]+[C,[A,B]]=0.
\end{equation}
   In accordance with Eqs.\ (\ref{2:1:uexp}) and (\ref{2:1:gmme1}),
elements of su(3) can be interpreted as tangent vectors in the identity of
SU(3).

   The structure of the Lie group is encoded in the commutation relations
of the Gell-Mann matrices,
\begin{equation}
\label{2:1:crgmm}
\left[\frac{\lambda_a}{2},\frac{\lambda_b}{2}\right]
=i f_{abc}\frac{\lambda_c}{2},
\end{equation}
where the totally antisymmetric real structure constants $f_{abc}$ are
obtained from Eq.\ (\ref{2:1:gmme3}) as
\begin{equation}
\label{2:1:fabc}
f_{abc}=\frac{1}{4i}\mbox{Tr}([\lambda_a,\lambda_b]\lambda_c).
\end{equation}
\begin{exercise}
\label{exercise_fabc}
\rm Verify Eq.\ (\ref{2:1:fabc}).

\noindent Hint: Multipy Eq.\ (\ref{2:1:crgmm}) by $\lambda_d$,
take the trace, make use of Eq.\ (\ref{2:1:gmme3}), and finally
rename $d\to c$.
\end{exercise}
\begin{exercise}
\label{exercise_fabc_antisymmetric}
\rm
Show that $f_{abc}$ is totally antisymmetric.

\noindent Hint:
Consider the symmetry properties of $\mbox{Tr}([A,B]C)$.
\end{exercise}
  The independent non-vanishing values are explicitly summarized in the
scheme of Table \ref{table:2:1:su3structurconstants}.
   Roughly speaking, these structure constants are a measure of
the non-commutativity of the group SU(3).

\begin{table}
\begin{center}
\begin{tabular}{|r|r|r|r|r|r|r|r|r|r|}
\hline
$abc$&123&147&156&246&257&345&367&458&678\\
\hline
$f_{abc}$&1&$\frac{1}{2}$&$-\frac{1}{2}$&$\frac{1}{2}$&
$\frac{1}{2}$&$\frac{1}{2}$&$-\frac{1}{2}$&$\frac{1}{2}\sqrt{3}$&
$\frac{1}{2}\sqrt{3}$\\
\hline
\end{tabular}
\caption{\label{table:2:1:su3structurconstants}
Totally antisymmetric non-vanishing structure constants of SU(3).}
\end{center}
\end{table}
   The anti-commutation relations of the Gell-Mann matrices read
 \begin{equation}
\label{2:1:acrgmm}
\{\lambda_a,\lambda_b\}
= \frac{4}{3}\delta_{ab} +2 d_{abc} \lambda_c,
\end{equation}
   where the totally symmetric $d_{abc}$ are given by
\begin{equation}
\label{2:1:dabc}
d_{abc}=\frac{1}{4}\mbox{Tr}(\{\lambda_a,\lambda_b\}\lambda_c),
\end{equation}
and are summarized in Table
\ref{table:2:1:su3dsymbols}.

\begin{exercise}
\label{exercise_dabc}
\rm Verify Eq.\ (\ref{2:1:dabc}) and show that $d_{abc}$ is totally
symmetric.
\end{exercise}
   Clearly, the anti-commutator of two Gell-Mann matrices is
not necessarily a Gell-Mann matrix.
   For example, the square of a (nontrivial) skew-Hermitian matrix
is not skew Hermitian.

\begin{table}
\begin{center}
\begin{tabular}{|r|r|r|r|r|r|r|r|r|}
\hline
$abc$&
118&
146&
157&
228&
247&
256&
338&
344\\
\hline
$d_{abc}$&
$\frac{1}{\sqrt{3}}$&
$\frac{1}{2}$&
$\frac{1}{2}$&
$\frac{1}{\sqrt{3}}$&
$-\frac{1}{2}$&
$\frac{1}{2}$&
$\frac{1}{\sqrt{3}}$&
$\frac{1}{2}$\\
\hline
$abc$&
355&
366&
377&
448&
558&
668&
778&
888\\
\hline
$d_{abc}$&
$\frac{1}{2}$&
$-\frac{1}{2}$&
$-\frac{1}{2}$&
$-\frac{1}{2\sqrt{3}}$&
$-\frac{1}{2\sqrt{3}}$&
$-\frac{1}{2\sqrt{3}}$&
$-\frac{1}{2\sqrt{3}}$&
$-\frac{1}{\sqrt{3}}$\\
\hline
\end{tabular}
\caption{\label{table:2:1:su3dsymbols}
Totally symmetric non-vanishing $d$ symbols of SU(3).}
\end{center}
\end{table}

  Moreover, it is convenient to introduce as a ninth matrix
$$\lambda_0 =\sqrt{2/3}\,\mbox{diag}(1,1,1),$$
such that Eqs.\ (\ref{2:1:gmme2}) and (\ref{2:1:gmme3}) are still
satisfied by the nine matrices $\lambda_a$.
   In particular, the set $\{i\lambda_a|a=0,\cdots, 8\}$ constitutes a
basis of the Lie algebra u(3) of U(3), i.e., the set of all
complex, skew-Hermitian, $3\times 3$ matrices.

\begin{itemize}
\item
   Many useful properties of the Gell-Mann matrices
can be found in Sect.\ 8 of CORE (Compendium of relations) by
V.~I.~Borodulin, R.~N.~Rogalyov, and S.~R.~Slabospitsky,
arXiv:hep-ph/9507456.
\end{itemize}

   Finally, an {\em arbitrary} $3\times 3$ matrix $M$ can be written as
\begin{equation}
\label{2:1:matrixa}
M=\sum_{a=0}^8 \lambda_a M_a,
\end{equation}
where $M_a$ are complex numbers given by
$$
M_a=\frac{1}{2}\mbox{Tr}(\lambda_a M).
$$

\section{The QCD Lagrangian}
\label{sec_qcdl}

   QCD is the gauge theory of the strong interactions
\cite{Gross:1973id:2,Weinberg:un:2,Fritzsch:pi:2} with color SU(3)
as the underlying gauge group.\footnote{Historically, the color
degree of freedom was introduced into the quark model to account
for the Pauli principle in the description of baryons as
three-quark states.}
   The matter fields of QCD are the so-called quarks which are
spin-1/2 fermions, with six different flavors in addition to their
three possible colors (see Table \ref{2:2:table:quarks}).
   Since quarks have not been observed as asymptotically free
states, the meaning of quark masses and their numerical values are
tightly connected with the method by which they are extracted from
hadronic properties (see Ref.\ \cite{Manohar_PDG:2} for a thorough
discussion).

\begin{table}
\begin{center}
\begin{tabular}{|l|c|c|c|}
\hline
flavor&u&d&s\\
\hline
charge [e] &$2/3$&$-1/3$&$-1/3$\\
\hline
mass [MeV]&$5.1\pm 0.9$ & $9.3\pm 1.4$ & $175\pm 25$\\
\hline
\hline
flavor&c&b&t\\
\hline
charge [e] &$2/3$&$-1/3$&$2/3$\\
\hline
mass [GeV] & $1.15-1.35$ & $4.0 - 4.4$ &$174.3\pm 3.2\pm 4.0$\\
\hline
\end{tabular}
\caption{\label{2:2:table:quarks} Quark flavors and their charges and masses.
   The absolute magnitude of $m_s$ is determined using QCD sum rules.
   The result is given for the $\overline{\mbox{MS}}$ running mass at scale
$\mu = 1$\, GeV.
   The light quark masses are obtained from the mass ratios found
using chiral perturbation theory, using the strange quark mass as input.
   The heavy-quark masses $m_c$ and $m_b$ are determined by the charmonium
and D masses, and the bottomium and B masses, respectively.
   The top quark mass $m_t$ results from the measurement of
lepton + jets and dilepton + jets channels in the D$\emptyset$ and CDF
experiments at Fermilab.}
\end{center}
\end{table}

   The QCD Lagrangian can be obtained from the
Lagrangian for free quarks by applying the gauge principle with
respect to the group SU(3).
   It reads
\begin{equation}
\label{2:3:lqcd}
{\cal L}_{\rm QCD}=\sum_{f={u,d,s, \atop c,b,t}}
\bar{q}_f(i D\hspace{-.6em}/ -m_f)q_f
-\frac{1}{4}{\cal G}_{\mu\nu,a}{\cal G}^{\mu\nu}_a.
\end{equation}
   For each quark flavor $f$ the quark field $q_f$ consists of a color triplet
(subscripts $r$, $g$, and $b$ standing for ``red,'' ``green,'' and ``blue''),
\begin{equation}
\label{2:3:qf}
q_f=\left(\begin{array}{c}q_{f,r}
\\q_{f,g}\\q_{f,b}\end{array}\right),
\end{equation}
   which transforms under a gauge transformation $g(x)$ described
by the set of parameters $\Theta(x)=[\Theta_1(x),\cdots,\Theta_8(x)]$
according to\footnote{For the sake of clarity, the Gell-Mann matrices
contain a superscript $C$, indicating the action in color space.}
\begin{equation}
\label{2:3:qft}
q_f\mapsto q_f'=\exp\left[-i\sum_{a=1}^8 \Theta_a(x)
\frac{\lambda_a^C}{2}\right]q_f=U[g(x)]q_f.
\end{equation}
   Technically speaking, each quark field $q_f$ transforms according to the
fundamental representation of color SU(3).
   Because SU(3) is an eight-parameter group, the covariant derivative
of Eq.\ (\ref{2:3:lqcd}) contains eight independent gauge potentials
${\cal A}_{\mu,a}$,
\begin{equation}
\label{2:3:ka}
D_\mu\left(\begin{array}{l}
q_{f,r}\\q_{f,g}\\q_{f,b}\end{array}
\right)
=\partial_\mu
\left(\begin{array}{l}
q_{f,r}\\q_{f,g}\\q_{f,b}\end{array}
\right)
-ig\sum_{a=1}^8
\frac{\lambda_a^C}{2}{\cal A}_{\mu,a} \left(\begin{array}{l}
q_{f,r}\\q_{f,g}\\q_{f,b}\end{array}
\right).
\end{equation}
   We note that the interaction between quarks and gluons is independent
of the quark flavors which can be seen from the fact that there
only appears one coupling constant $g$ in Eq.\ (\ref{2:3:ka}).
   Demanding gauge invariance of ${\cal L}_{\rm QCD}$
imposes the following transformation property of the gauge fields
(summation over $a$ implied)
\begin{equation}
\label{2:3:atraf} {\cal A}_\mu\equiv\frac{\lambda_a^C}{2} {\cal
A}_{\mu,a}(x)\mapsto U[g(x)]{\cal A}_\mu(x)U^\dagger[g(x)]
-\frac{i}{g}\partial_\mu U[g(x)]U^\dagger[g(x)].
\end{equation}
\begin{exercise}
\label{exercise_covariant_derivative}
\rm
Show that the covariant derivative $D_\mu q_f$
transforms as $q_f$, i.e. $D_\mu q_f\mapsto D'_\mu q'_f=U(g)D_\mu q_f$.
\end{exercise}
   Under a gauge transformation of the first kind, i.e., a global
SU(3) transformation, the second term on the right-hand side
of Eq.\ (\ref{2:3:atraf}) would vanish and the gauge fields would
transform according to the adjoint representation.

   So far we have only considered the matter-field part of ${\cal L}_{\rm
QCD}$ including its interaction with the gauge fields.
   Equation (\ref{2:3:lqcd}) also contains the generalization of
the field strength tensor to the non-Abelian case,
\begin{equation}
\label{2:3:gmunu}
{\cal G}_{\mu\nu,a}=\partial_\mu {\cal A}_{\nu,a}-\partial_\nu {\cal A}_{\mu,a}
+g f_{abc}{\cal A}_{\mu,b} {\cal A}_{\nu,c},
\end{equation}
   with the SU(3) structure constants given in Table
\ref{table:2:1:su3structurconstants} and a summation over repeated indices
implied.
   Given Eq.\ (\ref{2:3:atraf}) the field strength tensor transforms under
SU(3) as
\begin{equation}
\label{2:3:gtrafo}
{\cal G}_{\mu\nu}\equiv
\frac{\lambda_a^C}{2} {\cal G}_{\mu\nu,a}
\mapsto
U[g(x)]{\cal G}_{\mu\nu} U^\dagger[g(x)].
\end{equation}
\begin{exercise}
\label{exercise_gmunu_trans}
\rm Verify Eq.~(\ref{2:3:gtrafo}).

\noindent Hint: Introduce
${\cal A}_\mu\equiv \lambda_a^C {\cal A}_{\mu,a}/2$.
Equation (\ref{2:3:gmunu}) is then equivalent to
${\cal G}_{\mu\nu}=\partial_\mu{\cal A}_\nu-\partial_\nu{\cal A}_\mu
-ig[{\cal A}_\mu,{\cal A}_\nu]$.
\end{exercise}
   Using Eq.\ (\ref{2:1:gmme3}) the purely gluonic part
of ${\cal L}_{\rm QCD}$ can be written as
$$-\frac{1}{2}\mbox{Tr}_C({\cal G}_{\mu\nu} {\cal G}^{\mu\nu}),
$$
which, using the cyclic property of traces,
$\mbox{Tr}(AB)=\mbox{Tr}(BA)$, together
with $UU^\dagger=1$,
is easily seen to be invariant under the transformation of
Eq.\ (\ref{2:3:gtrafo}).

   In contradistinction to the Abelian case of quantum electrodynamics,
the squared field strength tensor gives rise to gauge-field self
interactions involving vertices with three and four gauge fields
of strength $g$ and $g^2$, respectively.
   Such interaction terms are characteristic of non-Abelian gauge
theories and make them much more complicated than Abelian theories.

   From the point of view of gauge invariance the strong-interaction
Lagrangian could also involve a term of the type
\begin{equation}
\label{2:3:ltheta}
{\cal L}_\theta=\frac{g^2\bar{\theta}}{64\pi^2}\epsilon^{\mu\nu\rho\sigma}
\sum_{a=1}^8{\cal G}^a_{\mu\nu}{\cal G}^a_{\rho\sigma},
\end{equation}
where $\epsilon_{\mu\nu\rho\sigma}$ denotes the totally antisymmetric
Levi-Civita tensor.\footnote{
\begin{displaymath}
\epsilon_{\mu\nu\rho\sigma}=\left\{
\begin{array}{rl}
+1& \mbox{if $\{\mu,\nu,\rho,\sigma\}$ is an even permutation of $\{0,1,2,3\}$}
\\
-1& \mbox{if $\{\mu,\nu,\rho,\sigma\}$ is an odd permutation of $\{0,1,2,3\}$}
\\
0& \mbox{otherwise}
\end{array}
\right.
\end{displaymath}}
   The so-called $\theta$ term of Eq.\ (\ref{2:3:ltheta}) implies an explicit
$P$ and $CP$ violation of the strong interactions which, for example, would
give rise to an electric dipole moment of the neutron.
   The present empirical information indicates that the $\theta$ term is
small and, in the following, we will omit Eq.\ (\ref{2:3:ltheta}) from our
discussion.

\section{Accidental, Global Symmetries of the QCD Lagrangian}
\label{sec_agsl}
\subsection{Light and Heavy Quarks}
\label{subsec_lhq}
   The six quark flavors are commonly divided into the three light quarks
$u$, $d$, and $s$ and the three heavy flavors $c$, $b$, and $t$,
\begin{equation}
\label{2:4:mq}
\left(\begin{array}{r}m_u=0.005\,\mbox{GeV}\\
m_d=0.009\,\mbox{GeV}\\
m_s=0.175\,\mbox{GeV}\end{array}\right)
\ll 1\, \mbox{GeV}\le
\left(\begin{array}{r}
m_c= (1.15 - 1.35)\, \mbox{GeV}\\
m_b= (4.0 - 4.4)\, \mbox{GeV}\\
m_t=174\,\mbox{GeV}\end{array}\right),
\end{equation}
   where the scale of 1 GeV is associated with the masses of the lightest
hadrons containing light quarks, e.g., \ $m_\rho$= 770 MeV,
which are not Goldstone bosons resulting from spontaneous
symmetry breaking.
   The scale associated with spontaneous symmetry breaking,
$4\pi F_\pi\approx$ 1170 MeV, is of the same order of magnitude.

   The masses of the lightest meson and baryon containing a charmed quark,
$D^+=c\bar{d}$ and $\Lambda^+_c=udc$, are $(1869.4\pm 0.5)\, \mbox{MeV}$
and $(2284.9\pm 0.6)\,\mbox{MeV}$, respectively.
   The threshold center-of-mass energy to produce, say, a $D^+ D^-$ pair
in $e^+ e^-$ collisions is approximately 3.74 GeV, and thus way beyond the
low-energy regime which we are interested in.
   In the following, we will approximate the full QCD Lagrangian by its
light-flavor version, i.e., we will ignore effects due to (virtual)
heavy quark-antiquark pairs $h\bar{h}$.

   Comparing the proton mass, $m_p$ = 938 MeV, with the sum of
two up and one down current-quark masses (see Table
\ref{2:2:table:quarks}),\footnote{The expression {\em
current-quark masses} for the light quarks is related to the fact
that they appear in the divergences of the vector and axial-vector
currents (see Section \ref{subsec_csbdqm}).}
\begin{equation}
\label{2:3:mp} m_p\gg 2m_u+m_d,
\end{equation}
shows that an interpretation of the proton mass in terms of
current-quark mass parameters must be very different from, say,
the situation in the hydrogen atom, where the mass is essentially
given by the sum of the electron and proton masses, corrected by a
small amount of binding energy.
   In this context we recall that the current-quark masses
must not be confused with the constituent quark masses of a (nonrelativistic)
quark model which are typically of the order of 350 MeV.
  In particular, Eq.\ (\ref{2:3:mp}) suggests that the Lagrangian
${\cal L}_{\rm QCD}^0$, containing only the light-flavor quarks in
the so-called chiral limit $m_u,m_d,m_s\to 0$,
might be a good starting point in the discussion
of low-energy QCD:
\begin{equation}
\label{2:4:lqcd0} {\cal L}^0_{\rm QCD}= \sum_{l=u,d,s}\bar{q}_l i
D\hspace{-.6em}/\hspace{.3em} q_l
-\frac{1}{4}{\cal G}_{\mu\nu,a} {\cal
G}^{\mu\nu}_a.
\end{equation}
   We repeat that the covariant derivative $D\hspace{-.6em}/\hspace{.3em}
q_{l}$ acts on color and Dirac indices only, but is independent of flavor.

\subsection{Left-Handed and Right-Handed Quark Fields}
\label{subsec_lhrhqf}

   In order to fully exhibit the global symmetries of Eq.\ (\ref{2:4:lqcd0}),
we consider the chirality matrix
$\gamma_5=\gamma^5=i\gamma^0\gamma^1
\gamma^2\gamma^3=\gamma_5^\dagger$, $\{\gamma^\mu,\gamma_5\}=0$,
$\gamma_5^2=1$, and introduce projection operators
\begin{equation}
\label{2:4:prpl}
P_R=\frac{1}{2}(1+\gamma_5)=P_R^\dagger,\quad
P_L=\frac{1}{2}(1-\gamma_5)=P_L^\dagger,
\end{equation}
  where the indices $R$ and $L$ refer to right-handed and
left-handed, respectively, as will become more clear below.
   Obviously, the
$4\times 4$ matrices $P_R$ and $P_L$ satisfy a completeness relation,
\begin{equation}
\label{2:4:prplcompleteness}
P_R+P_L=1,
\end{equation}
are idempotent,
\begin{equation}
\label{2:4:prplidempotent}
P_R^2=P_R,\quad P_L^2=P_L,
\end{equation}
and respect the orthogonality relations
\begin{equation}
\label{2:4:prplorthogonality}
P_R P_L=P_L P_R=0.
\end{equation}
\begin{exercise}
\label{exercise_projection_operators} \rm Verify the properties of
Eqs.\ (\ref{2:4:prpl}) -- (\ref{2:4:prplorthogonality}).
\end{exercise}
   The combined properties of Eqs.\ (\ref{2:4:prpl}) --
(\ref{2:4:prplorthogonality}) guarantee that $P_R$ and $P_L$ are
indeed projection operators which project from the Dirac field variable $q$ to
its chiral components $q_R$ and $q_L$,
\begin{equation}
\label{2:4:qlr}
q_R=P_R q,\quad
q_L=P_L q.
\end{equation}
   We recall in this context that a chiral (field) variable is one
which under parity is transformed into neither the original variable
nor its negative.\footnote{In case of fields,
a transformation of the argument $\vec{x}\to -\vec{x}$ is implied.}
   Under parity, the quark field is transformed into its parity conjugate,
$$
P:q(t,\vec{x})\mapsto \gamma_0 q(t,-\vec{x}),
$$
   and hence
$$
q_R(t,\vec{x})=P_R q(t,\vec{x})
\mapsto P_R \gamma_0 q (t,-\vec{x})
=\gamma_0 q_L (t,-\vec{x})
\neq \pm q_R(t,-\vec{x}),
$$
   and similarly for $q_L$.\footnote{Note that in the above sense,
also $q$ is a chiral variable.
However, the assignment of handedness
does not have such an intuitive meaning as in the case of $q_L$ and
$q_R$.}

   The terminology right-handed and left-handed fields can easily be
visualized in terms of the solution to the free Dirac equation.
   For that purpose, let us consider an extreme relativistic
positive-energy solution to the free Dirac equation with
three-momentum $\vec{p}$,\footnote{Here we adopt a covariant
normalization of the spinors,
$u^{(\alpha)\dagger}(\vec{p}\,)u^{(\beta)}(\vec{p}\,) = 2
E\delta_{\alpha\beta}$, etc.}
$$ u(\vec{p},\pm)=\sqrt{E+m}\left(\begin{array}{c}
\chi_\pm\\
\frac{\vec{\sigma}\cdot\vec{p}}{E+m}\chi_\pm\end{array}\right)
\stackrel{\mbox{$E\gg m$}}{\mapsto}
\sqrt{E}
\left(\begin{array}{r}\chi_\pm\\ \pm\chi_\pm\end{array}
\right)\equiv u_\pm(\vec{p}\,),
$$
   where we assume that the spin in the rest frame is either parallel
or antiparallel to the direction of momentum
$$
\vec{\sigma}\cdot \hat{p} \chi_{\pm}=\pm \chi_\pm.
$$
   In the standard representation of Dirac
   matrices\footnote{Unless stated otherwise, we use the
convention of J.~D.~Bjorken and S.~D.~Drell, {\em Relativistic
Quantum Mechanics} (McGraw-Hill, New York, 1964).}
    we find
$$ P_R=\frac{1}{2}\left(\begin{array}{rr}1_{2\times2}&
1_{2\times 2}\\1_{2\times 2}&1_{2\times 2}\end{array}\right),\quad
P_L=\frac{1}{2}\left(\begin{array}{rr}1_{2\times 2}&-1_{2\times 2}\\
-1_{2\times 2}&1_{2\times 2}\end{array}\right).
$$

\begin{exercise}
\label{exercise_plrupm}
\rm Show that
$$
P_R u_+=u_+,\quad
P_L u_+=0,\quad
P_R u_-=0,\quad
P_L u_-=u_-.
$$
\end{exercise}
    In the extreme relativistic limit (or better, in
the zero-mass limit), the operators $P_R$ and $P_L$ project to the
positive and negative helicity eigenstates, i.e., in this limit
chirality equals helicity.

   Our goal is to analyze the symmetry of the QCD Lagrangian with
respect to independent global transformations of the left- and right-handed
fields.
   There are 16 independent $4\times 4$ matrices, that can be expressed in
terms of the unit matrix, the Dirac matrices $\gamma^\mu$, the chirality matrix
$\gamma_5$, the products $\gamma^\mu \gamma_5$, and the six matrices
$\sigma^{\mu\nu}=i[\gamma^\mu,\gamma^\nu]/2$.
   In order to decompose the corresponding 16 quadratic forms into their
respective projections to right- and left-handed fields, we make
use of
\begin{equation}
\label{2:4:qgq}
\bar{q}\Gamma_i q=\left \{\begin{array}{lcl}
\bar{q}_R\Gamma_1 q_R+\bar{q}_L\Gamma_1 q_L&\mbox{for}&
\Gamma_1\in\{\gamma^\mu,\gamma^\mu\gamma_5\}\\
\bar{q}_R\Gamma_2 q_L +\bar{q}_L\Gamma_2 q_R&\mbox{for}& \Gamma_2
\in\{1,\gamma_5,\sigma^{\mu\nu}\}
\end{array}
\right.,
\end{equation}
where $\bar{q}_R=\bar{q}P_L$ and $\bar{q}_L=\bar{q}P_R.$
\begin{exercise}
\label{exercise_barq_gammai_q}
\rm
Verify Eq.\ (\ref{2:4:qgq}).

\noindent Hint: Insert unit matrices as
$$\bar{q}\Gamma_i q=\bar{q}(P_R+P_L)\Gamma_i(P_R+P_L)q,$$
and make use of $\{\Gamma_1,\gamma_5\}=0$ and $[\Gamma_2,\gamma_5]=0$
as well as the properties of the projection operators derived in
Exercise \ref{exercise_projection_operators}.
\end{exercise}
  We stress that the validity of Eq.\ (\ref{2:4:qgq}) is general
and does not refer to ``massless'' quark fields.

   We now apply Eq.\ (\ref{2:4:qgq}) to the term containing the contraction
of the covariant derivative with $\gamma^\mu$.
   This quadratic quark form decouples into the sum of two
terms which connect only left-handed with left-handed and right-handed
with right-handed quark fields.
   The QCD Lagrangian in the chiral limit can then be written as
\begin{equation}
\label{2:4:lqcd0lr}
{\cal L}^0_{\rm QCD}=\sum_{l=u,d,s}
(\bar{q}_{R,l}iD\hspace{-.6em}/\hspace{.3em}q_{R,l}+\bar{q}_{L,l}iD
\hspace{-.6em}/\hspace{.3em}
q_{L,l})-\frac{1}{4}{\cal G}_{\mu\nu,a} {\cal G}^{\mu\nu}_a.
\end{equation}
    Due to the flavor independence of the covariant derivative
${\cal L}^0_{\rm QCD}$ is invariant under
\begin{eqnarray}
\label{2:4:u3lu3r}
\left(\begin{array}{c}u_L\\d_L\\s_L\end{array}\right)
\mapsto U_L\left(\begin{array}{c}u_L\\d_L\\s_L\end{array}\right)
=\exp\left(-i\sum_{a=1}^8 \Theta^L_a \frac{\lambda_a}{2}\right)
e^{-i\Theta^L}\left(\begin{array}{c}u_L\\d_L\\s_L\end{array}\right),
\nonumber\\
\left(\begin{array}{c}u_R\\d_R\\s_R\end{array}\right)
\mapsto U_R\left(\begin{array}{c}u_R\\d_R\\s_R\end{array}\right)
=\exp\left(-i\sum_{a=1}^8 \Theta^R_a \frac{\lambda_a}{2}\right)
e^{-i\Theta^R}\left(\begin{array}{c}u_R\\d_R\\s_R\end{array}\right),
\end{eqnarray}
   where $U_L$ and $U_R$ are independent unitary $3\times 3$ matrices
and where we have extracted the factors $e^{-i\Theta^L}$ and
$e^{-i\Theta^R}$ for future convenience.
   Note that the Gell-Mann matrices act in flavor space.

   ${\cal L}^0_{\rm QCD}$ is said to have a classical
{\em global} $\mbox{U(3)}_L\times\mbox{U(3)}_R$ symmetry.
   Applying Noether's theorem
from such an invariance
one would expect a total of $2\times(8+1)=18$ conserved currents.

\subsection{Noether Theorem}
\label{subsec_nt}

   Noether's theorem establishes the connection between continuous symmetries
of a dynamical system and conserved quantities (constants of the motion).
   For simplicity we consider only internal symmetries.
(The method can also be used to discuss the consequences of Poincar\'e
invariance.)

   In order to identify the conserved currents associated with the
transformations of Eqs.\ (\ref{2:4:u3lu3r}), we briefly recall the
method of Gell-Mann and L{\'e}vy \cite{Gell-Mann:1960np:3:3},
which we will then apply to Eq.\ (\ref{2:4:lqcd0lr}).

   We start with a Lagrangian ${\cal L}$ depending on $n$ independent fields
$\Phi_i$ [typically $n\geq 2$ for bosons, and $n\geq 1$ for
fermions, e.g.~U(1)] and their first partial derivatives (the
extension to higher-order derivatives is also possible),
\begin{equation}
\label{2:3:l}
{\cal L}={\cal L}(\Phi_i,\partial_\mu\Phi_i),
\end{equation}
from which one obtains $n$ equations of motion:
\begin{equation}
\label{2:3:eom}
\frac{\partial \cal L}{\partial\Phi_i}-\partial_\mu
\frac{\partial\cal L}{\partial\partial_\mu\Phi_i}=0,\quad i=1,\cdots,n.
\end{equation}
   Suppose the Lagrangian of Eq.\ (\ref{2:3:l}) to be invariant
under a global symmetry transformation depending on $r$ real
parameters.
   The method of Gell-Mann and L{\'e}vy now consists of
promoting this global symmetry to a {\em local} one, from which we
will then be able to identify the Noether currents.
   To that end we consider transformations which depend on $r$ real local
parameters $\epsilon_a(x)$,\footnote{Note that the transformation
need not be realized linearly on the fields.}
\begin{equation}
\label{2:3:ltraf}
\Phi_i(x)\mapsto\Phi'_i(x)=\Phi_i(x)+\delta\Phi_i(x)
=\Phi_i(x)-i\epsilon_a(x) F^a_i[\Phi_j(x)],
\end{equation}
and obtain, neglecting terms of order $\epsilon^2$,
as the variation of the Lagrangian,
\begin{eqnarray}
\label{2:3:dl}
\delta{\cal L}&=& {\cal L}(\Phi'_i,\partial_\mu\Phi'_i)
-{\cal L}(\Phi_i,\partial_\mu\Phi_i)\nonumber\\
&=&\frac{\partial\cal L}{\partial\Phi_i}\delta\Phi_i
+\frac{\partial\cal L}{\partial\partial_\mu\Phi_i}
\hspace{-4.5em}\underbrace{\partial_\mu\delta\Phi_i}_{\mbox{$
-i[\partial_\mu\epsilon_a(x)] F^a_i-i\epsilon_a(x)
\partial_\mu F^a_i$}}\nonumber\\
&=&\epsilon_a(x)\left(-i\frac{\partial\cal L}{\partial\Phi_i}
F^a_i-i\frac{\partial\cal L}{\partial\partial_\mu\Phi_i}
\partial_\mu F^a_i\right)
+\partial_\mu\epsilon_a(x)\left(-i\frac{\partial\cal L}{
\partial\partial_\mu\Phi_i}F^a_i\right)\nonumber\\
&\equiv&\epsilon_a(x)\partial_\mu J^{\mu,a}
+\partial_\mu\epsilon_a(x)J^{\mu,a}.
\end{eqnarray}
   According to this equation we define for each infinitesimal
transformation a four-current density as
\begin{equation}
\label{2:3:strom}
   J^{\mu,a}=-i\frac{\partial\cal L}{\partial\partial_\mu\Phi_i}
F^a_i.
\end{equation}
     By calculating the divergence $\partial_\mu J^{\mu,a}$ of
Eq.\ (\ref{2:3:strom})
\begin{eqnarray*}
\partial_\mu J^{\mu,a}&=&-i\left(\partial_\mu\frac{\partial\cal L}{\partial
\partial_\mu\Phi_i}\right)F^a_i-i\frac{\partial\cal L}{\partial
\partial_\mu\Phi_i}\partial_\mu F^a_i\\
&=&
-i\frac{\partial\cal L}{\partial \Phi_i}
F^a_i-i\frac{\partial\cal L}{\partial\partial_\mu\Phi_i}
\partial_\mu F^a_i,
\end{eqnarray*}
where we made use of the equations of motion, Eq.\ (\ref{2:3:eom}),
we explicitly verify the consistency with the definition of
$\partial_\mu J^{\mu,a}$
according to Eq.\ (\ref{2:3:dl}).
   From Eq.\ (\ref{2:3:dl}) it is straightforward to obtain the four-currents
as well as their divergences as
\begin{eqnarray}
\label{2:3:strom2}
J^{\mu,a}&=&\frac{\partial \delta\cal L}{\partial \partial_\mu
\epsilon_a},\\
\label{2:3:divergenz}
\partial_\mu J^{\mu,a}&=&\frac{\partial \delta\cal L}{\partial
\epsilon_a}.
\end{eqnarray}
   We chose the parameters of the transformation to be local.
   However, the Langrangian of Eq.\ (\ref{2:3:l}) was only assumed
to be invariant under a {\em global} tranformation.
   In that case, the term $\partial_\mu\epsilon_a$ disappears, and
since the Lagrangian is invariant under such transformations, we
see from Eq.\ (\ref{2:3:dl}) that the current $J^{\mu,a}$ is
conserved, $\partial_\mu J^{\mu,a}=0$.
   For a conserved current the charge
\begin{equation}
\label{2:3:charge} Q^a(t)=\int d^3 x J^a_0(t,\vec{x})
\end{equation}
is time independent, i.e., a constant of the motion.
\begin{exercise}
\label{exercise_constant_motion}
\rm
By applying the divergence theorem for an infinite
volume with appropriate boundary conditions for $R\to \infty$,
show that $Q^a(t)$ is a constant of the motion for $\delta {\cal L}=0$.
\end{exercise}

   So far we have discussed Noether's theorem on the classical level, implying
that the charges $Q^a(t)$ can have any continuous real value.
   However, we also need to discuss the implications of a transition to
a quantum theory.

   To that end, let us first recall the transition from classical mechanics
to quantum mechanics.
   Consider a point mass $m$ in a central potential $V(\vec{r}\,)=V(r)$, i.e.,
the corresponding Lagrange and Hamilton functions are rotationally
invariant.
   As a result of this invariance, the angular momentum
$\vec{l}=\vec{r}\times\vec{p}\,$ is a constant of the motion
which, in classical mechanics, can have any continuous real value.
   In the transition to quantum mechanics, the components of $\vec{r}$
and $\vec{p}$ turn into Hermitian, linear operators, satisfying
the commutation relations
\begin{displaymath}
[\hat{x}_i,\hat{p}_j]=i\delta_{ij},\quad
[\hat{x}_i,\hat{x}_j]=0,\quad [\hat{p}_i,\hat{p}_j]=0.
\end{displaymath}
   The components  $\hat{l}_i=\epsilon_{ijk}
\hat{x}_j\hat{p}_k$ of the angular momentum operator satisfy the
commutation relations
\begin{displaymath}
[\hat{l}_i,\hat{l}_j]=i\epsilon_{ijk}\hat{l}_k,
\end{displaymath}
i.e., they cannot simultaneously be diagonalized.
   Rather, the states are organized as eigenstates of $\hat{\vec{l}^2}$ and
$\hat{l}_3$ with eigenvalues $l(l+1)$ and $m=-l,\cdots, l$
($l=0,1,2,\cdots$).
   Also note that the angular momentum operators are the
generators of rotations.
   The rotational invariance of the quantum system implies that the components
of the angular momentum operator commute with the Hamilton
operator,
\begin{displaymath}
[\hat{H},\hat{l}_i]=0,
\end{displaymath}
   i.e., they are still constants of the motion.
   One then simultaneously diagonalizes $\hat{H}$, $\hat{\vec{l}^2}$, and
$\hat{l}_3$.
   For example, the energy eigenvalues of the hydrogen atom are given by
\begin{displaymath}
E_n=-\frac{\alpha^2 m}{2n^2}\approx-\frac{13.6}{n^2}\,\mbox{eV},
\end{displaymath}
where $n=n'+l+1$, $n'\geq 0$ denotes the principal quantum number,
and the degeneracy of an energy level is given by $n^2$ (spin
neglected).
   The value $E_1$ and the spacing of the levels are determined by the
{\em dynamics} of the system, i.e., the specific form of the
potential, whereas the multiplicities of the energy levels are a
consequence of the underlying rotational {\em symmetry}. (In fact,
the accidental degeneracy for $n\geq 2$ is a result of an even
higher symmetry of the $1/r$ Hamiltonian, namely an O(4)
symmetry.)

   Having the example from quantum mechanics in mind, let us turn
to the analogous case in quantum field theory.
   After canonical quantization, the fields $\Phi_i$ and their conjugate
momenta $\Pi_i=\partial{\cal L}/\partial (\partial_0 \Phi_i)$
are considered as linear operators acting
on a Hilbert space which, in the Heisenberg picture, are subject to the
equal-time commutation relations
\begin{eqnarray}
\label{2:3:gzvr}
[\Phi_i(t,\vec{x}),\Pi_j(t,\vec{y})]&=&i\delta^3(\vec{x}-\vec{y})
\delta_{ij},\nonumber\\
{[}\Phi_i(t,\vec{x}),\Phi_j(t,\vec{y})]&=&0,\nonumber\\
{[}\Pi_i(t,\vec{x}),\Pi_j(t,\vec{y})]&=&0.
\end{eqnarray}
   As a special case of Eq.\ (\ref{2:3:ltraf}) let us consider infinitesimal
transformations which are {\em linear} in the fields,
\begin{equation}
\label{2:3:lt}
\Phi_i(x)\mapsto \Phi'_i(x)=\Phi_i(x)-i\epsilon_a(x)t^a_{ij}\Phi_j(x),
\end{equation}
where the $t^a_{ij}$ are constants generating a mixing of the fields.
   From Eq.\ (\ref{2:3:strom}) we then obtain\footnote{Normal ordering
symbols are suppressed.}
\begin{eqnarray}
\label{2:3:j}
J^{\mu,a}(x)&=&-it^a_{ij}\frac{\partial {\cal L}}{\partial
\partial_\mu\Phi_i}\Phi_j,\\
\label{2:3:q}
Q^{a}(t)&=&-i\int d^3x\, \Pi_i(x) t^a_{ij}\Phi_j(x),
\end{eqnarray}
where $J^{\mu,a}(x)$ and $Q^{a}(t)$ are now operators.
   In order to interpret the charge operators $Q^a(t)$, let us make use of
the equal-time commutation relations, Eqs.\ (\ref{2:3:gzvr}),
and calculate their commutators with the field operators,
\begin{eqnarray}
\label{2:3:qphi} [Q^a(t),\Phi_k(t,\vec{y})]&=&-it^a_{ij}\int
d^3x\,
[\Pi_i(t,\vec{x})\Phi_j(t,\vec{x}),\Phi_k(t,\vec{y})]\nonumber\\
&=&-t^a_{kj}\Phi_j(t,\vec{y}).
\end{eqnarray}
\begin{exercise}
\label{exercise_com_rel_ch_field} \rm Using the equal-time
commutation relations of Eqs.\ (\ref{2:3:gzvr}), verify Eq.
(\ref{2:3:qphi}).
\end{exercise}
   Note that we did not require the charge operators to be time
independent.

   On the other hand, for the transformation behavior of the Hilbert space
associated with a global infinitesimal transformation, we make an
ansatz in terms of an infinitesimal unitary transformation\footnote{
We have chosen to have the fields (field operators) rotate actively
and thus must transform the states of Hilbert space in the opposite
direction.}
\begin{equation}
\label{1:5:tz}
|\alpha'\rangle=[1+i\epsilon_a G^a(t)]|\alpha\rangle,
\end{equation}
with Hermitian operators $G^a$.
   Demanding
\begin{equation}
\label{2:3:at}
\langle\beta|A|\alpha\rangle=\langle\beta'|A'|\alpha'\rangle\quad
\forall\, |\alpha\rangle, |\beta\rangle, \epsilon_a,
\end{equation}
in combination with Eq.\ (\ref{2:3:lt}) yields the condition
\begin{eqnarray*}
\langle\beta|\Phi_i(x)|\alpha\rangle
&=&\langle\beta'|\Phi_i'(x)|\alpha'\rangle\nonumber\\
&=&\langle\beta|[1-i\epsilon_a G^a(t)][\Phi_i(x)-i\epsilon_b
t^b_{ij}\Phi_j(x)][1+i\epsilon_c G^c(t)]|\alpha\rangle.
\end{eqnarray*}
   By comparing the terms linear in $\epsilon_a$ on both sides,
\begin{equation}
0=-i\epsilon_a[G^a(t),\Phi_i(x)]
\underbrace{-i\epsilon_a t^a_{ij}\Phi_j(x)}_{\mbox{$
i\epsilon_a[Q^a(t),\Phi_i(x)]$}},
\end{equation}
   we see that the infinitesimal generators acting on the states of Hilbert
space which are associated with the transformation of the
fields are identical with the charge operators $Q^a(t)$ of
Eq.\ (\ref{2:3:q}).

   Finally, evaluating the commutation relations for the case of several
generators,
\begin{eqnarray}
\label{2:3:qaqbkom} [Q^a(t),Q^b(t)]
&=&-i(t^a_{ij}t^b_{jk}-t^b_{ij}t^a_{jk}) \int
d^3x\,\Pi_i(t,\vec{x})\Phi_k(t,\vec{x}),
\end{eqnarray}
we find the right-hand side of Eq.\ (\ref{2:3:qaqbkom}) to be
again proportional to a charge operator, if
\begin{equation}
\label{2:3:lrel}
t^a_{ij}t^b_{jk}-t^b_{ij}t^a_{jk}=iC_{abc}t^c_{ik},
\end{equation}
i.e., in that case the charge operators $Q^a(t)$ form a Lie algebra
\begin{equation}
\label{2:3:liealgebra}
[Q^a(t),Q^b(t)]=iC_{abc}Q^c(t)
\end{equation}
with structure constants $C_{abc}$.

\begin{exercise}
\label{exercise_commutation_relations_qa_qb}
\rm Using the canonical commutation relations of Eqs.\ (\ref{2:3:gzvr}),
verify Eq.\ (\ref{2:3:qaqbkom}).
\end{exercise}

   From now on we assume the validity of Eq.\ (\ref{2:3:lrel}) and interpret
the constants $t^a_{ij}$ as the entries in the $i$th row and $j$th column
of an $n\times n$ matrix $T^a$,
\begin{displaymath}
T^a=\left(\begin{array}{ccc}
t^a_{11}&\cdots&t^a_{1n}\\
\vdots & &\vdots\\
t^a_{n1}&\cdots &t^a_{nn}
\end{array}
\right).
\end{displaymath}
   Because of Eq.\ (\ref{2:3:lrel}), these matrices form an $n$-dimensional
representation of a Lie algebra,
$$[T^a,T^b]=iC_{abc}T^c.$$
   The infinitesimal, linear transformations of the fields $\Phi_i$ may
then be written in a compact form,
\begin{equation}
\label{2:3:ltrafo}
\left(\begin{array}{c}\Phi_1(x)\\\vdots\\ \Phi_n(x)\end{array}
\right)=\Phi(x)\mapsto\Phi'(x)=(1-i\epsilon_a T^a)\Phi(x).
\end{equation}
   In general, through an appropriate unitary transformation,
the matrices $T_a$ may be decomposed into their irreducible components, i.e.,
brought into block-diagonal form, such that only fields belonging
to the same multiplet transform into each other under the symmetry
group.

\begin{exercise}
\label{exercise_su2_symmetry_strong_interactions}
\renewcommand{\labelenumi}{(\alph{enumi})}
\rm
   In order to also deal with the case of fermions, we discuss
the isospin invariance of the strong interactions and consider, in
total, five fields.
   The commutation relations of the isospin algebra su(2) read
\begin{equation}
\label{1:5:su2}
[Q_i,Q_j]=i\epsilon_{ijk}Q_k.
\end{equation}
   A basis of the so-called fundamental representation ($N=2$) is given by
\begin{equation}
\label{1:5:su2f} T_i^{\rm f}=\frac{1}{2}\tau_i\quad\mbox{(f:
fundamental)}
\end{equation}
with the Pauli matrices
\begin{equation}
\label{1:5:pauli_matrices}
\tau_1=\left(\begin{array}{rr}0&1\\1&0\end{array}\right),\quad
\tau_2=\left(\begin{array}{rr}0&-i\\i&0\end{array}\right),\quad
\tau_3=\left(\begin{array}{rr}1&0\\0&-1\end{array}\right).
\end{equation}
   We substitute the fermion doublet for $\Phi_{4,5}$
\begin{equation}
\label{1:5:fd}
\Psi=\left(\begin{array}{c}p\\n\end{array}\right).
\end{equation}
   A basis of the so-called adjoint representation ($N=3$) is given by
\begin{equation}
\label{1:5:suad}
T_i^{\rm ad}=
\left(
\begin{array}{ccc}
t_{i,11}^{\rm ad}&
t_{i,12}^{\rm ad}&
t_{i,13}^{\rm ad}\\
t_{i,21}^{\rm ad}&
t_{i,22}^{\rm ad}&
t_{i,23}^{\rm ad}\\
t_{i,31}^{\rm ad}&
t_{i,32}^{\rm ad}&
t_{i,33}^{\rm ad}
\end{array}
\right),\quad
t^{\rm ad}_{i,jk}=-i\epsilon_{ijk},\quad
\mbox{(ad: adjoint)},
\end{equation}
i.e.
\begin{equation}
\label{1:5:adkon}
T_1^{\rm ad}
=\left(\begin{array}{rrr}0&0&0\\0&0&-i\\0&i&0\end{array}\right),\quad
T_2^{\rm ad}
=\left(\begin{array}{rrr}0&0&i\\0&0&0\\-i&0&0\end{array}\right),\quad
T_3^{\rm ad}
=\left(\begin{array}{rrr}0&-i&0\\i&0&0\\0&0&0\end{array}\right).
\end{equation}
With $\Phi_{1,2,3}\to\vec{\Phi}$
we consider the pseudoscalar pion-nucleon Lagrangian
\begin{equation}
\label{1:5:lps}
{\cal L}=\bar{\Psi}(i\partial\hspace{-.5em}/ -m_N)\Psi
+\frac{1}{2}\left(\partial_\mu\vec{\Phi}\cdot\partial^\mu\vec{\Phi}
-M_\pi^2\vec{\Phi}^2\right)
-ig\bar{\Psi}\gamma_5\vec{\tau}\cdot\vec{\Phi}\Psi.
\end{equation}
   As a specific application of the infinitesimal transformation of
Eq.\ (\ref{2:3:lt}) we take
\begin{equation}
\label{1:5:psipitr}
\left(\begin{array}{c}
\vec{\Phi}\\
\Psi
\end{array}
\right)
\mapsto
[1-i\epsilon_a(x)T_a]\left(\begin{array}{c}
\vec{\Phi}\\
\Psi
\end{array}
\right),\quad
T_a=\left(
\begin{array}{cc}
T_a^{\rm ad} & 0_{3\times 2}\\
0_{2\times 3} & T_a^{\rm f}
\end{array}
\right),
\end{equation}
($T_a$ block-diagonal, irreducible), i.e.
\begin{eqnarray}
\label{1:5:psip}
\Psi&\mapsto&\Psi'
=\left(1-\frac{i}{2}\vec{\tau}\cdot\vec{\epsilon}(x)\right)\Psi,
\\
\label{1:5:pip}
\vec{\Phi}&\mapsto&\left(1-i\vec{T}^{\rm ad}
\cdot\vec{\epsilon}(x)\right)\vec{\Phi}
\stackrel{\ast}{=}\vec{\Phi}+\vec{\epsilon}\times\vec{\Phi},
\end{eqnarray}
where in $\ast$ we made use of
$$-i\vec{T}^{\rm ad}\cdot\vec{\epsilon}\,\vec{\Phi}=
\left(\begin{array}{rrr}0&-\epsilon_3&\epsilon_2\\
\epsilon_3&0&-\epsilon_1\\
-\epsilon_2&\epsilon_1&0\end{array}\right)
\left(\begin{array}{c}\Phi_1\\{\Phi}_2\\{\Phi}_3\end{array}\right)
=\left(\begin{array}{l}-\epsilon_3\Phi_2+\epsilon_2\Phi_3\\
\epsilon_3\Phi_1-\epsilon_1\Phi_3\\
-\epsilon_2\Phi_1+\epsilon_1\Phi_2\end{array}\right)
=\vec{\epsilon}\times\vec{\Phi},
$$
i.e., the transformation acts on $\vec{\Phi}$ as an infinitesimal
rotation by the angle $|\vec{\epsilon}\,|$ about the axis $\hat{\epsilon}$
in isospin space.

\begin{enumerate}
\item
Show that the variation of the Lagrangian is given by
\begin{eqnarray}
\label{1:5:dl} \delta {\cal L}
&=&\partial_\mu\vec{\epsilon}\cdot(\bar{\Psi}\gamma^\mu\frac{\vec{\tau}}{2}\Psi
+\vec{\Phi}\times\partial^\mu\vec{\Phi}).
\end{eqnarray}
Hint: Make use of $\vec{\tau}\cdot\vec{a}\,\vec{\tau}\cdot
\vec{b}=\vec{a}\cdot \vec{b}\,1_{2\times 2}
+i\vec{\tau}\cdot\vec{a}\times\vec{b}$.

   From Eqs.\ (\ref{2:3:strom2}) and (\ref{2:3:divergenz}) we find
\begin{eqnarray}
\vec{J}^\mu&=&\frac{\partial \delta{\cal L}}{\partial \partial_\mu
\vec{\epsilon}}=\bar{\Psi}\gamma^\mu\frac{\vec{\tau}}{2}\Psi
+\vec{\Phi}\times \partial^\mu\vec{\Phi},\\
{\partial_\mu}\vec{J}^\mu&=&\frac{\partial \delta {\cal L}}{\partial
\vec{\epsilon}}=0.
\end{eqnarray}
   We obtain three time-independent charge operators
\begin{equation}
\vec{Q}=\int d^3x\left(\Psi^\dagger(x)\frac{\vec{\tau}}{2}\Psi(x)
+\vec{\Phi}(x)\times \vec{\Pi}(x)\right).
\end{equation}
   These operators are the infinitesimal generators of transformations
of the Hilbert space states.
\begin{itemize}
\item The generators decompose into a fermionic and a bosonic piece,
which commute with each other.
\end{itemize}
   Using the anti-commutation relations (fermions!)
\begin{eqnarray}
\label{1:5:komrelf1}
\{\Psi_{\alpha,r}(t,\vec{x}),\Psi^\dagger_{\beta,s}(t,\vec{y})\}&=&
\delta^3(\vec{x}-\vec{y})\delta_{\alpha\beta}\delta_{rs},\\
\label{1:5:komrelf2}
\{\Psi_{\alpha,r}(t,\vec{x}),\Psi_{\beta,s}(t,\vec{y})\}&=&0,\\
\label{1:5:komrelf3}
\{\Psi_{\alpha,r}^\dagger(t,\vec{x}),\Psi^\dagger_{\beta,s}(t,\vec{y})\}&=&0,
\end{eqnarray}
where $\alpha$ and $\beta$ denote Dirac indices, and $r$ and $s$ denote
isospin indices, and the commutation relations (bosons!)
\begin{eqnarray}
\label{1:5:komrelb1}
[\Phi_r(t,\vec{x}),\Pi_s(t,\vec{y})]&=&i\delta^3(\vec{x}-\vec{y})
\delta_{rs},\\
\label{1:5:komrelb2}
{[}\Phi_r(t,\vec{x}),\Phi_s(t,\vec{y})]&=&0,\\
\label{1:5:komrelb3}
{[}\Pi_r(t,\vec{x}),\Pi_s(t,\vec{y})]&=&0,
\end{eqnarray}
together with the fact that fermion fields and bosons fields commute, we
will verify:
\begin{equation}
\label{1:5:ver}
[Q_i,Q_j]=i\epsilon_{ijk}Q_k.
\end{equation}
\begin{itemize}
\item Proof:
\end{itemize}
\begin{eqnarray*}
[Q_i,Q_j]&=&\int d^3x d^3y\, [\Psi^\dagger(t,\vec{x})\frac{\tau_i}{2}
\Psi(t,\vec{x})+\epsilon_{ikl}\Phi_k(t,\vec{x})\Pi_l(t,\vec{x}),\\
&&\quad\quad\quad\Psi^\dagger(t,\vec{y})\frac{\tau_j}{2}
\Psi(t,\vec{y})+\epsilon_{jmn}\Phi_m(t,\vec{y})\Pi_n(t,\vec{y})]\\
&=&\int d^3x d^3y \Big([\Psi^\dagger(t,\vec{x})\frac{\tau_i}{2}
\Psi(t,\vec{x}),\Psi^\dagger(t,\vec{y})\frac{\tau_j}{2}
\Psi(t,\vec{y})]\\
&&\quad\quad+[\epsilon_{ikl}\Phi_k(t,\vec{x})\Pi_l(t,\vec{x}),
\epsilon_{jmn}\Phi_m(t,\vec{y})\Pi_n(t,\vec{y})]\Big)\\
&=& A+B.
\end{eqnarray*}
For the evaluation of $A$ we make use of
\begin{eqnarray}
\label{1:5:fk}
\lefteqn{[\Psi^\dagger_{\alpha,r}(t,\vec{x})\widehat{\cal O}_{\alpha\beta,rs}
\Psi_{\beta,s}(t,\vec{x}),\Psi^\dagger_{\gamma,t}(t,\vec{y})
\widehat{\cal O}_{\gamma\delta,tu}\Psi_{\delta,u}(t,\vec{y})]=}\nonumber\\
&=&\widehat{\cal O}_{\alpha\beta,rs}\widehat{\cal O}_{\gamma\delta,tu}
[\Psi^\dagger_{\alpha,r}(t,\vec{x})
\Psi_{\beta,s}(t,\vec{x}),\Psi^\dagger_{\gamma,t}(t,\vec{y})
\Psi_{\delta,u}(t,\vec{y})].
\end{eqnarray}
\item Verify
\begin{equation}
\label{1:5:abcdfk}
[ab,cd]=a\{b,c\}d-ac\{b,d\}+\{a,c\}db-c\{a,d\}b
\end{equation}
and express the commutator of fermion fields in terms of
anti-commu\-ta\-tors as
\begin{eqnarray*}
\lefteqn{[\Psi^\dagger_{\alpha,r}(t,\vec{x})
\Psi_{\beta,s}(t,\vec{x}),\Psi^\dagger_{\gamma,t}(t,\vec{y})
\Psi_{\delta,u}(t,\vec{y})]=}\\
&&\hspace{-2em}
\Psi^\dagger_{\alpha,r}(t,\vec{x})\Psi_{\delta,u}(t,\vec{y})\delta^3(
\vec{x}-\vec{y})\delta_{\beta\gamma}\delta_{st}
-\Psi^\dagger_{\gamma,t}(t,\vec{y})\Psi_{\beta,s}(t,\vec{x})
\delta^3(\vec{x}-\vec{y})\delta_{\alpha\delta}\delta_{ru}.
\end{eqnarray*}
   In a compact notation:
\begin{eqnarray}
\label{1:5:fkf}
\lefteqn{[\Psi^\dagger(t,\vec{x}) \Gamma_1 F_1 \Psi(t,\vec{x}),
\Psi^\dagger(t,\vec{y})\Gamma_2 F_2 \Psi(t,\vec{y})]=}\nonumber\\
&&\delta^3(\vec{x}-\vec{y})\left[
\Psi^\dagger(t,\vec{x})\Gamma_1\Gamma_2 F_1 F_2 \Psi(t,\vec{y})
-\Psi^\dagger(t,\vec{y})\Gamma_2 \Gamma_1 F_2 F_1 \Psi(t,\vec{x})\right],
\nonumber\\
\end{eqnarray}
where $\Gamma_i$ is one of the sixteen $4\times 4$ matrices
\begin{displaymath}
1_{4\times 4}, \gamma^\mu, \gamma_5, \gamma^\mu \gamma_5,
\sigma^{\mu\nu}=\frac{i}{2}[\gamma^\mu,\gamma^\nu],
\end{displaymath}
and $F_i$ one of the four $2\times 2$ matrices
\begin{displaymath}
1_{2\times 2}, \tau_i.
\end{displaymath}
\item Apply Eq.\ (\ref{1:5:fkf}) and integrate $\int d^3 y \cdots$ to obtain
\begin{eqnarray*}
A&=& i\epsilon_{ijk}\int d^3x\, \Psi^\dagger(x)\frac{\tau_k}{2}
\Psi(x).
\end{eqnarray*}
\item Verify
\begin{equation}
\label{1:5:abcdbk}
[ab,cd]=a[b,c]d+ac[b,d]+[a,c]db+c[a,d]b.
\end{equation}
\item Apply Eq.\ (\ref{1:5:abcdbk}) in combination with the
equal-time commutation relations to obtain
\begin{eqnarray}
\lefteqn{[\Phi_k(t,\vec{x})\Pi_l(t,\vec{x}),
\Phi_m(t,\vec{y})\Pi_n(t,\vec{y})]=}\nonumber\\
&&-i\Phi_k(t,\vec{x})\Pi_n(t,\vec{y})\delta^3(\vec{x}-\vec{y})\delta_{lm}
+i\Phi_m(t,\vec{y})\Pi_l(t,\vec{x})\delta^3(\vec{x}-\vec{y})\delta_{kn}.
\nonumber\\
\label{bkb}
\end{eqnarray}
\item Apply Eq.\ (\ref{bkb})  and integrate $\int d^3 y \cdots$ to obtain
for $B$
\begin{eqnarray*}
B&=& i\epsilon_{ijk}\int d^3x\, \epsilon_{klm}\Phi_l(x)\Pi_m(x).
\end{eqnarray*}
\end{enumerate}
\end{exercise}

\subsection{Global Symmetry Currents of the Light Quark Sector}
\label{subsec_gsclqs}
   The method of Gell-Mann and Lev\'y can now easily be applied to the
QCD Lagrangian by calculating the variation under the infinitesimal, local
form of Eqs.\ (\ref{2:4:u3lu3r}),
\begin{equation}
\label{2:4:dlqcd0}
\delta{\cal L}^0_{\rm QCD}=\bar{q}_R\left(\sum_{a=1}^8 \partial_\mu \Theta^R_a
\frac{\lambda_a}{2}+\partial_\mu \Theta^R\right)\gamma^\mu q_R
+\bar{q}_L\left(\sum_{a=1}^8 \partial_\mu \Theta^L_a
\frac{\lambda_a}{2}+\partial_\mu \Theta^L\right)\gamma^\mu q_L,
\end{equation}
from which, by virtue of Eqs.\ (\ref{2:3:strom2}) and (\ref{2:3:divergenz}),
one obtains the currents associated with the transformations of the
left-handed or right-handed quarks
\begin{eqnarray}
\label{2:4:str} L^{\mu,a}&=&\frac{\partial\delta{\cal L}^0_{\rm
QCD}}{\partial \partial_\mu \Theta^L_a}= \bar{q}_L\gamma^\mu
\frac{\lambda^a}{2}q_L,\quad
\partial_\mu L^{\mu,a}=
\frac{\partial\delta{\cal L}^0_{\rm QCD}}{\partial\Theta^L_a}=
0,\nonumber\\
R^{\mu,a}&=&\frac{\partial\delta{\cal L}^0_{\rm QCD}}{\partial
\partial_\mu \Theta^R_a}=\bar{q}_R\gamma^\mu
\frac{\lambda^a}{2}q_R,\quad
\partial_\mu R^{\mu,a}
=\frac{\partial\delta{\cal L}^0_{\rm QCD}}{\partial\Theta^R_a}
=0,\nonumber\\
L^{\mu}&=&\frac{\partial\delta{\cal L}^0_{\rm QCD}}{\partial
\partial_\mu \Theta^L}=\bar{q}_L\gamma^\mu q_L,\quad
\partial_\mu L^{\mu}=\frac{\partial\delta{\cal L}^0_{\rm QCD}}{\partial\Theta^L}=
0,\nonumber\\
R^{\mu}&=&\frac{\partial\delta{\cal L}^0_{\rm QCD}}{\partial
\partial_\mu \Theta^R}=\bar{q}_R\gamma^\mu q_R,\quad
\partial_\mu R^{\mu}=\frac{\partial\delta{\cal L}^0_{\rm QCD}}{\partial\Theta^R}=
0.
\end{eqnarray}
   The eight currents $L^{\mu,a}$ transform under
$\mbox{SU(3)}_L\times\mbox{SU(3)}_R$
as an $(8,1)$ multiplet, i.e., as octet and singlet under transformations
of the left- and right-handed fields, respectively.
   Similarly, the right-handed currents transform as a $(1,8)$ multiplet
under $\mbox{SU(3)}_L\times\mbox{SU(3)}_R$.
   Instead of these chiral currents one often uses linear combinations,
\begin{eqnarray}
\label{2:4:v}
V^{\mu,a}&=& R^{\mu,a}+L^{\mu,a}=\bar{q}\gamma^\mu\frac{\lambda^a}{2}q,\\
\label{2:4:a}
A^{\mu,a}&=&R^{\mu,a}-L^{\mu,a}=\bar{q}\gamma^\mu\gamma_5 \frac{\lambda^a}{2}q,
\end{eqnarray}
   transforming under parity as vector and axial-vector current
densities, respectively,
\begin{eqnarray}
\label{2:4:pv}
P:V^{\mu,a}(t,\vec{x})\mapsto V^a_\mu(t,-\vec{x}),\\
\label{2:4pa}
P: A^{\mu,a}(t,\vec{x})\mapsto -A_\mu^a(t,-\vec{x}).
\end{eqnarray}

   From Eq.\ (\ref{2:4:str}) one also obtains a conserved singlet
vector current resulting from a transformation of all left-handed
and right-handed quark fields by the {\em same} phase,
\begin{eqnarray}
\label{2:4:sv} V^\mu&=&R^{\mu}+L^{\mu}=\bar{q}\gamma^\mu q,\nonumber\\
\partial_\mu V^\mu&=&0.
\end{eqnarray}
   The singlet axial-vector current,
\begin{eqnarray}
\label{2:4:sa} A^\mu&=&
R^{\mu}-L^{\mu}=\bar{q}\gamma^\mu\gamma_5q,
\end{eqnarray}
originates from a transformation of all left-handed quark fields
with one phase and all right-handed with the {\em opposite} phase.
   However, such a singlet axial-vector current is only conserved on
the {\em classical} level.
    This symmetry is not preserved by quantization and there will be
extra terms, referred to as anomalies, resulting in\footnote{In
the large $N_C$ (number of colors) limit the singlet axial-vector
current is conserved, because the strong coupling constant behaves
as $g^2\sim N_C^{-1}$.}
\begin{equation}
\label{2:4:divsa}
\partial_\mu A^\mu=\frac{3 g^2}{32\pi^2}\epsilon_{\mu\nu\rho\sigma}
{\cal G}^{\mu\nu}_a {\cal G}^{\rho\sigma}_a,\quad \epsilon_{0123}=1,
\end{equation}
where the factor of 3 originates from the number of flavors.

\subsection{The Chiral Algebra}
\label{subsec_ca}
   The invariance of ${\cal L}^0_{\rm QCD}$ under global
$\mbox{SU(3)}_L\times\mbox{SU(3)}_R\times\mbox{U(1)}_V$ transformations
implies that also the QCD Hamilton operator in the chiral limit,
$H^0_{\rm QCD}$, exhibits a global
$\mbox{SU(3)}_L\times\mbox{SU(3)}_R\times\mbox{U(1)}_V$ symmetry.
   As usual, the ``charge operators'' are defined as the space integrals
of the charge densities,
\begin{eqnarray}
\label{2:3:ql}
Q^a_L(t)&=&\int d^3x\, q^\dagger_L(t,\vec{x})\frac{\lambda^a}{2}
q_L(t,\vec{x}),\quad a=1,\cdots, 8,\\
\label{2:3:qr}
Q^a_R(t)&=&\int d^3x\, q^\dagger_R(t,\vec{x})\frac{\lambda^a}{2}
q_R(t,\vec{x}),\quad a=1,\cdots, 8,\\
\label{2:3:qv}
Q_V(t)&=&\int d^3x\, \left[q^\dagger_L(t,\vec{x})q_L(t,\vec{x})+
q^\dagger_R(t,\vec{x})q_R(t,\vec{x})\right].
\end{eqnarray}
   For conserved symmetry currents, these operators are time independent,
i.e., they commute with the Hamiltonian,
\begin{equation}
\label{2:3:vrhq}
[Q_L^a,H^0_{\rm QCD}]=[Q_R^a,H^0_{\rm QCD}]=[Q_V,H^0_{\rm QCD}]=0.
\end{equation}
   The commutation relations of the charge operators with each other
are obtained by using Eq.\ (\ref{1:5:fkf}) applied to the quark fields
\begin{eqnarray}
\label{2:4:fkf}
\lefteqn{[q^\dagger(t,\vec{x}) \Gamma_1 F_1 q(t,\vec{x}),
q^\dagger(t,\vec{y})\Gamma_2 F_2 q(t,\vec{y})]=}\nonumber\\
&&\delta^3(\vec{x}-\vec{y})\left[
q^\dagger(t,\vec{x})\Gamma_1\Gamma_2 F_1 F_2 q(t,\vec{y})
-q^\dagger(t,\vec{y})\Gamma_2 \Gamma_1 F_2 F_1 q(t,\vec{x})\right],
\nonumber\\
\end{eqnarray}
   where $\Gamma_i$ and $F_i$ are $4\times 4$ Dirac matrices and
$3\times 3$ flavor matrices, respectively.\footnote{Strictly speaking,
we should
also include the color indices. However, since we are only discussing
color-neutral quadratic forms a summation over such indices is
always implied, with the net effect that one can completely omit them from
the discussion.}
   After inserting appropriate projectors $P_{L/R}$, Eq.\ (\ref{2:4:fkf})
is easily applied to the charge operators of Eqs.\ (\ref{2:3:ql}),
(\ref{2:3:qr}),
and (\ref{2:3:qv}),  showing that these operators
indeed satisfy the commutation relations corresponding to the Lie algebra
of $\mbox{SU(3)}_L\times\mbox{SU(3)}_R\times\mbox{U(1)}_V$,
\begin{eqnarray}
\label{2:4:crqll}
[Q_L^a,Q_L^b]&=&if_{abc}Q_L^c,\\
\label{2:4:crqrr}
{[Q_R^a,Q_R^b]}&=&if_{abc}Q_R^c,\\
\label{2:4:crqlr}
{[Q_L^a,Q_R^b]}&=&0,\\
\label{2:4:crqlvrv}
{[Q_L^a,Q_V]}&=&[Q_R^a,Q_V]=0.
\end{eqnarray}
   For example (recall $P_L^\dagger= P_L$ and $P_L^2=P_L$)
\begin{eqnarray*}
[Q_L^a,Q_L^b]&=&\int d^3 x d^3y [q^\dagger(t,\vec{x})P_L^\dagger
\frac{\lambda_a}{2} P_L q(t,\vec{x}),
q^\dagger(t,\vec{y})P_L^\dagger
\frac{\lambda_b}{2} P_L q(t,\vec{y})]\\
&=&\int d^3 x d^3y \delta^3(\vec{x}-\vec{y})
q^\dagger(t,\vec{x})\underbrace{P_L^\dagger P_L P_L^\dagger P_L}_{\mbox{$P_L$}}
\frac{\lambda_a}{2}\frac{\lambda_b}{2} q(t,\vec{y})
\\
&&-\int d^3 x d^3y\delta^3(\vec{x}-\vec{y})
q^\dagger(t,\vec{y}) P_L
\frac{\lambda_b}{2}\frac{\lambda_a}{2}
q(t,\vec{x})\\
&=&if_{abc}\int d^3 x q^\dagger(t,\vec{x}) \frac{\lambda_c}{2}P_L
q(t,\vec{x})=if_{abc}Q_L^c.
\end{eqnarray*}
\begin{exercise}
\label{exercise_chiral_algebra} \rm Verify the remaining
commutation relations, Eqs.\ (\ref{2:4:crqrr}), (\ref{2:4:crqlr}),
and (\ref{2:4:crqlvrv}).
\end{exercise}

  It should be stressed that, even without being able to explicitly solve
the equation of motion of the quark fields entering the charge
operators of Eqs.\ (\ref{2:4:crqll}) -- (\ref{2:4:crqlvrv}), we
know from the equal-time commutation relations and the symmetry of
the Lagrangian that these charge operators are the generators of
infinitesimal transformations of the Hilbert space associated with
$H^0_{\rm QCD}$.
   Furthermore, their commutation relations with a given operator,
specify the transformation behavior of the operator in question
under the group $\mbox{SU(3)}_L\times\mbox{SU(3)}_R\times\mbox{U(1)}_V$.

\subsection{Chiral Symmetry Breaking Due to Quark Masses}
\label{subsec_csbdqm}
   The finite $u$-, $d$-, and $s$-quark masses in the QCD Lagrangian
explicitly break the chiral symmetry, resulting in divergences of the symmetry
currents.
   As a consequence, the charge operators are, in general, no longer
time independent.
   However, as first pointed out by Gell-Mann, the equal-time-commutation
relations still play an important role even if the symmetry is
explicitly broken.
   As will be discussed later on in more detail, the symmetry currents will
give rise to chiral Ward identities relating various QCD Green functions
to each other.
   Equation (\ref {2:3:divergenz}) allows one to discuss the
divergences of the symmetry currents in the presence of quark
masses.
   To that end, let us consider the quark-mass matrix of the three
light quarks and project it on the nine $\lambda$ matrices of
Eq.\ (\ref{2:1:matrixa}),
\begin{eqnarray}
\label{2:4:qmm}
M&=&\left(\begin{array}{ccc}
m_u&0&0\\
0&m_d&0\\
0&0&m_s
\end{array}
\right).
\end{eqnarray}
\begin{exercise}
\label{exercise_quark_mass_matrix}
\rm
Express the quark mass matrix in terms of the $\lambda$ matrices
$\lambda_0$, $\lambda_3$, and $\lambda_8$.
\end{exercise}
   In particular, applying Eq.\ (\ref{2:4:qgq}) we see that the quark-mass
term mixes left- and right-handed fields,
\begin{equation}
\label{2:4:lm}
{\cal L}_M= -\bar{q}Mq=
-(\bar{q}_R M q_L +\bar{q}_L M q_R).
\end{equation}
   The symmetry-breaking term transforms under $\mbox{SU(3)}_L\times
\mbox{SU(3)}_R$ as a member of a $(3,3^\ast)+(3^\ast,3)$ representation,
i.e.,
$$
\bar{q}_{R,i} M_{ij} q_{L,j} +\bar{q}_{L,i} M_{ij} q_{R,j}
\mapsto
U_{L,jk}U^\ast_{R,il} \bar{q}_{R,l}M_{ij} q_{L,k}+ (L\leftrightarrow R),
$$
where $(U_L,U_R)\in\mbox{SU(3)}_L\times
\mbox{SU(3)}_R$.
   Such symmetry-breaking {\em patterns} were already discussed in
the pre-QCD era in Refs.\
\cite{Glashow:1967rx:3:6,Gell-Mann:rz:3:6}.

   From ${\cal L}_M$ one obtains as the variation $\delta{\cal L}_M$
under the transformations of Eqs.\ (\ref{2:4:u3lu3r}),
\begin{eqnarray}
\label{2:4:dlm}
\delta {\cal L}_M&=&
-i\left[ \bar{q}_R\left(\sum_{a=1}^8 \Theta^R_a \frac{\lambda_a}{2}
+ \Theta^R\right) M q_L
- \bar{q}_R M\left(\sum_{a=1}^8 \Theta^L_a \frac{\lambda_a}{2}
+\Theta^L\right) q_L\right.\nonumber\\
&&\left.+ \bar{q}_L\left(\sum_{a=1}^8 \Theta^L_a \frac{\lambda_a}{2}
+ \Theta^L\right) M q_R
- \bar{q}_L M\left(\sum_{a=1}^8 \Theta^R_a \frac{\lambda_a}{2}
+ \Theta^R\right) q_R \right]\nonumber\\
&=&
-i\left[ \sum_{a=1}^8 \Theta_a^R \left(
\bar{q}_R \frac{\lambda_a}{2}M q_L -\bar{q}_L M \frac{\lambda_a}{2} q_R
\right)
+\Theta^R  \left(\bar{q}_RMq_L-\bar{q}_LM q_R\right)\right.\nonumber\\
&&+\left.\sum_{a=1}^8 \Theta_a^L \left(
\bar{q}_L \frac{\lambda_a}{2}M q_R -\bar{q}_R M \frac{\lambda_a}{2} q_L
\right)
+\Theta^L  \left(\bar{q}_LMq_R-\bar{q}_RM q_L\right)\right],\nonumber\\
\end{eqnarray}
which results in the following divergences,\footnote{
   The divergences are proportional to the mass parameters which is
the origin of the expression current-quark mass.}
\begin{eqnarray}
\label{2:4:dslr}
\partial_\mu L^{\mu,a}&=&\frac{\partial \delta {\cal L}_M}{\partial \Theta^L_a}
=-i\left(\bar{q}_L\frac{\lambda_a}{2}M q_R -\bar{q}_R M \frac{\lambda_a}{2}
q_L\right),\nonumber\\
\partial_\mu R^{\mu,a}&=&\frac{\partial \delta {\cal L}_M}{\partial \Theta^R_a}
=-i\left(\bar{q}_R\frac{\lambda_a}{2}M q_L -\bar{q}_L M \frac{\lambda_a}{2}
q_R\right),\nonumber\\
\partial_\mu L^{\mu}&=&\frac{\partial \delta {\cal L}_M}{\partial \Theta^L}
=-i\left(\bar{q}_L M q_R -\bar{q}_R M q_L\right),\nonumber\\
\partial_\mu R^{\mu}&=&\frac{\partial \delta {\cal L}_M}{\partial \Theta^R}
=-i\left(\bar{q}_R M q_L -\bar{q}_L M q_R\right).
\end{eqnarray}
   The anomaly has not yet been considered.
   Applying Eq.\ (\ref{2:4:qgq}) to the case of the vector currents
and inserting projection operators for the axial-vector current,
the corresponding divergences read
\begin{eqnarray}
\label{2:4:dsva}
\partial_\mu V^{\mu,a}&=&
-i\bar{q}_R[\frac{\lambda_a}{2},M]q_L
-i\bar{q}_L[\frac{\lambda_a}{2},M]q_R
\stackrel{(\ref{2:4:qgq})}{=}
i\bar{q}[M,\frac{\lambda_a}{2}]q,\nonumber\\
\partial_\mu A^{\mu,a}&=&
-i\left(\bar{q}_R\frac{\lambda_a}{2}M q_L
-\bar{q}_L M\frac{\lambda_a}{2}q_R\right)
+i\left(\bar{q}_L\frac{\lambda_a}{2}M q_R
-\bar{q}_R M\frac{\lambda_a}{2}q_L\right)\nonumber\\
&=&
i\left(\bar{q}_L\{\frac{\lambda_a}{2},M\}q_R
-\bar{q}_R\{\frac{\lambda_a}{2},M\}q_L\right)
\nonumber\\
&=&i\left(\bar{q}\{\frac{\lambda_a}{2},M\}\frac{1}{2}(1+\gamma_5)q
-\bar{q}\{\frac{\lambda_a}{2},M\}\frac{1}{2}(1-\gamma_5)q\right)\nonumber\\
&=&
i\bar{q}\{\frac{\lambda_a}{2},M\}\gamma_5q,\nonumber\\
\partial_\mu V^\mu&=&0,\nonumber\\
\partial_\mu A^\mu&=&2i\bar{q}M\gamma_5 q+
\frac{3 g^2}{32\pi^2}\epsilon_{\mu\nu\rho\sigma}
{\cal G}^{\mu\nu}_a {\cal G}^{\rho\sigma}_a,\quad \epsilon_{0123}=1,
\end{eqnarray}
   where the axial anomaly has also been taken into account.

   We are now in the position to summarize the various (approximate)
symmetries of the strong interactions in combination with the corresponding
currents and their divergences.

\begin{itemize}
\item In the limit of massless quarks, the sixteen currents $L^{\mu,a}$
and $R^{\mu,a}$ or, alternatively, $V^{\mu,a}$ and $A^{\mu,a}$ are
conserved.
   The same is true for the singlet vector current $V^\mu$, whereas the
singlet axial-vector current $A^\mu$ has an anomaly. \item For any
value of quark masses, the individual flavor currents
$\bar{u}\gamma^\mu u$, $\bar{d}\gamma^\mu d$, and
$\bar{s}\gamma^\mu s$ are always conserved in the strong
interactions reflecting the flavor independence of the strong
coupling and the diagonality of the quark-mass matrix.
   Of course, the singlet vector current $V^\mu$, being the sum of
the three flavor currents, is always conserved.
\item In addition to the anomaly, the singlet axial-vector current
has an explicit divergence due to the quark masses.
\item For equal quark masses, $m_u=m_d=m_s$, the eight vector currents
$V^{\mu,a}$ are conserved, because $[\lambda_a,1]=0$.
   Such a scenario is the origin of the SU(3) symmetry
originally proposed by Gell-Mann and Ne'eman
\cite{EightfoldWay:3:6}.
   The eight axial-vector currents $A^{\mu,a}$ are not conserved.
   The divergences of the octet axial-vector currents of Eq.\ (\ref{2:4:dsva})
are proportional to pseudoscalar quadratic forms.
   This can be interpreted as the microscopic origin of the
PCAC relation (partially conserved axial-vector current)
\cite{Gell-Mann:1964tf:3:6,Adler:1968:3:6} which states that the
divergences of the axial-vector currents are proportional to
renormalized field operators representing the lowest-lying
pseudoscalar octet.
\end{itemize}

\section{Green Functions and Ward Identities *}
\label{sec_gfwi}

  In this section we will show how to derive Ward identities for Green
functions in the framework of canonical quantization on the one hand,
and quantization via the Feynman path integral on the other hand,
by means of an explicit example.
   In order to keep the discussion transparent, we will concentrate on
a simple scalar field theory with a global O(2) or U(1) invariance.
   To that end, let us consider the Lagrangian
\begin{eqnarray}
\label{2:5:gfwi:lphi4}
{\cal L}&=&\frac{1}{2}(\partial_\mu \Phi_1\partial^\mu \Phi_1
+\partial_\mu \Phi_2\partial^\mu \Phi_2)
-\frac{m^2}{2} (\Phi_1^2+\Phi_2^2)
-\frac{\lambda}{4}(\Phi_1^2+\Phi_2^2)^2\nonumber\\
&=&
\partial_\mu \Phi^\dagger \partial^\mu \Phi -m^2 \Phi^\dagger\Phi
-\lambda (\Phi^\dagger\Phi)^2,
\end{eqnarray}
where
\begin{displaymath}
\Phi(x)=\frac{1}{\sqrt{2}}[\Phi_1(x)+i\Phi_2(x)],
\quad
\Phi^\dagger(x)=\frac{1}{\sqrt{2}}[\Phi_1(x)-i\Phi_2(x)],
\end{displaymath}
with real scalar fields $\Phi_1$ and $\Phi_2$.
   Furthermore, we assume $m^2>0$ and $\lambda>0$, so there is
no spontaneous symmetry breaking (see Chapter \ref{chap_ssbgt}) and the energy is
bounded from below.
   Equation (\ref{2:5:gfwi:lphi4}) is invariant under the global (or rigid)
transformations
\begin{equation}
\label{2:5::gfwi:inftrans1}
\Phi'_1=\Phi_1-\epsilon \Phi_2,\quad
\Phi'_2=\Phi_2+\epsilon \Phi_1,
\end{equation}
or, equivalently,
\begin{equation}
\label{2:5:gfwi:inftrans2}
\Phi'=(1+i\epsilon)\Phi,\quad
\Phi'^\dagger=(1-i\epsilon)\Phi^\dagger,
\end{equation}
where $\epsilon$ is an infinitesimal real parameter.
   Applying the method of Gell-Mann and L{\'e}vy,
we obtain for a {\em local}
parameter $\epsilon(x)$,
\begin{equation}
\label{2:5:gfwi:dlphi4}
\delta{\cal L}=\partial_\mu\epsilon(x)(i\partial^\mu\Phi^\dagger \Phi
-i\Phi^\dagger\partial^\mu\Phi),
\end{equation}
from which, via Eqs.\ (\ref{2:3:strom2}) and (\ref{2:3:divergenz}),
we derive for the current corresponding to the global symmetry,
\begin{eqnarray}
J^{\mu}&=&\frac{\partial \delta\cal L}{\partial \partial_\mu
\epsilon}=(i\partial^\mu\Phi^\dagger \Phi
-i\Phi^\dagger\partial^\mu\Phi),\\
\partial_\mu J^{\mu}&=&\frac{\partial \delta\cal L}{\partial
\epsilon}=0.
\end{eqnarray}
   Recall that the identification of Eq.\ (\ref{2:3:divergenz}) as
the divergence of the current is only true for fields
satisfying the Euler-Lagrange equations of motion.

   We now extend the analysis to a {\em quantum} field theory.
   In the framework of canonical quantization, we first define
conjugate momenta,
\begin{equation}
\label{2:5:gfwi:conjmom}
\Pi_i(x)=\frac{\partial \cal L}{\partial \partial_0\Phi_i},\quad
\Pi(x)=\frac{\partial \cal L}{\partial \partial_0\Phi},\quad
\Pi^\dagger(x)=\frac{\partial \cal L}{\partial \partial_0\Phi^\dagger},
\end{equation}
and interpret the fields and their conjugate momenta
as operators which, in the Heisenberg
picture, are subject to the equal-time commutation relations
\begin{equation}
[\Phi_i(t,\vec{x}),\Pi_j(t,\vec{y})]=i\delta_{ij}\delta^3(\vec{x}-\vec{y}),
\end{equation}
and
\begin{equation}
\label{2:5:gfwi:eqtcr}
{[}\Phi(t,\vec{x}),\Pi(t,\vec{y})]=
[\Phi^\dagger(t,\vec{x}),\Pi^\dagger(t,\vec{y})]
=i\delta^3(\vec{x}-\vec{y}).
\end{equation}
   The remaining equal-time commutation relations, involving fields or momenta
only, vanish.
   For the quantized theory, the current operator then reads
\begin{equation}
J^\mu(x)=:(i\partial^\mu\Phi^\dagger \Phi
-i\Phi^\dagger\partial^\mu\Phi):,
\end{equation}
where $:\quad :$ denotes normal or Wick ordering, i.e., annihilation
operators appear to the right of creation operators.
   For a conserved current, the charge operator, i.e., the space integral of
the charge density, is time independent and serves as the generator of
infinitesimal transformations of the Hilbert space states,
\begin{equation}
Q=\int d^3 x J^0(t,\vec{x}).
\end{equation}
   Applying Eq.\ (\ref{2:5:gfwi:eqtcr}), it is
straightforward to calculate the equal-time commutation
relations\footnote{The
transition to normal ordering involves an (infinite)
constant which does not contribute to the commutator.}
\begin{eqnarray}
\label{2:5:gfwi:j0comrel}
[J^0(t,\vec{x}),\Phi(t,\vec{y})]&=&\delta^3(\vec{x}-\vec{y})\Phi(t,\vec{x}),
\nonumber\\
{[}J^0(t,\vec{x}),\Pi(t,\vec{y})]&=&-\delta^3(\vec{x}-\vec{y})
\Pi(t,\vec{x}),\nonumber\\
{[}J^0(t,\vec{x}),\Phi^\dagger(t,\vec{y})]&=&-\delta^3(\vec{x}-\vec{y})
\Phi^\dagger(t,\vec{x}),\nonumber\\
{[}J^0(t,\vec{x}),\Pi^\dagger(t,\vec{y})]&=&\delta^3(\vec{x}-\vec{y})
\Pi^\dagger(t,\vec{x}).
\end{eqnarray}
   In particular, performing the space integrals in Eqs.\
(\ref{2:5:gfwi:j0comrel}), one obtains
\begin{eqnarray}
\label{2:5:gfwi:Qcomrel}
[Q,\Phi(x)]&=&\Phi(x),
\nonumber\\
{[}Q,\Pi(x)]&=&-\Pi(x),\nonumber\\
{[}Q,\Phi^\dagger(x)]&=&-\Phi^\dagger(x),\nonumber\\
{[}Q,\Pi^\dagger(x)]&=&\Pi^\dagger(x).
\end{eqnarray}
   In order to illustrate the implications of Eqs.\ (\ref{2:5:gfwi:Qcomrel}),
let us take an eigenstate $|\alpha\rangle$ of $Q$ with eigenvalue
$q_\alpha$ and consider, for example, the action of $\Phi(x)$ on that
state,
\begin{eqnarray*}
Q\left(\Phi(x)|\alpha\rangle\right)=\left([Q,\Phi(x)]+\Phi(x)Q\right)
|\alpha\rangle
=(1+q_\alpha)\left(\Phi(x)|\alpha\rangle\right).
\end{eqnarray*}
   We conclude that the operators $\Phi(x)$ and $\Pi^\dagger(x)$
[$\Phi^\dagger(x)$ and $\Pi(x)$] increase (decrease) the Noether charge
of a system by one unit.

   We are now in the position to discuss the consequences of the U(1)
symmetry of Eq.\ (\ref{2:5:gfwi:lphi4}) for the Green functions of the theory.
   To that end, let us consider as our prototype the Green function
\begin{equation}
\label{2:5:gfwi:gmuxyz}
G^\mu(x,y,z)=\langle 0|T[\Phi(x) J^\mu(y) \Phi^\dagger(z)]|0\rangle,
\end{equation}
   which describes the transition amplitude for the creation of a quantum
of Noether charge $+1$ at $x$, propagation to $y$, interaction at $y$
via the current operator, propagation to $z$ with annihilation at $z$.
   First of all we observe that under the global infinitesimal transformations
of Eq.\ (\ref{2:5:gfwi:inftrans2}), $J^\mu(x)\mapsto J'^\mu(x)=J^\mu(x)$, or in
other words $[Q,J^\mu(x)]=0$.
   We thus obtain
\begin{eqnarray}
\label{2:5:gfwi:gmutrans}
G^\mu(x,y,z)\mapsto G'^\mu(x,y,z)&=&
\langle 0|T[(1+i\epsilon)\Phi(x) J'^\mu(y) (1-i\epsilon)\Phi^\dagger(z)]
|0\rangle
\nonumber\\
&=&\langle 0|T[\Phi(x) J^\mu(y) \Phi^\dagger(z)]|0\rangle\nonumber\\
&=&G^\mu(x,y,z),
\end{eqnarray}
   the Green function remaining invariant under
the U(1) transformation.
   (In general, the transformation behavior of a Green function depends on the
irreducible representations under which the fields transform.
   In particular, for more complicated groups such as SU($N$), standard tensor
methods of group theory may be applied to reduce the product
representations into irreducible components.
   We also note that for
U(1), the symmetry current is charge neutral, i.e.\ invariant,
which for more complicated groups, in general, is not the case.)

   Moreover, since $J^\mu(x)$ is the Noether current of the underlying
U(1) there are further restrictions on the Green function beyond
its transformation behavior under the group.
   In order to see this, we consider the divergence of
Eq.\ (\ref{2:5:gfwi:gmuxyz})
   and apply the equal-time commutation relations of Eqs.\
(\ref{2:5:gfwi:j0comrel}) to obtain
\begin{eqnarray}
\label{2:5:gfwi:gmuwt}
\partial_\mu^y G^\mu(x,y,z)&=&[\delta^4(x-y)-\delta^4(z-y)]
\langle 0|T[\Phi(x)\Phi^\dagger(z)]|0\rangle,
\end{eqnarray}
   where we made use of $\partial_\mu J^\mu=0$.
   Equation (\ref{2:5:gfwi:gmuwt}) is the analogue of the Ward identity of QED
[see Eq.\ (\ref{2:6:qedwardidentity})].
   In other words, the underlying symmetry not only determines the
transformation behavior of Green functions under the group, but also
relates $n$-point Green functions containing a symmetry current
to $(n-1)$-point Green functions [see Eq.\ (\ref{2:6:gendmug})].
   In principle, calculations similar to those leading to Eqs.\
(\ref{2:5:gfwi:gmutrans}) and (\ref{2:5:gfwi:gmuwt}), can be performed for any
Green function of the theory.

\begin{exercise}
\label{exercise_ward_takahashi_identity}
\renewcommand{\labelenumi}{(\alph{enumi})}
\rm
Consider the three-point Green function
\begin{displaymath}
G^\mu(x,y,z)=\langle 0|T[J^\mu(x)\pi^+(y) \pi^-(z)]|0\rangle,
\end{displaymath}
where $J^\mu(x)$ is the electromagnetic current operator, and
$\pi^{+/-}(x)$ are field operators destroying a $\pi^{+/-}$ or creating
a $\pi^{-/+}$. The time ordering is defined as
\begin{eqnarray*}
T[J^\mu(x)\pi^+(y) \pi^-(z)]&=&
J^\mu(x)\pi^+(y) \pi^-(z)\Theta(x_0-y_0)\Theta(y_0-z_0)\\
&&+J^\mu(x)\pi^-(z)\pi^+(y)\Theta(x_0-z_0)\Theta(z_0-y_0)
+\cdots.
\end{eqnarray*}
   All in all there exist $3!=6$ distinct orderings.
   The equal-time commutation relations between the charge density operator
$J^0$ and the field operators $\pi^{+/-}$,
\begin{eqnarray*}
[J^0(x),\pi^-(y)]\delta(x_0-y_0)&=&\delta^4(x-y) \pi^-(y),\\
{[}J^0(x),\pi^+(y)]\delta(x_0-y_0)&=&-\delta^4(x-y) \pi^+(y),
\end{eqnarray*}
in combination with current conservation $\partial_\mu J^\mu(x)=0$ are the
main ingredients of obtaining the Ward-Takahashi identity.
   To that end consider
\begin{displaymath}
\partial_\mu^x \langle 0|T[J^\mu(x)\pi^+(y) \pi^-(z)]|0\rangle.
\end{displaymath}
   Note that the $x$ dependence resides in both $J^\mu(x)$ and the
$\Theta$ functions of the time ordering.
\begin{enumerate}
\item
Make use of
\begin{eqnarray*}
\partial_\mu^x \Theta(x_0-y_0)&=&g_{\mu 0}\delta(x_0-y_0),\\
\partial_\mu^x \Theta(y_0-x_0)&=&-g_{\mu 0}\delta(y_0-x_0),
\end{eqnarray*}
   to obtain
\begin{eqnarray*}
\partial_\mu^x T[J^\mu(x)\pi^+(y) \pi^-(z)]&=&T[
\underbrace{\partial_\mu^x J^\mu(x)}_{\mbox{0}}
 \pi^+(y) \pi^-(z)] \\
&&+J^0(x)\pi^+(y)\pi^-(z)\delta(x_0-y_0)\Theta(y_0-z_0)\\
&&-\pi^+(y)J^0(x)\pi^-(z)\delta(y_0-x_0)\Theta(x_0-z_0)\\
&&+J^0(x)\pi^-(z)\pi^+(y)\delta(x_0-z_0)\Theta(z_0-y_0)\\
&&-\pi^-(z)J^0(x)\pi^+(y)\delta(z_0-x_0)\Theta(x_0-y_0)\\
&&+\pi^+(y)J^0(x)\pi^-(z)\Theta(y_0-x_0)\delta(x_0-z_0)\\
&&-\pi^+(y)\pi^-(z)J^0(x)\Theta(y_0-z_0)\delta(z_0-x_0)\\
&&+\pi^-(z)J^0(x)\pi^+(y)\Theta(z_0-x_0)\delta(x_0-y_0)\\
&&-\pi^-(z)\pi^+(y)J^0(x)\Theta(z_0-y_0)\delta(y_0-x_0).
\end{eqnarray*}
\item
   Apply the equal-time commutation relations and combine the result
to obtain
\begin{eqnarray*}
\partial_\mu^x G^\mu(x,y,z)=[\delta^4(x-z)-\delta^4(x-y)]
\langle 0|T[\pi^+(y)\pi^-(z)]|0\rangle.
\end{eqnarray*}
\end{enumerate}
\noindent Remarks:
\begin{itemize}
\item Usually, the WT identity is expressed in momentum space.
\item It does not rely on perturbation theory!
\item The generalization to $n$-point functions ($n\geq 4$) is straightforward
(proof via induction).
\item It can also be applied to currents which are not conserved (e.g.,
PCAC).
\end{itemize}
\end{exercise}

   We will now show that the symmetry constraints imposed by the Ward identities
can be compactly summarized in terms of an invariance property of
a generating functional.
   The generating functional is defined as the vacuum-to-vacuum transition
amplitude in the presence of external fields,
\begin{eqnarray}
\label{2:5:gfwi:genfunc1}
\lefteqn{W[j,j^\ast,j_\mu]}\nonumber\\
&=& \langle 0 ;{\rm out}|0 ;{\rm in}\rangle_{j,j^\ast,j_\mu}
\nonumber\\
&=&\exp(iZ[j,j^\ast,j_\mu])\nonumber\\
&=&\langle 0|T\left(\exp\left\{i\int d^4x[j(x)\Phi^\dagger(x)+j^\ast(x) \Phi(x)
+j_\mu(x) J^\mu(x)]\right\}\right)|0\rangle,\nonumber\\
\end{eqnarray}
   where $\Phi$ and $\Phi^\dagger$ are the field operators and $J^\mu(x)$
is the Noether current.
   Note that the field operators and the conjugate momenta are subject
to the equal-time commutation relations and, in addition, must satisfy
the Heisenberg equations of motion.
   Via this second condition and implicitly through the ground state,
the generating functional depends on the dynamics
of the system which is determined by the Lagrangian of
Eq.\ (\ref{2:5:gfwi:lphi4}).
   The Green functions of the theory involving $\Phi$, $\Phi^\dagger$, and
$J^\mu$ are obtained through functional derivatives of
Eq.\ (\ref{2:5:gfwi:genfunc1}).
   For example, the Green function of Eq.\ (\ref{2:5:gfwi:gmuxyz}) is given by
\begin{equation}
\label{2:5:gfwi:gmuxyz1}
G^\mu(x,y,z)=(-i)^3 \left.\frac{\delta^3 W[j,j^\ast,j_\mu]}{\delta j^\ast(x)
\delta j_\mu(y) \delta j(z)}\right|_{j=0,j^\ast=0,j_\mu=0}.
\end{equation}

   In order to discuss the constraints imposed on the generating functional
via the underlying symmetry of the theory, let us consider its path integral
rep\-re\-sentation,\footnote{Up to an irrelevant constant
the measure $[d\Phi_1][d\Phi_2]$ is equivalent to $[d\Phi][d\Phi^\ast]$,
with $\Phi$ and $\Phi^\ast$ considered as independent variables of integration.
}
\begin{eqnarray}
\label{2:5:gfwi:genfunc2}
W[j,j^\ast,j_\mu]&=&\int [d\Phi_1][d\Phi_2] e^{iS[\Phi,\Phi^\ast,j,j^\ast,
j_\mu]},
\end{eqnarray}
where
\begin{equation}
\label{2:5:gfwi:actionsources}
S[\Phi,\Phi^\ast,j,j^\ast,j_\mu]=S[\Phi,\Phi^\ast]+
\int d^4 x [\Phi(x)j^\ast(x)+\Phi^\ast(x) j(x)+J^\mu(x) j_\mu(x)]
\end{equation}
denotes the action corresponding to the Lagrangian of
Eq.\ (\ref{2:5:gfwi:lphi4})
in combination with a coupling to the external sources.
    Let us now consider a {\em local} infinitesimal transformation of the
fields [see Eqs.\ (\ref{2:5:gfwi:inftrans2})] together
with a {\em simultaneous}
transformation of the external sources,
\begin{equation}
\label{2:5:gfwi:inftranssources}
j'(x)=[1+i\epsilon(x)]j(x),\quad
j'^\ast(x)=[1-i\epsilon(x)] j^\ast(x),\quad
j_\mu'(x)=j_\mu(x)-\partial_\mu\epsilon(x).
\end{equation}
   The action of Eq.\ (\ref{2:5:gfwi:actionsources}) remains invariant under
such a transformation,
\begin{equation}
S[\Phi',\Phi'^\ast,j',j'^\ast,j_\mu']=
S[\Phi,\Phi^\ast,j,j^\ast,j_\mu].
\end{equation}
   We stress that the transformation of the external current $j_\mu$ is
necessary to cancel a term resulting from the kinetic term in
the Lagrangian.
   Also note that the {\em global} symmetry of the Lagrangian
determines the explicit form of the transformations of Eq.\
(\ref{2:5:gfwi:inftranssources}).
   We can now verify the invariance of the generating functional as follows,
\begin{eqnarray}
\label{2:5:gfwi:winvariance}
W[j,j^\ast,j_\mu]&=&\int [d\Phi_1][d\Phi_2] e^{iS[\Phi,\Phi^\ast,j,j^\ast,
j_\mu]}\nonumber\\
&=&\int [d\Phi_1][d\Phi_2] e^{iS[\Phi',\Phi'^\ast,j',j'^\ast,
j'_\mu]}\nonumber\\
&=&\int [d\Phi_1'][d\Phi_2'] \left|\left(\frac{\partial\Phi_i}{\partial
\Phi_j'}\right)\right|
e^{iS[\Phi',\Phi'^\ast,j',j'^\ast,
j'_\mu]}\nonumber\\
&=&\int [d\Phi_1][d\Phi_2]
e^{iS[\Phi,\Phi^\ast,j',j'^\ast,
j'_\mu]}\nonumber\\
&=&W[j',j'^\ast,j'_\mu].
\end{eqnarray}
   We made use of the fact that the Jacobi determinant is one
and renamed the integration variables.
   In other words, given the {\em global} U(1) symmetry of the Lagrangian,
Eq.\ (\ref{2:5:gfwi:lphi4}), the generating functional is invariant under the
{\em local} transformations of Eq.\ (\ref{2:5:gfwi:inftranssources}).
   It is this observation which, for the more general case of the
chiral group SU($N$)$\times$SU($N$), was used by Gasser and Leutwyler
as the starting point of chiral perturbation theory.

   We still have to discuss how this invariance allows us to collect the Ward
identities in a compact formula.
   We start from Eq.\ (\ref{2:5:gfwi:winvariance}),
\begin{eqnarray*}
0&=&\int[d\Phi_1] [d\Phi_2]\left(e^{iS[\Phi,\Phi^\ast,j',j'^\ast,j'_\mu]}
-e^{iS[\Phi,\Phi^\ast,j,j^\ast,j_\mu]}\right)\\
&=&\int[d\Phi_1] [d\Phi_2]\int d^4x \left\{
\epsilon[\Phi j^\ast-\Phi^\ast j]
-iJ^\mu\partial_\mu\epsilon\right\}
e^{iS[\Phi,\Phi^\ast,j,j^\ast,j_\mu]}.
\end{eqnarray*}
   Observe that
\begin{displaymath}
\Phi(x) e^{iS[\Phi,\Phi^\ast,j,j^\ast,j_\mu]}
=-i\frac{\delta}{\delta j^\ast(x)} e^{iS[\Phi,\Phi^\ast,j,j^\ast,j_\mu]},
\end{displaymath}
and similarly for the other terms, resulting in
\begin{eqnarray*}
0&=&\int[d\Phi_1] [d\Phi_2]\int d^4 x\left\{
\epsilon(x)\left[-ij^\ast(x)\frac{\delta}{\delta j^\ast(x)}
+ij(x)\frac{\delta}{\delta j(x)}\right]\right.\\
&&\left.
-\partial_\mu\epsilon(x)\frac{\delta}{\delta j_\mu(x)}\right\}
e^{iS[\Phi,\Phi^\ast,j,j^\ast,j_\mu]}.
\end{eqnarray*}
   Finally we interchange the order of integration, make use of
partial integration, and apply the divergence theorem:
\begin{equation}
\label{2:5:gfwi:0fl}
0=\int d^4 x \epsilon(x)\left[i j(x)\frac{\delta}{\delta j(x)}
-i j^\ast(x)\frac{\delta}{\delta j^\ast(x)}
+\partial_\mu ^x \frac{\delta}{\delta j_\mu(x)}\right]
W[j,j^\ast,j_\mu].
\end{equation}
   Since Eq.\ (\ref{2:5:gfwi:0fl}) must hold for any $\epsilon(x)$ we obtain
as the master equation for deriving Ward identities,
\begin{equation}
\label{2:5:gfwi:mewi}
\left[j(x)\frac{\delta}{\delta j(x)}-
j^\ast(x)\frac{\delta}{\delta j^\ast(x)}
-i\partial_\mu^x \frac{\delta}{\delta j_\mu(x)}\right] W[j,j^\ast,j_\mu]
=0.
\end{equation}
   We note that Eqs.\ (\ref{2:5:gfwi:winvariance}) and
(\ref{2:5:gfwi:mewi}) are equivalent.

   As an illustration let us re-derive the Ward identity of
Eq.\ (\ref{2:5:gfwi:gmuwt}) using Eq.\ (\ref{2:5:gfwi:mewi}).
   For that purpose we start from Eq.\ (\ref{2:5:gfwi:gmuxyz1}),
\begin{displaymath}
\partial_\mu^y G^\mu(x,y,z)=
(-i)^3 \partial_\mu^y
\left.\frac{\delta^3 W}{\delta j^\ast(x)\delta j_\mu (y)\delta j(z)},
\right|_{j=0,j^\ast=0,j_\mu=0},
\end{displaymath}
apply Eq.\ (\ref{2:5:gfwi:mewi}),
\begin{displaymath}
=(-i)^2\left\{
\frac{\delta^2}{\delta j^\ast(x)\delta j(z)}\left[
j^\ast(y)\frac{\delta}{\delta j^\ast(y)}-j(y)\frac{\delta}{\delta j(y)}
\right] W\right\}_{j=0,j^\ast=0,j_\mu=0},
\end{displaymath}
make use of
$\delta j^\ast(y)/\delta j^\ast(x)=
\delta^4(y-x)$ and $\delta j(y)/\delta j(z)=\delta^4(y-z)$
for the functional derivatives,
\begin{displaymath}
=(-i)^2\left\{\delta^4(x-y)\frac{\delta^2 W}{\delta j^\ast(y)
\delta j(z)}
-\delta^4(z-y)\frac{\delta^2 W}{\delta j^\ast(x)
\delta j(y)}\right\}_{j=0,j^\ast=0,j_\mu=0},
\end{displaymath}
and, finally, use the definition of Eq.\ (\ref{2:5:gfwi:genfunc1}),
\begin{displaymath}
\partial_\mu^y G^\mu(x,y,z)
=[\delta^4(x-y)-\delta^4(z-y)]\langle 0| T\left[\Phi(x)\Phi^\dagger(z)
\right]|0\rangle
\end{displaymath}
   which is the same as Eq.\ (\ref{2:5:gfwi:gmuwt}).
   In principle, any Ward identity can be obtained by taking appropriate
higher functional derivatives of $W$ and then using Eq.\ (\ref{2:5:gfwi:mewi}).

\section{Green Functions and Chiral Ward Identities}
\label{sec_gfcwi}

\subsection{Chiral Green Functions}
\label{subsec_cgf}
   For conserved currents, the spatial integrals of the
charge densities are time independent, i.e., in a quantized theory
the corresponding charge operators commute with the Hamilton operator.
   These operators are generators of infinitesimal transformations
on the Hilbert space of the theory.
   The mass eigenstates
should organize themselves in
degenerate multiplets with dimensionalities corresponding to
irreducible representations of the Lie group in
question.\footnote{Here we assume that the dynamical system described
by the Hamiltonian does not lead to a spontaneous symmetry breakdown.
   We will come back to this point later.}
   Which irreducible representations ultimately appear, and what
the actual energy eigenvalues are, is determined by the dynamics of
the Hamiltonian.
   For example, SU(2) isospin symmetry of the strong interactions reflects
itself in degenerate SU(2) multiplets such as the nucleon doublet,
the pion triplet, and so on.
   Ultimately, the actual masses of the nucleon and the pion should
follow from QCD.

   It is also well-known that symmetries imply relations between $S$-matrix
elements. For example, applying the Wigner-Eckart theorem to pion-nucleon
scattering, assuming the strong-interaction Hamiltonian to be an isoscalar,
it is sufficient to consider two isospin amplitudes describing transitions
between states of total isospin $I=1/2$ or $I=3/2$.
   All the dynamical information is contained in these isospin amplitudes and
the results for physical processes can be expressed in terms of
these amplitudes together with geometrical coefficients, namely,
the Clebsch-Gordan coefficients.

   In quantum field theory, the objects of interest are the
Green functions which are vacuum expectation values of time-ordered
products.\footnote{Later on, we will also refer to matrix elements
of time-ordered products between states other than the vacuum as
Green functions.}
   Pictorially, these Green functions can be understood as vertices
and are related to physical scattering amplitudes through the
Lehmann-Symanzik-Zimmermann (LSZ) reduction formalism.
   Symmetries provide strong constraints not only for scattering amplitudes,
i.e.~their transformation behavior, but, more generally speaking, also for
 Green functions and, in particular,
{\em among} Green functions.
   The famous example in this context is, of course, the Ward identity
of QED associated with U(1) gauge invariance
\begin{equation}
\label{2:6:qedwardidentity}
\Gamma^\mu(p,p)=-\frac{\partial}{\partial p_\mu}\Sigma(p),
\end{equation}
which relates the electromagnetic vertex of an electron at zero
momentum transfer, $\Gamma^\mu(p,p)$,
to the electron self energy, $\Sigma(p)$.

   Such symmetry relations can be extended to non-vanishing momentum transfer
and also to more complicated groups and are referred to as
Ward-Fradkin-Takahashi identities
(or Ward identities for short).
   Furthermore, even if a symmetry is broken, i.e., the infinitesimal
generators are time dependent, conditions related to the symmetry
breaking terms can still be obtained using equal-time commutation
relations.

   At first, we are interested in
time-ordered products of color-neutral, Hermitian quadratic forms involving
the light quark fields evaluated between the vacuum of QCD.
   Using the LSZ reduction formalism
such Green functions can be related to physical processes
involving mesons as well as their interactions with the
electroweak gauge fields of the Standard Model.
   The interpretation depends on the transformation properties
and quantum numbers of the quadratic forms, determining for which mesons
they may serve as an interpolating field.
  In addition to the vector and axial-vector currents of
Eqs.\ (\ref{2:4:v}), (\ref{2:4:a}), and (\ref{2:4:sv})
we want to investigate scalar and pseudoscalar
densities,\footnote{The singlet axial-vector current involves an
anomaly such that the Green functions involving this current operator
are related to Green functions containing the contraction of the
gluon field-strength tensor with its dual.}
\begin{equation}
\label{2:6:quadraticforms}
S_a(x)=\bar{q}(x)\lambda_a q(x),\quad
P_a(x)=i\bar{q}(x)\gamma_5 \lambda_a  q(x),\quad
a=0,\cdots, 8,
\end{equation}
   which enter, for example, in Eqs.\ (\ref{2:4:dsva}) as the divergences of
the vector and axial-vector currents for nonzero quark masses.
   Whenever it is more convenient, we will also use
\begin{equation}
\label{2:6:SP}
S(x)=\bar{q}(x) q(x),\quad
P(x)=i\bar{q}(x)\gamma_5 q(x),
\end{equation}
instead of $S_0$ and $P_0$.

   One may also consider similar time-ordered products
evaluated between a single nucleon in the initial and final states
in addition to the vacuum Green functions.
   This allows one to discuss properties of the nucleon as well as
dynamical processes involving a single nucleon.

   Generally speaking, a chiral Ward identity relates the divergence
of a Green function containing at least one factor of $V^{\mu,a}$ or
$A^{\mu,a}$ [see Eqs.\ (\ref{2:4:v}) and (\ref{2:4:a})] to some linear
combination of other Green functions.
   The terminology {\em chiral} refers to the underlying $\mbox{SU(3)}_L\times
\mbox{SU(3)}_R$ group.
   To make this statement more precise, let us consider as a simple example
the two-point Green function involving an axial-vector current and a
pseudoscalar density,\footnote{The time ordering of $n$ points
$x_1,\cdots,x_n$ gives rise to $n$! distinct orderings, each
involving products of $n-1$ theta functions.}
\begin{eqnarray}
\label{2:6:gfaav}
G^{\mu,ab}_{AP}(x,y)&=&\langle 0| T[A^\mu_a(x) P_b(y)]|0\rangle\nonumber\\
&=&\Theta(x_0-y_0)\langle 0|A^\mu_a(x) P_b(y)|0\rangle
+\Theta(y_0-x_0)\langle 0|P_b(y) A^\mu_a(x)|0\rangle,\nonumber\\
\end{eqnarray}
   and evaluate the divergence
\begin{eqnarray*}
\lefteqn{\partial_\mu^x G^{\mu,ab}_{AP}(x,y)}\\
&=&
 \partial_\mu^x [\Theta(x_0-y_0)\langle 0| A^\mu_a(x) P_b(y)|0\rangle
+\Theta(y_0-x_0)\langle 0|  P_b(y)A^\mu_a(x)|0\rangle ]\\
&=&\delta(x_0-y_0)\langle 0| A_0^a(x) P_b(y)|0\rangle
-\delta(x_0-y_0)\langle 0|  P_b(y)A_0^a (x)|0\rangle\\
&&
+\Theta(x_0-y_0)\langle 0|\partial_\mu^x A^\mu_a(x) P_b(y)|0\rangle
+\Theta(y_0-x_0)\langle 0| P_b(y)\partial_\mu^x A^\mu_a(x)|0\rangle\\
&=&\delta(x_0-y_0)\langle 0|[A^a_0(x),P_b(y)]|0\rangle
+\langle 0|T[\partial_\mu^x A^\mu_a(x) P_b(y)]|0\rangle,
\end{eqnarray*}
   where we made use of $\partial_\mu^x \Theta(x_0-y_0)=\delta(x_0-y_0)
g_{0\mu}=-\partial_\mu^x \Theta(y_0-x_0)$.
   This simple example already shows the main features of (chiral) Ward
identities.
   From the differentiation of the theta functions
one obtains equal-time commutators between a charge density and the
remaining quadratic forms.
   The results of such commutators are a reflection of the underlying symmetry,
as will be shown below.
   As a second term, one obtains the divergence of the current operator in
question.
   If the symmetry is perfect, such terms vanish identically.
   For example, this is always true for the electromagnetic case with its
U(1) symmetry.
   If the symmetry is only approximate, an additional term involving the
symmetry breaking appears.
   For a soft breaking such a divergence can be treated as a perturbation.

   Via induction, the generalization of the above simple example to an
$(n+1)$-point Green function is symbolically of the form
\begin{eqnarray}
\label{2:6:gendmug}
\lefteqn{\partial_\mu^x \langle 0|T\{J^\mu(x) A_1(x_1)\cdots A_n(x_n)
\}|0\rangle=}\nonumber\\
&&\langle 0|T\{[\partial_\mu^x J^\mu(x)]
A_1(x_1)\cdots A_n(x_n)\}|0\rangle\nonumber\\
&&+\delta(x^0-x_1^0)\langle 0|T\{[J_0(x),A_1(x_1)] A_2(x_2)\cdots A_n(x_n)\}|
0\rangle\nonumber\\
&&+\delta(x^0-x_2^0)\langle 0|T\{A_1(x_1)[J_0(x),A_2(x_2)]
\cdots A_n(x_n)\}|0\rangle\nonumber\\
&&+\cdots+\delta(x^0-x_n^0)
\langle 0|T\{A_1(x_1)\cdots [J_0(x),A_n(x_n)]\}|0\rangle,
\end{eqnarray}
where $J^\mu$ stands generically for any of the Noether currents.

\subsection{The Algebra of Currents *}
\label{subsec_ac}
   In the above example, we have seen that chiral Ward identities
depend on the equal-time commutation relations of the {\em charge densities}
of the symmetry currents with the relevant quadratic quark forms.
   Unfortunately, a naive application of Eq.\ (\ref{2:4:fkf}) may lead
to erroneous results.
   Let us illustrate this by means of a simplified example,
the equal-time commutator of the time and space components
of the ordinary electromagnetic current in QED.
   A naive use of the canonical commutation relations leads to
\begin{eqnarray}
\label{2:6:schwinger}
[J_0(t,\vec{x}), J_i(t,\vec{y})]&=&
[\Psi^\dagger(t,\vec{x})\Psi(t,\vec{x}),\Psi^\dagger(t,\vec{y})\gamma_0\gamma_i
\Psi(t,\vec{y})]\nonumber\\
&=&\delta^3(\vec{x}-\vec{y})\Psi^\dagger(t,\vec{x})[1,\gamma_0\gamma_i]
\Psi(t,\vec{x})=0,
\end{eqnarray}
   where we made use of the delta function to evaluate the fields at
$\vec{x}=\vec{y}$.
   It was noticed a long time ago by Schwinger that this result
cannot be true \cite{Schwinger:xd:5:2}.
   In order to see this, consider the commutator
\begin{displaymath}
[J_0(t,\vec{x}),\vec{\nabla}_y\cdot \vec{J}(t,\vec{y})]=
-[J_0(t,\vec{x}),\partial_t J_0(t,\vec{y})],
\end{displaymath}
   where we made use of current conservation, $\partial_\mu J^\mu=0$.
   If Eq.\ (\ref{2:6:schwinger}) were true, one would necessarily also have
\begin{displaymath}
0=[J_0(t,\vec{x}),\partial_t J_0(t,\vec{y})],
\end{displaymath}
which we evaluate for $\vec{x}=\vec{y}$ between the ground state,
\begin{eqnarray*}
0&=&\langle 0|[J_0(t,\vec{x}),\partial_t J_0(t,\vec{x})]|0\rangle\\
&=&\sum_n \Big(\langle 0|J_0(t,\vec{x})|n\rangle\langle n|
\partial_t J_0(t,\vec{x})|0\rangle
-\langle 0|\partial_t J_0(t,\vec{x})|n\rangle\langle n |
J_0(t,\vec{x})|0\rangle\Big)\\
&=&2i\sum_n(E_n-E_0)|\langle 0|J_0(t,\vec{x})|n\rangle|^2.
\end{eqnarray*}
   Here, we inserted a complete set of states and made use of
\begin{displaymath}
\partial_t J_0(t,\vec{x})=i[H,J_0(t,\vec{x})].
\end{displaymath}
   Since every individual term in the sum is non-negative, one would
need $\langle 0|J_0(t,\vec{x})|n\rangle=0$ for any intermediate
state which is obviously unphysical.
   The solution is that the starting point,  Eq.\ (\ref{2:6:schwinger}),
is not true.
   The corrected version of Eq.\ (\ref{2:6:schwinger}) picks up an
additional, so-called Schwinger term containing a derivative of the
delta function.

   Quite generally, by evaluating commutation relations with the component
$\Theta^{00}$ of the energy-momentum tensor one can show that the
equal-time commutation relation between a charge density and a
current density can be determined up to one derivative of the
$\delta$ function \cite{Jackiw:1972:5:2},
\begin{equation}
\label{2:6:j0jigen}
[J_0^a(0,\vec{x}),J_i^b(0,\vec{y})]=iC_{abc} J_i^c(0,\vec{x})\delta^3
(\vec{x}-\vec{y})+S_{ij}^{ab}(0,\vec{y})\partial^j\delta^3(\vec{x}-\vec{y}),
\end{equation}
where the Schwinger term possesses the symmetry
\begin{displaymath}
S_{ij}^{ab}(0,\vec{y})=S_{ji}^{ba}(0,\vec{y}),
\end{displaymath}
and $C_{abc}$ denote the structure constants of the group in question.

    However, in our above derivation of the chiral Ward identity,
we also made use of the {\em naive} time-ordered product ($T$) as opposed to
the {\em covariant} one ($T^\ast$) which, typically, differ by another
non-covariant term which is called a seagull.
   Feynman's conjecture \cite{Jackiw:1972:5:2}
states that there is a cancelation between Schwinger terms
and seagull terms such that a Ward identity obtained by
using the naive T product and by simultaneously omitting Schwinger terms
ultimately yields the correct result to be satisfied by the Green
function (involving the covariant $T^\ast$ product).
   Although this will not be true in general, a sufficient condition for it
to happen is that the time component algebra of the full theory remains
the same as the one derived canonically and does not posses a Schwinger
term.

   Keeping the above discussion in mind, the complete list of equal-time
commutation relations, omitting Schwinger terms, reads
\begin{eqnarray}
\label{2:6:letcr}
[V^a_0(t,\vec{x}),V^\mu_b(t,\vec{y})]
&=&\delta^3(\vec{x}-\vec{y})if_{abc} V^\mu_c(t,\vec{x}),\nonumber\\
{[}V^a_0(t,\vec{x}),V^\mu(t,\vec{y})]&=&0,\nonumber\\
{[}V^a_0(t,\vec{x}),A^\mu_b(t,\vec{y})]
&=&\delta^3(\vec{x}-\vec{y})if_{abc} A^\mu_c(t,\vec{x}),\nonumber\\
{[}V^a_0(t,\vec{x}),S_b(t,\vec{y})]
&=&\delta^3(\vec{x}-\vec{y})if_{abc} S_c(t,\vec{x}),\quad b=1,\cdots,8,
\nonumber\\
{[}V^a_0(t,\vec{x}),S_0(t,\vec{y})]&=&0,\nonumber\\
{[}V^a_0(t,\vec{x}),P_b(t,\vec{y})]
&=&\delta^3(\vec{x}-\vec{y})if_{abc} P_c(t,\vec{x}),\quad b=1,\cdots,8,
\nonumber\\
{[}V^a_0(t,\vec{x}),P_0(t,\vec{y})]&=&0,\nonumber\\
{[}A^a_0(t,\vec{x}),V^\mu_b(t,\vec{y})]
&=&\delta^3(\vec{x}-\vec{y})if_{abc} A^\mu_c(t,\vec{x}),\nonumber\\
{[}A^a_0(t,\vec{x}),V^\mu(t,\vec{y})]&=&0,\nonumber\\
{[}A^a_0(t,\vec{x}),A^\mu_b(t,\vec{y})]
&=&\delta^3(\vec{x}-\vec{y})if_{abc} V^\mu_c(t,\vec{x}),\nonumber\\
{[}A^a_0(t,\vec{x}),S_b(t,\vec{y})]
&=&i\delta^3(\vec{x}-\vec{y})\left[\sqrt{\frac{2}{3}}\delta_{ab}P_0(t,\vec{x})
+d_{abc}P_c(t,\vec{x})\right],\nonumber\\
&&b=1,\cdots,8,
\nonumber\\
{[}A^a_0(t,\vec{x}),S_0(t,\vec{y})]&=&
i\delta^3(\vec{x}-\vec{y})
\sqrt{\frac{2}{3}}P_a(t,\vec{x}),\nonumber\\
{[}A^a_0(t,\vec{x}),P_b(t,\vec{y})]
&=&-i\delta^3(\vec{x}-\vec{y})\left[\sqrt{\frac{2}{3}}\delta_{ab}S_0(t,\vec{x})
+d_{abc}S_c(t,\vec{x})\right],\nonumber\\
&&b=1,\cdots,8,
\nonumber\\
{[}A^a_0(t,\vec{x}),P_0(t,\vec{y})]&=&
-i\delta^3(\vec{x}-\vec{y})
\sqrt{\frac{2}{3}}S_a(t,\vec{x}).
\end{eqnarray}
   For example,
\begin{eqnarray*}
\lefteqn{[V^0_a(t,\vec{x}),V^\mu_b(t,\vec{y})]}\\
&=&
[q^\dagger(t,\vec{x}) 1 \frac{\lambda_a}{2}q(t,\vec{x}),
q^\dagger(t,\vec{y})\gamma_0\gamma^\mu \frac{\lambda_b}{2}q(t,\vec{y})]\\
&=&\delta^3(\vec{x}-\vec{y})\left[
q^\dagger(t,\vec{x})\gamma_0\gamma^\mu \frac{\lambda_a}{2}
\frac{\lambda_b}{2}q(t,\vec{y})
-q^\dagger(t,\vec{y})\gamma_0 \gamma^\mu
\frac{\lambda_b}{2}\frac{\lambda_a}{2}
q(t,\vec{x})\right]\\
&=&\delta^3(\vec{x}-\vec{y})i f_{abc} V^\mu_c(t,\vec{x}).
\end{eqnarray*}
   The remaining expressions are obtained analogously.

\subsection{QCD in the Presence of External Fields and the Generating
Functional}
\label{subsec_qcdpefgf}
   Here, we want to consider the consequences of Eqs.\ (\ref{2:6:letcr})
for the Green functions of QCD (in particular, at low energies).
   In principle, using the techniques of the last section, for each
Green function one can {\em explicitly} work out the chiral Ward identity
which, however, becomes more and more tedious as the number $n$ of quark
quadratic forms increases.
   However, there exists an elegant way of formally combining all Green
functions in a generating functional.
   The (infinite) set of {\em all} chiral Ward identities is encoded
as an invariance property of that functional.
    To see this, one has to consider a coupling to external c-number fields
such that through functional methods one can, in principle, obtain all
Green functions from a generating functional.
   The rationale behind this approach is that, in the absence of anomalies,
the Ward identities obeyed by the Green functions are equivalent
to an invariance of the generating functional under a {\em local}
transformation of the external fields \cite{Leutwyler:1993iq:5}.
   The use of local transformations allows one to also consider divergences
of Green functions.
   For an illustration of this statement, the reader is referred to
Section \ref{sec_gfwi}.

   Following the procedure of Gasser and Leutwyler
\cite{Gasser:1983yg:5,Gasser:1984gg:5}, we introduce into the
Lagrangian of QCD the couplings of the nine vector currents and
the eight axial-vector currents as well as the scalar and
pseudoscalar quark densities to external c-number fields $v^\mu
(x)$, $v^\mu_{(s)}$, $a^\mu (x)$, $s(x)$, and $p(x)$,
\begin{equation}
\label{2:6:lqcds}
{\cal L}={\cal L}^0_{\rm QCD}+{\cal L}_{\rm ext}
={\cal L}^0_{\rm QCD}+\bar{q}\gamma_\mu (v^\mu +\frac{1}{3}v^\mu_{(s)}
+\gamma_5 a^\mu )q
-\bar{q}(s-i\gamma_5 p)q.
\end{equation}
     The external fields are color-neutral, Hermitian $3\times 3$ matrices,
where the matrix character, with respect to the (suppressed) flavor indices
$u$, $d$, and $s$ of the quark fields, is\footnote{
We omit the coupling to the
singlet axial-vector current which has an anomaly,
but include a singlet vector current $v^\mu_{(s)}$ which is of some physical
relevance in the two-flavor sector.}
\begin{equation}
\label{2:6:mch}
v^\mu=\sum_{a=1}^8\frac{\lambda_a}{2}v_a^\mu,\quad
a^\mu=\sum_{a=1}^8\frac{\lambda_a}{2}a_a^\mu,\quad
s=\sum_{a=0}^8 \lambda_a s_a,\quad
p=\sum_{a=0}^8\lambda_a p_a.
\end{equation}
The ordinary three flavor QCD Lagrangian is recovered by setting
$v^\mu=v^\mu_{(s)}=a^\mu=p=0$ and $s=\mbox{diag}(m_u,m_d,m_s)$ in
Eq.\ (\ref{2:6:lqcds}).

   If one defines the generating functional\footnote{Many
books on Quantum Field Theory reserve the symbol $Z[v,a,s,p]$ for
the generating functional of all Green functions as opposed to the
argument of the exponential which denotes the generating
functional of connected Green functions.}
\begin{eqnarray}
\label{2:6:genfun}
\lefteqn{\exp[i Z(v,a,s,p)] =\langle
0|T\exp\left[i\int d^4 x {\cal L}_{\rm ext}(x)
\right]|0\rangle}\nonumber\\
&=&\langle 0|T\exp\left( i\int d^4 x
\bar{q}(x)\{\gamma_\mu[v^\mu(x)+\gamma_5 a^\mu(x)]
-s(x)+i\gamma_5 p(x)\}q(x)\right)|0\rangle,\nonumber\\
\end{eqnarray}
then any Green function consisting of the time-ordered product of
color-neutral, Hermitian quadratic forms can be obtained from Eq.\
(\ref{2:6:genfun}) through a functional derivative with respect to
the external fields.
   The quark fields are operators in the Heisenberg picture and
have to satisfy the equation of motion and the canonical
anti-commutation relations.
   The actual value of the generating functional for a given configuration
of external fields $v$, $a$, $s$, and $p$ reflects the dynamics generated
by the QCD Lagrangian.
   The generating functional is related to the vacuum-to-vacuum transition
amplitude in the presence of external fields,
\begin{equation}
\label{2:6:genfunvv} \exp[i Z(v,a,s,p)]= \langle 0 ;{\rm out}|0
;{\rm in}\rangle_{v,a,s,p}.
\end{equation}

   For example,\footnote{In order to obtain Green functions from the
generating functional the simple rule
$$\frac{\delta f(x)}{\delta f(y)}=\delta(x-y)$$
is extremely useful. Furthermore, the functional derivative satisfies
properties similar to the ordinary differentiation, namely linearity,
the product and chain rules.}
the $\bar{u} u$ component of the scalar quark condensate in the chiral limit,
$\langle 0| \bar{u}u|0\rangle_0$, is given by
\begin{eqnarray}
\label{2:6:sqc}
\lefteqn{\langle 0|\bar{u}(x) u(x)|0\rangle_0 =}\nonumber\\
&&\left.\frac{i}{2}\left[\sqrt{\frac{2}{3}}\frac{\delta}{\delta s_0(x)}
+\frac{\delta}{\delta s_3(x)}
+\frac{1}{\sqrt{3}}
\frac{\delta}{\delta s_8(x)}\right]
\exp(iZ [v,a,s,p])\right|_{v=a=s=p=0},\nonumber\\
\end{eqnarray}
where we made use of Eq.\ (\ref{2:1:matrixa}).
   Note that both the quark field operators and the ground state are considered
in the chiral limit, which is denoted by the subscript 0.

    As another example, let us consider the two-point function of the
axial-vector currents of Eq.\ (\ref{2:4:a}) of the ``real world,''
i.e., for $s=\mbox{diag}(m_u,m_d,m_s)$, and the ``true vacuum'' $|0\rangle$,
\begin{eqnarray}
\label{2:4:tpfavc}
\lefteqn{\langle 0|T[A_\mu^a(x) A_\nu^b(0)]|0\rangle =}\nonumber\\
&&\left.
(-i)^2 \frac{\delta^2}{\delta a^\mu_a(x)\delta a^\nu_b(0)}
\exp(iZ[v,a,s,p])\right|_{v=a=p=0,s=\mbox{diag}(m_u,m_d,m_s)}.\nonumber\\
\end{eqnarray}

   Requiring the total Lagrangian of Eq.\ (\ref{2:6:lqcds}) to be Hermitian
and invariant under $P$, $C$, and $T$ leads to constraints on the
transformation behavior of the external fields.
   In fact, it is sufficient to consider $P$ and $C$, only, because
$T$ is then automatically incorporated owing to the $CPT$ theorem.

   Under parity, the quark fields transform as
\begin{equation}
\label{2:6:qtrafop}
q_f(t,\vec{x})\stackrel{\mbox{$P$}}{\mapsto}\gamma^0 q_f(t,-\vec{x}),
\end{equation}
   and the requirement of parity conservation,
\begin{equation}
\label{2:6:parinv}
{\cal L}(t,\vec{x}) \stackrel{\mbox{$P$}}{\mapsto} {\cal L}(t,-\vec{x}),
\end{equation}
leads, using the results of Table \ref{2:6:parity}, to the following
constraints for the external fields,
\begin{equation}
\label{2:6:eftrafop}
v^\mu\stackrel{\mbox{$P$}}{\mapsto}v_\mu,\quad
v^\mu_{(s)}\stackrel{\mbox{$P$}}{\mapsto}v_\mu^{(s)},\quad
a^\mu\stackrel{\mbox{$P$}}{\mapsto}-a_\mu,\quad
s\stackrel{\mbox{$P$}}{\mapsto}s,\quad
p\stackrel{\mbox{$P$}}{\mapsto}-p.
\end{equation}
   In Eq.\ (\ref{2:6:eftrafop}) it is understood that the arguments change
from $(t,\vec{x})$ to $(t,-\vec{x})$.
   Let us verify Eq.\ (\ref{2:6:eftrafop}) by means of an example:
\begin{eqnarray*}
\bar{q}(t,\vec{x})\gamma^\mu v_\mu(t,\vec{x}) q(t,\vec{x})
&\stackrel{\mbox{$P$}}{\mapsto}&
\bar{q}(t,-\vec{x})\gamma^0\gamma^\mu \tilde{v}_\mu(t,-\vec{x})
\gamma^0 q(t,-\vec{x})=\cdots,
\end{eqnarray*}
where the tilde denotes the transformed external field. With the
help of $\gamma^0\gamma^\mu\gamma^0=\gamma_\mu$ we find
$$
\cdots
=\bar{q}(t,-\vec{x})\gamma_\mu \tilde{v}_\mu(t,-\vec{x})
q(t,-\vec{x})
\stackrel{\mbox{!}}{=}
\bar{q}(t,-\vec{x})\gamma_\mu v^\mu(t,-\vec{x})
q(t,-\vec{x}).
$$
   We thus obtain
$$v_\mu(t,\vec{x})\stackrel{\mbox{$P$}}{\mapsto} v^\mu(t,-\vec{x}).$$

\begin{table}
\begin{center}
\begin{tabular}{|c|c|c|c|c|c|}
\hline
$\Gamma$&
$1$&
$\gamma^\mu$&
$\sigma^{\mu\nu}$&
$\gamma_5$&
$\gamma^\mu\gamma_5$\\
\hline
$\gamma_0 \Gamma \gamma_0$&
$1$&
$\gamma_\mu$&
$\sigma_{\mu\nu}$&
$-\gamma_5$&
$-\gamma_\mu\gamma_5$
\\
\hline
\end{tabular}
\end{center}
\caption{\label{2:6:parity} Transformation properties of the Dirac
matrices $\Gamma$ under parity.}
\end{table}

   Similarly, under charge conjugation the quark fields transform as
\begin{equation}
\label{2:6:qtrafc}
q_{\alpha,f}\stackrel{\mbox{$C$}}{\mapsto}C_{\alpha\beta}\bar{q}_{\beta,f},
\quad
\bar{q}_{\alpha,f}\stackrel{\mbox{$C$}}{\mapsto}
-q_{\beta,f}C^{-1}_{\beta\alpha},
\end{equation}
   where the subscripts $\alpha$ and $\beta$ are Dirac
spinor indices,
\begin{displaymath}
C=i\gamma^2\gamma^0
=\left(\begin{array}{cccc}0&0&0&-1\\0&0&1&0\\
0&-1&0&0\\
1&0&0&0
\end{array}\right)=-C^{-1}=-C^\dagger=-C^T
\end{displaymath}
is the usual charge conjugation matrix and $f$ refers to flavor.
Using
\begin{eqnarray*}
\bar{q}\Gamma F q&=&\bar{q}_{\alpha,f}\Gamma_{\alpha\beta}F_{ff'}q_{\beta,f'}\\
&\stackrel{\mbox{$C$}}{\mapsto}&-q_{\gamma,f}
C^{-1}_{\gamma\alpha}\Gamma_{\alpha\beta}
F_{ff'} C_{\beta\delta} \bar{q}_{\delta,f'}\\
&\stackrel{\mbox{Fermi statistics}}{=}
&\bar{q}_{\delta,f'}\underbrace{F_{ff'}}_{\mbox{$F^T_{f'f}$}}
\underbrace{C^{-1}_{\gamma\alpha}\Gamma_{\alpha\beta}C_{\beta\delta}}_{
\mbox{$(C^{-1}\Gamma C)^T_{\delta\gamma}$}}
q_{\gamma,f}
\\
&=&\bar{q} F^T \underbrace{(C^{-1}\Gamma C)^T}_{\mbox{$
C^T\Gamma^T {C^{-1}}^T$}} q\\
&=& -\bar{q} C \Gamma^T C F^T q
\end{eqnarray*}
in combination with Table
\ref{2:6:chargeconjugation} it is straightforward to show that invariance
of ${\cal L}_{\rm ext}$ under charge conjugation requires the transformation
properties
\begin{equation}
\label{2:6:eftrafoc}
v_\mu\stackrel{C}{\rightarrow}-v_\mu^T,\quad
v_\mu^{(s)}\stackrel{C}{\rightarrow}-v_\mu^{(s)T},\quad
a_\mu\stackrel{C}{\rightarrow}a_\mu^T,\quad
s,p\stackrel{C}{\rightarrow}s^T,p^T,
\end{equation}
where the transposition refers to the flavor space.

\begin{table}
\begin{center}
\begin{tabular}{|c|c|c|c|c|c|}
\hline
$\Gamma$&$1$&$\gamma^\mu$&$\sigma^{\mu\nu}$&$\gamma_5$&$\gamma^\mu\gamma_5$\\
\hline
$-C\Gamma^TC
$&$1$&$-\gamma^\mu$&$-\sigma^{\mu\nu}$&$\gamma_5$&$\gamma^\mu\gamma_5$
\\
\hline
\end{tabular}
\end{center}
\caption{\label{2:6:chargeconjugation}
Transformation properties of the Dirac matrices $\Gamma$
under charge conjugation.}
\end{table}

   Finally, we need to discuss the requirements to be met by the external
fields under local $\mbox{SU(3)}_L\times\mbox{SU(3)}_R\times\mbox{U}(1)_V$
transformations.
   In a first step, we write Eq.\ (\ref{2:6:lqcds}) in terms of the
left- and right-handed quark fields.
\begin{exercise}
\label{exercise_rewrite_external_lagrangian}
\renewcommand{\labelenumi}{(\alph{enumi})}
\rm We first define
\begin{equation}
\label{2:6:vrlarl} r_\mu=v_\mu+a_\mu,\quad l_\mu=v_\mu-a_\mu.
\end{equation}
\begin{enumerate}
\item
   Make use of the projection operators $P_L$ and $P_R$ and verify
\begin{displaymath}
\bar{q}\gamma^\mu(v_\mu+\frac{1}{3}v_\mu^{(s)} +\gamma_5 a_\mu)q
=\bar{q}_R\gamma^\mu \left(r_\mu +\frac{1}{3}v_\mu^{(s)}\right)q_R
+\bar{q}_L\gamma^\mu \left(l_\mu+\frac{1}{3}v_\mu^{(s)}\right)
q_L.
\end{displaymath}
\item
Also verify
\begin{displaymath}
\bar{q}(s-i\gamma_5 p)q= \bar{q}_L(s-ip)q_R+\bar{q}_R(s+ip)q_L.
\end{displaymath}
\end{enumerate}
\end{exercise}

   We obtain for the Lagrangian of Eq.\ (\ref{2:6:lqcds})
\begin{eqnarray}
\label{2:6:lqcdsn}
{\cal L}&=&{\cal L}_{\rm QCD}^0
+\bar{q}_L\gamma^\mu\left(l_\mu+\frac{1}{3}v^{(s)}_\mu\right)q_L
+\bar{q}_R\gamma^\mu\left(r_\mu+\frac{1}{3}v^{(s)}_\mu\right)q_R\nonumber\\
&&-\bar{q}_R(s+ip)q_L-\bar{q}_L(s-ip)q_R.
\end{eqnarray}
   Equation (\ref{2:6:lqcdsn}) remains invariant under {\em local}
transformations
\begin{eqnarray}
\label{2:6:qrl}
q_R&\mapsto&\exp\left(-i\frac{\Theta(x)}{3}\right) V_R(x) q_R,\nonumber\\
q_L&\mapsto&\exp\left(-i\frac{\Theta(x)}{3}\right) V_L(x) q_L,
\end{eqnarray}
where $V_R(x)$ and $V_L(x)$ are independent space-time-dependent SU(3)
matrices, provided the external fields are subject
to the transformations
\begin{eqnarray}
\label{2:6:sg}
r_\mu&\mapsto& V_R r_\mu V_R^{\dagger}
+iV_R\partial_\mu V_R^{\dagger},\nonumber\\
l_\mu&\mapsto& V_L l_\mu V_L^{\dagger}
+iV_L\partial_\mu V_L^{\dagger},
\nonumber\\
v_\mu^{(s)}&\mapsto&v_\mu^{(s)}-\partial_\mu\Theta,\nonumber\\
s+ip&\mapsto& V_R(s+ip)V_L^{\dagger},\nonumber\\
s-ip&\mapsto& V_L(s-ip)V_R^{\dagger}.
\end{eqnarray}
   The derivative terms in Eq.\ (\ref{2:6:sg}) serve the same purpose as
in the construction of gauge theories, i.e., they cancel analogous
terms originating from the kinetic part of the quark Lagrangian.

   There is another, yet, more practical aspect of the local invariance,
namely: such a procedure allows one to also discuss a coupling to external
gauge fields in the transition to the effective theory to be discussed later.
   For example, a coupling of the electromagnetic field to point-like
fundamental particles results from
gauging a U(1) symmetry.
  Here, the corresponding U(1) group is to be understood
as a subgroup of a local $\mbox{SU(3)}_L\times\mbox{SU(3)}_R$.
   Another example deals with the interaction of the light quarks
with the charged and neutral gauge bosons of the weak interactions.

   Let us consider both examples explicitly. The coupling of quarks
to an external electromagnetic field ${\cal A}_\mu$ is given by
\begin{equation}
\label{2:6:rla}
r_\mu=l_\mu=-e Q {\cal A}_\mu,
\end{equation}
where $Q=\mbox{diag}(2/3,-1/3,-1/3)$ is the quark charge matrix:
\begin{eqnarray*}
{\cal L}_{\rm ext}&=&-e {\cal A}_\mu(\bar{q}_L Q\gamma^\mu q_L
+\bar{q}_R Q \gamma^\mu q_R)\\
&=&-e {\cal A}_\mu \bar{q}Q\gamma^\mu q\\
&=&-e {\cal A}_\mu\left(\frac{2}{3}\bar{u}\gamma^\mu u
-\frac{1}{3} \bar{d}\gamma^\mu d -\frac{1}{3}\bar{s}\gamma^\mu s\right)\\
&=&-e {\cal A}_\mu J^\mu.
\end{eqnarray*}
   On the other hand, if one considers only the SU(2) version of ChPT one
has to insert for the external fields
\begin{equation}
\label{2:6:rlasu2}
r_\mu=l_\mu=-e\frac{\tau_3}{2}{\cal A}_\mu,\quad
v_\mu^{(s)}=-\frac{e}{2}{\cal A}_\mu.
\end{equation}

   In the description of semileptonic interactions such as
$\pi^-\to \mu^-\bar{\nu}_\mu$,  $\pi^-\to\pi^0e^-\bar{\nu}_e$, or
neutron decay $n\to p e^-\bar{\nu}_e$ one needs the interaction of quarks with
the massive charged weak bosons
${\cal W}^\pm_\mu=({\cal W}_{1\mu}\mp i {\cal W}_{2\mu})/\sqrt{2}$,
\begin{equation}
\label{2:6:rlw} r_\mu=0,\quad l_\mu=-\frac{g}{\sqrt{2}} ({\cal W}^+_\mu T_+ +
\mbox{H.c.}),
\end{equation}
where H.c.~refers to the Hermitian conjugate and
$$
T_+=\left(\begin{array}{rrr}0&V_{ud}&V_{us}\\0&0&0\\0&0&0\end{array}\right).
$$
   Here, $V_{ij}$ denote the elements of the
Cabibbo-Kobayashi-Maskawa quark-mixing matrix describing the
transformation between the mass eigenstates of QCD and the weak
eigenstates,
$$|V_{ud}|=0.9735\pm 0.0008,\quad
|V_{us}|=0.2196\pm 0.0023.
$$
   At lowest order in perturbation theory, the Fermi constant is related
to the gauge coupling $g$ and the $W$ mass as
$$
G_F=\sqrt{2} \frac{g^2}{8 M^2_W}=1.16639(1)\times 10^{-5}\,\mbox{GeV}^{-2}.
$$
   Making use of
\begin{eqnarray*}
\bar{q}_L\gamma^\mu {\cal W}_\mu^+ T_+ q_L&=&
{\cal W}_\mu^+ (\bar{u}\,\,\bar{d}\,\, \bar{s}) P_R\gamma^\mu
\left(\begin{array}{rrr}0&V_{ud}&V_{us}\\0&0&0\\0&0&0\end{array}
\right)
P_L
\left(\begin{array}{c}u\\ d\\ s\end{array}\right)\\
&=&{\cal W}_\mu^
+(\bar{u}\,\,\bar{d}\,\,\bar{s})\gamma^\mu \frac{1}{2}(1-\gamma_5)
\left(\begin{array}{c}V_{ud} d+ V_{us} s\\0\\0\end{array}
\right)\\
&=&\frac{1}{2}{\cal W}_\mu^+[V_{ud}\bar{u}\gamma^\mu(1-\gamma_5)d
+V_{us}\bar{u}\gamma^\mu(1-\gamma_5)s],
\end{eqnarray*}
   we see that inserting Eq.\ (\ref{2:6:rlw}) into Eq.\ (\ref{2:6:lqcdsn})
leads to the standard charged-current weak interaction in the light
quark sector,
\begin{eqnarray*}
{\cal L}_{\rm ext}&=&-\frac{g}{2\sqrt{2}}\left\{{\cal W}^+_\mu[
V_{ud}\bar{u}\gamma^\mu(1-\gamma_5)d+V_{us}\bar{u}\gamma^\mu(1-\gamma_5)s]
+\mbox{H.c.}\right\}.
\end{eqnarray*}

   The situation is slightly different for the neutral weak interaction.
Here, the SU(3) version requires a coupling to the singlet axial-vector
current which, because of the anomaly of Eq.\ (\ref{2:4:divsa}), we have
dropped from our discussion.
   On the other hand, in the SU(2) version the axial-vector current part
is traceless and we have
\begin{eqnarray}
\label{2:6:rlz}
r_\mu&=&e \tan(\theta_W) \frac{\tau_3}{2} {\cal Z}_\mu,\nonumber\\
l_\mu&=&-\frac{g}{\cos(\theta_W)}\frac{\tau_3}{2} {\cal Z}_\mu+
e \tan(\theta_W) \frac{\tau_3}{2} {\cal Z}_\mu,
\nonumber\\
v_\mu^{(s)}&=&\frac{e\tan(\theta_W)}{2}{\cal Z}_\mu,
\end{eqnarray}
where  $\theta_W$ is the weak angle.
   With these external fields, we obtain the standard weak neutral-current
interaction
\begin{eqnarray*}
{\cal L}_{\rm ext}&=&-\frac{g}{2\cos(\theta_W)}{\cal Z}_\mu\left(
\bar{u}\gamma^\mu\left\{\left[\frac{1}{2}-\frac{4}{3}\sin^2(\theta_W)\right]
-\frac{1}{2}\gamma_5\right\}u\right.\nonumber\\
&&\left.+\bar{d}\gamma^\mu\left\{\left[-\frac{1}{2}
+\frac{2}{3}\sin^2(\theta_W)\right]
+\frac{1}{2}\gamma_5\right\}d\right),
\end{eqnarray*}
where we made use of $e=g\sin(\theta_W)$.

\subsection{PCAC in the Presence of an External Electromagnetic Field *}
\label{subsec_pcacpeef}
   Finally, the technique of coupling the QCD Lagrangian to external fields
also allows us to determine the current divergences for rigid external fields,
i.e., fields which are {\em not} simultaneously transformed.
   For the sake of simplicity we restrict ourselves to the SU(2) sector.
   (The generalization to the SU(3) case is straightforward.)

\begin{exercise}
\label{exercise_divergence_currents}
\rm
   Consider a {\em global} chiral transformation only and assume that
the external fields are {\em not} simultaneously transformed.
   Show that the divergences of the currents
read [see Eq.\ (\ref{2:3:divergenz})]
\begin{eqnarray}
\label{2:6:divv}
\partial_\mu V^\mu_i&=&i\bar{q}\gamma^\mu[\frac{\tau_i}{2},v_\mu]q
+i\bar{q}\gamma^\mu\gamma_5[\frac{\tau_i}{2},a_\mu]q
-i\bar{q}[\frac{\tau_i}{2},s]q-\bar{q}\gamma_5[\frac{\tau_i}{2},p]q,
\nonumber\\
\label{2:6:diva}\\
\partial_\mu A^\mu_i&=&i\bar{q}\gamma^\mu\gamma_5[\frac{\tau_i}{2},v_\mu]q
+i\bar{q}\gamma^\mu[\frac{\tau_i}{2},a_\mu]q
+i\bar{q}\gamma_5\{\frac{\tau_i}{2},s\}q
+\bar{q}\{\frac{\tau_i}{2},p\}q.\nonumber\\
\end{eqnarray}
\end{exercise}

\begin{exercise}
\label{exercise_pcac_em_field}
\rm
   As an example, let us consider the QCD Lagrangian for a finite light quark
mass $m_q$ in combination with a coupling to an external
electromagnetic field ${\cal A}_\mu$ [see Eq.\ (\ref{2:6:rlasu2}),
$a_\mu=0=p$].
   Show that the expressions for the divergence of the vector and
axial-vector currents, respectively, are given by
\begin{eqnarray}
\label{2:6:divvsc}
\partial_\mu V^\mu_i&=&-\epsilon_{3ij}e{\cal A}_\mu \bar{q}\gamma^\mu
\frac{\tau_j}{2}q=-\epsilon_{3ij}e{\cal A}_\mu V^\mu_j,\\
\label{2:6:divasc}
\partial_\mu A^\mu_i
&=&-e {\cal A}_\mu \epsilon_{3ij} \bar{q}\gamma^\mu
\gamma_5 \frac{\tau_j}{2} q+2m_q i\bar{q}\gamma_5 \frac{\tau_i}{2}q
\nonumber\\
&=&-e {\cal A}_\mu \epsilon_{3ij} A^\mu_j+m_q P_i,
\end{eqnarray}
where we have introduced the isovector pseudoscalar density
\begin{equation}
\label{2:6:psd}
P_i=i\bar{q}\gamma_5 \tau_i q.
\end{equation}

In fact, Eq.\ (\ref{2:6:divasc}) is incomplete, because the third
component of the axial-vector current, $A^\mu_3$, has an anomaly
which is related to the decay $\pi^0\to\gamma\gamma$.
   The full equation reads
\begin{equation}
\label{2:6:divascfull}
\partial_\mu A^\mu_i
=m_q P_i-e {\cal A}_\mu \epsilon_{3ij} A^\mu_j
+\delta_{i3}
\frac{e^2}{32\pi^2}\epsilon_{\mu\nu\rho\sigma}{\cal F}^{\mu\nu}
{\cal F}^{\rho\sigma},
\end{equation}
where ${\cal F}_{\mu\nu}=\partial_\mu{\cal A}_\nu-\partial_\nu{\cal A}_\mu$ is
the electromagnetic field strength tensor.
\end{exercise}

   We emphasize the formal similarity of Eq.\ (\ref{2:6:divasc}) to the
(pre-QCD) PCAC (Partially Conserved Axial-Vector Current) relation
obtained by Adler \cite{Adler:1965:5:5} through the inclusion of the
electromagnetic interactions
with minimal electromagnetic coupling.\footnote{
   In Adler's version, the right-hand side of Eq.\ (\ref{2:6:divascfull})
contains a renormalized field operator creating and destroying pions instead of
$m_q P_i$.  From a modern point of view, the combination
$m_q P_i/(M_\pi^2 F_\pi)$  serves as an interpolating
pion field.
    Furthermore, the anomaly term is not yet present in Ref.\
\cite{Adler:1965:5:5}.}
   Since in QCD the quarks are taken as truly elementary, their interaction
with an (external) electromagnetic field is of such a minimal type.

\chapter{Spontaneous Symmetry Breaking and the Goldstone Theorem}
\label{chap_ssbgt}

   So far we have concentrated on the chiral symmetry of the QCD Hamiltonian
and the {\em explicit} symmetry breaking through the quark masses.
   We have discussed the importance of chiral symmetry for the properties
of Green functions with particular emphasis on the relations
{\em among} different Green functions as expressed through the chiral Ward
identities.
   Now it is time to address a second aspect which, for the low-energy
structure of QCD, is equally important, namely, the concept of
{\em spontaneous} symmetry breaking.
   A (continuous) symmetry is said to be spontaneously broken or hidden,
if the ground state of the system is no longer invariant under the
full symmetry group of the Hamiltonian.
\section{Spontaneous Breakdown of a Global, Continuous,
Non-Abelian Symmetry}
\label{sec_sbgcnas}
\begin{figure}
\begin{center}
\epsfig{file=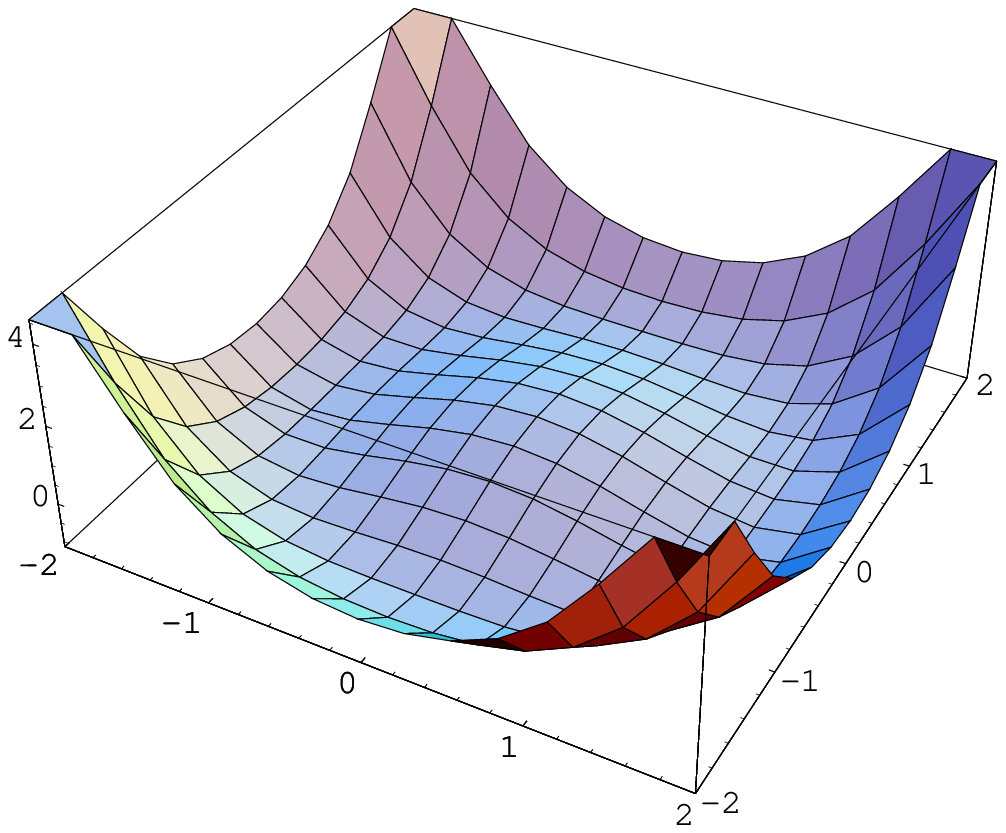,width=6cm}
\caption{\label{3:2:pot2dim}}
Two-dimensional rotationally invariant potential:
${\cal V}(x,y)=-(x^2+y^2)+\frac{(x^2+y^2)^2}{4}$.
\end{center}
\end{figure}

   Using the example of the familiar O(3) sigma model we recall a few
aspects relevant to our subsequent discussion of spontaneous symmetry breaking.
   To that end, we consider the Lagrangian
\begin{eqnarray}
\label{3:2:lphi}
{\cal L}(\vec{\Phi},\partial_\mu\vec{\Phi})
&=&{\cal L}(\Phi_1,\Phi_2,\Phi_3,\partial_\mu\Phi_1,
\partial_\mu\Phi_2,\partial_\mu\Phi_3)\nonumber\\
&=&\frac{1}{2}\partial_\mu \Phi_i\partial^\mu \Phi_i-\frac{m^2}{2}\Phi_i\Phi_i
-\frac{\lambda}{4}(\Phi_i\Phi_i)^2,
\end{eqnarray}
where $m^2<0$, $\lambda>0$, with Hermitian fields $\Phi_i$.
The Lagrangian of Eq.\ (\ref{3:2:lphi}) is invariant under a global
``isospin'' rotation,\footnote{The Lagrangian is invariant under
the full group O(3) which can be decomposed into its two components: the
proper rotations connected to the identity, SO(3), and the
rotation-reflections. For our purposes it is sufficient to
discuss SO(3).}
\begin{equation}
\label{3:2:phitrafo}
g\in \mbox{SO(3)}:\,\,\Phi_i\to\Phi_i'=D_{ij}(g)\Phi_j=
(e^{-i\alpha_k T_k})_{ij}\Phi_j.
\end{equation}
   For the $\Phi_i'$ to also be Hermitian, the Hermitian $T_k$ must be purely
imaginary and thus antisymmetric.
   The $iT_k$ provide the basis of a representation of the so(3) Lie algebra
and satisfy the commutation relations $[T_i,T_j]=i\epsilon_{ijk} T_k$.
   We will use the representation with the matrix elements
given by $t^i_{jk}=-i\epsilon_{ijk}$.
   We now look for a minimum of the potential which does
not depend on $x$.
\begin{exercise}
\label{exercise_pot_min}
\rm    Determine the minimum of the potential
\begin{displaymath}
{\cal V}(\Phi_1,\Phi_2,\Phi_3)
= \frac{m^2}{2}\Phi_i\Phi_i+\frac{\lambda}{4}(\Phi_i\Phi_i)^2.
\end{displaymath}
\end{exercise}
We find
\begin{equation}
\label{3:2:phimin}
|\vec{\Phi}_{\rm min}|=\sqrt{\frac{-m^2}{\lambda}}\equiv v,
\quad |\vec{\Phi}|=\sqrt{\Phi_1^2+\Phi_2^2+\Phi_3^2}.
\end{equation}
   Since $\vec{\Phi}_{\rm min}$ can point in any direction in isospin
space we have a non-countably infinite number of degenerate
vacua.
   Any infinitesimal external perturbation which is not invariant under SO(3)
will select a particular direction which, by an appropriate orientation of
the internal coordinate frame, we denote as the 3 direction,
\begin{equation}
\label{3:2:phimin3}
\vec{\Phi}_{\rm min}=v \hat{e}_3.
\end{equation}
   Clearly, $\vec{\Phi}_{\rm min}$ of Eq.\ (\ref{3:2:phimin3}) is {\em not}
invariant under the full group $G=\mbox{SO(3)}$ since rotations about
the 1 and 2 axis change $\vec{\Phi}_{\rm min}$.\footnote{We say, somewhat
loosely, that $T_1$ and $T_2$ do not annihilate the ground state or,
 equivalently, finite group elements generated by $T_1$ and $T_2$ do not
leave the ground state invariant. This should become clearer later on.}
   To be specific, if
$$
\vec{\Phi}_{\rm min}=v\left(\begin{array}{r}0\\0\\1\end{array}\right),
$$
we obtain
\begin{equation}
\label{3:2:t12phimin}
T_1 \vec{\Phi}_{\rm min}=
v\left(\begin{array}{r}0\\-i\\0\end{array}\right),\quad
T_2 \vec{\Phi}_{\rm min}=
v\left(\begin{array}{r}i\\0\\0\end{array}\right),
\quad
T_3 \vec{\Phi}_{\rm min}=0.
\end{equation}
   Note that the set of transformations which do not leave
$\vec{\Phi}_{\rm min}$ invariant does {\em not} form a group, because it does
not contain the identity.
   On the other hand, $\vec{\Phi}_{\rm min}$ is invariant under a
subgroup $H$ of $G$, namely, the rotations about the 3 axis:
\begin{equation}
\label{3:2:phimintrafoh}
h\in H:\quad \vec{\Phi}'=D(h)\vec{\Phi}=e^{-i\alpha_3 T_3}\vec{\Phi},
\quad
D(h)\vec{\Phi}_{\rm min}=\vec{\Phi}_{\rm min}.
\end{equation}
\begin{exercise}
\label{exercise_eta_expansion}
\rm
  Expand $\Phi_3$ with respect
to $v$,
\begin{equation}
\label{3:2:entw} \Phi_3(x)=v+\eta(x),
\end{equation}
where $\eta(x)$ is a new field replacing $\Phi_3(x)$, and express
the Lagrangian in terms of the fields $\Phi_1$, $\Phi_2$, and
$\eta$, where $v=\sqrt{-m^2/\lambda}$.
\end{exercise}
   The new expression for the potential is given by
\begin{eqnarray}
\label{3:2:ventw} \tilde{\cal V} &=&\frac{1}{2}(-2m^2)\eta^2
+\lambda v\eta (\Phi_1^2+\Phi_2^2+\eta^2)
+\frac{\lambda}{4}(\Phi_1^2+\Phi_2^2+\eta^2)^2-\frac{\lambda}{4}v^4.
\nonumber\\
\end{eqnarray}
   Upon inspection of the terms quadratic in the fields, one finds
after spontaneous symmetry breaking two massless Goldstone bosons and one
massive boson:
\begin{eqnarray}
\label{3:2:masses}
m_{\Phi_1}^2=m_{\Phi_2}^2&=&0,\nonumber\\
m_\eta^2&=&-2m^2.
\end{eqnarray}
   The model-independent feature of the above example is given by the
fact that for each of the two generators $T_1$ and $T_2$ which do not
annihilate the ground state one obtains a {\em massless} Goldstone
boson.
   By means of a two-dimensional simplification (see the ``Mexican hat''
potential shown in Fig.\ \ref{3:2:pot2dim}) the mechanism at hand can easily
be visualized.
    Infinitesimal variations orthogonal to the circle of the minimum
of the potential generate quadratic terms, i.e., ``restoring forces
linear in the displacement,'' whereas tangential variations experience
restoring forces only of higher orders.

   Now let us generalize the model to the case of an arbitrary compact
Lie group $G$ of order $n_G$ resulting in $n_G$ infinitesimal
generators.\footnote{The restriction to compact groups
allows for a complete decomposition into finite-dimensional irreducible
unitary representations.}
   Once again, we start from a Lagrangian of the form \cite{Goldstone:es}
\begin{equation}
\label{3:2:lallg}
{\cal L}(\vec{\Phi},\partial_\mu\vec{\Phi})
=\frac{1}{2}\partial_\mu \vec{\Phi}\cdot \partial^\mu
\vec{\Phi}- {\cal V}(\vec{\Phi}),
\end{equation}
where $\vec{\Phi}$ is a multiplet of scalar (or pseudoscalar)
Hermitian fields.
   The Lagrangian ${\cal L}$ and thus also ${\cal V}(\vec{\Phi})$ are
supposed to be globally invariant under $G$, where the infinitesimal
transformations of the fields are given by
\begin{equation}
\label{3:2:symmtrans}
g\in G:\quad \Phi_i\to\Phi_i+\delta\Phi_i,\quad
\delta \Phi_i=-i\epsilon_a t^a_{ij}\Phi_j.
\end{equation}
   The Hermitian representation matrices $T^a=(t^a_{ij})$
are again antisymmetric and purely imaginary.
   We now assume that, by choosing an appropriate form of ${\cal V}$,
the Lagrangian generates a spontaneous symmetry breaking resulting in
a ground state with a vacuum expectation
value $\vec{\Phi}_{\rm min}=\langle\vec{\Phi}\rangle$
which is invariant under a continuous subgroup $H$ of $G$.
   We expand
${\cal V}(\vec{\Phi})$ with respect to  $\vec{\Phi}_{\rm min}$,
$|\vec{\Phi}_{\rm min}|=v$, i.e.,
$\vec{\Phi}=\vec{\Phi}_{\rm min}+\vec{\chi}$,
\begin{equation}
\label{3:2:venta}
{\cal V}(\vec{\Phi})={\cal V}(\vec{\Phi}_{\rm min})
+\underbrace{\frac{\partial {\cal V}(\vec{\Phi}_{\rm min})}{\partial \Phi_i}}_{
\mbox{$0$}}\chi_i
+\frac{1}{2}\underbrace{\frac{\partial^2 {\cal V}(\vec{\Phi}_{\rm
min})}{\partial
\Phi_i\partial \Phi_j}}_{\mbox{$m^2_{ij}$}}\chi_i\chi_j+\cdots.
\end{equation}
  The matrix $M^2=(m^2_{ij})$
must be symmetric and, since one is expanding around
a minimum,  positive semidefinite, i.e.,
\begin{equation}
\label{3:2:m2}
\sum_{i,j}m^2_{ij}x_i x_j\ge 0\quad \forall \quad \vec{x}.
\end{equation}
   In that case, all eigenvalues of $M^2$ are nonnegative.
   Making use of the invariance of  ${\cal V}$ under the symmetry group
$G$,
\begin{eqnarray}
\label{3:2:vinv}
{\cal V}(\vec{\Phi}_{\rm min})&=&{\cal V}(D(g)\vec{\Phi}_{\rm min})
={\cal V}(\vec{\Phi}_{\rm min}+\delta\vec{\Phi}_{\rm min})\nonumber\\
&\stackrel{\mbox{(\ref{3:2:venta})}}{=}&
{\cal V}(\vec{\Phi}_{\rm min})+\frac{1}{2}m^2_{ij}\delta\Phi_{\rm min,i}
\delta{\Phi}_{\rm min,j}+\cdots,
\end{eqnarray}
one obtains, by comparing coefficients,
\begin{equation}
\label{3:2:kv}
m^2_{ij}\delta\Phi_{\rm min,i}\delta\Phi_{\rm min,j}=0.
\end{equation}
   Differentiating Eq.\ (\ref{3:2:kv}) with respect to
$\delta\Phi_{\rm min,k}$ and using $m^2_{ij}=m^2_{ji}$
results in the matrix equation
\begin{equation}
\label{3:2:rel1}
M^2\delta\vec{\Phi}_{\rm min}=\vec{0}.
\end{equation}
   Inserting the variations of Eq.\ (\ref{3:2:symmtrans}) for arbitrary
$\epsilon_a$,
$\delta\vec{\Phi}_{\rm min}=-i\epsilon_a T^a\vec{\Phi}_{\rm min}$,
we conclude
\begin{equation}
\label{3:2:result}
M^2 T^a \vec{\Phi}_{\rm min}=\vec{0}.
\end{equation}
   The solutions of Eq.\ (\ref{3:2:result}) can be classified into
two categories:
\begin{enumerate}
\item $T^a$, $a=1,\cdots, n_H$, is a representation of an element of
the Lie algebra belonging to the subgroup $H$ of $G$, leaving
the selected ground state invariant.
   In that case one has
$$ T^a \vec{\Phi}_{\rm min}=\vec{0}, \quad a=1,\cdots,n_H,$$
such that Eq.\ (\ref{3:2:result}) is automatically satisfied without
any knowledge of $M^2$.

\item $T^a$, $a=n_H+1,\cdots, n_G,$ is {\em not} a representation of an
element of the Lie algebra belonging to the subgroup $H$. In that case
$T^a\vec{\Phi}_{\rm min}\neq\vec{0}$, and $T^a\vec{\Phi}_{\rm min}$ is
an eigenvector of $M^2$ with eigenvalue 0.
   To each such eigenvector corresponds a massless Goldstone boson.
   In particular, the different $T^a\vec{\Phi}_{\rm min}\neq \vec{0}$ are
linearly independent, resulting in $n_G-n_H$ independent
Goldstone bosons.
   (If they were not linearly independent, there would exist a nontrivial
linear combination
$$\vec{0}=\sum_{a=n_H+1}^{n_G}c_a (T^a\vec{\Phi}_{\rm min})=
\underbrace{\left(\sum_{a=n_H+1}^{n_G}c_a T^a\right)}_{\mbox{$:=T$}}
\vec{\Phi}_{\rm min},
$$
such that $T$ is an element of the Lie algebra of $H$ in contradiction
to our assumption.)
\end{enumerate}

\noindent Remark: It may be necessary to perform a similarity transformation
on the fields in order to diagonalize the mass matrix.

   Let us check these results by reconsidering the example of
Eq.\ (\ref{3:2:lphi}).
   In that case $n_G=3$ and $n_H=1$, generating 2 Goldstone bosons
[see Eq.\ (\ref{3:2:masses})].

   We conclude this section with two remarks. First, the number of
Goldstone bosons is determined by the structure of the symmetry groups.
   Let $G$ denote the symmetry group of the Lagrangian, with $n_G$ generators
and $H$ the subgroup with $n_H$ generators
which leaves the ground state after spontaneous symmetry
breaking invariant.
   For each generator which does not annihilate the vacuum one obtains
a massless Goldstone boson, i.e., the total number of
Goldstone bosons equals $n_G-n_H$.
   Second, the Lagrangians used in {\em motivating} the phenomenon
of a spontaneous symmetry breakdown are typically constructed in such a
fashion that the degeneracy of the ground states is built into
the potential at the classical level (the prototype being the ``Mexican hat''
potential of Fig.\ \ref{3:2:pot2dim}).
   As in the above case, it is then argued that an
{\em elementary} Hermitian field of a multiplet transforming non-trivially
under the symmetry group $G$ acquires a vacuum expectation
value signaling a spontaneous symmetry breakdown.
   However, there also exist theories such as QCD where one cannot infer
from inspection of the Lagrangian whether the theory exhibits spontaneous
symmetry breaking.
   Rather, the criterion for spontaneous symmetry breaking is a non-vanishing
vacuum expectation value of some Hermitian operator, not an elementary field,
which is generated through the dynamics of the underlying theory.
   In particular, we will see that the quantities developing a vacuum
expectation value may also be local Hermitian operators composed of more
fundamental degrees of freedom of the theory.
   Such a possibility was already emphasized in the derivation of Goldstone's
theorem in Ref.\ \cite{Goldstone:es}.

\section{Goldstone Theorem}
\label{sec_gt}
   By means of the above example, we motivate another approach to
Goldstone's theorem without delving into all the subtleties of a
quantum field-theoretical approach (for further reading, see
Chapter 2 of Ref.\ \cite{Bernstein}).
   Given a Hamilton operator with a global symmetry group $G=\mbox{SO(3)}$,
let $\vec{\Phi}(x)=(\Phi_1(x),\Phi_2(x),\Phi_3(x))$ denote a triplet
of local Hermitian operators transforming as a vector
under $G$,
\begin{eqnarray}
\label{3:3:phitrafo}
g\in G:&&\vec{\Phi}(x)\mapsto \vec{\Phi}'(x)=
e^{i\sum_{k=1}^3\alpha_k Q_k}\vec{\Phi}(x)e^{-i\sum_{l=1}^3\alpha_l Q_l}
\nonumber\\
&&=e^{-i\sum_{k=1}^3 \alpha_k T_k}
\vec{\Phi}(x)
\neq \vec{\Phi}(x),
\end{eqnarray}
where the $Q_i$ are the generators of the SO(3) transformations on the Hilbert
space satisfying $[Q_i,Q_j]=i\epsilon_{ijk}Q_k$
and the $T_i=(t^i_{jk})$ are the matrices of the three dimensional
representation satisfying $t^i_{jk}=-i\epsilon_{ijk}$.
   We assume that one component of the multiplet acquires a non-vanishing
vacuum expectation value:
\begin{equation}
\label{3:3:phi3vac}
\langle0|\Phi_1(x)|0\rangle
=\langle0|\Phi_2(x)|0\rangle=0,\quad \langle0|\Phi_3(x)|0\rangle=v\neq 0.
\end{equation}
  Then the two generators $Q_1$ and $Q_2$ do not annihilate the ground state,
and to each such generator corresponds a massless Goldstone boson.

   In order to prove these two statements
let us expand Eq.\ (\ref{3:3:phitrafo})
to first order in the $\alpha_k$:
$$\vec{\Phi}'=\vec{\Phi}+i\sum_{k=1}^3\alpha_k[Q_k,\vec{\Phi}]
=\left(1-i\sum_{k=1}^3\alpha_k T_k\right)\vec{\Phi}
=\vec{\Phi}+\vec{\alpha}\times\vec{\Phi}.
$$
   Comparing the terms linear in the  $\alpha_k$
$$i[\alpha_k Q_k,\Phi_l]=\epsilon_{lkm}\alpha_k\Phi_m$$
and noting that all
three $\alpha_k$ can be chosen independently, we obtain
$$i[Q_k,\Phi_l]=-\epsilon_{klm}\Phi_m,$$
   which, of course, simply expresses the fact that the field operators
$\Phi_i$ transform as a vector.
   Using $\epsilon_{klm}\epsilon_{kln}=2\delta_{mn}$, we find
$$-\frac{i}{2}\epsilon_{kln}[Q_k,\Phi_l]=\delta_{mn}\Phi_m=\Phi_n.$$
   In particular,
\begin{equation}
\label{3:3:phi3zykl}
\Phi_3=-\frac{i}{2}([Q_1,\Phi_2]-[Q_2,\Phi_1]),
\end{equation}
with cyclic permutations for the other two cases.

   In order to prove that $Q_1$ and $Q_2$ do not annihilate the ground
state, let us consider Eq.\ (\ref{3:3:phitrafo})
for $\vec{\alpha}=(0,\pi/2,0)$,
\begin{eqnarray*}
e^{-i\frac{\pi}{2} T_2}\vec{\Phi}&=&
\left(\begin{array}{ccc} \cos(\pi/2)&0&\sin(\pi/2)\\
0&1&0\\
-\sin(\pi/2)&0&\cos(\pi/2)\end{array}\right)
\left(\begin{array}{c}
\Phi_1\\ \Phi_2 \\ \Phi_3\end{array}\right)
=\left(\begin{array}{c}\Phi_3\\ \Phi_2 \\ -\Phi_1\end{array}\right)\\
&=&e^{i\frac{\pi}{2}Q_2}\left(\begin{array}{c}
\Phi_1\\ \Phi_2 \\ \Phi_3\end{array}\right)
e^{-i\frac{\pi}{2}Q_2}.
\end{eqnarray*}
   From the first row we obtain
$$\Phi_3=e^{i\frac{\pi}{2}Q_2}\Phi_1 e^{-i\frac{\pi}{2}Q_2}.$$
   Taking the vacuum expectation value
$$v=\langle 0| e^{i\frac{\pi}{2}Q_2}\Phi_1 e^{-i\frac{\pi}{2}Q_2}|0\rangle
$$
   and using Eq.\ (\ref{3:3:phi3vac}) clearly $Q_2|0\rangle\neq 0$,
since otherwise the exponential operator could be replaced by unity
and the right-hand side would vanish.
   A similar argument shows $Q_1|0\rangle\neq 0$.

   At this point let us make two remarks.
   \begin{itemize}
   \item
   The ``states''  $Q_{1(2)}|0\rangle$ cannot be normalized.
In a more rigorous derivation  one makes use of integrals of the form
$$\int d^3 x \langle0|[J^{0,b}(t,\vec{x}),\Phi_c(0)]|0\rangle,$$
and first determines the commutator before evaluating the integral
\cite{Bernstein}.
\item
   Some derivations of Goldstone's theorem right away start by
assuming $Q_{1(2)}|0\rangle$ $\neq 0$.
   However, for the discussion of spontaneous symmetry breaking in
the framework of QCD it is advantageous to establish the
connection between the existence of Goldstone bosons and a
non-vanishing expectation value (see Section \ref{sec_ssbqcd}).
\end{itemize}

   Let us now turn to the existence of Goldstone bosons,
taking the vacuum expectation value of Eq.\ (\ref{3:3:phi3zykl}):
$$0\neq v=\langle0|\Phi_3(0)|0\rangle
=
-\frac{i}{2}\langle0|\left([Q_1,\Phi_2(0)]-[Q_2,\Phi_1(0)]\right)|0\rangle
\equiv -\frac{i}{2}(A-B).
$$
   We will first show $A=-B$. To that end we perform a rotation of the
fields as well as the generators by $\pi/2$ about the 3 axis
[see Eq.\ (\ref{3:3:phitrafo}) with $\vec{\alpha}=(0,0,\pi/2)$]:
\begin{eqnarray*}
e^{-i\frac{\pi}{2}T_3}\vec{\Phi}
&=&\left(\begin{array}{r}-\Phi_2\\ \Phi_1\\ \Phi_3\end{array}\right)=
e^{i\frac{\pi}{2}Q_3}
\left(\begin{array}{c}\Phi_1\\ \Phi_2\\ \Phi_3\end{array}\right)
e^{-i\frac{\pi}{2}Q_3},
\end{eqnarray*}
and analogously for the charge operators
$$
\left(\begin{array}{r} -Q_2\\Q_1\\Q_3\end{array}\right)
=e^{i\frac{\pi}{2}Q_3}\left(\begin{array}{r}Q_1\\Q_2\\Q_3\end{array}
\right)e^{-i\frac{\pi}{2}Q_3}.
$$
   We thus obtain
\begin{eqnarray*}
B=\langle0|[Q_2, \Phi_1(0)]|0\rangle&=&
\langle0|\Big(e^{i\frac{\pi}{2}Q_3}(- Q_1)
\underbrace{e^{-i\frac{\pi}{2}Q_3}e^{i\frac{\pi}{2}Q_3}}_{\mbox{1}}
\Phi_2(0) e^{-i\frac{\pi}{2}Q_3}\\
&&- e^{i\frac{\pi}{2}Q_3}\Phi_2(0)
e^{-i\frac{\pi}{2}Q_3}e^{i\frac{\pi}{2}Q_3}
(-Q_1) e^{-i\frac{\pi}{2}Q_3}\Big)|0\rangle\\
&=&-\langle0|[Q_1,\Phi_2(0)]|0\rangle=-A,
\end{eqnarray*}
where we made use of $Q_3|0\rangle=0$, i.e., the vacuum is invariant
under rotations about the 3 axis.
   In other words, the non-vanishing vacuum expectation value $v$
can also be written as
\begin{eqnarray}
\label{3:3:vq1phi2}
0\neq v&=&\langle0|\Phi_3(0)|0\rangle=-i\langle0|[Q_1,\Phi_2(0)]|0\rangle
\nonumber\\
&=&-i\int d^3 x\langle0|[J^1_0(t,\vec{x}),\Phi_2(0)]|0\rangle.
\end{eqnarray}
   We insert a complete set of states $1=\sum_n\hspace{-1.4em}\int
\hspace{0.5em} |n\rangle
\langle n|$ into the commutator\footnote{The abbreviation
$\sum_n\hspace{-1.4em}\int\hspace{0.5em} |n\rangle\langle n|$
includes an integral over the total momentum $\vec{p}$ as well
as all other quantum numbers necessary to fully specify the states.}
\begin{eqnarray*}
v&=&-i\sum_n\hspace{-1.1em}\int
\int d^3x
\left(\langle0|J^{1}_0(t,\vec{x})|n\rangle
\langle n|\Phi_2(0)|0\rangle-\langle0|\Phi_2(0)|n\rangle
\langle n|J^{1}_0(t,\vec{x})|0\rangle\right),
\end{eqnarray*}
and make use of translational invariance
\begin{eqnarray*}
&=&-i\sum_n\hspace{-1.1em}\int \int d^3x\left(e^{-iP_n\cdot x}
\langle0|J^{1}_0(0)|n\rangle\langle n|\Phi_2(0)|0\rangle
-\cdots\right)\\
&=&-i\sum_n\hspace{-1.1em}\int (2\pi)^3\delta^3(\vec{P}_n)
\left(e^{-iE_n t}
\langle0|J^{1}_0(0)|n\rangle\langle n|\Phi_2(0)|0\rangle\right.\nonumber\\
&&\left.-e^{iE_n t}\langle 0|\Phi_2(0)|n\rangle
\langle n|J^{1}_0(0)|0\rangle\right).
\end{eqnarray*}
Integration with respect to the momentum of the inserted intermediate
states yields an expression of the form
$$
=-i(2\pi)^3 \sum_n'
 \left(e^{-iE_n t}\cdots
-e^{iE_n t}\cdots\right),
$$
   where the prime indicates that only states with $\vec{P}=0$
need to be considered.
   Due to the Hermiticity of the symmetry current operators
$J^{\mu,a}$ as well as the $\Phi_l$, we have
$$c_n:=\langle 0|J^{1}_0(0)|n\rangle\langle n|\Phi_2(0)|0\rangle
=\langle n|J^{1}_0(0)|0\rangle^\ast \langle0|\Phi_2(0)|n\rangle^\ast,$$
such that
\begin{equation}
\label{3:3:vresult}
v=-i(2\pi)^3\sum_n'
 \left(c_n e^{-iE_n t}-c_n^\ast e^{iE_n t}\right).
\end{equation}
   From Eq.\ (\ref{3:3:vresult}) we draw the following conclusions.
\begin{enumerate}
\item Due to our assumption of a non-vanishing vacuum expectation
value $v$, there must exist states $|n\rangle$ for which both
$\langle0|J^{0}_{1(2)}(0)|n\rangle$ and\linebreak $\langle
n|\Phi_{1(2)}(0)|0\rangle$ do not vanish.
   The vacuum itself cannot contribute to Eq.\ (\ref{3:3:vresult})
because $\langle0|\Phi_{1(2)}(0)|0\rangle=0$.
\item States with $E_n>0$ contribute ($\varphi_n$ is the phase of $c_n$)
\begin{eqnarray*}
\frac{1}{i}\left(c_n e^{-iE_n t}-c_n^\ast e^{iE_n t}\right)
&=&\frac{1}{i}|c_n|\left(e^{i\varphi_n}e^{-iE_n t}
-e^{-i\varphi_n}e^{iE_n t}\right)\\
&=&2|c_n|\sin(\varphi_n-E_n t)
\end{eqnarray*}
to the sum.
   However, $v$ is time-independent and therefore the sum over states
with $(E_n>0,\vec{0})$ must vanish.
\item The right-hand side of Eq.\ (\ref{3:3:vresult}) must therefore
contain the contribution from states with zero energy as well as
zero momentum thus zero mass.
   These zero-mass states are the Goldstone bosons.
\end{enumerate}

\section{Explicit Symmetry Breaking: A First Look *
}
\label{sec_esbfl}
  Finally, let us illustrate the consequences of adding to our Lagrangian
of Eq.\ (\ref{3:2:lphi}) a small perturbation which {\em explicitly} breaks
the symmetry.
   To that end, we modify the potential of Eq.\ (\ref{3:2:lphi})
by adding a term $a\Phi_3$,
\begin{equation}
\label{3:4:pot}
{\cal V}(\Phi_1,\Phi_2,\Phi_3)=
\frac{m^2}{2}\Phi_i\Phi_i
+\frac{\lambda}{4}(\Phi_i\Phi_i)^2 + a\Phi_3,
\end{equation}
where $m^2<0$, $\lambda>0$, and $a>0$, with Hermitian fields $\Phi_i$.
   Clearly, the potential no longer has the original O(3) symmetry but
is only invariant under O(2).
   The conditions for the new minimum, obtained from
$\vec{\nabla}_\Phi {\cal V}=0$, read
\begin{displaymath}
\Phi_1=\Phi_2=0,\quad \lambda \Phi_3^3+m^2\Phi_3+a=0.
\end{displaymath}
\begin{exercise}
\label{exercise_solcubeq}
\rm
Solve the cubic equation for $\Phi_3$ using the perturbative ansatz
\begin{equation}
\label{3:4:phi3ansatz} \langle \Phi_3
\rangle=\Phi^{(0)}_3+a\Phi^{(1)}_3+{\cal O}(a^2).
\end{equation}
\end{exercise}
The solution reads
\begin{displaymath}
\Phi^{(0)}_3=\pm \sqrt{-\frac{m^2}{\lambda}},\quad
\Phi^{(1)}_3=\frac{1}{2m^2}.
\end{displaymath}
   Of course, $\Phi^{(0)}_3$ corresponds to our result without explicit
perturbation.
   The condition for a {\em minimum} [see Eq.\ (\ref{3:2:m2})] excludes
$\Phi^{(0)}_3=+\sqrt{-\frac{m^2}{\lambda}}$.
   Expanding the potential with $\Phi_3=\langle\Phi_3\rangle +\eta$ we obtain,
after a short calculation, for the masses
\begin{eqnarray}
\label{3:4:masses}
m_{\Phi_1}^2=m_{\Phi_2}^2&=& a\sqrt{\frac{\lambda}{-m^2}},\nonumber\\
m_\eta^2&=&-2m^2+3 a \sqrt{\frac{\lambda}{-m^2}}.
\end{eqnarray}
   The important feature here is that the original Goldstone bosons
of Eq.\ (\ref{3:2:masses}) are now massive.
   The squared masses are proportional to the symmetry breaking parameter $a$.
   Calculating {\em quantum} corrections to observables in terms of
Goldstone-boson loop diagrams will generate corrections which are non-analytic
in the symmetry breaking parameter such as $a\ln(a)$.
   Such so-called chiral logarithms originate from the mass terms in the
Goldstone boson propagators entering the calculation of loop integrals.
   We will come back to this point in the next Chapter when we discuss the
masses of the pseudoscalar octet in terms of the quark masses which, in QCD,
represent the analogue to the parameter $a$ in the above example.

\chapter{Chiral Perturbation Theory for Mesons}
\label{chap_cptm}
   Chiral perturbation theory provides a systematic method for discussing
the consequences of the global flavor symmetries of QCD at low energies
by means of an {\em effective field theory}.
   The effective Lagrangian is expressed in terms of those hadronic
degrees of freedom which, at low energies, show up as observable asymptotic
states.
   At very low energies these are just the members of the pseudoscalar octet
($\pi,K,\eta$)
which are regarded as the Goldstone bosons of the {\em spontaneous} breaking
of the chiral $\mbox{SU(3)}_L\times\mbox{SU}(3)_R$ symmetry down to
$\mbox{SU}(3)_V$.
   The non-vanishing masses of the light pseudoscalars in the ``real'' world
are related to the explicit symmetry breaking in QCD due to the light quark
masses.
\section{Effective Field Theory}
\label{sec_EFT}

   Chiral perturbation theory is an example of an \emph{effective
field theory} (EFT).
   Before discussing chiral perturbation theory
in detail we want to briefly outline some of the main features of
the effective field theory approach, as it finds a wide range of
applications in physics.

   The basic idea of an EFT is that one does not need to know everything
in order to make a sensible description of the particular part of physics one is
interested in.
   In general, effective field theories are low-energy
approximations to more fundamental theories.
   Instead of solving the underlying theory, low-energy physics is
described with a set of variables that is suited for the
particular energy region you are interested in.
   The effective field theory can then be used to calculate
physical quantities in terms of an expansion in $p/\Lambda$, where
$p$ stands for momenta or masses that are smaller than a certain
momentum scale $\Lambda$.
   By suited we mean that the description in the low-energy
degrees of freedom is more convenient for actual calculations.
   For example, we will use pions and nucleons instead of the more
fundamental quarks and gluons as the degrees of freedom in
low-energy processes in hadronic physics.
   This is more convenient since so far we do not know how to solve
QCD, and standard perturbation theory cannot be applied for
energies well below $1\ \rm{GeV}$.
   It should be noted that an effective field theory only has a
limited range of applicability, since it gives the wrong
high-energy behavior.
   After all, it is not the same as the underlying theory.
   But as long as one stays well below the momentum scale
$\Lambda$, the EFT is designed to give an appropriate description
up to \emph{finite} accuracy, as in actual calculations only a
finite number of terms in the expansion in $p/\Lambda$ has to be
considered.

   In order to construct an EFT one has to write down the
effective Lagrangian, which includes \emph{all} terms that are
compatible with the symmetries of the underlying theory.
   This means that the Lagrangian actually consists of an
\emph{infinite} number of terms.
   The coefficients of these terms should in principle be calculable
from the underlying theory.
   In the case of chiral perturbation theory, however, we cannot yet solve
the underlying theory, QCD, and the parameters are taken as free parameters that
are fitted to experimental data.
   The second important ingredient together with the effective
Lagrangian is a method that allows to decide which terms
contribute in a calculation up to a certain accuracy.
   We will see an example of such a method when considering
Weinberg's power counting in Section \ref{sec_elwpcs}.

   Effective field theories are non-renormalizable in the
traditional sense.
   However, as long as one considers all terms that are allowed by
the symmetries, divergences that occur in calculations up to any
given order of $p/\Lambda$ can be renormalized by redefining
fields and parameters of the Lagrangian of the effective field
theory.

\section{Spontaneous Symmetry Breaking in QCD}
\label{sec_ssbqcd}
   While the toy model of Section \ref{sec_sbgcnas} by construction led to
a spontaneous symmetry breaking, it is not fully understood
theoretically why QCD should exhibit this phenomenon.
   We will first motivate why experimental input, the hadron spectrum
of the ``real'' world, indicates that spontaneous symmetry breaking
happens in QCD.
   Secondly, we will show that a non-vanishing singlet scalar quark
condensate is a sufficient condition for a spontaneous symmetry breaking in
QCD.

\subsection{The Hadron Spectrum}
\label{subsec_hs}
   We saw in Section \ref{sec_agsl} that the QCD Lagrangian possesses
an $\mbox{SU(3)}_L\times\mbox{SU(3)}_R\times \mbox{U(1)}_V$
symmetry in the chiral limit in which the light quark masses
vanish.
   From symmetry considerations involving the Hamiltonian $H^0_{\rm QCD}$
only, one would naively expect that hadrons organize themselves into
approximately degenerate multiplets fitting the dimensionalities
of irreducible representations of the group
$\mbox{SU(3)}_L\times\mbox{SU(3)}_R\times\mbox{U(1)}_V$.
   The $\mbox{U(1)}_V$ symmetry results in
baryon number conservation and leads to a classification of
hadrons into mesons ($B=0$) and baryons ($B=1$).
   The linear combinations $Q^a_V=Q^a_R+Q^a_L$ and $Q^a_A=Q^a_R-Q^a_L$
of the left- and right-handed charge operators commute with
$H^0_{\rm QCD}$, have opposite parity, and thus for any state of positive
parity one would expect the existence of a degenerate state of negative
parity (parity doubling) which can be seen as follows.
   Let $|i,+\rangle$ denote an eigenstate of $H^0_{\rm QCD}$ with eigenvalue
$E_i$,
$$H^0_{\rm QCD}|i,+\rangle=E_i|i,+\rangle,$$
having positive parity,
$$P|i,+\rangle=+ |i,+\rangle,$$
such as, e.g., a member of the ground state baryon octet (in the chiral limit).
   Defining $|\phi\rangle= Q_A^a|i,+\rangle$, because of
$[H^0_{\rm QCD},Q_A^a]=0$, we have
\begin{displaymath}
H^0_{\rm QCD}|\phi\rangle
=H^0_{\rm QCD} Q_A^a|i,+\rangle
= Q_A^a H^0_{\rm QCD}|i,+\rangle
= E_i Q_A^a|i,+\rangle
= E_i |\phi\rangle,
\end{displaymath}
i.e, the new state $|\phi\rangle$ is also an eigenstate of
$H^0_{\rm QCD}$ with the same eigenvalue $E_i$ but of
opposite parity:
$$P|\phi\rangle= PQ_A^a P^{-1} P|i,+\rangle=-Q_A^a(+|i,+\rangle)
=-|\phi\rangle.
$$
   The state $|\phi\rangle$ can be expanded in terms of the members of the
multiplet with negative parity,
\begin{displaymath}
|\phi\rangle=Q^a_A|i,+\rangle=-t^a_{ij}|j,-\rangle.
\end{displaymath}
   However, the low-energy spectrum of baryons does not contain a degenerate
baryon octet of negative parity.
   Naturally the question arises whether the above chain of arguments is
incomplete.
   Indeed, we have tacitly assumed that the ground state of QCD is annihilated
by $Q^a_A$.

   Let $a^\dagger_i$ symbolically denote an operator which creates quanta
with the quantum numbers of the state $|i,+\rangle$, whereas
$b_i^\dagger$ creates degenerate quanta of opposite parity.
   Let us assume the states $|i,+\rangle$ and $|i,-\rangle$ to be
linear combinations which are constructed from members of two
bases carrying irreducible representations of
$\mbox{SU(3)}_L\times\mbox{SU(3)}_R$.
   In analogy to Eq.\ (\ref{2:3:qphi}), we assume that under
$\mbox{SU(3)}_L\times\mbox{SU(3)}_R$ the creation operators are related
by
$$
[Q^a_A,a^\dagger_i]= -t^a_{ij} b_j^\dagger.
$$
   The usual chain of arguments then works as
\begin{eqnarray}
\label{4:1:pardoub}
Q^a_A|i,+\rangle&=&Q^a_A a^\dagger_i|0\rangle
=\Big([Q^a_A,a^\dagger_i]+a_i^\dagger \underbrace{Q_A^a}_{
\mbox{$\hookrightarrow 0$}}\Big)|0\rangle
= -t^a_{ij} b_j^\dagger |0\rangle.
\end{eqnarray}
   However, if the ground state is {\em not} annihilated by $Q_A^a$, the
reasoning of Eq.\ (\ref{4:1:pardoub}) does no longer apply.
   In that case the ground state is not invariant under the full
symmetry group of the Lagrangian resulting in a spontaneous
symmetry breaking.
   In other words, the non-existence of degenerate multiplets
of opposite parity points to the fact that $\mbox{SU(3)}$ instead
of $\mbox{SU(3)}_L\times\mbox{SU(3)}_R$ is approximately realized
as a symmetry of the hadrons.
   Furthermore the octet of the pseudoscalar mesons is special in
the sense that the masses of its members are small in comparison with
the corresponding $1^-$ vector mesons.
   They are candidates for the Goldstone bosons of a spontaneous
symmetry breaking.

   In order to understand the origin of the SU(3) symmetry let us consider
the vector charges $Q^a_V=Q_R^a+Q^a_L$ [see Eq.\ (\ref{2:4:v})].
   They satisfy the commutation relations of an SU(3) Lie algebra
[see Eqs.\ (\ref{2:4:crqll}) - (\ref{2:4:crqlr})],
\begin{equation}
\label{4:1:su3v}
[Q_R^a+Q_L^a,Q_R^b+Q_L^b]=[Q_R^a,Q_R^b]+[Q_L^a,Q_L^b]
=if_{abc} Q_R^c+if_{abc}Q_L^c=if_{abc}Q^c_V.
\end{equation}
  It was shown by Vafa and Witten \cite{Vafa:tf:1:1}
that, in the chiral limit, the ground state is necessarily
invariant under $\mbox{SU(3)}_V\times\mbox{U(1)}_V$, i.e., the
eight vector charges $Q^a_V$ as well as the baryon number
operator\footnote{Recall that each quark is assigned a baryon
number 1/3.} $Q_V/3$ annihilate the ground state,
\begin{equation}
Q^a_V|0\rangle =Q_V|0\rangle =0.
\end{equation}
   If the vacuum is invariant under $\mbox{SU(3)}_V\times\mbox{U(1)}_V$,
then so is the Hamiltonian (but not vice versa) (Coleman's theorem
\cite{Coleman:1966:1:1}).
   Moreover, the invariance of the ground state {\em and} the Hamiltonian
implies that the physical states of the spectrum of $H^0_{\rm QCD}$
can be organized according to irreducible representations of
$\mbox{SU(3)}_V\times\mbox{U(1)}_V$.
   The index $V$ (for vector) indicates that the generators result from
integrals of the zeroth component of vector current operators and
thus transform with a positive sign under parity.

   Let us now turn to the linear combinations $Q^a_A=Q^a_R-Q^a_L$
satisfying the commutation relations
[see Eqs.\ (\ref{2:4:crqll}) - (\ref{2:4:crqlr})]
\begin{eqnarray}
\label{4:1:crqaa}
[Q^a_A,Q^b_A]&=&[Q^a_R-Q^a_L,Q^b_R-Q^b_L]
=[Q^a_R,Q^b_R]+[Q^a_L,Q^b_L]\nonumber\\
&=&if_{abc}Q^c_R+if_{abc}Q^c_L=
if_{abc}Q^c_V,\nonumber\\
\label{4:1:crqva}
{[Q_V^a,Q^b_A]}&=&[Q_R^a+Q_L^a,Q_R^b-Q_L^b]=
[Q_R^a,Q^b_R]-[Q^a_L,Q^b_L]\nonumber\\
&=&if_{abc}Q^c_R-if_{abc}Q^c_L=
if_{abc}Q^c_A.
\end{eqnarray}
   Note that these charge operators do {\em not} form a closed algebra,
i.e., the commutator of two axial charge operators is not again an
axial charge operator.
   Since the parity doubling is not observed for the low-lying states,
one assumes that the $Q_A^a$ do {\em not} annihilate the ground
state,
\begin{equation}
\label{4:1:qav}
Q^a_A|0\rangle\neq 0,
\end{equation}
i.e., the ground state of QCD is not invariant under ``axial'' transformations.
   According to Goldstone's theorem,
to each axial generator $Q^a_A$, which does not annihilate the ground state,
corresponds a massless Goldstone boson field $\phi^a(x)$ with spin 0,
whose symmetry properties are tightly connected to the generator
in question.
   The Goldstone bosons have the same transformation behavior under
parity,
\begin{equation}
\label{4:1:parityphi}
\phi^a(t,\vec{x})\stackrel{P}{\mapsto}-\phi^a(t,-\vec{x}),
\end{equation}
i.e., they are pseudoscalars, and transform under the subgroup
$H=\mbox{SU(3)}_V$, which leaves the vacuum invariant, as
an octet [see Eq.\ (\ref{4:1:crqva})]:
 \begin{equation}
\label{4:1:transformationphiqv}
[Q^a_V,\phi^b(x)]=if_{abc}\phi^c(x).
\end{equation}
   In the present case, $G=\mbox{SU(3)}_L\times\mbox{SU(3)}_R$ with $n_G=16$
and $H=\mbox{SU(3)}_V$ with $n_H=8$ and we expect eight
Goldstone bosons.

\subsection{The Scalar Quark Condensate *}
\label{subsec_sqc}
   In the following, we will show that a non-vanishing scalar quark condensate
in the chiral limit is a sufficient (but not a necessary) condition for a
spontaneous symmetry breaking in QCD.\footnote{In this Section all physical
quantities such as the ground state, the quark operators etc.\ are considered
in the chiral limit.}
   The subsequent discussion will parallel that of the toy model in
Section \ref{sec_gt} after replacement of the elementary fields
$\Phi_i$ by appropriate composite Hermitian operators of QCD.

   Let us first recall the definition of the nine scalar and pseudoscalar
quark densities:
\begin{eqnarray}
\label{4:1:sqd}
S_a(y)&=&\bar{q}(y)\lambda_a q(y),
\quad a=0,\cdots,8,\\
\label{4:1:psqd} P_a(y)&=&i\bar{q}(y)\gamma_5\lambda_a  q(y),
\quad a=0,\cdots,8.
\end{eqnarray}
\begin{exercise}
\label{exercise_trans_sp}
\rm    Show that $S_a$ and $P_a$ transform under
$\mbox{SU(3)}_L\times\mbox{SU(3)}_R$, i.e.,
$q_L\mapsto q_L'=U_L q_L$ and $q_R \mapsto q_R'= U_R q_R$, as
\begin{eqnarray*}
S_a&\mapsto & S_a'=\bar{q}_L U_L^\dagger \lambda_a U_R q_R
+\bar{q}_R U_R^\dagger \lambda_a U_L q_L,\\
P_a&\mapsto & P_a'=i\bar{q_L}U_L^\dagger \lambda_a U_R q_R
-i\bar{q}_R U_R^\dagger \lambda_a U_L q_L.
\end{eqnarray*}
Hint: Express $S_a$ and $P_a$ in terms of left- and right-handed quark fields.
What are $\gamma_5 P_R$ and $\gamma_5 P_L$?

In technical terms: The components $S_a$ transform as members of a
$(3^\ast,3)+(3,3^\ast)$ representation.
\end{exercise}

   The equal-time commutation relation of two quark operators of the form
$A_i(x)=q^\dagger(x)\hat{A}_i q(x)$,
where $\hat{A}_i$ symbolically denotes Dirac- and flavor matrices and
a summation over color indices is implied, can compactly be written
as [see Eq.\ (\ref{2:4:fkf})]
\begin{equation}
\label{4:1:comrel}
[A_1(t,\vec{x}),A_2(t,\vec{y})]=\delta^3(\vec{x}-\vec{y})
q^\dagger(x)[\hat{A}_1,\hat{A}_2]q(x).
\end{equation}
   With the definition
\begin{displaymath}
Q_V^a(t)=\int d^3 x q^\dagger(t,\vec{x}) \frac{\lambda^a}{2} q(t,\vec{x}),
\end{displaymath}
   and using
\begin{eqnarray*}
[\frac{\lambda_a}{2},\gamma_0\lambda_0]&=&0,\\
{[}\frac{\lambda_a}{2},\gamma_0\lambda_b]&=&
\gamma_0 i f_{abc} \lambda_c,
\end{eqnarray*}
we see, after integration of Eq.\ (\ref{4:1:comrel})
over $\vec{x}$, that the scalar quark densities of
Eq.\ (\ref{4:1:sqd}) transform under
$\mbox{SU(3)}_V$ as a singlet and as an octet, respectively,
\begin{eqnarray}
\label{4:1:sitr}
[Q^a_V(t),S_0(y)]&=&0,\quad a=1,\cdots,8,\\
\label{4:1:octr}
{[Q^a_V(t),S_b(y)]}&=&i\sum_{c=1}^8f_{abc}S_c(y),
\quad a,b=1,\cdots,8,
\end{eqnarray}
with analogous results for the pseudoscalar quark densities.
   In the $\mbox{SU(3)}_V$ limit and, of course, also in the even more
restrictive chiral limit, the charge operators in Eqs.\ (\ref{4:1:sitr})
and (\ref{4:1:octr})
are actually time independent.\footnote{
The commutation relations also remain valid for {\em equal} times if the
symmetry is
explicitly broken.}
   Using the relation
\begin{equation}
\sum_{a,b=1}^8 f_{abc}f_{abd}=3\delta_{cd}
\end{equation}
for the structure constants of SU(3), we re-express the octet components of
the scalar quark densities as
\begin{equation}
\label{4:1:soktett}
S_a(y)=-\frac{i}{3}\sum_{b,c=1}^8f_{abc}[Q_V^b(t),S_c(y)],
\end{equation}
which represents the analogue of Eq.\ (\ref{3:3:phi3zykl}) in the
discussion of Goldstone's theorem.

   In the chiral limit the ground state is necessarily invariant under
$\mbox{SU(3)}_V$, i.e., $Q_V^a|0\rangle=0$, and we
obtain from Eq.\ (\ref{4:1:soktett})
\begin{equation}
\label{4:1:saun}
\langle 0|S_a(y)|0\rangle
=\langle 0|S_a(0)|0\rangle
\equiv\langle S_a\rangle =0,\quad a=1,\cdots,8,
\end{equation}
where we made use of translational invariance of the ground state.
   In other words, the octet components of the scalar quark condensate
{\em must} vanish in the chiral limit.
   From Eq.\ (\ref{4:1:saun}), we obtain for $a=3$
$$\langle\bar{u}u\rangle-\langle\bar{d}d\rangle=0,$$
i.e.\ $\langle\bar{u}u\rangle=\langle\bar{d}d\rangle$
and for $a=8$
$$\langle\bar{u}u\rangle+\langle\bar{d}d\rangle
-2\langle\bar{s}s\rangle=0,
$$
i.e.\ $\langle\bar{u}u\rangle=\langle\bar{d}d\rangle=
\langle\bar{s}s\rangle$.

   Because of Eq.\ (\ref{4:1:sitr}) a similar argument cannot be used
for the singlet condensate, and if we assume a non-vanishing
singlet scalar quark condensate in the chiral limit, we find using
$\langle\bar{u}u\rangle=\langle\bar{d}d\rangle=
\langle\bar{s}s\rangle$:
\begin{equation}
\label{4:1:cqc}
0\neq \langle \bar{q}q\rangle
=\langle\bar{u}u+\bar{d}d+\bar{s}s\rangle
=3\langle\bar{u}u\rangle =
3\langle\bar{d}d\rangle
=3\langle \bar{s}s\rangle.
\end{equation}
   Finally, we make use of (no summation implied!)
$$ (i)^2 [\gamma_5 \frac{\lambda_a}{2},\gamma_0\gamma_5\lambda_a]
=\lambda^2_a\gamma_0$$ in combination with
\begin{eqnarray*}
\lambda_1^2=\lambda_2^2=\lambda_3^2&=&
\left(
\begin{array}{rrr}
1&0&0\\
0&1&0\\
0&0&0
\end{array}
\right),\\
\lambda_4^2=\lambda_5^2&=&
\left(
\begin{array}{rrr}
1&0&0\\
0&0&0\\
0&0&1
\end{array}
\right),\\
\lambda_6^2=\lambda_7^2&=&
\left(
\begin{array}{rrr}
0&0&0\\
0&1&0\\
0&0&1
\end{array}
\right),\\
\lambda_8^2&=&
\frac{1}{3}
\left(
\begin{array}{rrr}
1&0&0\\
0&1&0\\
0&0&4
\end{array}
\right)
\end{eqnarray*}
to obtain
\begin{equation}
\label{4:1:crqapsqd}
i[Q_a^A(t), P_a(y)]
=
\left \{\begin{array}{cl}
\bar{u}u+\bar{d}d, & a=1,2,3\\
\bar{u}u+\bar{s}s, & a=4,5\\
\bar{d}d+\bar{s}s, & a=6,7\\
\frac{1}{3}(\bar{u}u+\bar{d}d+4\bar{s}s), & a=8
\end{array}
\right.
\end{equation}
where we have suppressed the $y$ dependence on the right-hand side.
   We evaluate Eq.\ (\ref{4:1:crqapsqd}) for a ground state which is
invariant under $\mbox{SU(3)}_V$, assuming a non-vanishing singlet scalar
quark condensate,
\begin{equation}
\label{4:1:crqc}
\langle 0|i[Q_a^A(t),P_a(y)]|0\rangle
=\frac{2}{3}\langle\bar{q}q\rangle,\quad a=1,\cdots,8,
\end{equation}
   where, because of translational invariance, the right-hand side
is independent of $y$.
   Inserting a complete set of states into the commutator of
Eq.~(\ref{4:1:crqc}) yields, in complete analogy to Section
\ref{sec_gt} [see the discussion following Eq.\
(\ref{3:3:vq1phi2})] that both the pseudoscalar density $P_a(y)$
as well as the axial charge operators $Q^a_A$ must have a
non-vanishing matrix element between the vacuum and massless one
particle states $|\phi^b\rangle$.
   In particular, because of Lorentz covariance, the
matrix element of the axial-vector current operator between the
vacuum and these massless states,
appropriately normalized,
 can be written as
\begin{equation}
\label{4:1:acc}
\langle 0|A^a_\mu(0)|\phi^b(p)\rangle=ip_\mu F_0 \delta^{ab},
\end{equation}
where $F_0\approx 93$ MeV denotes the ``decay'' constant of
the Goldstone bosons in the chiral limit.
   From Eq.\ (\ref{4:1:acc}) we see that a non-zero value of $F_0$ is a necessary
and sufficient criterion for spontaneous chiral symmetry breaking.
   On the other hand, because of Eq.\ (\ref{4:1:crqc})
a non-vanishing scalar quark condensate $\langle \bar{q}
q\rangle$ is a sufficient (but not a necessary) condition for
a spontaneous symmetry breakdown in QCD.

   Table \ref{table:4:1:comparison} contains a summary of the patterns of
spontaneous symmetry breaking as discussed in Section
\ref{sec_gt}, the generalization of Section \ref{sec_sbgcnas} to
the so-called O($N$) linear sigma model, and QCD.
\begin{table}
\begin{center}
\begin{tabular}{|c|c|c|c|}
\hline
&Section \ref{sec_gt}&O($N$) linear  &QCD\\
&&sigma model&\\
\hline
Symmetry group $G$ of&O(3)&O($N$)&$\mbox{SU(3)}_L\times\mbox{SU(3)}_R$\\
the Lagrangian density&&&\\
\hline
Number of &3&$N(N-1)/2$&16\\
generators $n_G$&&&\\
\hline
Symmetry group $H$& O(2)&O($N-1$)&SU(3)$_V$\\
of the ground state&&&\\
\hline
Number of &1&$(N-1)(N-2)/2$&8\\
generators $n_H$&&&\\
\hline
Number of &&&\\
Goldstone bosons&2&$N-1$&8\\
$n_G-n_H$&&&\\
\hline
Multiplet of
&$(\Phi_1(x),\Phi_2(x))$
&$(\Phi_1(x),\cdots,\Phi_{N-1}(x))$&
$i\bar{q}(x)\gamma_5\lambda_a q(x)$\\
Goldstone boson fields&&&\\
\hline
Vacuum expectation & $v=\langle\Phi_3\rangle$&$v=\langle\Phi_N\rangle$&
$v=\langle\bar{q}q\rangle$\\
value&&&\\
\hline
\end{tabular}
\end{center}
\caption{\label{table:4:1:comparison}Comparison of spontaneous symmetry
breaking.}
\end{table}

\section{Transformation Properties of the Goldstone Bosons}
\label{sec_tpgb}

   The purpose of this section is to discuss the transformation properties
of the field variables describing the Goldstone bosons.
   We will need the concept of a {\em nonlinear realization} of a group in
addition to a {\em representation} of a group which one usually encounters in
Physics.
   We will first discuss a few general group-theoretical properties before
specializing to QCD.

\subsection{General Considerations *}
\label{subsec_gc}
   Let us consider a physical system with a Hamilton operator $\hat{H}$ which
is invariant under a compact Lie group $G$.
   Furthermore we assume the ground state of the system to be invariant under
only a subgroup $H$ of $G$, giving rise to $n=n_G-n_H$ Goldstone bosons.
   Each of these Goldstone bosons will be described by an independent
field $\phi_i$ which is a continuous real function on Minkowski space
$M^4$.\footnote{Depending on the equations of motion, we will
require more restrictive properties of the functions $\phi_i$.
}
   We collect these fields in an $n$-component vector $\Phi$ and define the
vector space
\begin{equation}
\label{4:2:m1}
M_1\equiv\{\Phi:M^4\to R^n|\phi_i:M^4\to R\,\,\mbox{continuous}\}.
\end{equation}
   Our aim is to find a mapping $\varphi$ which uniquely associates with each
pair $(g,\Phi)\in G\times M_1$ an element $\varphi(g,\Phi)\in M_1$ with
the following properties:
\begin{eqnarray}
\label{4:2:condmap1}
&&\varphi(e,\Phi)=\Phi\,\,\forall\,\,\Phi\in M_1,\, e\,\,
\mbox{identity of}\,\, G,\\
\label{4:2:condmap2}
&&\varphi(g_1,\varphi(g_2,\Phi))=\varphi(g_1 g_2,\Phi)\,\,\forall\,\,
g_1,g_2\in G,\,\forall\,\Phi\in M_1.
\end{eqnarray}
   Such a mapping defines an {\em operation} of the group $G$ on $M_1$.
   The second condition is the so-called group-homomorphism property.
   The mapping will, in general, {\em not} define a {\em representation} of
the group $G$, because we do not require the mapping to be linear, i.e.,
$\varphi(g,\lambda \Phi)\neq \lambda\varphi(g,\Phi)$.

   Let $\Phi=0$ denote the ``origin'' of $M_1$  which,
in a theory containing Goldstone bosons only, loosely speaking corresponds
to the ground state configuration.
   Since the ground state is supposed to be invariant under the subgroup
$H$ we require the mapping $\varphi$ to be such that all elements
$h\in H$ map the origin onto itself.
   In this context the subgroup $H$ is also known as the little group of
$\Phi=0$.
   Given that such a mapping indeed exists, we need to verify for
infinite groups that:
\begin{enumerate}
\item $H$ is not empty, because the identity $e$ maps the origin onto itself.
\item If $h_1$ and $h_2$ are elements satisfying $\varphi(h_1,0)=
\varphi(h_2,0)=0$, so does $\varphi(h_1 h_2,0)=
\varphi(h_1,\varphi(h_2,0))=\varphi(h_1,0)=0$,
i.e., because of the homomorphism property also the product $h_1 h_2\in H$.
\item For $h\in H$ we have
$$\varphi(h^{-1},0)=\varphi(h^{-1},\varphi(h,0))
=\varphi(h^{-1}h,0)=\varphi(e,0),
$$
i.e., $h^{-1}\in H$.
\end{enumerate}

   We will establish a connection
between the Goldstone boson fields and the set of all left cosets
$\{gH|g\in G\}$ which is also referred to as the quotient $G/H$.
   For a subgroup $H$ of $G$ the set $gH=\{gh|h\in H\}$ defines the left
coset of $g$ (with an analogous definition for the right coset) which is one
element of $G/H$.\footnote{
An {invariant} subgroup has the additional property that the left and right
cosets coincide for each $g$ which allows for a definition of the
factor group $G/H$ in terms of the complex product.
However, here we do not need this property.}
   For our purposes we need the property
that cosets either completely overlap or are
completely disjoint,
i.e, the quotient is a
set whose elements themselves are sets of group elements, and these sets
are completely disjoint.
\begin{itemize}
\item As an illustration, consider the symmetry group $C_4$ of a
square with directed sides:
\end{itemize}
\begin{center}
\epsfig{file=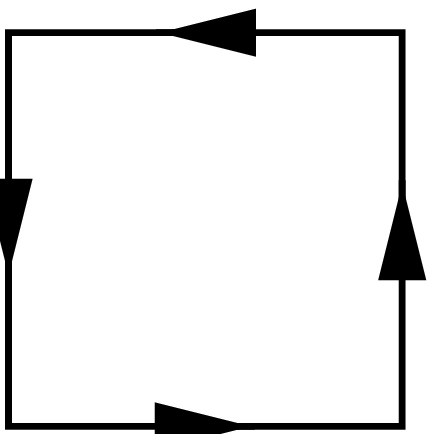,width=2cm}
\end{center}
\begin{eqnarray*}
G=C_4&=&\{e,a,a^2,a^3\},\quad\mbox{$a$ rotation by 90$^\circ$},\quad a^4=e,\\
H&=&\{e,a^2\}.\\
\end{eqnarray*}
$$eH=\{e,a^2\},\,\,aH=\{a,a^3\},\,\,
a^2H=\{e,a^2\},\,\,a^3H=\{a,a^3\}.
$$
$$G/H=\{gH|g\in G\}=\{\{e,a^2\},\{a,a^3\}\}.
$$

   Let us first show that for all elements of a given coset,
$\varphi$ maps the origin onto the same vector in $R^n$:
$$\varphi(gh,0)=\varphi(g,\varphi(h,0))
=\varphi(g,0)\,\,\forall\,\, g\in G\,
\mbox{and}\, h\in H.$$
   Secondly, the mapping is injective with respect to the elements
of $G/H$, which can be proven as follows.
Consider two elements $g$ and $g'$ of $G$ where $g'\not\in g H$.
   We need to show $\varphi(g,0)\neq \varphi(g',0)$.
   Let us assume $\varphi(g,0)=\varphi(g',0)$:
$$0=\varphi(e,0)=\varphi(g^{-1}g,0)
=\varphi(g^{-1},\varphi(g,0))
=\varphi(g^{-1},\varphi(g',0))=\varphi(g^{-1}g',0).$$
   However, this implies $g^{-1}g'\in H$ or $g'\in gH $
in contradiction to the assumption.
   Thus $\varphi(g,0)=\varphi(g',0)$ cannot be true.
   In other words, the mapping can be inverted on the image of
$\varphi(g,0)$.
   The conclusion is that there exists an {\em isomorphic mapping}
between the quotient $G/H$ and and the Goldstone boson
fields.\footnote{Of course, the Goldstone boson fields are not
constant vectors in $R^n$ but functions on Minkowski space
[see Eq.\ (\ref{4:2:m1})].
   This is accomplished by allowing the cosets $gH$ to also
depend on $x$.}

   Now let us discuss the transformation behavior of the Goldstone boson
fields under an arbitrary $g\in G$ in terms of the
isomorphism established above.
   To each $\Phi$ corresponds a  coset $\tilde{g}H$ with appropriate
$\tilde{g}$.
   Let $f=\tilde{g}h\in \tilde{g}H$ denote a representative of this
coset such that
$$\Phi=\varphi(f,0)=\varphi(\tilde{g}h,0).$$
   Now apply the mapping $\varphi(g)$ to $\Phi$:
$$\varphi(g,\Phi)=\varphi(g,\varphi(\tilde{g}h,0))
=\varphi(g\tilde{g}h,0)=\varphi(f',0)=\Phi',\quad
f'\in g(\tilde{g}H).
$$
   In other words, in order to obtain the transformed $\Phi'$ from a given
$\Phi$ we simply need to multiply the left coset $\tilde{g}H$ representing
$\Phi$ by $g$  in order to obtain the new left coset representing $\Phi'$.
   This procedure uniquely determines the transformation behavior of
the Goldstone bosons up to an appropriate choice of variables
parameterizing the elements of the quotient $G/H$.

\subsection{Application to QCD}
\label{subsec_aqcd}

   Now let us apply the above general considerations to the specific case
relevant to QCD and consider the group
$G=\mbox{SU($N$)}\times\mbox{SU($N$)}=\{(L,R)|
L\in \mbox{SU($N$)}, R\in
\mbox{SU($N$)}\}$ and $H=\{(V,V)|V\in \mbox{SU($N$)}\}$ which is
isomorphic to $\mbox{SU($N$)}$.
Let $\tilde{g}=(\tilde{L},\tilde{R})\in G$.
   We may uniquely characterize the left coset of $\tilde{g}$,
$\tilde{g}H=\{(\tilde{L}V,\tilde{R}V)|V\in \mbox{SU($N$)}\}$,
through the SU($N$) matrix $U=\tilde{R}\tilde{L}^\dagger$
\cite{Balachandran:zj:2:2},
$$
(\tilde{L}V,\tilde{R}V)=(\tilde{L}V,\tilde{R}\tilde{L}^\dagger\tilde{L}V)
=(1,\tilde{R}\tilde{L}^\dagger)\underbrace{(\tilde{L}V,\tilde{L}V)}_{
\mbox{$\in H$}},\quad
\mbox{i.e.} \quad \tilde{g}H=(1,\tilde{R}\tilde{L}^\dagger)H,
$$
   if we follow the convention that we choose the representative of the
coset such that the unit matrix stands in its first argument.
   According to the above derivation, $U$ is isomorphic to a $\Phi$.
   The transformation behavior of $U$ under $g=(L,R)\in G$ is obtained by
multiplication in the left coset:
$$g \tilde{g}H=(L, R\tilde{R}\tilde{L}^\dagger)H=
(1,R\tilde{R}\tilde{L}^\dagger L^\dagger)(L,L)H=
(1,R(\tilde{R}
\tilde{L}^\dagger)L^\dagger)H,$$
i.e.\
\begin{equation}
\label{4:2:utrafo}
U=\tilde{R}\tilde{L}^\dagger \mapsto U'=R(\tilde{R}\tilde{L}^\dagger)L^\dagger
=RUL^\dagger.
\end{equation}
   As mentioned above, we finally need to introduce an $x$ dependence so
that
\begin{equation}
\label{4:2:utrfafo}
U(x)\mapsto R U(x) L^\dagger.
\end{equation}

    Let us now restrict ourselves to the physically relevant cases of
$N=2$ and $N=3$ and define
$$
M_1\equiv\left\{
\begin{array}{l}
\{\Phi: M^4\to  R^3|\phi_i: M^4\to
R\,\,\mbox{continuous}\}\,\,\mbox{for $N=2$,}\\
\{\Phi: M^4\to R^8|\phi_i: M^4 \to
R\,\,\mbox{continuous}\}\,\,\mbox{for $N=3$}.
\end{array}
\right.
$$
   Furthermore let $\tilde{\cal H}(N)$ denote the set of all
Hermitian and traceless $N\times N$ matrices,
\begin{displaymath}
\tilde{\cal H}(N)\equiv\{A\in
\mbox{gl}(N, C)|A^\dagger=A\wedge \mbox{Tr}(A)=0\},
\end{displaymath}
   which under addition of matrices defines a real vector space.
   We define a second set $M_2:=\{\phi:M^4\to\tilde{\cal H}(N)|\phi\,\,
\mbox{continuous}\}$, where the entries are continuous functions.
   For $N=2$ the elements of $M_1$ and $M_2$ are related to each
other according to
\begin{eqnarray*}
\phi(x)&=&\sum_{i=1}^3\tau_i\phi_i(x)
=\left(\begin{array}{cc} \phi_3 & \phi_1-i\phi_2\\
\phi_1+i\phi_2&-\phi_3
\end{array}\right)
\equiv
\left(\begin{array}{cc}
\pi^0&\sqrt{2}\pi^+\\
\sqrt{2}\pi^-&-\pi^0
\end{array}\right),\\
\end{eqnarray*}
where the $\tau_i$ are the usual Pauli matrices and $\phi_i(x)=\frac{1}{2}
\mbox{Tr}[\tau_i \phi(x)]$.
   Analogously for $N=3$ ,
\begin{eqnarray*}
\phi(x)=\sum_{a=1}^8 \lambda_a \phi_a(x) \equiv
\left(\begin{array}{ccc}
\pi^0+\frac{1}{\sqrt{3}}\eta &\sqrt{2}\pi^+&\sqrt{2}K^+\\
\sqrt{2}\pi^-&-\pi^0+\frac{1}{\sqrt{3}}\eta&\sqrt{2}K^0\\
\sqrt{2}K^- &\sqrt{2}\bar{K}^0&-\frac{2}{\sqrt{3}}\eta
\end{array}\right),\\
\end{eqnarray*}
with the Gell-Mann matrices $\lambda_a$ and
$\phi_a(x)=\frac{1}{2}\mbox{Tr}[\lambda_a \phi(x)]$.
   Again, $M_2$ forms a real vector space.
\begin{exercise}
\label{exercise_physical_fields} \rm Make use of the Gell-Mann
matrices of Eq.\ (\ref{2:1:gmm}) and express the physical fields
in terms of the Cartesian components, e.g.,
$$\pi^+(x)=\frac{1}{\sqrt{2}}[\phi_1(x)-i\phi_2(x)].$$
\end{exercise}

   Let us finally define
\begin{displaymath}
M_3\equiv\left\{U:M^4\to \mbox{SU}(N)|U(x)
=\exp\left(i\frac{\phi(x)}{F_0}\right),
\phi\in M_2\right\}.
\end{displaymath}
   At this point it is important to note that $M_3$ does not define
a vector space because the sum of two SU($N$) matrices is not
an SU($N$) matrix.

   We are now in the position to discuss the so-called nonlinear
realization of $\mbox{SU($N$)}\times\mbox{SU($N$)}$ on $M_3$.
   The homomorphism
\begin{displaymath}
\varphi: G\times M_3 \to M_3\quad\mbox{with}\quad
\varphi[(L,R),U](x)\equiv R U (x)L^\dagger,
\end{displaymath}
defines an operation of $G$ on $M_3$, because
\begin{enumerate}
\item  $RUL^\dagger\in M_3$, since $U\in M_3$ and
$R, L^\dagger\in \mbox{SU}(N)$.
\item $\varphi[(1_{N\times N},1_{N\times N}),U](x)=1_{N\times N}U(x)
 1_{N\times N}=U(x).$
\item Let $g_i=(L_i,R_i)\in G$ and thus $g_1 g_2=(L_1 L_2,R_1 R_2)\in G$.
\begin{eqnarray*}
\varphi[g_1,\varphi[g_2,U]](x)
&=&\varphi[g_1,(R_2 U L_2^\dagger)](x)=R_1 R_2 U(x) L_2^\dagger
L_1^\dagger,\\
\varphi[g_1g_2,U](x)&=&R_1 R_2 U(x)(L_1 L_2)^\dagger= R_1 R_2 U(x) L_2^\dagger
L_1^\dagger.
\end{eqnarray*}
\end{enumerate}
   The mapping $\varphi$ is called a nonlinear realization, because $M_3$ is
{\em not} a vector space.

   The origin $\phi(x)=0$, i.e.\ $U_0=1$, denotes the ground state of
the system.
   Under transformations of the subgroup $H=\{(V,V)|V\in \mbox{SU($N$)}\}$
corresponding to rotating both left- and right-handed quark fields in
QCD by the same $V$, the ground state remains invariant,
$$\varphi[g=(V,V),U_0]=VU_0 V^\dagger=V V^\dagger= 1=U_0.$$
   On the other hand, under ``axial transformations,'' i.e.\ rotating
the left-handed quarks by $A$ and the right-handed quarks by $A^\dagger$,
the ground state does {\em not} remain invariant,
$$\varphi[g=(A,A^\dagger),U_0]=A^\dagger U_0 A^\dagger=A^\dagger A^\dagger
\neq U_0,$$
   which, of course, is consistent with the assumed spontaneous symmetry
breakdown.

   Let us finally discuss the transformation behavior of $\phi(x)$ under
the subgroup $H=\{(V,V)|V\in \mbox{SU($N$)}\}$.
   Expanding
$$U=1+i\frac{\phi}{F_0}-\frac{\phi^2}{2F_0^2}+\cdots,$$
we immediately see that the realization restricted to the subgroup $H$,
\begin{equation}
\label{4:2:uhtrafo}
1+i\frac{\phi}{F_0}-\frac{\phi^2}{2F_0^2}+\cdots\mapsto
V(1+i\frac{\phi}{F_0}-\frac{\phi^2}{2F_0^2}+\cdots)V^\dagger
=1+i\frac{V \phi V^\dagger}{F_0}-\frac{V\phi V^\dagger V\phi V^\dagger}{2F_0^2}
+\cdots,
\end{equation}
defines a representation on $M_2 \ni \phi\mapsto V\phi V^\dagger
\in M_2$, because
\begin{eqnarray*}
&&
(V\phi V^\dagger)^\dagger= V\phi V^\dagger,\quad
\mbox{Tr}(V\phi V^\dagger)=\mbox{Tr}(\phi)=0,\\
&&V_1 (V_2 \phi V_2^\dagger) V_1^\dagger=(V_1 V_2) \phi (V_1 V_2)^\dagger.
\end{eqnarray*}
   Let us consider the SU(3) case and parameterize
$$V=\exp\left(-i\Theta^V_a\frac{\lambda_a}{2}\right),$$
from which we obtain, by comparing both sides of Eq.\ (\ref{4:2:uhtrafo}),
\begin{equation}
\label{4:2:phihtrafo}
\phi=\lambda_b\phi_b\stackrel{\mbox{$h\in \mbox{SU}(3)_V$}}{\mapsto}
V\phi V^\dagger=\phi-i\Theta^V_a
\underbrace{[\frac{\lambda_a}{2},\phi_b\lambda_b]}_{
\mbox{$\phi_b if_{abc}\lambda_c$}}
+\cdots
=\phi+f_{abc}\Theta^V_a\phi_b\lambda_c +\cdots.
\end{equation}
   However, this corresponds exactly to the adjoint representation,
i.e., in SU(3) the fields $\phi_a$ transform as an octet which is
also consistent with the transformation behavior we discussed in Eq.\
(\ref{4:1:transformationphiqv}):
\begin{eqnarray}
\label{4:2:phivergf}
e^{i\Theta^V_a Q^a_V}\lambda_b\phi_b e^{-i\Theta^V_a Q^a_V}
&=&\lambda_b\phi_b +i\Theta^V_a\lambda_b\underbrace{[Q^a_V,\phi_b]}_{
\mbox{$if_{abc}\phi_c$}}+\cdots\nonumber\\
&=&\phi+f_{abc}\Theta^V_a\phi_b\lambda_c +\cdots.
\end{eqnarray}

   For group elements of $G$ of the form $(A,A^\dagger)$ one may
proceed in a completely analogous fashion.
   However, one finds that the fields $\phi_a$ do {\em not} have
a simple transformation behavior under these group elements.
   In other words, the commutation relations of the fields with
the {\em axial} charges are complicated nonlinear functions of
the fields.

\section{The Lowest-Order Effective Lagrangian}
\label{sec_loel}
   Our goal is the construction of the most general theory describing the
dynamics of the Goldstone bosons associated with the spontaneous
symmetry breakdown in QCD.
   In the chiral limit, we want the effective Lagrangian to be invariant under
 $\mbox{SU(3)}_L\times\mbox{SU(3)}_R\times\mbox{U(1)}_V$.
   It should contain exactly eight pseudoscalar degrees of freedom transforming
as an octet under the subgroup $H=\mbox{SU(3)}_V$.
   Moreover, taking account of spontaneous symmetry breaking, the ground
state should only be invariant under $\mbox{SU(3)}_V\times\mbox{U(1)}_V$.

   Following the discussion of Section \ref{subsec_aqcd} we collect the
dynamical variables in the SU(3) matrix $U(x)$,
\begin{eqnarray}
\label{4:3:upar}
U(x)&=&\exp\left(i\frac{\phi(x)}{F_0}\right),\nonumber\\
\phi(x)&=&\sum_{a=1}^8 \lambda_a \phi_a(x)\equiv
\left(\begin{array}{ccc}
\pi^0+\frac{1}{\sqrt{3}}\eta &\sqrt{2}\pi^+&\sqrt{2}K^+\\
\sqrt{2}\pi^-&-\pi^0+\frac{1}{\sqrt{3}}\eta&\sqrt{2}K^0\\
\sqrt{2}K^- &\sqrt{2}\bar{K}^0&-\frac{2}{\sqrt{3}}\eta
\end{array}\right).
\end{eqnarray}
   The most general, chirally invariant, effective Lagrangian density with the
minimal number of derivatives reads
\begin{equation}
\label{4:3:l2}
{\cal L}_{\rm eff}
=\frac{F^2_0}{4}\mbox{Tr}\left(\partial_\mu U \partial^\mu U^\dagger
\right),
\end{equation}
   where $F_0\approx 93$ MeV is a free parameter which later on will be
related to the pion decay $\pi^+\to\mu^+\nu_\mu$. (see Section
\ref{sec_alo_pion_decay}).

   First of all, the Lagrangian is invariant under the {\em global}\,
$\mbox{SU(3)}_L\times\mbox{SU(3)}_R$ transformations of
Eq.\ (\ref{4:2:utrafo}):
\begin{eqnarray*}
U&\mapsto& R U L^\dagger,\\
\partial_\mu U&\mapsto&\partial_\mu(R U L^\dagger)=
\underbrace{\partial_\mu R}_{\mbox{0}}UL^\dagger+R\partial_\mu U L^\dagger
+RU\underbrace{\partial_\mu L^\dagger}_{\mbox{0}}=R\partial_\mu U L^\dagger,\\
U^\dagger&\mapsto& L U^\dagger R^\dagger,\\
\partial_\mu U^\dagger&\mapsto&L\partial_\mu U^\dagger R^\dagger,
\end{eqnarray*}
because
\begin{displaymath}
{\cal L}_{\rm eff}
\mapsto\frac{F^2_0}{4}\mbox{Tr}\Big(R\partial_\mu U
\underbrace{L^\dagger L}_{\mbox{1}}\partial^\mu U^\dagger R^\dagger\Big)
=\frac{F^2_0}{4}\mbox{Tr}\Big(\underbrace{R^\dagger R}_{\mbox{1}}
\partial_\mu U
\partial^\mu U^\dagger\Big)
={\cal L}_{\rm eff},
\end{displaymath}
   where we made use of the trace property $\mbox{Tr}(AB)=\mbox{Tr}(BA)$.
   The global $\mbox{U(1)}_V$ invariance is trivially satisfied, because
the Goldstone bosons have baryon number zero, thus transforming
as $\phi\mapsto\phi$ under $\mbox{U(1)}_V$ which also implies $U \mapsto U$.

   The substitution $\phi_a(t,\vec{x})\mapsto -\phi_a(t,\vec{x})$ or,
equivalently, $U(t,\vec{x})\mapsto U^\dagger(t,\vec{x})$ provides a simple
method of testing, whether an expression is of so-called even or odd
{\em intrinsic} parity,\footnote{
Since the Goldstone bosons are pseudoscalars, a true parity transformation
is given by $\phi_a(t,\vec{x})\mapsto -\phi_a(t,-\vec{x})$ or,
equivalently, $U(t,\vec{x})\mapsto U^\dagger(t,-\vec{x})$.}
i.e., even or odd in the number of Goldstone boson fields.
   For example, it is easy to show, using the trace property, that
the Lagrangian of Eq.\ (\ref{4:3:l2}) is even.

   The purpose of the multiplicative constant $F^2_0/4$ in Eq.\ (\ref{4:3:l2})
is to generate the standard form of the kinetic term
$\frac{1}{2}\partial_\mu \phi_a\partial^\mu \phi_a$, which can be seen by
expanding the exponential
$U=1+i\phi/F_0+\cdots$, $\partial_\mu U=i\partial_\mu\phi/F_0
+\cdots$, resulting in
\begin{eqnarray*}
{\cal L}_{\rm eff}&=&
\frac{F^2_0}{4}\mbox{Tr}\left[\frac{i\partial_\mu\phi}{F_0}
\left(-\frac{i\partial^\mu\phi}{F_0}\right)\right]+\cdots
=\frac{1}{4}\mbox{Tr}(\lambda_a\partial_\mu\phi_a\lambda_b\partial^\mu
\phi_b)+\cdots\\
&=&\frac{1}{4}\partial_\mu\phi_a\partial^\mu
\phi_b\mbox{Tr}(\lambda_a\lambda_b)+\cdots
=\frac{1}{2}\partial_\mu\phi_a\partial^\mu\phi_a +{\cal L}_{\rm int},
\end{eqnarray*}
where we made use of $\mbox{Tr}(\lambda_a \lambda_b)=2\delta_{ab}$.
   In particular, since there are no other terms containing only two fields
(${\cal L}_{\rm int}$ starts with interaction terms
containing at least four Goldstone bosons)
the eight fields $\phi_a$ describe eight independent {\em massless}
particles.\footnote{At this stage this is only a tree-level argument.
However, the Goldstone bosons remain massless in the chiral limit
even after loop corrections have been included.}

   A term of the type $\mbox{Tr}[(\partial_\mu\partial^\mu U) U^\dagger]$ may
be re-expressed as\footnote{In the present case
$\mbox{Tr}(\partial^\mu U U^\dagger)=0.$}
$$\mbox{Tr}[(\partial_\mu\partial^\mu U) U^\dagger]
=\partial_\mu[\mbox{Tr}(\partial^\mu U U^\dagger)]
-\mbox{Tr}(\partial^\mu U \partial_\mu U^\dagger),$$
i.e., up to a total derivative it is proportional to the Lagrangian
of Eq.\ (\ref{4:3:l2}).
   However, in the present context, total derivatives do not have a dynamical
significance, i.e.\ they leave the equations of motion unchanged and can thus
be dropped.
   The product of two invariant traces is excluded at lowest order,
because $\mbox{Tr}(\partial_\mu U U^\dagger)=0$.
\begin{exercise}
\label{exercise_tr_dmuuudagger}
\rm
   Prove the general SU($N$) case by considering an
SU($N$)-valued field
$$U=\exp\left(i \frac{\Lambda_a\phi_a(x)}{F_0}\right),$$
with $N^2-1$ Hermitian, traceless matrices $\Lambda_a$ and real
fields $\phi_a(x)$.
   Defining $\phi=\Lambda_a\phi_a$, expand
the exponential
$$U=1+i\frac{\phi}{F_0}+\frac{1}{2F_0^2}(i\phi)^2+\frac{1}{3!F_0^3}(i\phi)^3+\cdots$$
and consider the derivative
$$\partial_\mu U=i\frac{\partial_\mu \phi}{F_0}+\frac{i\partial_\mu \phi i\phi
+i\phi i\partial_\mu \phi}{2 F_0^2}+\frac{i\partial_\mu\phi
(i\phi)^2+i\phi i\partial_\mu\phi
i\phi+(i\phi)^2i\partial_\mu\phi}{3! F_0^3}+ \cdots.$$ Remark:
$\phi$ and $\partial_\mu\phi$ are matrices which, in general, do
not commute!

\noindent Verify
\begin{equation}
\label{4:3:trpmuuud}
\mbox{Tr}(\partial_\mu U U^\dagger)=0.
\end{equation}
Hint: $[\phi, U^\dagger]=0$.
\end{exercise}

   Let us turn to the vector and axial-vector currents associated with
the global $\mbox{SU(3)}_L\times\mbox{SU(3)}_R$ symmetry of the effective
Lagrangian of Eq.\ (\ref{4:3:l2}).
   To that end, we parameterize
\begin{eqnarray}
\label{4:3:l}
L&=&\exp\left(-i\Theta^L_a\frac{\lambda_a}{2}\right),\\
\label{4:3:r}
R&=&\exp\left(-i\Theta^R_a\frac{\lambda_a}{2}\right).
\end{eqnarray}
   In order to construct $J^{\mu,a}_L$, set $\Theta^R_a=0$ and choose
$\Theta^L_a=\Theta^L_a(x)$ (see Section \ref{subsec_nt}).
   Then, to first order in $\Theta^L_a$,
\begin{eqnarray}
\label{dul}
U&\mapsto& U'=R U L^\dagger=U\left(1+i\Theta^L_a\frac{\lambda_a}{2}\right),
\nonumber\\
U^\dagger&\mapsto&U'^\dagger=
\left(1-i\Theta^L_a\frac{\lambda_a}{2}\right)U^\dagger,
\nonumber\\
\partial_\mu U&\mapsto&\partial_\mu U'
=\partial_\mu U \left(1+i\Theta^L_a\frac{\lambda_a}{2}\right) +U
i\partial_\mu\Theta_a^L\frac{\lambda_a}{2},
\nonumber\\
\partial_\mu U^\dagger&\mapsto&\partial_\mu U'^\dagger
=\left(1-i\Theta_a^L\frac{\lambda_a}{2}\right)\partial_\mu
U^\dagger -i\partial_\mu \Theta_a^L\frac{\lambda_a}{2} U^\dagger,
\end{eqnarray}
from which we obtain for $\delta \cal L_{\rm eff}$:
\begin{eqnarray}
\label{4:3:dll}
\delta{\cal L}_{\rm eff}&=&\frac{F^2_0}{4}\mbox{Tr}\left[
U i\partial_\mu \Theta_a^L\frac{\lambda_a}{2}\partial^\mu U^\dagger
+\partial_\mu U \left(-i\partial^\mu \Theta_a^L\frac{\lambda_a}{2}U^\dagger
\right)\right]
\nonumber\\
&=& \frac{F^2_0}{4}i\partial_\mu \Theta^L_a\mbox{Tr}\left[
\frac{\lambda_a}{2}(\partial^\mu U^\dagger U-U^\dagger\partial^\mu U)
\right]\nonumber\\
&=&\frac{F^2_0}{4}i\partial_\mu \Theta^L_a\mbox{Tr}\left(\lambda_a
\partial^\mu U^\dagger U\right).
\end{eqnarray}
   (In the last step we made use of
$$
\partial^\mu U^\dagger U=-U^\dagger \partial^\mu U,
$$
   which follows from differentiating $U^\dagger U=1$.)
   We thus obtain for the left currents
\begin{equation}
\label{4:3:jl}
J^{\mu,a}_L=\frac{\partial \delta {\cal L}_{\rm eff}}{
\partial \partial_\mu \Theta_a^L}
=i\frac{F^2_0}{4}\mbox{Tr}\left(\lambda_a \partial^\mu U^\dagger U\right),
\end{equation}
   and, completely analogously, choosing $\Theta_a^L=0$ and
$\Theta_a^R=\Theta_a^R(x)$,
\begin{equation}
\label{4:3:jr}
J^{\mu,a}_R=\frac{\partial \delta {\cal L}_{\rm eff}
}{\partial \partial_\mu \Theta^R_a}
=-i\frac{F^2_0}{4}\mbox{Tr}\left(\lambda_a U \partial^\mu U^\dagger\right)
\end{equation}
   for the right currents.
   Combining Eqs.\ (\ref{4:3:jl}) and (\ref{4:3:jr}) the vector and
axial-vector currents read
\begin{eqnarray}
\label{4:3:jv}
J^{\mu,a}_V&=&J^{\mu,a}_R+J^{\mu,a}_L=-i\frac{F^2_0}{4}
\mbox{Tr}\left(\lambda_a[U,\partial^\mu U^\dagger]\right),\\
\label{4:3:ja}
J^{\mu,a}_A&=&J^{\mu,a}_R-J^{\mu,a}_L=-i\frac{F^2_0}{4}
\mbox{Tr}\left(\lambda_a\{U,\partial^\mu U^\dagger\}\right).
\end{eqnarray}
   Furthermore, because of the symmetry of ${\cal L}_{\rm eff}$ under
$\mbox{SU(3)}_L\times\mbox{SU(3)}_R$, both vector and axial-vector
currents are conserved.
   The vector current densities $J^{\mu,a}_V$ of Eq.\ (\ref{4:3:jv})
contain only terms with an even number of Goldstone bosons,
\begin{eqnarray*}
J^{\mu,a}_V&\stackrel{\mbox{$\phi\mapsto-\phi$}}{\mapsto}&
-i\frac{F^2_0}{4}\mbox{Tr}[\lambda_a(U^\dagger\partial^\mu
U-\partial^\mu U U^\dagger)]\\
&=&-i\frac{F^2_0}{4}\mbox{Tr}[\lambda_a(-\partial^\mu U^\dagger U
+U \partial^\mu U^\dagger)]
=J^{\mu,a}_V.
\end{eqnarray*}
   On the other hand, the expression for the axial-vector currents is
{\em odd} in the number of Goldstone bosons,
\begin{eqnarray*}
J^{\mu,a}_A&\stackrel{\mbox{$\phi\mapsto-\phi$}}{\mapsto}&
-i\frac{F^2_0}{4}\mbox{Tr}[\lambda_a(U^\dagger\partial^\mu
U+\partial^\mu U U^\dagger)]\\
&=&i\frac{F^2_0}{4}\mbox{Tr}[\lambda_a(\partial^\mu U^\dagger U
+U \partial^\mu U^\dagger)]
=-J^{\mu,a}_A.
\end{eqnarray*}
   To find the leading term let us expand Eq.\ (\ref{4:3:ja}) in the fields,
\begin{displaymath}
J^{\mu,a}_A=-i\frac{F^2_0}{4}\mbox{Tr}\left(\lambda_a\left\{1+\cdots,
-i\frac{\lambda_b\partial^\mu \phi_b}{F_0}+\cdots\right\}\right)=
-F_0\partial^\mu\phi_a+\cdots
\end{displaymath}
   from which we conclude that the axial-vector current has a non-vanishing
matrix element when evaluated between the vacuum and a one-Goldstone boson
state [see  Eq.\ (\ref{4:1:acc})]:
\begin{eqnarray*}
\langle 0|J^{\mu,a}_A(x)|\phi^b(p)\rangle
&=&\langle 0|-F_0\partial^\mu\phi_a(x)|\phi^b(p)\rangle\\
&=&-F_0\partial^\mu \exp(-ip\cdot x)\delta^{ab}
=ip^\mu F_0\exp(-ip\cdot x)\delta^{ab}.
\end{eqnarray*}

    So far we have assumed a perfect
$\mbox{SU(3)}_L\times\mbox{SU(3)}_R$ symmetry.
   However, in Section \ref{sec_esbfl} we saw, by means of a simple example,
how an explicit symmetry breaking may lead to finite masses of the
Goldstone bosons.
   As has been discussed in Section \ref{subsec_csbdqm},
the quark mass term of QCD results
in such an explicit symmetry breaking,
\begin{equation}
\label{4:3:qmt}
{\cal L}_M=-\bar{q}_RM q_L-\bar{q}_L M^\dagger q_R,\quad
M=\left(\begin{array}{ccc}m_u&0&0\\0&m_d&0\\0&0&m_s\end{array}\right).
\end{equation}
   In order to incorporate the consequences of Eq.\ (\ref{4:3:qmt})
into the effective-Lagrangian framework, one makes use of the
following argument \cite{Georgi:3}:
   Although $M$ is in reality just a constant matrix and does not
transform along with the quark fields, ${\cal L}_M$ of Eq.\ (\ref{4:3:qmt})
{\em would be} invariant {\em if} $M$ transformed as
\begin{equation}
\label{4:3:mgtrafo}
M\mapsto R M L^\dagger.
\end{equation}
   One then constructs the most general Lagrangian ${\cal L}(U,M)$ which is
invariant under Eqs.\ (\ref{4:2:utrfafo}) and
(\ref{4:3:mgtrafo}) and expands this function in powers of $M$.
   At lowest order in $M$ one obtains
\begin{equation}
\label{4:3:lqm}
{\cal L}_{\rm s.b.}=\frac{F^2_0 B_0}{2}\mbox{Tr}(MU^\dagger+UM^\dagger),
\end{equation}
   where the subscript s.b.\ refers to symmetry breaking.
   In order to interpret the new parameter $B_0$ let us consider the
energy density of the ground state ($U=U_0=1$),
\begin{equation}
\label{4:3:Heff}
\langle{\cal H}_{\rm eff}\rangle=-F_0^2 B_0(m_u+m_d+m_s),
\end{equation}
   and compare its derivative with respect to (any of) the light quark masses
$m_q$ with the corresponding quantity in QCD,
\begin{displaymath}
\left.\frac{\partial \langle 0|{\cal H}_{\rm QCD}|0\rangle}{\partial m_q}
\right|_{m_u=m_d=m_s=0}=\frac{1}{3}\langle 0|\bar{q}{q}|0\rangle_0
=\frac{1}{3}\langle \bar{q}q\rangle,
\end{displaymath}
   where $\langle\bar{q}{q}\rangle$ is the chiral quark condensate of
Eq.\ (\ref{4:1:cqc}).
   Within the framework of the lowest-order effective
Lagrangian, the constant $B_0$ is thus related to the chiral quark condensate
as
\begin{equation}
\label{4:3:b0}
3 F^2_0B_0=-\langle\bar{q}q\rangle.
\end{equation}

   Let us add a few remarks.
\begin{enumerate}
\item A term $\mbox{Tr}(M)$ by itself is not invariant.
\item The combination $\mbox{Tr}(MU^\dagger-U M^\dagger)$ has the wrong
behavior under parity $\phi(t,\vec{x})\to-\phi(t,-\vec{x})$, because
\begin{eqnarray*}
\mbox{Tr}[M U^\dagger(t,\vec{x})-U(t,\vec{x})M^\dagger]
&\stackrel{P}{\mapsto}
&\mbox{Tr}[M U(t,-\vec{x})-U^\dagger(t,-\vec{x})M^\dagger]
\\
&\stackrel{M=M^\dagger}{=}&-\mbox{Tr}[M U^\dagger(t,-\vec{x})-
U(t,-\vec{x}) M^\dagger].
\end{eqnarray*}
\item Because $M=M^\dagger$, ${\cal L}_{\rm s.b.}$ contains only terms even
in $\phi$.
\end{enumerate}
   In order to determine the masses of the Goldstone bosons, we identify the
terms of second order in the fields in ${\cal L}_{\rm s.b.}$,
\begin{equation}
\label{4:3:lmzo}
{\cal L}_{\rm s.b}=-\frac{B_0}{2}\mbox{Tr}(\phi^2M) +\cdots.
\end{equation}
\begin{exercise}
\label{exercise_mass_term_l2}
\rm
Expand the mass term to second order in the fields and determine the mass
squares of the Goldstone bosons.
\end{exercise}
   Using Eq.\ (\ref{4:3:upar}) we find
\begin{eqnarray*}
\mbox{Tr}(\phi^2M) &=&2(m_u+m_d)\pi^+\pi^- +2(m_u+m_s)K^+ K^-
+2(m_d+m_s)K^0\bar{K}^0\\
&&+(m_u+m_d)\pi^0\pi^0 +\frac{2}{\sqrt{3}}(m_u-m_d)\pi^0\eta
+\frac{m_u+m_d+4m_s}{3}\eta^2.
\end{eqnarray*}
   For the sake of simplicity we consider the isospin-symmetric limit
$m_u=m_d=m$ so that the $\pi^0\eta$ term vanishes and there is no
$\pi^0$-$\eta$ mixing.
   We then obtain for the masses of the Goldstone bosons, to lowest order in
the quark masses,
\begin{eqnarray}
\label{4:3:mpi2}
M^2_\pi&=&2 B_0 m,\\
\label{4:3:mk2}
M^2_K&=&B_0(m+m_s),\\
\label{4:3:meta2}
M^2_\eta&=&\frac{2}{3} B_0\left(m+2m_s\right).
\end{eqnarray}
   These results, in combination with Eq.\ (\ref{4:3:b0}),
$B_0=-\langle\bar{q}q\rangle/(3 F_0^2)$, correspond to relations
obtained in Ref.\ \cite{Gell-Mann:rz:3} and are referred to as the
Gell-Mann, Oakes, and Renner relations.
   Furthermore, the masses of Eqs.\ (\ref{4:3:mpi2}) - (\ref{4:3:meta2})
satisfy the Gell-Mann-Okubo relation
\begin{equation}
\label{4:3:gmof}
4M^2_K=4B_0(m+m_s)=2B_0(m+2m_s)+2B_0m=3M^2_\eta+M^2_\pi
\end{equation}
   independent of the value of $B_0$.
   Without additional input regarding the numerical value of $B_0$,
Eqs.\ (\ref{4:3:mpi2}) - (\ref{4:3:meta2}) do not allow for an extraction of
the  absolute values of the quark masses $m$ and $m_s$, because rescaling
$B_0\to \lambda B_0$ in combination with $m_q\to m_q/\lambda$ leaves the
relations invariant.
   For the ratio of the quark masses one obtains, using the empirical
values of the pseudoscalar octet,
\begin{eqnarray}
\frac{M^2_K}{M^2_\pi}=\frac{m+m_s}{2m}&\Rightarrow&\frac{m_s}{m}=25.9,
\nonumber\\
\frac{M^2_\eta}{M^2_\pi}=\frac{2m_s+m}{3m}&\Rightarrow&\frac{m_s}{m}=24.3.
\end{eqnarray}

   Let us conclude this section with the following remark.
   We saw in Section \ref{subsec_sqc} that a non-vanishing quark condensate in
the chiral
limit is a sufficient but not a necessary condition for a spontaneous
chiral symmetry breaking.
   The effective Lagrangian term of Eq.\ (\ref{4:3:lqm}) not only results in
a shift of the vacuum energy but also in finite Goldstone boson
masses and both effects are proportional to the parameter
$B_0$.\footnote{In Exercise \ref{exercise_pion_pion_scattering} we
will also see that the $\pi\pi$ scattering amplitude is effected
by ${\cal L}_{\rm s.b.}$.}
   We recall that it was
a symmetry argument which excluded a term $\mbox{Tr}(M)$ which,
at leading order in $M$, would decouple the vacuum energy shift
from the Goldstone boson masses.
   The scenario underlying ${\cal L}_{\rm s.b.}$ of Eq.\ (\ref{4:3:lqm})
is similar to that of a Heisenberg ferromagnet which exhibits a
spontaneous magnetization $\langle \vec{M}\rangle$, breaking the
O(3) symmetry of the Heisenberg Hamiltonian down to O(2).
   In the present case the analogue of the order parameter
$\langle \vec{M}\rangle$ is the quark condensate $\langle \bar{q} q\rangle$.
   In the case of the ferromagnet, the interaction with an external magnetic
field is given by $-\langle \vec{M}\rangle\cdot \vec{H}$, which corresponds
to Eq.\ (\ref{4:3:Heff}), with the quark masses playing the role of the
external field $\vec{H}$.
   However, in principle, it is also possible that $B_0$ vanishes or is
rather small.
   In such a case  the quadratic masses of the Goldstone bosons might
be dominated by terms which are nonlinear in the quark masses, i.e., by
higher-order terms in the expansion of ${\cal L}(U,M)$.
   Such a scenario is the origin of the so-called generalized chiral
perturbation theory \cite{Knecht:1995tr:3}.
   The analogue would be an antiferromagnet which shows a spontaneous
symmetry breaking but with $\langle \vec{M}\rangle=0$.

    The analysis of recent data on $K^+\to \pi^+\pi^-e^+\nu_e$
\cite{Pislak:2001bf:3} in terms of the isoscalar $s$-wave
scattering length $a_0^0$ \cite{Colangelo:2001sp:3} supports the
conjecture that the quark condensate is indeed the leading order
parameter of the spontaneously broken chiral symmetry.

\section{Effective Lagrangians and Weinberg's Power Counting Scheme}
\label{sec_elwpcs}

   An essential prerequisite for the construction of effective field theories
is a ``theorem'' of Weinberg stating that a perturbative
description in terms of the most general effective Lagrangian
containing all possible terms compatible with assumed symmetry
principles yields the most general $S$ matrix consistent with the
fundamental principles of quantum field theory and the assumed
symmetry principles \cite{Weinberg:1978kz:4}.
   The corresponding effective Lagrangian will contain an infinite number of
terms with an infinite number of free parameters.
   Turning Weinberg's theorem into a practical tool requires two
steps: one needs some scheme to organize the effective Lagrangian and
a systematic method of assessing the importance of diagrams generated by
the interaction terms of this Lagrangian when calculating a physical matrix
element.

    In the framework of mesonic chiral perturbation theory, the most general
chiral Lagrangian describing the dynamics of the Goldstone bosons is organized
as a string of terms with an increasing number of derivatives and quark mass
terms,
\begin{equation}
\label{4:4:ll2l4}
{\cal L}_{\rm eff}={\cal L}_2 + {\cal L}_4 + {\cal L}_6 +\cdots,
\end{equation}
   where the subscripts refer to the order in a momentum and
quark mass expansion.
   The index 2, for example, denotes either two derivatives or one quark
mass term.
   In the context of Feynman rules, derivatives generate four-momenta,
whereas the convention of counting quark-mass terms as being of
the same order as two derivatives originates from Eqs.\
(\ref{4:3:mpi2}) - (\ref{4:3:meta2}) in conjunction with the
on-shell condition $p^2=M^2$.
   In an analogous fashion, ${\cal L}_4$ and ${\cal L}_6$ denote more
complicated terms of so-called chiral orders ${\cal O}(p^4)$ and
${\cal O}(p^6)$ with corresponding numbers of derivatives and quark mass terms.
   With such a counting scheme, the chiral orders in the mesonic sector are
always even [${\cal O}(p^{2n})$] because Lorentz indices of derivatives always
have to be contracted with either the metric tensor $g^{\mu\nu}$ or the
Levi-Civita tensor $\epsilon^{\mu\nu\rho\sigma}$ to generate scalars,
and the quark mass terms are counted as ${\cal O}(p^2)$.

   Weinberg's power counting scheme analyzes
the behavior of a given diagram under a linear rescaling of all the
{\em external} momenta, $p_i\mapsto t p_i$, and a quadratic rescaling of the
light quark masses, $m_q\mapsto t^2 m_q$, which, in terms of the Goldstone
boson masses, corresponds to $M^2\mapsto t^2 M^2$.
   The chiral dimension $D$ of a given diagram with amplitude
${\cal M}(p_i,m_q)$ is defined by
\begin{equation}
\label{4:4:mr1}
{\cal M}(tp_i, t^2 m_q)=t^D {\cal M}(p_i,m_q),
\end{equation}
and thus
\begin{equation}
\label{4:4:mr2}
D=2+\sum_{n=1}^\infty2(n-1)N_{2n} +2N_L,
\end{equation}
where $N_{2n}$ denotes the number of vertices originating from
${\cal L}_{2n}$, and $N_L$ is the number of independent loops.
   Going to smaller momenta and masses corresponds to a rescaling
with $0<t<1$.
   Clearly, for small enough momenta and masses diagrams with small $D$, such
as $D=2$ or $D=4$, should dominate.
   Of course, the rescaling of Eq.\ (\ref{4:4:mr1}) must be viewed as
a mathematical tool.
   While external three-momenta can, to a certain extent, be made arbitrarily
small, the rescaling of the quark masses is a theoretical instrument only.
   Note that loop diagrams are always suppressed due to the term $2N_L$ in
Eq.\ (\ref{4:4:mr2}).
   It may happen, though, that the leading-order tree diagrams vanish and
therefore that the lowest-order contribution to a certain process is
a one-loop diagram.
   An example is the reaction
$\gamma\gamma\to\pi^0\pi^0$.

    For the purpose of actually determining the chiral dimension $D$ of a given
diagram it is more convenient to use the expression
\begin{equation}
\label{4:4:D2}
 D=N_L-2N_I+\sum_{n=1}^\infty2nN_{2n},
\end{equation}
where $N_I$ denotes the number of internal lines.
   The equivalence with Eq.\ (\ref{4:4:mr2}) is shown by using
a relation among the number of independent loops, total number of
vertices, and number of internal lines:\footnote{Note that the
number of independent momenta is {\em not} the number of faces or
closed circuits that may be drawn on the internal lines of a
diagram. This may, for example, be seen using a diagram with the
topology of a tetrahedron which has four faces but $N_L=6-(4-1)=3$
(see, e.g., Chapter 6-2 of C.~Itzykson and J.-B.~Zuber, {\em
Quantum Field Theory}).} \begin{equation} \label{4:4:NL}
N_L=N_I-(N_V-1).\end{equation}
   Each of the $N_V$ vertices generates a delta function which,
after extracting one overall delta function, yields $N_V-1$
conditions for the internal momenta.
   Finally, the total number of vertices is
   given by $N_V=\sum_{n} N_{2n}$.

   In order to prove Eq.\ (\ref{4:4:mr2}) we start from the usual Feynman
rules for evaluating an $S$-matrix element (see, e.g., Ref.\
\cite{Itzykson:rh:4}).
   Each internal meson line contributes a factor
\begin{eqnarray}
\label{4:4:intlines}
\int\frac{d^4k}{(2\pi)^4} \frac{i}{k^2-M^2+i\epsilon}
&\stackrel{\mbox{$(M^2\mapsto t^2 M^2)$}}{\mapsto}&
t^{-2}\int\frac{d^4k}{(2\pi)^4} \frac{i}{k^2/t^2-M^2+i\epsilon}\nonumber\\
&\stackrel{\mbox{$(k=tl)$}}{=}&
t^2 \int\frac{d^4l}{(2\pi)^4} \frac{i}{l^2-M^2+i\epsilon}.
\end{eqnarray}
   For each vertex, originating from ${\cal L}_{2n}$, we obtain symbolically
a factor $p^{2n}$ together with a four-momentum conserving delta function
resulting in $t^{2n}$ for the vertex factor and $t^{-4}$ for the delta
function.
   At this point one has to take into account the fact that, although
Eq.\ (\ref{4:4:mr1}) refers to a rescaling of {\em external} momenta,
a substitution $k=tl$ for internal momenta as in Eq.\ (\ref{4:4:intlines})
acts in exactly the same way as a rescaling of external momenta:
\begin{eqnarray*}
\delta^4(p+k)&\stackrel{\mbox{$p\mapsto tp, k=tl$}}{\mapsto}t^{-4}
\delta^4(p+l),\\
p^{2n-m}k^m&\stackrel{\mbox{$p\mapsto tp, k=tl$}}{\mapsto}t^{2n}p^{2n-m}l^m,
\end{eqnarray*}
where $p$ and $k$ denote external and internal momenta, respectively.

   So far we have discussed the rules for determining the power $D_S$
referring to the $S$-matrix element which is related to the invariant amplitude
through a four-momentum conserving delta function,
$$S\sim \delta^4(P_f-P_i){\cal M}.$$
   The delta function contains external momenta only, and thus re-scales
under $p_i\mapsto tp_i$ as $t^{-4}$, so
$$t^{D_S}=t^{-4}t^D.$$
    We thus find as an intermediate result
\begin{equation}
\label{4:4:d}
D=4+2N_I+\sum_{n=1}^\infty N_{2n}(2n-4),
\end{equation}
which, using Eq.\ (\ref{4:4:NL}), we bring into the form of Eq.\
(\ref{4:4:mr2}):
$$
 D=4+2(N_L+N_V-1)+\sum_{n=1}^\infty N_{2n}(2n-4)
=2+2 N_L +\sum_{n=1}^\infty N_{2n}(2n-2).
$$
  In the discussion of loop integrals we need to
address the question of convergence, since applying the
substitution $tl=k$ in Eq.\ (\ref{4:4:intlines}) is well-defined
only for convergent integrals.
   Later on we will regularize the integrals by use of the method of
dimensional regularization, introducing a renormalization scale $\mu$ which
also has to be rescaled linearly.
   However, at a given chiral order, the sum of all diagrams will, by
construction, not depend on the renormalization scale.

\section{Construction of the Effective Lagrangian}
\label{sec_cel}
   In Section \ref{sec_loel} we have derived the lowest-order effective
Lagrangian for a {\em global} $\mbox{SU(3)}_L\times\mbox{SU(3)}_R$ symmetry.
   On the other hand, the Ward identities originating in the global
$\mbox{SU(3)}_L\times\mbox{SU(3)}_R$ symmetry of QCD are obtained
from a {\em locally} invariant generating functional involving a
coupling to external fields (see Sections \ref{sec_gfwi} and
\ref{subsec_qcdpefgf}).
   Our goal is to approximate the ``true'' generating functional
$Z_{\rm QCD}[v,a,s,p]$ of Eq.\ (\ref{2:6:genfun}) by a sequence
$Z^{(2)}_{\rm eff}[v,a,s,p] + Z^{(4)}_{\rm eff}[v,a,s,p]
+\cdots$, where the effective generating
functionals are obtained using the effective field theory.
   Therefore, we need to promote the global symmetry of the effective
Lagrangian to a local one and introduce a coupling to the {\em same} external
fields $v$, $a$, $s$, and $p$ as in QCD.

   In the following we will outline the principles entering the construction
of the effective Lagrangian for a local
$G=\mbox{SU(3)}_L\times\mbox{SU(3)}_R$
symmetry.\footnote{In principle, we could also ``gauge'' the
U(1)$_V$ symmetry. However, this is primarily of relevance to the SU(2)
sector in order to fully incorporate the coupling to the electromagnetic
field [see Eq.\ (\ref{2:6:rlasu2})].
Since in SU(3), the quark-charge matrix is traceless, this important
case is included in our considerations.}
   The matrix $U$ transforms as $U \mapsto U'=V_R U V_L^{\dagger}$,
where $V_L(x)$ and $V_R(x)$ are independent space-time-dependent SU(3)
matrices.
   As in the case of gauge theories, we need external fields
$l_\mu^a(x)$ and $r_\mu^a(x)$
[see Eqs.\ (\ref{2:6:mch}), (\ref{2:6:vrlarl}), and
(\ref{2:6:sg}) and Table \ref{4:5:table_trafprop}]
 corresponding to the parameters $\Theta^L_a(x)$
and $\Theta^R_a(x)$ of $V_L(x)$ and $V_R(x)$, respectively.
   For any object $A$ transforming as $V_R A V_L^\dagger$ such as, e.g., $U$
we define the covariant derivative $D_\mu A$ as
\begin{equation}
\label{4:5:kaa}
D_\mu A\equiv\partial_\mu A -i r_\mu A+iA l_\mu.
\end{equation}
\begin{exercise}
\label{exercise_covariant_derivative_a} \rm Verify the
transformation behavior
$$
D_\mu A\mapsto V_R (D_\mu A) V_L^\dagger.$$
Hint: Make use of
$V_R\partial_\mu V_R^\dagger=-\partial_\mu V_R V_R^\dagger$.
\end{exercise}
   Again, the defining property for the covariant derivative is that it
should transform in the same way as the object it acts on.\footnote{Under
certain circumstances it is advantageous to introduce for each object with
a well-defined transformation behavior a separate covariant derivative.
   One may then use a product rule similar to the one
of ordinary differentiation.}
   Since the effective Lagrangian will ultimately contain arbitrarily high
powers of derivatives we also need the field strength tensors
$f^L_{\mu\nu}$ and $f^R_{\mu\nu}$ corresponding to the external
fields $r_{\mu}$ and $l_{\mu}$,
\begin{eqnarray}
\label{4:5:fr}
f_{\mu\nu}^R&\equiv&\partial_\mu r_\nu-\partial_\nu r_\mu-i{[r_\mu,r_\nu]},\\
\label{4:5:fl}
f_{\mu\nu}^L&\equiv&\partial_\mu l_\nu-\partial_\nu l_\mu-i{[l_\mu,l_\nu]}.
\end{eqnarray}
   The field strength tensors are traceless,
\begin{equation}
\label{4:5:trflfr}
\mbox{Tr}(f^L_{\mu\nu})=\mbox{Tr}(f^R_{\mu\nu})=0,
\end{equation}
because $\mbox{Tr}(l_\mu)=\mbox{Tr}(r_\mu)=0$ and the trace of any
commutator vanishes.
   Finally, following the convention of Gasser and Leutwyler we introduce
the linear combination $\chi\equiv 2B_0(s+ip)$ with the scalar and pseudoscalar
external fields of Eq.\ (\ref{2:6:mch}),  where $B_0$ is defined in
Eq.\ (\ref{4:3:b0}).
   Table \ref{4:5:table_trafprop} contains the transformation
properties of all building blocks under the group ($G$), charge
conjugation ($C$), and parity ($P$).

\begin{table}
\begin{center}
\begin{tabular}{|c|c|c|c|}
\hline
element&$G$&$C$&$P$\\
\hline
$U$&$V_R U V_L^\dagger$&$U^T$&$U^\dagger$\\
\hline
$D_{\lambda_1}\cdots D_{\lambda_n}U$&
$V_R D_{\lambda_1}\cdots D_{\lambda_n}U V_L^\dagger$&
$(D_{\lambda_1}\cdots D_{\lambda_n}U)^T$&
$(D^{\lambda_1}\cdots D^{\lambda_n}U)^\dagger$\\
\hline
$\chi$&$V_R \chi V_L^\dagger$&$\chi^T$&$\chi^\dagger$\\
\hline
$D_{\lambda_1}\cdots D_{\lambda_n}\chi$&
$V_R D_{\lambda_1}\cdots D_{\lambda_n}\chi V_L^\dagger$&
$(D_{\lambda_1}\cdots D_{\lambda_n}\chi)^T$&
$(D^{\lambda_1}\cdots D^{\lambda_n}\chi)^\dagger$\\
\hline
$r_\mu$&$V_R r_\mu V_R^\dagger+iV_R\partial_\mu
V^\dagger_R$&$-l_\mu^T$&$l^\mu$\\
\hline
$l_\mu$&$V_L l_\mu V_L^\dagger+iV_L\partial_\mu
V^\dagger_L$&$-r_\mu^T$&$r^\mu$\\
\hline
$f^R_{\mu\nu}$&$V_R f^R_{\mu\nu}V_R^\dagger$&
$-(f_{\mu\nu}^L)^T$&$f_L^{\mu\nu}$\\
\hline
$f^L_{\mu\nu}$&$V_L f^L_{\mu\nu}V_L^\dagger$&
$-(f_{\mu\nu}^R)^T$&$f_R^{\mu\nu}$\\
\hline
\end{tabular}
\end{center}
\caption[test]{\label{4:5:table_trafprop} Transformation
properties under the group ($G$), charge conjugation ($C$), and
parity ($P$).
    The expressions for adjoint matrices are trivially obtained
by taking the Hermitian conjugate of each entry.
   In the parity transformed
expression it is understood that the argument is $(t,-\vec{x})$
and that partial derivatives $\partial_{\mu}$ act with respect to
$x$ and not with respect to the argument of the corresponding
function.}
\end{table}

  In the chiral counting scheme of chiral perturbation theory the elements
are counted as:
\begin{equation}
\label{4:5:powercounting}
U =  {\cal O}(p^0),\, D_{\mu} U  = {\cal  O}(p),\, r_{\mu},l_{\mu}
= {\cal O}(p),\,
f^{L/R}_{\mu\nu}  =  {\cal O}(p^2),\, \chi  =  {\cal O}(p^2).
\end{equation}
   The external fields $r_{\mu}$ and $l_{\mu}$ count as ${\cal O}(p)$
to match $\partial_\mu A$, and $\chi$ is of ${\cal O}(p^2)$ because
of Eqs.\ (\ref{4:3:mpi2}) - (\ref{4:3:meta2}).
   Any additional covariant derivative counts as ${\cal O}(p)$.

   The construction of the effective Lagrangian in terms of the building blocks
of Eq.\ (\ref{4:5:powercounting}) proceeds as follows.\footnote{There is
a certain freedom in the choice of the elementary building blocks.
   For example, by a suitable multiplication with $U$ or $U^\dagger$ any
building block can be made to transform as $V_R \cdots V_R^\dagger$ without
changing its chiral order.
   The present approach most naturally leads to the Lagrangian of Gasser
and Leutwyler.}
   Given objects $A,B,\dots$, all of which transform as
\mbox{$A'=V_R A V_L^{\dagger},$} \mbox{$B'= V_R B V_L^{\dagger},\,\dots,$}
one can form invariants by taking the trace of products of the
type $A B^{\dagger}$:
\begin{eqnarray*}
\mbox{Tr}(A B^\dagger)&\mapsto& \mbox{Tr}[V_R A V_L^\dagger
(V_R B V_L^\dagger)^\dagger]=\mbox{Tr}(V_R A V_L^\dagger V_L B^\dagger
V_R^\dagger)=\mbox{Tr}(A B^\dagger V_R^\dagger V_R)\\
&=&\mbox{Tr}(A B^\dagger).
\end{eqnarray*}
The generalization to more terms is obvious and, of course, the product of
invariant traces is invariant:
\begin{equation}
\label{4:5:chains}
\mbox{Tr}(AB^\dagger C D^\dagger),\quad
\mbox{Tr}(A B^\dagger)\mbox{Tr}(C D^\dagger),\quad \cdots.
\end{equation}
   The complete list of relevant elements up to and including order ${\cal O}(p^2)$
transforming as $V_R\cdots V_L^\dagger$ reads
\begin{equation}
\label{4:5:lis}
U, D_\mu U, D_\mu D_\nu U,\chi, U f_{\mu\nu}^L, f^R_{\mu\nu}U.
\end{equation}
   For the invariants up to ${\cal O}(p^2)$ we then obtain
\begin{eqnarray}
\label{4:5:invariants}
{\cal O}(p^0)&:& \mbox{Tr}(UU^\dagger)=3,\nonumber\\
{\cal O}(p)&:& \mbox{Tr}(D_\mu U U^\dagger)
\stackrel{\ast}{=}-\mbox{Tr}[U (D_\mu U)^\dagger]
\stackrel{\ast}{=}0,\nonumber
\\
{\cal O}(p^2)&:& \mbox{Tr}(D_\mu D_\nu U U^\dagger)
\stackrel{\ast\ast}{=}-\mbox{Tr}[D_\nu U (D_\mu U)^\dagger]
\stackrel{\ast\ast}{=}\mbox{Tr}[U(D_\nu D_\mu U)^\dagger],\nonumber\\
&&\mbox{Tr}(\chi U^\dagger),\nonumber\\
&&\mbox{Tr}(U \chi^\dagger),\nonumber\\
&&\mbox{Tr}(U f^L_{\mu\nu}U^\dagger)=\mbox{Tr}(f^L_{\mu\nu})=0,\nonumber\\
&&\mbox{Tr}(f^R_{\mu\nu})=0.
\end{eqnarray}
   In $\ast$ we made use of two important properties of the covariant
derivative $D_\mu U$:
\begin{eqnarray}
\label{4:5:kauprop1}
D_\mu U U^\dagger &=& -U (D_\mu U)^\dagger,\\
\label{4:5:kauprop2}
\mbox{Tr}(D_\mu U U^\dagger)&=&0.
\end{eqnarray}
   The first relation results from the unitarity of $U$ in combination
with the definition of the covariant derivative, Eq.\ (\ref{4:5:kaa}).
\begin{eqnarray*}
D_\mu U U^\dagger&=& \underbrace{\partial_\mu U U^\dagger}_{
\mbox{$-U\partial_\mu U^\dagger$}}-ir_\mu \underbrace{U U^\dagger}_{\mbox{1}}
+i Ul_\mu U^\dagger,\\
-U (D_\mu U)^\dagger&=&-U\partial_\mu U^\dagger -\underbrace{U U^\dagger}_{
\mbox{1}} i r_\mu
-U (-i l_\mu U^\dagger).
\end{eqnarray*}
   Equation (\ref{4:5:kauprop2}) is shown using
$\mbox{Tr}(r_\mu)=\mbox{Tr}(l_\mu)=0$
together with Eq.\ (\ref{4:3:trpmuuud}),
$\mbox{Tr}(\partial_\mu U U^\dagger)=0$:
\begin{eqnarray*}
\mbox{Tr}(D_\mu U U^\dagger)&=&\mbox{Tr}(\partial_\mu U U^\dagger
-ir_\mu U U^\dagger
+iUl_\mu U^\dagger)=0.
\end{eqnarray*}

\begin{exercise}
\label{exercise_total_derivative} \rm Verify $\ast\ast$
$$\mbox{Tr}(D_\mu D_\nu U U^\dagger)
=-\mbox{Tr}[D_\nu U (D_\mu U)^\dagger]=\mbox{Tr}[U(D_\nu D_\mu
U)^\dagger]$$
by explicit calculation.
\end{exercise}

   Finally, we impose Lorentz invariance, i.e., Lorentz indices
have to be contracted, resulting in three candidate terms:
\begin{eqnarray}
\label{4:5:lis1}
&&\mbox{Tr}[D_\mu U (D^\mu U)^\dagger],\\
\label{4:5:lis2}
&&\mbox{Tr}(\chi U^\dagger\pm U \chi^\dagger).
\end{eqnarray}
   The term in Eq.\ (\ref{4:5:lis2}) with the minus sign is excluded because
it has the wrong sign under parity (see Table \ref{4:5:table_trafprop}),
and we end up with the most general, {\em locally} invariant,
effective Lagrangian at lowest chiral order,\footnote{At ${\cal O}(p^2)$
invariance under $C$ does not
provide any additional constraints.}
\begin{equation}
\label{4:5:l2}
{\cal L}_2=\frac{F_0^2}{4}\mbox{Tr}[D_\mu U (D^\mu U)^\dagger]
+\frac{F^2_0}{4}\mbox{Tr}(\chi U^\dagger + U\chi^\dagger).
\end{equation}
   Note that ${\cal L}_2$ contains two free parameters:
the pion-decay constant $F_0$ and $B_0$ of Eq.\ (\ref{4:3:b0})
(hidden in the definition of $\chi$).

\begin{exercise}
\renewcommand{\labelenumi}{(\alph{enumi})}
\label{exercise_charge_conjugation}
\rm Under charge conjugation fields describing
particles are mapped on fields describing antiparticles, i.e.,
$\pi^0\mapsto\pi^0$, $\eta\mapsto\eta$, $\pi^+\leftrightarrow\pi^-$,
$K^+\leftrightarrow K^-$, $K^0\leftrightarrow \bar{K}^0$.
\begin{enumerate}
\item
What does that mean for the matrix
$$\phi=
\left(\begin{array}{ccc}
\pi^0+\frac{1}{\sqrt{3}}\eta &\sqrt{2}\pi^+&\sqrt{2}K^+\\
\sqrt{2}\pi^-&-\pi^0+\frac{1}{\sqrt{3}}\eta&\sqrt{2}K^0\\
\sqrt{2}K^- &\sqrt{2}\bar{K}^0&-\frac{2}{\sqrt{3}}\eta
\end{array}\right)?
$$
\item Using $A^TB^T=(BA)^T$ show by induction
${(A^T)}^n={(A^n)}^T$. In combination with (a) show that
$U=\exp(i\phi/F_0)\stackrel{C}{\mapsto} U^T$.
\item Under charge conjugation the external fields transform
as
$$v_\mu\mapsto -v_\mu^T,\quad a_\mu\mapsto a_\mu^T,\quad
s\mapsto s^T,\quad p\mapsto p^T.
$$
   Derive the transformation behavior of $r_\mu=
v_\mu+a_\mu$, $l_\mu=v_\mu-a_\mu$, $\chi=2B_0(s+ip)$, and
$\chi^\dagger$.
\item Using (b) and (c) show that the covariant
derivative of $U$ under charge conjugation transforms as
$$
D_\mu U\mapsto (D_\mu U)^T.
$$
\item Show that
\begin{displaymath}
{\cal L}_2=\frac{F_0^2}{4}\mbox{Tr}[D_\mu U (D^\mu U)^\dagger]
+\frac{F^2_0}{4}\mbox{Tr}(\chi U^\dagger + U\chi^\dagger)
\end{displaymath}
is invariant under charge conjugation. Note that ${(A^T)}^\dagger
={(A^\dagger)}^T$ and $\mbox{Tr}(A^T)=\mbox{Tr}(A)$.
\item As an example, show the invariance of the $L_3$ term
of ${\cal L}_4$ under charge conjugation:
$$L_3 \mbox{Tr}\left
[D_{\mu}U (D^{\mu}U)^{\dagger}D_{\nu}U (D^{\nu}U)^{\dagger}
\right ].
$$
Hint: At the end you will need $(D_\mu U)^\dagger=-U^\dagger
D_\mu U U^\dagger$ and $U^\dagger D_\mu U U^\dagger=-(D_\mu U)^\dagger$.
\end{enumerate}
\end{exercise}

\section{Application at Lowest Order:
Pion Decay} \label{sec_alo_pion_decay}
 As an example of a tree-level calculation
 we discuss the weak decay $\pi^+\to \mu^+\nu_\mu$ which
will allow us to relate the free parameter $F_0$ of ${\cal L}_2$
to the pion-decay constant.
   According to Eq.\ (\ref{4:4:mr2}) we only need to consider tree-level
diagrams with vertices of ${\cal L}_2$.

\begin{figure}
\epsfig{file=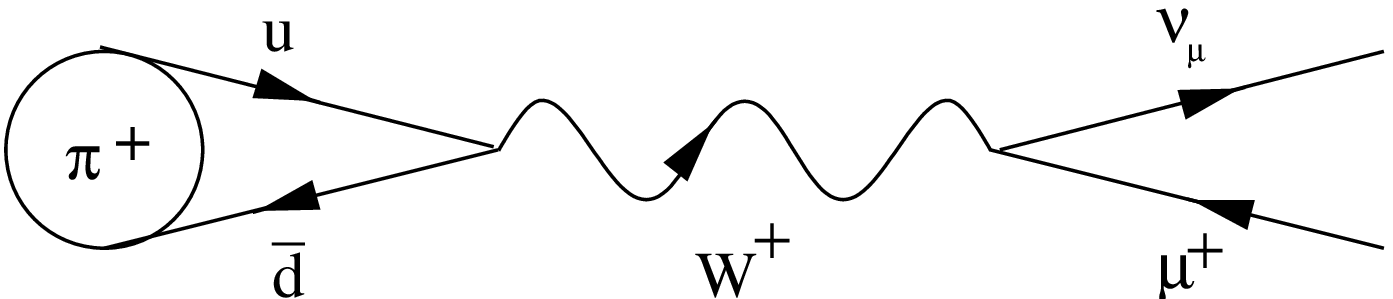,width=12cm}
\caption{\label{4:6:piondecay}
Pion decay $\pi^+\to \mu^+\nu_\mu$.}
\end{figure}

   At the level of the degrees of freedom of the Standard Model,
pion decay is described by the annihilation of a $u$ quark and a
$\bar{d}$ antiquark, forming the $\pi^+$, into a $W^+$ boson,
propagation of the intermediate $W^+$, and creation of the leptons
$\mu^+$ and $\nu_\mu$ in the final state (see Figure
\ref{4:6:piondecay}).
    The coupling of the $W$ bosons to the leptons is given by
\begin{equation}
\label{4:6:lwl} {\cal L}=-\frac{g}{2\sqrt{2}}\left[{\cal W}^+_\alpha\bar{\nu}_\mu
\gamma^\alpha(1-\gamma_5)\mu+{\cal W}^-_\alpha\bar{\mu}\gamma^\alpha
(1-\gamma_5)\nu_\mu\right],
\end{equation}
    whereas their interaction with the quarks forming the Goldstone
bosons is effectively taken into account by inserting Eq.\ (\ref{2:6:rlw})
into the Lagrangian of Eq.\ (\ref{4:5:l2}).
   Let us consider the first term of Eq.\ (\ref{4:5:l2}) and set $r_\mu=0$
with, at this point, still arbitrary $l_\mu$.
\begin{exercise}
\label{exercise_lagrangian_pion_decay} \rm Using $D_\mu
U=\partial_\mu U+iUl_\mu$ derive
\begin{eqnarray*}
\frac{F^2_0}{4}\mbox{Tr}[D_\mu U (D^\mu U)^\dagger]
&=&i\frac{F^2_0}{2}\mbox{Tr}(l_\mu \partial^\mu U^\dagger
U)+\cdots,
\end{eqnarray*}
   where only the term linear in $l_\mu$ is shown.
\end{exercise}

   If we parameterize
$$l_\mu =\sum_{a=1}^8\frac{\lambda_a}{2}l^a_\mu,$$
the interaction term linear in $l_\mu$ reads
\begin{equation}
\label{4:6:lwli}
{\cal L}_{\rm int}=\sum_{a=1}^8l_\mu^a\left[i \frac{F_0^2}{4}\mbox{Tr}(
\lambda_a \partial^\mu U^\dagger U)\right]=
\sum_{a=1}^8l^a_\mu J^{\mu,a}_L,
\end{equation}
   where we made use of Eq.\ (\ref{4:3:jl}) defining $J^{\mu,a}_L$.
   Again, we expand $J^{\mu,a}_L$ by using Eq.\ (\ref{4:3:upar}) to first
order in $\phi$,
\begin{eqnarray}
\label{4:6:jlent} J^{\mu,a}_L&=& \frac{F_0}{2}\partial^\mu \phi^a+{\cal
O}(\phi^2),
\end{eqnarray}
from which we obtain the matrix element
\begin{equation}
\label{4:6:lpi}
\langle 0|J^{\mu,a}_L(0)|\phi^b(p)\rangle
=\frac{F_0}{2}\langle 0|\partial^\mu \phi^a(0)|\phi^b(p)\rangle
=-ip^\mu \frac{F_0}{2}\delta^{ab}.
\end{equation}
   Inserting $l_\mu$ of Eq.\ (\ref{2:6:rlw}), we find for the interaction
term of a single Goldstone boson with a $W$
$$
{\cal L}_{W\phi}=\frac{F_0}{2}\mbox{Tr}(l_\mu \partial^\mu \phi)
= -\frac{g}{\sqrt{2}}\frac{F_0}{2}\mbox{Tr}
[({\cal W}_\mu^+T_+ + {\cal W}^-_\mu T_-)\partial^\mu\phi].
$$
   Thus, we need to calculate\footnote{Recall that the entries $V_{ud}$
and $V_{us}$ of the Cabibbo-Kobayashi-Maskawa matrix are real.}
\begin{eqnarray*}
\lefteqn{\mbox{Tr}(T_+\partial^\mu\phi)}\\
&=&
\mbox{Tr}\left[
\left(\begin{array}{ccc}0&V_{ud}&V_{us}\\0&0&0\\0&0&0\end{array}
\right)
\partial^\mu
\left(\begin{array}{ccc}
\pi^0+\frac{1}{\sqrt{3}}\eta &\sqrt{2}\pi^+&\sqrt{2}K^+\\
\sqrt{2}\pi^-&-\pi^0+\frac{1}{\sqrt{3}}\eta&\sqrt{2}K^0\\
\sqrt{2}K^- &\sqrt{2}\bar{K}^0&-\frac{2}{\sqrt{3}}\eta
\end{array}\right)\right]\\
&=&
V_{ud}\sqrt{2}\partial^\mu\pi^-+V_{us}\sqrt{2}\partial^\mu K^-,\\
\lefteqn{\mbox{Tr}(T_-\partial^\mu\phi)}\\
&=&\mbox{Tr}\left[
\left(\begin{array}{ccc}0&0&0\\
V_{ud}&0&0\\
V_{us}&0&0
\end{array}
\right)
\partial^\mu
\left(\begin{array}{ccc}
\pi^0+\frac{1}{\sqrt{3}}\eta &\sqrt{2}\pi^+&\sqrt{2}K^+\\
\sqrt{2}\pi^-&-\pi^0+\frac{1}{\sqrt{3}}\eta&\sqrt{2}K^0\\
\sqrt{2}K^- &\sqrt{2}\bar{K}^0&-\frac{2}{\sqrt{3}}\eta
\end{array}\right)\right]\\
&=&V_{ud} \sqrt{2}\partial^\mu\pi^++V_{us}\sqrt{2}\partial^\mu K^+.
\end{eqnarray*}
   We then obtain for the interaction term
\begin{equation}
\label{4:6:lwphi}
{\cal L}_{W\phi}=
-g \frac{F_0}{2}
[{\cal W}_\mu^+(V_{ud}\partial^\mu\pi^-+V_{us}\partial^\mu K^-)
+{\cal W}_\mu^-(V_{ud}\partial^\mu \pi^++V_{us}\partial^\mu K^+)].
\end{equation}
   Expanding the Feynman propagator for $W$ bosons,
\begin{equation}
\label{4:6:wprop} \frac{-g_{\mu\nu}+\frac{k_\mu k_\nu}{M^2_W}}{k^2-M^2_W}
=\frac{g_{\mu\nu}}{M^2_W}+{\cal O}\left(\frac{kk}{M^4_W}\right),
\end{equation}
and neglecting terms which are of higher order in $(\mbox{momentum}/M_W)^2$, the
Feynman rule for the invariant amplitude for the weak pion decay reads
\begin{eqnarray}
\label{4:6:mpionzerfall} {\cal M}&=&i\left[-\frac{g}{2\sqrt{2}}\bar{u}_{\nu_\mu}
\gamma^\beta (1-\gamma_5)v_{\mu^+}\right] \frac{ig_{\beta\alpha}}{M^2_W}
i\left[-g\frac{F_0}{2}V_{ud}
(-ip^\alpha)\right]\nonumber\\
&=&-G_F V_{ud} F_0 \bar{u}_{\nu_\mu} p\hspace{-.4em}/(1-\gamma_5)
v_{\mu^+},
\end{eqnarray}
where  $p$ denotes the four-momentum of the pion and
$$G_F=\frac{g^2}{4\sqrt{2}M^2_W}=
1.16639(1)\times
10^{-5}\,\mbox{GeV}^{-2}
$$
is the Fermi constant.
   The evaluation of the decay rate is a standard textbook exercise
and we only quote the final result
\begin{equation}
\label{4:6:zr}
\frac{1}{\tau}=\frac{G^2_F |V_{ud}|^2}{4\pi} F^2_0 M_\pi m_\mu^2
\left(1-\frac{m_\mu^2}{M_\pi^2}\right)^2.
\end{equation}
   The constant $F_0$ is referred to as the pion-decay constant in the
chiral limit.\footnote{Of course, in the chiral limit, the pion is
massless and, in such a world, the massive leptons would decay into Goldstone
bosons, e.g., $e^-\to\pi^-\nu_e$.
   However, at ${\cal O}(p^2)$, the symmetry breaking term of
Eq.\ (\ref{4:3:lqm}) gives rise to Goldstone-boson masses, whereas
the decay constant is not modified at ${\cal O}(p^2)$.}
   It measures the strength of the matrix element of the axial-vector
current operator between a one-Goldstone-boson state and the vacuum
[see Eq.\ (\ref{4:1:acc})].
   Since the interaction of the $W$ boson with the quarks is of the type
$l_\mu^a L^{\mu,a}=l_\mu^a(V^{\mu,a}-A^{\mu,a})/2$ [see Eq.\ (\ref{2:6:rlw})]
and the vector current operator does not contribute to the
matrix element between a single pion and the vacuum, pion decay
is completely determined by the axial-vector current.
   The degeneracy of a single constant $F_0$ in Eq.\ (\ref{4:1:acc}) is
lifted at ${\cal O}(p^4)$ \cite{Gasser:1984gg:6} once SU(3)
symmetry breaking is taken into account.
   The empirical numbers for $F_\pi$ and $F_K$ are $92.4$ MeV and
$113$ MeV, respectively.\footnote{In the analysis of D.~E.~Groom
{\it et al.} [Particle Data Group Collaboration], Eur.\ Phys.\ J.\
C {\bf 15}, 1 (2000) $f_\pi=\sqrt{2} F_\pi$ is used.}

\begin{exercise}
\label{exercise_pion_decay}
\renewcommand{\labelenumi}{(\alph{enumi})}
\rm
The differential decay rate for
$\pi^+(p_\pi)\to \nu_\mu(p_\nu)+\mu^+(p_\mu)$ in the pion rest frame
is given by
\begin{displaymath}
d\omega=\frac{1}{2M_\pi}
\overline{|{\cal M}|^2}
\frac{d^3 p_\nu}{2E_\nu (2\pi)^3}
\frac{d^3 p_\mu}{2E_\mu (2\pi)^3}
(2\pi)^4 \delta^4 (p_\pi-p_\nu-p_\mu).
\end{displaymath}
   Here, we assume the neutrino to be massless and make use of the
normalization $u^\dagger u=2 E=v^\dagger v$.
   The invariant amplitude is given by Eq.\ (\ref{4:6:mpionzerfall}).
   The neutrino spinors satisfy
\begin{eqnarray*}
\frac{1-\gamma_5}{2} u_{\nu_\mu}(p_\nu) &=& u_{\nu_\mu}(p_\nu),\\
\frac{1+\gamma_5}{2} u_{\nu_\mu}(p_\nu) &=& 0.
\end{eqnarray*}
\begin{enumerate}
\item[(a)]
Make use of the Dirac equation
\begin{eqnarray*}
\bar{u}_{\nu_\mu}(p_\nu) p_\nu \hspace{-.9em}/&=&0,\\
p_\mu\hspace{-.9em}/\hspace{.5em}
v_{\mu^+}(p_\mu,s_\mu)&=&-m_\mu v_{\mu^+}(p_\mu,s_\mu),
\end{eqnarray*}
and show
\begin{displaymath}
\bar{u}_{\nu_\mu}(p_\nu) (p_\nu+p_\mu)_\alpha\gamma^\alpha (1-\gamma_5)
v_{\mu^+}(p_\mu,s_\mu) =-m_\mu \bar{u}_{\nu_\mu}(p_\nu) (1+\gamma_5)
v_{\mu^+}(p_\mu,s_\mu).
\end{displaymath}
Hint: $\{\gamma_\alpha,\gamma_5\}=0$.
\item[(b)]
Verify using trace techniques
\begin{eqnarray*}
\lefteqn{ [\bar{u}_{\nu_\mu}(p_\nu) (p_\nu+p_\mu)_\alpha\gamma^\alpha
(1-\gamma_5) v_{\mu^+}(p_\mu,s_\mu)]}\\
&&\times [\bar{u}_{\nu_\mu}(p_\nu) (p_\nu+p_\mu)_\beta\gamma^\beta
(1-\gamma_5) v_{\mu^+}(p_\mu,s_\mu)]^\ast\\
&=&m_\mu^2\bar{u}_{\nu_\mu}(p_\nu) (1+\gamma_5)v_{\mu^+}(p_\mu,s_\mu)
\bar{v}_{\mu^+}(p_\mu,s_\mu)(1-\gamma_5)u_{\nu_\mu}(p_\nu)\\
&=&m_\mu^2\mbox{Tr}[u_{\nu_\mu}(p_\nu)\bar{u}_{\nu_\mu}(p_\nu) (1+\gamma_5)v_{\mu^+}(p_\mu,s_\mu)
\bar{v}_{\mu^+}(p_\mu,s_\mu)(1-\gamma_5)]\\
&=&\cdots\\
&=& 4m_\mu^2 M_\pi^2\left[\frac{1}{2}\left(1-\frac{m_\mu^2}{m_\pi^2}\right)
-\frac{m_\mu p_\nu\cdot s_\mu}{M_\pi^2}\right].
\end{eqnarray*}

\noindent Hints:
\begin{eqnarray*}
(1-\gamma_5)u_{\nu_\mu}(p_\nu)\bar{u}_{\nu_\mu}(p_\nu)(1+\gamma_5)
&=&(1-\gamma_5)p_\nu \hspace{-.9em}/\hspace{.5em} (1+\gamma_5),\\
v_{\mu^+}(p_\mu,s_\mu)\bar{v}_{\mu^+}(p_\mu,s_\mu)&=&
(p_\mu\hspace{-.9em}/\hspace{.5em}-m_\mu)\frac{1+\gamma_5
s_\mu\hspace{-.9em}/\hspace{.5em}}{2},\\
\mbox{Tr(odd \# of gamma matrices)}&=&0,\\
\gamma_5&=& \mbox{product of 4 gamma matrices},\\
\gamma_5^2&=&1,\\
\mbox{Tr}(a\hspace{-.5em}/ \hspace{.45em} b\hspace{-.5em}/
\hspace{.1em})&=&4 a\cdot b,\\
\mbox{Tr}(\gamma_5 a\hspace{-.5em}/ \hspace{.5em}
b\hspace{-.45em}/\hspace{.1em})&=&0.
\end{eqnarray*}
\item[(c)] Sum over the spin projections of the muon and integrate
with respect to the (unobserved) neutrino
\begin{displaymath}
d\omega=\frac{1}{8\pi^2}G_F^2|V_{ud}|^2 F_0^2 m_\mu^2 M_\pi
\left(1-\frac{m_\mu^2}{M_\pi^2}\right)
\int \frac{d^3 p_\mu}{E_\mu E_\nu}\delta(M_\pi-E_\mu - E_\nu).
\end{displaymath}
Make use of
\begin{displaymath}
d^3 p_\mu=p_\mu^2 dp_\mu d\Omega_\mu
\end{displaymath}
and note that the argument of the delta function implicitly depends
on $p_\mu=|\vec{p}_\mu|$.
  Moreover,
\begin{eqnarray*}
E_\nu+E_\mu&=&M_\pi,\\
E_\nu&=&|\vec{p}_\nu|=|\vec{p}_\mu|.
\end{eqnarray*}
   The final result reads
\begin{eqnarray*}
\omega=\frac{1}{\tau}&=&
\underbrace{G_F^2|V_{ud}|^2 F_0^2 4m^2_\mu M^2_\pi
\left(1-\frac{m^2_\mu}{M^2_\pi}\right)}_{\mbox{$\overline{|{\cal M}|^2}$}}
\underbrace{\frac{1}{16\pi M_\pi}\left(1-\frac{m^2_\mu}{M^2_\pi}\right)}_{
\mbox{phase space}}\\
&=&\frac{1}{4\pi}G_F^2|V_{ud}|^2 F_0^2 m_\mu^2 M_\pi
 \left(1-\frac{m^2_\mu}{M^2_\pi}\right)^2.
\end{eqnarray*}
\end{enumerate}
\end{exercise}

\section{Application at Lowest Order:
Pion-Pion Scattering}
\label{sec_pipiscat}
   Our second example deals with the prototype of a Goldstone boson reaction:
$\pi\pi$ scattering.

\begin{exercise}
\renewcommand{\labelenumi}{(\alph{enumi})}
\label{exercise_pion_pion_scattering}
\rm
   Consider the Lagrangian
\begin{displaymath}
{\cal L}_2=\frac{F^2}{4}\mbox{Tr}
\left(\partial_\mu U \partial^\mu U^\dagger\right)
+\frac{F^2}{4}\mbox{Tr}\left(\chi U^\dagger+ U \chi^\dagger\right)
\end{displaymath}
in SU(2) with
\begin{displaymath}
\chi=2B\underbrace{\left(\!\!\begin{array}{cc}
m &0\\0&m\end{array}
\!\!\right)}_{\mbox{$M$}}
\end{displaymath}
and $U$ given by
\begin{displaymath}
U(x)=\exp\left(i\frac{\phi(x)}{F}\right),\quad
\phi=\sum_{a=1}^3 \tau_a \phi_a\equiv
\left(\begin{array}{cc}
\pi^0 &\sqrt{2}\pi^+\\
\sqrt{2}\pi^-&-\pi^0
\end{array}\right).
\end{displaymath}
\begin{enumerate}
\item
   Show that ${\cal L}_2$ contains only even powers of $\phi$,
\begin{displaymath}
{\cal L}_2= {\cal L}_2^{2\phi} +{\cal L}_2^{4\phi} +\cdots.
\end{displaymath}
\item
Since ${\cal L}_2$ does not produce a three-Goldstone-boson vertex,
the scattering of two Goldstone bosons is described by a
4-Goldstone-boson contact interaction.
   Verify
$$
{\cal L}_2^{4\phi}=\frac{1}{24 F^2}\left[\mbox{Tr}(
[\phi,\partial_\mu \phi]\phi \partial^\mu \phi)
+B\mbox{Tr}(M\phi^4)\right]
$$
by using the expansion
$$
U=1+i\frac{\phi}{F}-\frac{1}{2}\frac{\phi^2}{F^2}-\frac{i}{6}
\frac{\phi^3}{F^3}+\frac{1}{24}\frac{\phi^4}{F^4}+\cdots.
$$
\noindent Remark: An analogous formula would be obtained in SU(3)
with the corresponding replacements.
\item Show that the interaction Lagrangian can be written as
\begin{displaymath}
{\cal L}^{4\pi}_2=\frac{1}{6F^2}\left(\vec{\phi}\cdot\partial_\mu\vec{\phi}
\vec{\phi}\cdot\partial^\mu\vec{\phi}
-\vec{\phi}^2\partial_\mu\vec{\phi}\cdot\partial^\mu\vec{\phi}\right)
+\frac{M^2_\pi}{24 F^2}(\vec{\phi}^2)^2,
\end{displaymath}
where $M_\pi^2=2 B m$ at ${\cal O}(p^2)$.
\item Derive the Feynman rule for
$\pi^a(p_a)+\pi^b(p_b)\to\pi^c(p_c)+\pi^d(p_d)$:
\begin{eqnarray*}
{\cal M}&=& i\left[\delta^{ab}\delta^{cd}\frac{s-M^2_\pi}{F^2}
+\delta^{ac}\delta^{bd}\frac{t-M^2_\pi}{F^2}
+\delta^{ad}\delta^{bc}\frac{u-M^2_\pi}{F^2}\right]\nonumber\\
&&-\frac{i}{3F^2}
\left(\delta^{ab}\delta^{cd}+\delta^{ac}\delta^{bd}+\delta^{ad}\delta^{bc}
\right)
\left(\Lambda_a+\Lambda_b+\Lambda_c+\Lambda_d\right),
\end{eqnarray*}
where we introduced $\Lambda_k=p_k^2-M^2_\pi$.
\item
Using four-momentum conservation, show that the so-called Mandelstam
variables $s=(p_a+p_b)^2$, $t=(p_a-p_c)^2$, and $u=(p_a-p_d)^2$ satisfy
the relation
$$s+t+u=p_a^2+p_b^2+p_c^2+p_d^2.$$
\item The $T$-matrix element (${\cal M}=iT$) of the scattering process
$\pi^a(p_a)+\pi^b(p_b)\to\pi^c(p_c)+\pi^d(p_d)$ can be parameterized as
\begin{displaymath}
T^{ab;cd}(p_a,p_b;p_c,p_d)=\delta^{ab}\delta^{cd}A(s,t,u)
                  +\delta^{ac}\delta^{bd}A(t,s,u)
                  +\delta^{ad}\delta^{bc}A(u,t,s),
\end{displaymath}
where the function $A$ satisfies
$A(s,t,u)=A(s,u,t)$.
   Since the pions form an isospin triplet, the two isovectors of both
the initial and final states may be coupled to $I=0,1,2$.
   For $m_u=m_d=m$ the strong interactions are invariant under isospin
transformations, implying that scattering matrix elements can be decomposed
as
\begin{displaymath}
\langle I',I_3'|T|I,I_3\rangle=T^I \delta_{II'}\delta_{I_3 I_3'}.
\end{displaymath}
    For the case of $\pi\pi$ scattering the three isospin amplitudes
are given in terms of the invariant amplitude $A$ by
\begin{eqnarray*}
\label{4:10:isk}
T^{I=0}&=&3A(s,t,u)+A(t,u,s)+A(u,s,t),\nonumber\\
T^{I=1}&=&A(t,u,s)-A(u,s,t),\nonumber\\
T^{I=2}&=&A(t,u,s)+A(u,s,t).
\end{eqnarray*}
   For example, the physical $\pi^+\pi^+$ scattering process is described by
$T^{I=2}$.
   Other physical processes are obtained using the appropriate Clebsch-Gordan
coefficients.
   Evaluating the $T$ matrices at threshold, one obtains the $s$-wave
$\pi\pi$-scattering lengths \footnote{The definition differs by a
factor of $(-M_\pi)$ \cite{Gasser:1983yg:7} from the conventional
definition of scattering lengths in the effective range expansion
(see, e.g., Ref.\ \cite{Preston:1962:7}).}
\begin{displaymath}
T^{I=0}|_{\rm thr}=32\pi a^0_0,\quad T^{I=2}|_{\rm thr}=32\pi a^2_0,
\end{displaymath}
   where the subscript $0$ refers to the $s$ wave and the superscript to
the isospin.
   ($T^{I=1}|_{\rm thr}$ vanishes because of Bose symmetry).
   Using the results of (d) verify the famous current-algebra prediction
for the scattering lengths
\begin{displaymath}
a_0^0=\frac{7 M_\pi^2}{32 \pi F_\pi^2}=0.156,\quad
a_0^2=-\frac{M_\pi^2}{16 \pi F_\pi^2}=-0.045,
\end{displaymath}
   where we replaced $F$ by $F_\pi$ and made use of the numerical values
$F_\pi=93.2$ MeV and $M_\pi=139.57$ MeV.

Conclusion: Given that we know the value of $F$, the Lagrangian ${\cal L}_2$
{\em predicts} the low-energy scattering amplitude.
\item
   Sometimes it is more convenient to use a different parameterization of $U$ which is
very popular in SU(2) calculations:
\begin{displaymath}
U(x)=\frac{1}{F}\left[\sigma(x)+i\vec{\tau}\cdot\vec{\pi}(x)\right],
\quad\sigma(x)=\sqrt{F^2-\vec{\pi}\,^2(x)}.
\end{displaymath}
    The fields of the two parameterizations are non-linearly related
by a field transformation,
\begin{displaymath}
\frac{\vec{\pi}}{F}=\hat{\phi}\sin\left(\frac{|\vec{\phi}|}{F}\right)
=\frac{\vec{\phi}}{F}\left(1-\frac{1}{6}\frac{\vec{\phi}\,^2}{F^2}
+\cdots\right).
\end{displaymath}
   Repeat the above steps with the new parameterization.
   Because of the equivalence theorem of field theory, the results for
observables (such as, e.g., $S$-matrix elements) do not depend on the
parameterization.
   On the other hand, intermediate building blocks such as Feynman rules with
off-mass-shell momenta depend on the parameterization chosen.
\item You may also consider the SU(3) calculation which proceeds analogously.
   Using the properties of the Gell-Mann matrices show that
\begin{eqnarray*}
{\cal L}^{4\phi}_2&=&
-\frac{1}{6 F_0^2}\phi_a\partial_\mu\phi_b\phi_c\partial^\mu \phi_d
f_{abe}f_{cde}\\
&&+\frac{(2m+m_s)B_0}{36 F_0^2}(\phi_a\phi_a)^2\\
&&
+\frac{(m-m_s)B_0}{12 \sqrt{3}F_0^2}\left(\frac{2}{3}\phi_8\phi_a\phi_b
\phi_c d_{abc}+\phi_a\phi_a\phi_b\phi_c d_{bc8}\right).
\end{eqnarray*}

\end{enumerate}

\end{exercise}

\section{Application at Lowest Order:
Compton Scattering}
\label{sec_comptonscat}
\begin{exercise}
\renewcommand{\labelenumi}{(\alph{enumi})}
\label{exercise_compton_scattering}
\rm
We will investigate the reaction
$\gamma(q) +\pi^+(p)\to\gamma(q')+\pi^+(p')$ at lowest order in the
momentum expansion [${\cal O}(p^2)$].
\begin{enumerate}
\item Consider the first term of ${\cal L}_2$ and substitute
$$r_\mu=l_\mu=-eQ {\cal A}_\mu,\quad
Q=\left(\begin{array}{rrr} \frac{2}{3}&0&0\\
0&-\frac{1}{3}&0\\
0&0&-\frac{1}{3}
\end{array}
\right),\quad e>0,\quad \frac{e^2}{4\pi}\approx\frac{1}{137},
$$
where ${\cal A}_\mu$ is a Hermitian (external) electromagnetic field.
    Show that
\begin{eqnarray*}
D_\mu U&=&\partial_\mu U+ie {\cal A}_\mu[Q,U],\\
(D^\mu U)^\dagger&=&\partial^\mu U^\dagger +ie {\cal A}^\mu[Q,U^\dagger].
\end{eqnarray*}
Using the substitution $U\leftrightarrow U^\dagger$, show that the
resulting Lagrangian consists of terms involving only even numbers of
Goldstone boson fields.
\item Insert the result of (a) into ${\cal L}_2$
and verify
\begin{eqnarray*}
\frac{F^2_0}{4}\mbox{Tr}[D_\mu U (D^\mu U)^\dagger]
&=&\frac{F^2_0}{4}\mbox{Tr}[\partial_\mu U \partial^\mu U^\dagger]\\
&&-ie {\cal A}_\mu \frac{F^2_0}{2}
\mbox{Tr}[Q(\partial^\mu U U^\dagger-U^\dagger
\partial^\mu U)]\\
&&-e^2 {\cal A}_\mu {\cal A}^\mu \frac{F^2_0}{4}\mbox{Tr}([Q,U][Q,U^\dagger]).
\end{eqnarray*}
Hint: $U\partial^\mu U^\dagger=-\partial^\mu U U^\dagger$
and $\partial^\mu U^\dagger U=-U^\dagger\partial^\mu U$.\\
   The second term describes interactions with a single photon and
the third term with two photons.
\item
Using $U=\exp(i\phi/F_0)=1+i\phi/F_0-\phi^2/(2F_0^2)+\cdots$,
identify those interaction terms
which contain exactly two Goldstone bosons:
\begin{eqnarray*}
{\cal L}_2^{A-2\phi}&=&-e {\cal A}_\mu
\frac{i}{2}\mbox{Tr}(Q[\partial^\mu\phi,\phi]),
\\
{\cal L}_2^{2A-2\phi}&=&-\frac{1}{4}e^2 {\cal A}_\mu {\cal A}^\mu \mbox{Tr}(
[Q,\phi][Q,\phi]).
\end{eqnarray*}
\item Insert $\phi$ of Exercise \ref{exercise_physical_fields}.
Verify the intermediate steps
\begin{eqnarray*}
([\partial^\mu \phi,\phi])_{11}&=&2(\partial^\mu\pi^+\pi^--\pi^+\partial^\mu
\pi^-+\partial^\mu K^+ K^--K^+\partial^\mu K^-),\\
([\partial^\mu \phi,\phi])_{22}&=&2(\partial^\mu\pi^-\pi^+-\pi^-\partial^\mu
\pi^++\partial^\mu K^0\bar{K}^0-K^0\partial^\mu\bar{K}^0),\\
([\partial^\mu \phi,\phi])_{33}&=&2(\partial^\mu K^- K^+-K^-\partial^\mu
K^++\partial^\mu\bar{K}^0 K^0-\bar{K}^0\partial^\mu K^0),\\
{[}Q,\phi]&=&\sqrt{2}\left(\begin{array}{ccc} 0&\pi^+&K^+\\
-\pi^-&0&0\\
-K^-&0&0
\end{array}
\right),\\
{[}Q,\phi][Q,\phi]&=&-2\left(\begin{array}{ccc} \pi^+\pi^-+K^+K^-&0&0\\
0&\pi^-\pi^+&\pi^-K^+\\
0&K^-\pi^+&K^-K^+
\end{array}
\right).
\end{eqnarray*}
Now show
\begin{eqnarray*}
{\cal L}_2^{A-2\phi}&=&
-{\cal A}_\mu ie(\partial^\mu\pi^+\pi^--\pi^+\partial^\mu\pi^-
+\partial^\mu K^+ K^--K^+\partial^\mu K^-),
\\
{\cal L}_2^{2A-2\phi}&=&e^2{\cal A}_\mu {\cal A}^\mu(\pi^+\pi^-+K^+ K^-).
\end{eqnarray*}
\item The corresponding
Feynman rules read
\begin{displaymath}
{\cal L}_2^{A-2\phi}\,\Rightarrow\,\mbox{vertex for}\,\,
\gamma(q,\epsilon)+\pi^\pm(p)\to\pi^\pm(p'):\, \mp ie\epsilon\cdot(p+p'),
\end{displaymath}
\begin{displaymath}
{\cal L}_2^{2A-2\phi}\,\Rightarrow\,\mbox{vertex for}\,\,
\gamma(q,\epsilon)+\pi^\pm(p)\to\gamma(q',\epsilon')+\pi^\pm(p'):\,
2ie^2\epsilon'^\ast\cdot\epsilon,
\end{displaymath}
and analogously for charged kaons.
   An internal line of momentum $p$ is described by the propagator
$i/(p^2-M^2+i0^+)$.
   Determine the Compton scattering amplitude for
$\gamma(q,\epsilon)
+\pi^+(p)\to\gamma(q',\epsilon')+\pi^+(p')$:
\begin{center}
\epsfig{file=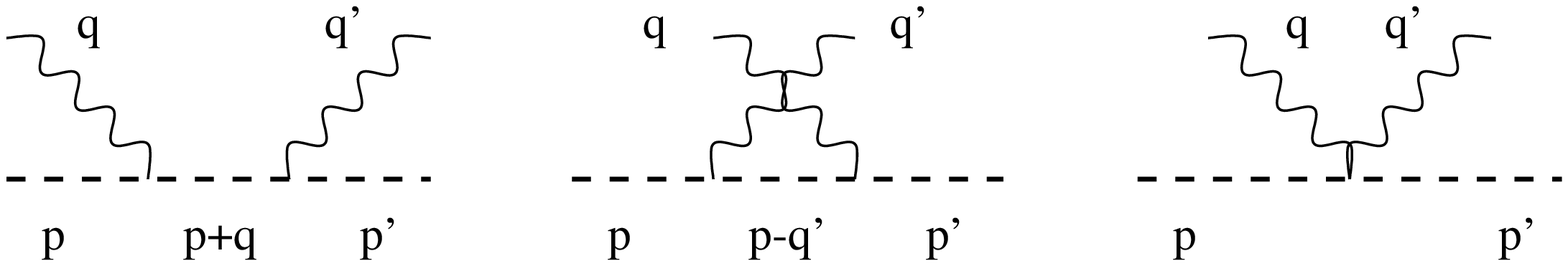,width=12cm}
\end{center}
What is the scattering amplitude for $\gamma(q,\epsilon)
+\pi^-(p)\to\gamma(q',\epsilon')+\pi^-(p')$?
\item Verify gauge invariance in terms of the substitution
$q\to\epsilon$.
\item Verify the invariance of the matrix element under
the substitution $(q,\epsilon)\leftrightarrow (-q',\epsilon'^\ast)$
(photon crossing).
\end{enumerate}

\end{exercise}

\section{The Chiral Lagrangian at Fourth Order}
\label{sec_clop4}
   Applying the ideas outlined in Section \ref{sec_cel} it is possible to
construct the most general Lagrangian at ${\cal O}(p^4)$.
   Here we only quote the result of
Gasser and Leutwyler \cite{Gasser:1984gg:9}:
\begin{eqnarray}
\label{4:8:l4gl}
\lefteqn{{\cal L}_4=
L_1 \left\{\mbox{Tr}[D_{\mu}U (D^{\mu}U)^{\dagger}] \right\}^2
+ L_2 \mbox{Tr} \left [D_{\mu}U (D_{\nu}U)^{\dagger}\right]
\mbox{Tr} \left [D^{\mu}U (D^{\nu}U)^{\dagger}\right]}\nonumber\\
& & + L_3 \mbox{Tr}\left[
D_{\mu}U (D^{\mu}U)^{\dagger}D_{\nu}U (D^{\nu}U)^{\dagger}
\right ]
+ L_4 \mbox{Tr} \left [ D_{\mu}U (D^{\mu}U)^{\dagger} \right ]
\mbox{Tr} \left( \chi U^{\dagger}+ U \chi^{\dagger} \right )
\nonumber \\
& & +L_5 \mbox{Tr} \left[ D_{\mu}U (D^{\mu}U)^{\dagger}
(\chi U^{\dagger}+ U \chi^{\dagger})\right]
+ L_6 \left[ \mbox{Tr} \left ( \chi U^{\dagger}+ U \chi^{\dagger} \right )
\right]^2
\nonumber \\
& & + L_7 \left[ \mbox{Tr} \left ( \chi U^{\dagger} - U \chi^{\dagger} \right )
\right]^2
+ L_8 \mbox{Tr} \left ( U \chi^{\dagger} U \chi^{\dagger}
+ \chi U^{\dagger} \chi U^{\dagger} \right )
\nonumber \\
& & -i L_9 \mbox{Tr} \left [ f^R_{\mu\nu} D^{\mu} U (D^{\nu} U)^{\dagger}
+ f^L_{\mu\nu} (D^{\mu} U)^{\dagger} D^{\nu} U \right ]
+ L_{10} \mbox{Tr} \left ( U f^L_{\mu\nu} U^{\dagger} f_R^{\mu\nu} \right )
\nonumber \\
& & + H_1 \mbox{Tr} \left ( f^R_{\mu\nu} f^{\mu\nu}_R +
f^L_{\mu\nu} f^{\mu\nu}_L \right )
+ H_2 \mbox{Tr} \left ( \chi \chi^{\dagger} \right ).
\end{eqnarray}
   The numerical values of the low-energy coupling constants $L_i$ are not
determined by chiral symmetry.
   In analogy to $F_0$ and $B_0$ of ${\cal L}_2$ they are parameters
containing information on the underlying dynamics and should,
in principle, be calculable in terms of the (remaining) parameters of
QCD, namely, the heavy-quark masses and the QCD scale $\Lambda_{\rm QCD}$.
   In practice, they parameterize our inability to solve the
dynamics of QCD in the non-perturbative regime.
   So far they have either been fixed using empirical input
or theoretically using QCD-inspired models, meson-resonance
saturation, and lattice QCD.

   From a practical point of view the coefficients are also required for
another purpose.
   When calculating one-loop graphs, using vertices from ${\cal L}_2$ of
Eq.\ (\ref{4:5:l2}), one generates infinities (so-called ultraviolet
divergences).
      In the framework of dimensional regularization (see below)
these divergences appear as poles at space-time dimension $n=4$.
   The loop diagrams are renormalized by absorbing the infinite parts
into the redefinition of the fields and the parameters of the most
general Lagrangian (see the end of this section and Section
\ref{sec_aop4}).
   Since ${\cal L}_2$ of Eq.\ (\ref{4:5:l2}) is not renormalizable in
the traditional sense, the infinities cannot be absorbed by a renormalization
of the coefficients $F_0$ and $B_0$.
   However, to quote from Ref.\ \cite{Weinberg:1995mt}:
   ``... the cancellation of ultraviolet divergences does
not really depend on renormalizability; as long as we include every one of the
infinite number of interactions allowed by symmetries, the so-called
non-renormalizable theories are actually just as renormalizable as renormalizable
theories.''
   According to Weinberg's power counting of Eq.\ (\ref{4:4:mr2}),
one loop graphs with vertices from ${\cal L}_2$ are of ${\cal O}(p^4)$.
   The conclusion is that one needs to adjust (renormalize) the parameters
of ${\cal L}_4$ to cancel one-loop infinities.

   By construction Eq.\ (\ref{4:8:l4gl}) represents the most general
Lagrangian at ${\cal O}(p^4)$, and it is thus possible to absorb the one-loop
divergences by an appropriate renormalization of the coefficients
$L_i$ and $H_i$:
\begin{eqnarray}
\label{4:8:li}
L_i&=&L_i^r+\frac{\Gamma_i}{32\pi^2}R, \quad i=1,\cdots,10,\\
\label{4:8:hi}
H_i&=&H^r_i+\frac{\Delta_i}{32\pi^2}R,\quad i=1,2,
\end{eqnarray}
   where $R$ is defined as
\begin{equation}
\label{4:8:R}
R=\frac{2}{n-4}-[\mbox{ln}(4\pi)-\gamma_E+1],
\end{equation}
   with $n$ denoting the number of space-time dimensions and
$\gamma_E=-\Gamma'(1)$ being Euler's constant.
   The constants $\Gamma_i$ and $\Delta_i$ are given in Table
\ref{4:8:tableli}.
   Except for $L_3$ and $L_7$, the low-energy coupling constants $L_i$ and the
``contact terms''---i.e., pure external field terms---$H_1$ and $H_2$
are required in the renormalization of the one-loop graphs.
   Since $H_1$ and $H_2$ contain only external fields, they are of no
physical relevance.
   The renormalized coefficients $L_i^r$ depend on the scale $\mu$
introduced by dimensional regularization [see Eq.\ (\ref{app:drb:im22})] and
their values at two different scales $\mu_1$ and $\mu_2$
are related by
\begin{equation}
\label{4:8:limu1mu2}
L^r_i(\mu_2)=L^r_i(\mu_1)
+\frac{\Gamma_i}{16\pi^2}\ln\left(\frac{\mu_1}{\mu_2}\right).
\end{equation}
   We will see that the scale dependence of the coefficients and
the finite part of the loop-diagrams compensate each other in
such a way that physical observables are scale independent.

\begin{table}
\begin{center}
\begin{tabular}{|c|r|r|}
\hline
Coefficient &Empirical Value &$\Gamma_i$\\
\hline
$L_1^r$ &    $ 0.4\pm 0.3$  &$\frac{3}{32}$\\
$L_2^r$ &    $ 1.35\pm 0.3$ &$\frac{3}{16}$\\
$L_3^r$ &    $-3.5\pm 1.1$  &$0$\\
$L_4^r$ &    $-0.3\pm 0.5$  &$\frac{1}{8}$\\
$L_5^r$ &    $ 1.4\pm 0.5$  &$\frac{3}{8}$\\
$L_6^r$ &    $-0.2\pm 0.3$  &$\frac{11}{144}$\\
$L_7^r$ &    $-0.4\pm 0.2$  &$0$\\
$L_8^r$ &    $ 0.9\pm 0.3$  &$\frac{5}{48}$\\
$L_9^r$ &    $ 6.9\pm 0.7$  &$\frac{1}{4}$\\
$L_{10}^r$ & $-5.5\pm 0.7$  &$-\frac{1}{4}$\\
\hline
\end{tabular}
\end{center}
\caption[test]{\label{4:8:tableli} Renormalized low-energy
coupling constants $L_i^r$ in units of $10^{-3}$ at the scale
$\mu=M_\rho$, see J.\ Bijnens, G.\ Ecker, and J. Gasser, {\em The
Second DA$\Phi$NE Physics Handbook}, Vol.\ I, Chapter 3.
$\Delta_1=-1/8$, $\Delta_2=5/24$. }
\end{table}

\section{Brief Introduction to Dimensional Regularization}
\label{sec_bidr}

   For the sake of completeness we provide a simple illustration of
the method of dimensional regularization.

   Let us consider the integral
\begin{equation}
\label{app:drb:int}
I=\int\frac{d^4k}{(2\pi)^4}\frac{i}{k^2-M^2+i0^+}.
\end{equation}
   We introduce
$$a\equiv\sqrt{\vec{k}^2+M^2}>0$$
so that
\begin{eqnarray*}
k^2-M^2+i0^+
&=&k_0^2-\vec{k}^2-M^2+i0^+\\
&=&k_0^2-a^2+i0^+\\
&=&k_0^2-(a-i0^+)^2\\
&=&
[k_0+(a-i0^+)][k_0-(a-i0^+)],
\end{eqnarray*}
   and define
$$f(k_0)=\frac{1}{[k_0+(a-i0^+)][k_0-(a-i0^+)]}.$$
   In order to determine $\int_{-\infty}^{\infty} dk_0 f(k_0)$ as part of
the calculation of $I$, we consider
$f$ in the complex $k_0$ plane and make use of Cauchy's theorem
\begin{equation}
\label{app:drb:cauchy}
\oint_C dz f(z)=0
\end{equation}
for functions which are differentiable in every point inside the closed
contour $C$.
\begin{figure}
\begin{center}
\epsfig{file=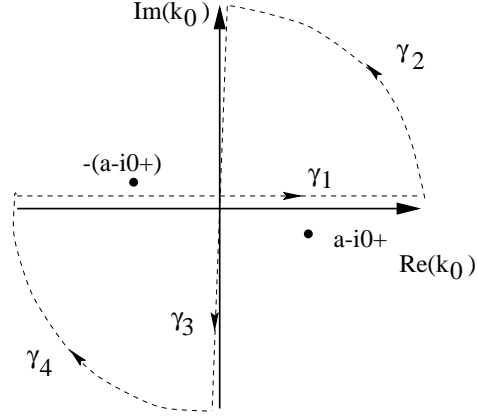,width=6cm}
\caption{\label{app:drb:wickrotation_fig}
Path of integration in the complex $k_0$ plane.}
\end{center}
\end{figure}
   We choose the contour as shown in Figure \ref{app:drb:wickrotation_fig},
$$
0=\sum_{i=1}^4 \int_{\gamma_i} dz f(z),
$$
   and make use of
$$\int_\gamma f(z)dz=\int_a^b f[\gamma(t)]\gamma'(t)dt
$$
to obtain for the individual integrals:
$$\gamma_1(t)=t,\,\,\gamma_1'(t)=1,\,\,a=-\infty,\,\,
b=\infty:\quad
\int_{\gamma_1} f(z) dz= \int_{-\infty}^\infty f(t)dt,
$$
$$
\gamma_2(t)=R e^{it},\,\,\gamma_2'(t)=iR e^{it},\,\,a=0,\,\,
b=\frac{\pi}{2}:$$
$$\int_{\gamma_2} f(z) dz=
\lim_{R\to\infty} \int_{0}^\frac{\pi}{2} f(Re^{it})iRe^{it}dt =0,
\,\,\mbox{because}\,\,
\lim_{R\to\infty} \underbrace{Rf(Re^{it})}_{\mbox{$\sim \frac{1}{R}$}}=0,
$$
$$\gamma_3(t)=it,\,\,\gamma_3'(t)=i,\,\,a=+\infty,\,\,
b=-\infty:\quad
\int_{\gamma_3} f(z) dz= \int_{\infty}^{-\infty} f(it)idt,
$$

$$
\gamma_4(t)=R e^{it},\,\,\gamma_4'(t)=iR e^{it},\,\,a=\frac{3}{2}\pi,\,\,
b=\pi:$$
$$\int_{\gamma_4} f(z) dz=0\,\,\mbox{analogous to  $\gamma_2$}.
$$
   In combination with Eq.\ (\ref{app:drb:cauchy}) we obtain the so-called
Wick rotation
\begin{equation}
\label{app:drb:wickrotation}
\int_{-\infty}^\infty f(t)dt=-i\int_{\infty}^{-\infty}dt f(it)
=i\int_{-\infty}^\infty dt f(it).
\end{equation}
   As an intermediate result the integral of
Eq.\ (\ref{app:drb:int}) reads
$$I=\frac{1}{(2\pi)^4}i\int_{-\infty}^\infty dk_0\int d^3 k
\frac{i}{(ik_0)^2-\vec{k}^2 -M^2 +i0^+}
=\int \frac{d^4 l}{(2\pi)^4} \frac{1}{l^2+M^2-i0^+},
$$
   where $l^2=l_1^2+l_2^2+l_3^2+l_4^2$ denotes a Euclidian scalar
product.
   In this {\em special} case, the integrand does not have
a pole and we can thus omit the $-i0^+$ which gave the positions of the
poles in the original integral consistent with the boundary conditions.
   Introducing polar coordinates in 4 dimensions, $d^4l=d\Omega
   l^3dl$, we see that the integral diverges.
   The degree of divergence can be estimated by simply counting the powers of
momenta.
   If the integral behaves asymptotically as $\int d^4 l /l^2$,
$\int d^4 l /l^3$, $\int d^4 l /l^4$
the integral is said to diverge quadratically, linearly, and
logarithmically, respectively.
   Thus, our example $I$ diverges quadratically.

   Various methods have been devised to regularize divergent integrals.
   We will make use of {\em dimensional} regularization,  because it preserves
algebraic relations among Green functions (Ward identities) if the
underlying symmetries do not depend on the number of dimensions of
space-time.

   In dimensional regularization, we generalize the integral from 4 to $n$
dimensions and introduce polar coordinates
\begin{eqnarray}
\label{app:drb:polkoord}
l_1&=& l\cos(\theta_1),\nonumber\\
l_2&=& l\sin(\theta_1)\cos(\theta_2),\nonumber\\
l_3&=& l\sin(\theta_1)\sin(\theta_2)\cos(\theta_3),\nonumber\\
&\vdots&\nonumber\\
l_{n-1}&=&l\sin(\theta_1)\sin(\theta_2)\cdots\cos(\theta_{n-1}),\nonumber\\
l_{n}&=&l\sin(\theta_1)\sin(\theta_2)\cdots\sin(\theta_{n-1}),
\end{eqnarray}
where $0\leq l$, $\theta_i\in[0,\pi], i=1,\cdots,n-2$, $\theta_{n-1}\in
[0,2\pi]$.
   A general integral is then symbolically of the form
\begin{equation}
\label{app:drb:volumenelement}
\int d^n l\cdots = \int_0^\infty l^{n-1}dl\int_0^{2\pi}d\theta_{n-1}
\int_0^\pi d\theta_{n-2}\sin(\theta_{n-2})\cdots\int_0^\pi d\theta_1
\sin^{n-2}(\theta_1)\cdots .
\end{equation}
   If the integrand does not depend on the angles, the angular integration
can explicitly be carried out.
   To that end one makes use of
$$
\int_0^\pi \sin^m(\theta) d\theta=\frac{\sqrt{\pi} \Gamma\left(\frac{m+1}{2}
\right)}{\Gamma\left(\frac{m+2}{2}\right)}
$$
which can be shown by induction (see Exercise
\ref{exercise_induction}).
   We then obtain for the angular integration
\begin{eqnarray}
\label{app:drb:winkelintegration}
\int_0^{2\pi}d\theta_{n-1}\cdots \int_0^\pi d\theta_1
\sin^{n-2}(\theta_1)&=&
2\pi\underbrace{\frac{\sqrt{\pi}\Gamma(1)}{\Gamma\left(\frac{3}{2}\right)}
\frac{\sqrt{\pi}\Gamma\left(\frac{3}{2}\right)}{\Gamma(2)}\cdots
\frac{\sqrt{\pi}\Gamma\left(\frac{n-1}{2}\right)}{\Gamma\left(\frac{n}{2}
\right)}}_{\mbox{$(n-2)$ factors}}\nonumber\\
&=&
2\pi \frac{\pi^{\frac{n-2}{2}}}{\Gamma\left(\frac{n}{2}\right)}
=2\frac{\pi^\frac{n}{2}}{\Gamma\left(\frac{n}{2}\right)}.
\end{eqnarray}
   We define the integral for $n$ dimensions ($n$ integer) as
\begin{equation}
\label{app:drb:im2}
I_n(M^2,\mu^2)=\mu^{4-n}\int\frac{d^nk}{(2\pi)^n}\frac{i}{k^2-M^2+i0^+},
\end{equation}
   where for convenience we have introduced the renormalization scale $\mu$
so that the integral has the same dimension for arbitrary $n$.
   (The integral of Eq.\ (\ref{app:drb:im2}) is convergent only for
$n=1$.)
   After the Wick rotation of Eq.\ (\ref{app:drb:wickrotation}) and
the angular integration of Eq.\ (\ref{app:drb:winkelintegration}) the
integral formally reads
$$
I_n(M^2,\mu^2)=\mu^{4-n}2\frac{\pi^\frac{n}{2}}{\Gamma\left(\frac{n}{2}\right)}
\frac{1}{(2\pi)^n}
\int_0^\infty dl \frac{l^{n-1}}{l^2+M^2}.
$$
  For later use, we investigate the (more general) integral
\begin{equation}
\label{app:drb:mgint}
\int_0^\infty \frac{l^{n-1}dl}{(l^2+M^2)^\alpha}
=\frac{1}{(M^2)^\alpha}\int_0^\infty \frac{l^{n-1}dl}{(\frac{l^2}{M^2}+1
)^\alpha}
=\frac{1}{2}(M^2)^{\frac{n}{2}-\alpha} \int_0^\infty \frac{
t^{\frac{n}{2}-1}dt}{(t+1)^\alpha},
\end{equation}
where we made use of the substitution $t\equiv l^2/M^2$.
   We then make use of the Beta function
\begin{equation}
\label{app:drb:betafunktion}
B(x,y)=\int_0^\infty \frac{t^{x-1}dt}{(1+t)^{x+y}}=\frac{\Gamma(x)\Gamma(y)}{
\Gamma(x+y)},
\end{equation}
where the {\em integral} converges for $x>0$, $y>0$ and diverges if
$x\leq 0$ or $y\leq 0$.
   For non-positive values of $x$ or $y$ we make use of the analytic
continuation in terms of the Gamma function to define the Beta
function and thus the integral of Eq.\ (\ref{app:drb:mgint}).\footnote{
Recall that
$\Gamma(z)$ is single valued and analytic over the entire complex plane, save
for the points $z=-n$, $n=0,1,2,\cdots$, where it possesses simple poles with
residue $(-1)^n/n!$.}
   Putting $x=n/2$, $x+y=\alpha$ and $y=\alpha-n/2$
our (intermediate) integral reads
\begin{equation}
\label{app:drb:allgint}
\int_0^\infty \frac{l^{n-1}dl}{(l^2+M^2)^\alpha}=
\frac{1}{2}(M^2)^{\frac{n}{2}-\alpha}\frac{\Gamma\left(\frac{n}{2}\right)
\Gamma\left(\alpha-\frac{n}{2}\right)}{\Gamma(\alpha)}
\end{equation}
which, for $\alpha=1$, yields for our original integral
\begin{eqnarray}
\label{app:drb:iint}
I_n(M^2,\mu^2)&=&\mu^{4-n}\underbrace{2\frac{\pi^\frac{n}{2}}{\Gamma\left(
\frac{n}{2}\right)}}_{\mbox{angular integration}}
\frac{1}{(2\pi)^n} \frac{1}{2} (M^2)^{\frac{n}{2}-1}\frac{\Gamma
\left(\frac{n}{2}\right)\Gamma\left(1-\frac{n}{2}\right)}{
\underbrace{\Gamma(1)}_{\mbox{1}}}\nonumber\\
&=&\frac{\mu^{4-n}}{(4\pi)^{\frac{n}{2}}} (M^2)^{\frac{n}{2}-1}
\Gamma\left(1-\frac{n}{2}\right).
\end{eqnarray}
   Since $\Gamma(z)$ is an analytic function in the complex plane except
for poles of first order in $0,-1,-2,\cdots$, and
$a^z=\exp[\ln(a)z]$, $a\in R^+$ is an analytic function in $C$,
the right-hand side of Eq.\ (\ref{app:drb:iint}) can be thought of
as a function of a {\em complex} variable $n$ which is
analytic in $C$ except for poles of first order
for $n=2,4,6,\cdots$.
   Making use of
$$\mu^{4-n}=(\mu^2)^{2-\frac{n}{2}},\quad
(M^2)^{\frac{n}{2}-1}=M^2 (M^2)^{\frac{n}{2}-2},
\quad
(4\pi)^\frac{n}{2}=(4\pi)^2(4\pi)^{\frac{n}{2}-2},
$$
we define (for complex n)
$$
I(M^2,\mu^2,n)=\frac{M^2}{(4\pi)^2}\left(\frac{4\pi\mu^2}{M^2}\right)^{2-
\frac{n}{2}} \Gamma\left(1-\frac{n}{2}\right).
$$
   Of course, for $n\to 4$ the Gamma function has a pole and we want
to investigate how this pole is approached.
   The property $\Gamma(z+1)=z\Gamma(z)$ allows one to rewrite
\begin{displaymath}
\Gamma\left(1-\frac{n}{2}\right)=
\frac{\Gamma\left(1-\frac{n}{2}+1\right)}{1-\frac{n}{2}}
=\frac{\Gamma\left(2-\frac{n}{2}+1\right)}{\left(1-\frac{n}{2}\right)
\left(2-\frac{n}{2}\right)}=\frac{\Gamma\left(1+\frac{\epsilon}{2}\right)}{
(-1)\left(1-\frac{\epsilon}{2}\right)\frac{\epsilon}{2}},
\end{displaymath}
where we defined $\epsilon\equiv 4-n$.
   Making use of $a^x=\exp[\ln(a)x]=1+\ln(a)x+O(x^2)$ we expand
the integral for small $\epsilon$
\begin{eqnarray*}
I(M^2,\mu^2,n)
&=&\frac{M^2}{16\pi^2}\left[1+\frac{\epsilon}{2}\ln\left(
\frac{4\pi\mu^2}{M^2}\right)+O(\epsilon^2)\right]\\
&&\times
\left(-\frac{2}{\epsilon}\right)\left[1+\frac{\epsilon}{2}+O(\epsilon^2)\right]
\left[\underbrace{\Gamma(1)}_{\mbox{1}}
+\frac{\epsilon}{2}\Gamma'(1)+O(\epsilon^2)\right]
\\
&=&\frac{M^2}{16\pi^2}\left[-\frac{2}{\epsilon}-\Gamma'(1)
-1-\ln(4\pi)+\ln\left(\frac{M^2}{\mu^2}\right)
+O(\epsilon)\right],
\end{eqnarray*}
   where $-\Gamma'(1)=\gamma_E=0.5772\cdots$ is Euler's constant.
   We finally obtain
  \begin{equation}
\label{app:drb:im22}
I(M^2,\mu^2,n)=\frac{M^2}{16\pi^2}\left[
R+\ln\left(\frac{M^2}{\mu^2}\right)\right]+O(n-4),
\end{equation}
where
\begin{equation}
\label{app:constantR}
R=\frac{2}{n-4}-[\mbox{ln}(4\pi)+\Gamma'(1)+1].
\end{equation}

    Using the result of Eq.\ (\ref{app:drb:im22}), we are now in a position
to motivate why we assign the scale $4\pi F_0$ to the parameter that
characterizes the convergence of the momentum and quark-mass expansion.
   In a loop correction every endpoint of an internal line is multiplied
by a factor $1/F_0$, since the SU($N$) matrix of Eq.\ (\ref{4:3:upar}) contains
the Goldstone boson fields in the combination $\phi/F_0$.
   On the other hand, external momenta $q$ or Goldstone boson masses produce
factors of $q^2$ or $M^2$ as, e.g., in Eq.\ (\ref{app:drb:im22}) such that they
combine to corrections of the order of $[q/(4\pi F_0)]^2$ for each independent
loop.

   Using the same techniques one can easily derive a very useful expression
for the more general integral (see Exercise
\ref{exercise_general_integral})
\begin{eqnarray}
\label{app:drb:moregenint}
\lefteqn{\int \frac{d^n k}{(2\pi)^n} \frac{(k^2)^p}{(k^2-M^2+i0^+)^q}=}
\nonumber\\
&&i(-)^{p-q} \frac{1}{(4\pi)^{\frac{n}{2}}}(M^2)^{p+\frac{n}{2}-q}
\frac{\Gamma\left(p+\frac{n}{2}\right)\Gamma\left(q-p-\frac{n}{2}\right)}{
\Gamma\left(\frac{n}{2}\right)\Gamma(q)}.
\end{eqnarray}

  In the context of combining propagators by using Feynman's trick
one encounters integrals of the type of Eq.\
(\ref{app:drb:moregenint})
with $M^2$ replaced by $A-i0^+$, where $A$ is a real number.
   In this context it is important to consistently deal with the
boundary condition $-i0^+$.
   For example, let us consider a term of the type
$\ln(A-i0^+)$.
   To that end one expresses a complex number $z$ in its polar form
$z=|z|\exp(i\varphi),$
where the argument $\varphi$ of $z$ is uniquely determined if, in addition,
we demand $-\pi\leq\varphi < \pi$.
   For $A>0$ one simply has $\ln(A-i0^+)=\ln(A)$.
   For $A<0$ the infinitesimal imaginary part indicates that  $-|A|$
is reached in the third quadrant from below the real axis
so that we have to use the $-\pi$.
   We then make use of $\ln(ab)=\ln(a)+\ln(b)$ and obtain
$$
\ln(A-i0^+)=\ln(|A|)+\ln(e^{-i\pi})=\ln(|A|)-i\pi,\quad
A<0.
$$
   Both cases can be summarized in a single expression
\begin{equation}
\label{app:drb:lna}
\ln(A-i0^+)=\ln(|A|)-i\pi\Theta(-A)\quad\mbox{for}\,A\in R.
\end{equation}
   The preceding discussion is of importance for consistently determining
imaginary parts of loop integrals.

   Let us conclude with the general observation that
(ultraviolet) divergences of
one-loop integrals in dimensional regularization always show up
as single poles in $\epsilon=4-n$.

The following 5 exercises are related to dimensional regularization.

\begin{exercise}
\label{exercise_standard_integral}
\rm
We consider the integral
\begin{displaymath}
I=\int\frac{d^4 k}{(2\pi)^4}\frac{i}{k^2-M^2+i0^+}.
\end{displaymath}
Introduce $a\equiv\sqrt{\vec{k}^2+M^2}$ and define
\begin{displaymath}
f(k_0)=\frac{1}{[k_0+(a-i0^+)][k_0-(a-i0^+)]}.
\end{displaymath}
   In order to determine $\int_{-\infty}^{\infty} dk_0 f(k_0)$ as part of
the calculation of $I$, we consider
$f$ in the complex $k_0$ plane and choose the paths
\begin{displaymath}
\gamma_1(t)=t,\quad t_1=-\infty,\quad t_2=+\infty\quad\mbox{and}\quad
\gamma_2(t)=R e^{it},\quad t_1=0,\quad t_2=\pi.
\end{displaymath}
\renewcommand{\labelenumi}{(\alph{enumi})}
\begin{enumerate}
\item Using the residue theorem determine
\begin{displaymath}
\oint_C f(z)dz =\int_{\gamma_1} f(z) dz+\lim_{R\to \infty}\int_{\gamma_2}
f(z)dz=2\pi i \mbox{Res}[f(z),-(a+i0^+)].
\end{displaymath}
Verify
\begin{displaymath}
\int_{-\infty}^\infty dk_0 f(k_0)=\frac{-i\pi}{\sqrt{\vec{k}^2+M^2}-i0^+}.
\end{displaymath}
\item Using (a) show
\begin{displaymath}
\int\frac{d^4 k}{(2\pi)^4}\frac{i}{k^2-M^2+i0^+}
=\frac{1}{2}\int \frac{d^3 k}{(2\pi)^3}\frac{1}{\sqrt{\vec{k}^2+M^2}-i0^+}.
\end{displaymath}
\item Now consider the generalization from $4 \to n$
dimensions:
\begin{displaymath}
\int \frac{d^{n-1} k}{(2\pi)^{n-1}}\frac{1}{\sqrt{\vec{k}^2+M^2}},
\quad \vec{k}^2=k_1^2+k_2^2+\cdots+ k_{n-1}^2.
\end{displaymath}
   We can omit the $-i0^+$, because the integrand no longer has a pole.
   Introduce polar coordinates in $n-1$ dimensions and perform the angular
integration to obtain
\begin{displaymath}
\int \frac{d^{n-1} k}{(2\pi)^{n-1}}\frac{1}{\sqrt{\vec{k}^2+M^2}}
=\frac{1}{2^{n-2}}\pi^{-\frac{n-1}{2}}
\frac{1}{\Gamma\left(\frac{n-1}{2}\right)}
\int_0^\infty dr \frac{r^{n-2}}{\sqrt{r^2+M^2}}.
\end{displaymath}
\item Using the substitutions $t=r/M$ and $y=t^2$ show
\begin{displaymath}
\int_0^\infty dr \frac{r^{n-2}}{\sqrt{r^2+M^2}}
=\frac{1}{2} M^{n-2}\frac{\Gamma\left(\frac{n-1}{2}\right)
\Gamma\left(1-\frac{n}{2}\right)}{\underbrace{\Gamma\left(\frac{1}{2}\right)}_{
\sqrt{\pi}}}.
\end{displaymath}
Hint: Make use of the Beta function
\begin{displaymath}
B(x,y)=\int_0^\infty \frac{t^{x-1}dt}{(1+t)^{x+y}}=\frac{\Gamma(x)\Gamma(y)}{
\Gamma(x+y)}.
\end{displaymath}
\item Now put the results together to obtain
\begin{displaymath}
\int \frac{d^n k}{(2\pi)^n}\frac{i}{k^2-M^2+i0^+}=
\frac{1}{(4\pi)^\frac{n}{2}}M^{n-2}\Gamma\left(1-\frac{n}{2}\right),
\end{displaymath}
which agrees with the result of the lecture.
\end{enumerate}
\end{exercise}
\begin{exercise}
\label{exercise_polar_coordinates}
\rm
Consider polar coordinates in 4 dimensions:
\begin{eqnarray*}
l_1&=&l\cos(\theta_1),\quad \theta_1\in[0,\pi],\\
l_2&=&l\sin(\theta_1)\cos(\theta_2),\quad \theta_2\in[0,\pi],\\
l_3&=&l\sin(\theta_1)\sin(\theta_2)\cos(\theta_3),\quad \theta_3\in[0,2\pi],\\
l_4&=&l\sin(\theta_1)\sin(\theta_2)\sin(\theta_3),
\end{eqnarray*}
where $l=\sqrt{l_1^2+l_2^2+l_3^2+l_4^2}$.
   The transition from four-dimensional Cartesian coordinates to polar
coordinates introduces the determinant of the Jacobi or functional matrix
\begin{displaymath}
J=\left(
\begin{array}{ccc}
\frac{\partial l_1}{\partial l}&\cdots&\frac{\partial l_1}{\partial \theta_3}\\
\vdots  &&\vdots\\
\frac{\partial l_4}{\partial l}&\cdots&\frac{\partial l_4}{\partial \theta_3}
\end{array}
\right).
\end{displaymath}
   Show that
\begin{displaymath}
\mbox{det}(J)=l^3\sin^2(\theta_1)\sin(\theta_2)
\end{displaymath}
and thus
\begin{displaymath}
dl_1 dl_2 dl_3 dl_4=l^3dl\underbrace{\sin^2(\theta_1)\sin(\theta_2)d\theta_1
d\theta_2d\theta_3}_{d\Omega}
\end{displaymath}
with
\begin{displaymath}
\int d \Omega=2\pi^2.
\end{displaymath}
\end{exercise}

\begin{exercise}
\label{exercise_induction}
\rm
Show by induction
$$
\int_0^\pi \sin^m(\theta) d\theta=\frac{\sqrt{\pi} \Gamma\left(\frac{m+1}{2}
\right)}{\Gamma\left(\frac{m+2}{2}\right)}
$$
for $m \geq 1$.

Hints:
Make use of partial integration. $\Gamma(1)=1$, $\Gamma(1/2)=\sqrt{\pi}$,
$x\Gamma(x)=\Gamma(x+1)$.

\end{exercise}

\begin{exercise}
\label{exercise_general_integral}
\rm
\renewcommand{\labelenumi}{(\alph{enumi})}
Show that in dimensional regularization
\begin{eqnarray*}
\lefteqn{\int \frac{d^n k}{(2\pi)^n} \frac{(k^2)^p}{(k^2-M^2+i0^+)^q}=}
\nonumber\\
&&i(-)^{p-q} \frac{1}{(4\pi)^{\frac{n}{2}}}(M^2)^{p+\frac{n}{2}-q}
\frac{\Gamma\left(p+\frac{n}{2}\right)\Gamma\left(q-p-\frac{n}{2}\right)}{
\Gamma\left(\frac{n}{2}\right)\Gamma(q)}.
\end{eqnarray*}
   We first assume $M^2>0$, $p=0,1,\cdots$, $q=1,2,\cdots$, and $p<q$.
   The last condition is used in the Wick rotation to guarantee that the
quarter circles at infinity do not contribute to the integral.
\begin{enumerate}
\item Show that the transition to the Euclidian metric produces
the factor $i(-)^{p-q}$. \item Perform the angular integration in
$n$ dimensions.
   Perform the remaining radial integration using
\begin{displaymath}
\int_0^\infty \frac{l^{n-1}dl}{(l^2+M^2)^\alpha}=
\frac{1}{2}(M^2)^{\frac{n}{2}-\alpha}\frac{\Gamma\left(\frac{n}{2}\right)
\Gamma\left(\alpha-\frac{n}{2}\right)}{\Gamma(\alpha)}.
\end{displaymath}
   What do you have to substitute for $n-1$ and $\alpha$, respectively?
\end{enumerate}
Now put the results together.
The analytic continuation of the right-hand side is used to also define
expressions with (integer) $q\leq p$ in dimensional regularization.
\end{exercise}

\begin{exercise}
\label{exercise_cr} \rm Consider the complex function
\renewcommand{\labelenumi}{(\alph{enumi})}
\begin{displaymath}
f(z)=a^z=\exp(\ln(a)z)\equiv u(x,y)+iv(x,y),\quad a\in R,\quad
z=x+iy.
\end{displaymath}
\begin{enumerate}
\item Determine $u(x,y)$ and $v(x,y)$. Note that
$u,v\in C^\infty(R^2)$.
\item Determine $\partial u/\partial x$,
$\partial u/\partial y$, $\partial v/\partial x$, and $\partial v/\partial y$.
Show that the Cauchy-Riemann differential equations $\partial u/\partial x
=\partial v/\partial y$ and $\partial u/\partial y=-\partial v/\partial x$
are satisfied. The complex function $f$ is thus holomorphic in $C$.
We made use of this fact when discussing $I(M^2,\mu^2,n)$ as a function
of the complex variable $n$ in the context of dimensional regularization.
\end{enumerate}
\end{exercise}

\section{Application at Fourth Order:
Masses of the Goldstone Bosons}
\label{sec_aop4}
   A discussion of the masses at ${\cal O}(p^4)$ will allow us to illustrate
various properties typical of chiral perturbation theory:
\begin{enumerate}
\item The relation between the bare low-energy coupling constants $L_i$ and
the renormalized coefficients $L_i^r$ in Eq.~(\ref{4:8:li}) is such that
the divergences of one-loop diagrams are canceled.
\item Similarly, the scale dependence of the coefficients $L^r_i(\mu)$ on the
one hand and of the finite contributions of the one-loop diagrams on the other
hand lead to scale-independent predictions for physical observables.
\item A perturbation expansion in the explicit symmetry breaking
with respect to a symmetry that is realized in the Nambu-Goldstone mode
generates corrections which are non-analytic in the symmetry breaking
parameter, here the quark masses.
\end{enumerate}
   Let us consider ${\cal L}_2 + {\cal L}_4$ for QCD with finite quark
masses but in the absence of external fields.
   We restrict ourselves to the limit of isospin symmetry, i.e.,
$m_u=m_d=m$.
   In order to determine the masses we calculate the so-called self energies
$\Sigma(p^2)$ of the Goldstone bosons.

   The propagator of a (pseudo-) scalar field is defined as the
Fourier transform of the two-point Green function:
\begin{equation}
\label{4:9:propdef}
i\Delta(p)=\int d^4 x e^{-ip\cdot x}\langle
0|T\left[\Phi_0(x)\Phi_0(0)\right]|0\rangle,
\end{equation}
   where the index 0  refers to the fact that we still deal with the
bare unrenormalized field---not to be confused with a free field without
interaction.
   At lowest order ($D=2$) the propagator simply reads
\begin{equation}
\label{4:9:prop}
i\Delta(p)=\frac{i}{p^2-M^2_0+i0^+},
\end{equation}
where the lowest-order masses $M_0$ are given in
Eqs.\ (\ref{4:3:mpi2}) - (\ref{4:3:meta2}):
\begin{eqnarray*}
M^2_{\pi,2}&=&2 B_0 m,\\
M^2_{K,2}&=&B_0(m+m_s),\\
M^2_{\eta,2}&=&\frac{2}{3} B_0\left(m+2m_s\right).
\end{eqnarray*}
   (The subsrcipt 2 refers to chiral order 2.)
   The loop diagrams with ${\cal L}_2$ and the contact diagrams with
${\cal L}_4$ result in so-called proper self-energy insertions
$-i\Sigma(p^2)$, which may be summed using a geometric series (see
Figure \ref{4:9:fullprop}):
\begin{figure}
\begin{center}
\epsfig{file=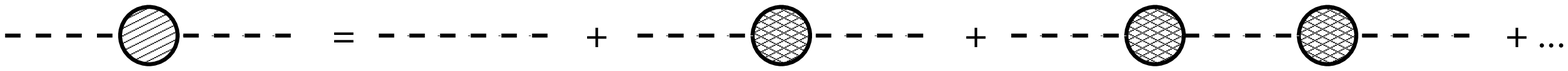,width=12cm}
\caption{\label{4:9:fullprop}
Unrenormalized propagator as the sum of irreducible self-energy diagrams.
Hatched and cross-hatched ``vertices'' denote one-particle-reducible
and one-particle-irreducible contributions, respectively.}
\end{center}
\end{figure}
\begin{eqnarray}
\label{4:9:prop1}
i\Delta(p)&=&\frac{i}{p^2-M^2_0+i0^+}+\frac{i}{p^2-M^2_0+i0^+}
[-i\Sigma(p^2)]\frac{i}{p^2-M^2_0+i0^+}+\cdots\nonumber\\
&=&\frac{i}{p^2-M^2_0+i0^+}\frac{1}{1+i\Sigma(p^2) \frac{i}{p^2-M^2_0
+i0^+}}\nonumber\\
&=&\frac{i}{p^2-M^2_0-\Sigma(p^2)+i0^+}.
\end{eqnarray}
   Note that $-i\Sigma(p^2)$ consists of one-particle-irreducible
diagrams only, i.e., diagrams which do not fall apart into
two separate pieces when cutting an arbitrary internal line.
   The physical mass, including the interaction, is defined as the position
of the pole of Eq.\ (\ref{4:9:prop1}),
\begin{equation}
\label{4:9:mdef}
M^2-M^2_0-\Sigma(M^2)\stackrel{!}{=}0.
\end{equation}

\begin{figure}
\begin{center}
\epsfig{file=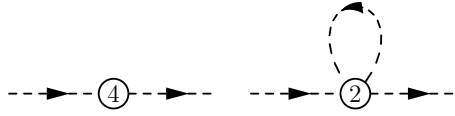,width=6cm}
\caption{\label{4:9:selfenergy}
Self-energy diagrams at ${\cal O}(p^4)$.
   Vertices derived from ${\cal L}_{2n}$ are denoted by $2n$ in the
interaction blobs.}
\end{center}
\end{figure}

   Let us now turn to the calculation within the framework of ChPT.
   Since ${\cal L}_2$ and ${\cal L}_4$ without external fields generate
vertices with an even number of Goldstone bosons only, the
candidate terms at $D=4$ contributing to the self energy are those
shown in Figure \ref{4:9:selfenergy}.
   For our particular application with exactly two external meson lines,
the relevant interaction Lagrangians can be written as
\begin{equation}
{\cal L}_{\rm int}={\cal L}_2^{4\phi}+{\cal L}_4^{2\phi},
\end{equation}
   where ${\cal L}_2^{4\phi}$ is given by
\begin{equation}
\label{4:9:l24phi}
{\cal L}^{4\phi}_2=\frac{1}{24 F^2_0}\left\{\mbox{Tr}(
[\phi,\partial_\mu \phi]\phi \partial^\mu \phi)
+B_0\mbox{Tr}(M\phi^4)\right\}.
\end{equation}
   The terms of ${\cal L}_4$ proportional to  $L_9$, $L_{10}$, $H_1$, and
$H_2$ do not contribute, because they either contain field-strength tensors or
external fields only.
   Since $\partial_\mu U={\cal O}(\phi)$, the $L_1$, $L_2$,
and $L_3$ terms of Eq.\ (\ref{4:8:l4gl}) are ${\cal O}(\phi^4)$ and need not be
considered.
   The only candidates are the $L_4$ - $L_8$ terms, of which we consider
the $L_4$ term as an explicit example,\footnote{For pedagogical reasons, we make
use of the physical fields. From a technical point of view, it is often
advantageous to work with the Cartesian fields and, at the end of the
calculation, express physical processes in terms of the Cartesian components.}
\begin{eqnarray*}
\lefteqn{L_4\mbox{Tr}(\partial_\mu U \partial^\mu U^\dagger)
\mbox{Tr}(\chi U^\dagger
+U \chi^\dagger)=}\\
&&L_4 \frac{2}{F_0^2}[\partial_\mu \eta \partial^\mu \eta
+\partial_\mu \pi^0 \partial^\mu \pi^0
+2\partial_\mu \pi^+\partial^\mu\pi^-+2\partial_\mu K^+\partial^\mu K^-\\
&&+2\partial_\mu K^0\partial^\mu \bar{K}^0+{\cal O}(\phi^4)]
[4B_0(2m+m_s)+{\cal O}(\phi^2)].
\end{eqnarray*}
   The remaining terms are treated analogously and we obtain for
${\cal L}_4^{2\phi}$
\begin{eqnarray}
\label{4:9:l42phi} {\cal L}_4^{2\phi}&=& -\frac{1}{2}\left(a_\eta\eta^2
+b_\eta\partial_\mu\eta\partial^\mu\eta\right)
\nonumber\\
&&-\frac{1}{2}\left(a_\pi\pi^0\pi^0+b_\pi
\partial_\mu\pi^0\partial^\mu\pi^0\right)\nonumber\\
&&-a_\pi\pi^+\pi^--b_\pi\partial_\mu\pi^+\partial^\mu\pi^-\nonumber\\
&&-a_K K^+ K^- - b_K \partial_\mu K^+\partial^\mu K^-\nonumber\\
&&-a_K K^0\bar{K}^0 - b_K \partial_\mu K^0\partial^\mu\bar{K}^0,
\end{eqnarray}
where the constants $a_\phi$ and $ b_\phi$ are given by
\begin{eqnarray}
\label{4:9:ab} a_\eta&=&\frac{64
B^2_0}{3F^2_0}\left[(2m+m_s)(m+2m_s)L_6+2(m-m_s)^2L_7 +(m^2+2m_s^2)L_8\right],
\nonumber\\
b_\eta&=&-\frac{16B_0}{F^2_0}\left[(2m+m_s)L_4+\frac{1}{3}(m+2m_s)L_5\right],
\nonumber\\
a_\pi&=&\frac{64 B^2_0}{F^2_0}\left[(2m+m_s)mL_6+m^2L_8\right],
\nonumber\\
b_\pi&=&-\frac{16 B_0}{F^2_0}\left[(2m+m_s)L_4+mL_5\right],
\nonumber\\
a_K&=&\frac{32B^2_0}{F^2_0}\left[(2m+m_s)(m+m_s)L_6+\frac{1}{2}(m+m_s)^2L_8
\right],
\nonumber\\
b_K&=&-\frac{16 B_0}{F^2_0}\left[(2m+m_s)L_4+\frac{1}{2}(m+m_s)L_5\right].
\end{eqnarray}
   At ${\cal O}(p^4)$ the self energies are of the form
\begin{equation}
\label{4:9:sigmaphi}
\Sigma_\phi(p^2)=A_\phi+B_\phi p^2,
\end{equation}
   where the constants $A_\phi$ and $B_\phi$ receive a tree-level
contribution from ${\cal L}_4$ and a one-loop contribution with a
vertex from ${\cal L}_2$ (see Fig.\ \ref{4:9:selfenergy}).
   For the tree-level contribution of ${\cal L}_4$ this is easily seen, because
the Lagrangians of Eq.\ (\ref{4:9:l42phi}) contain either exactly
two derivatives of the fields or no derivatives at all.
   For example, the contact contribution for the $\eta$ reads
$$-i\Sigma_\eta^{\rm contact}(p^2)=i 2\left[-\frac{1}{2}a_\eta
-b_\eta\frac{1}{2}(ip_\mu)(-ip^\mu)\right]=-i(a_\eta + b_\eta p^2),$$ where, as
usual, $\partial_\mu \phi$ generates $-ip_\mu$ and $ip_\mu$ for initial and final
lines, respectively, and the factor two takes account of two combinations of
contracting the fields with external lines.

   For the one-loop contribution the argument is as follows.
   The Lagrangian ${\cal L}_2^{4\phi}$ contains either two derivatives
or no derivatives at all which, symbolically, can be written as
$\phi\phi\partial\phi\partial\phi$ and $\phi^4$, respectively.
   The first term results in $M^2$ or $p^2$, depending on whether the
$\phi$ or the $\partial \phi$ are contracted with the external fields.
   The ``mixed'' situation vanishes upon integration.
   The second term, $\phi^4$, does not generate a momentum dependence.

\begin{figure}
\begin{center}
\epsfig{file=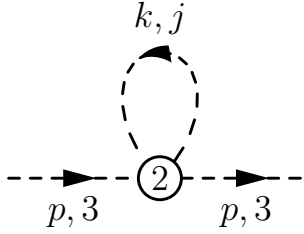,width=4cm}
\caption{\label{4:9:pi0seloop} Contribution of the pion loops to
the $\pi^0$ self energy.}
\end{center}
\end{figure}

   As a specific example, we evaluate the pion-loop contribution to the
$\pi^0$ self energy (see Figure \ref{4:9:pi0seloop}) by applying
the Feynman rule of Exercise \ref{exercise_pion_pion_scattering}
for $a=c=3$, $p_a=p_c=p$, $b=d=j$, and $p_b=p_d=k$:\footnote{Note
that we work in SU(3) and thus with the exponential
parameterization of $U$.}
\begin{eqnarray}
\label{4:9:diag} && \frac{1}{2}\int \frac{d^4k}{(2\pi)^4}i \left[
\underbrace{\delta^{3j}\delta^{3j}}_{\mbox{$1$}}
\frac{(p+k)^2-M_{\pi,2}^2}{F_0^2}
+\underbrace{\delta^{33}\delta^{jj}}_{\mbox{$3$}}
\frac{-M_{\pi,2}^2}{F_0^2} \right.\nonumber\\
&& +\underbrace{\delta^{3j}\delta^{j3}}_{\mbox{$1$}}
\frac{(p-k)^2-M_{\pi,2}^2}{F_0^2}\nonumber\\
&&\left.-\frac{1}{3 F_0^2}
\underbrace{(\delta^{3j}\delta^{3j}+\delta^{33}\delta^{jj}
+\delta^{3j}\delta^{j3})}_{\mbox{5}}(2p^2+2k^2-4M_{\pi,2}^2)\right]
\frac{i}{k^2-M_{\pi,2}^2+i0^+}\nonumber\\
&&=\frac{1}{2}\int \frac{d^4k}{(2\pi)^4}\frac{i}{3F_0^2}
[-4p^2-4k^2+5 M^2_{\pi,2}]
\frac{i}{k^2-M_{\pi,2}^2+i0^+},
\end{eqnarray}
   where $1/2$ is a symmetry factor.\footnote{When deriving the
Feynman rule of Exercise \ref{exercise_pion_pion_scattering}, we
took account of 24 distinct combinations of contracting four field
operators with four external lines. However, the Feynman diagram
of Eq.\ (\ref{4:9:diag}) involves only 12 possibilities to
contract two fields with each other and the remaining two fields
with two external lines.}
   The integral of Eq.\ (\ref{4:9:diag}) diverges and we thus consider
its extension to $n$ dimensions in order to make use of the
dimensional-regularization technique described in Section
\ref{sec_bidr}.
   In addition to the loop-integral of Eq.\ (\ref{app:drb:im22}),
\begin{eqnarray}
\label{4:9:im22}
I(M^2,\mu^2,n)&=&\mu^{4-n}\int\frac{d^nk}{(2\pi)^n}\frac{i}{k^2-M^2+i0^+}
\nonumber\\
&=&\frac{M^2}{16\pi^2}\left[
R+\ln\left(\frac{M^2}{\mu^2}\right)\right]+O(n-4),
\end{eqnarray}
where $R$ is given in Eq.\ (\ref{4:8:R}),
we need
\begin{displaymath}
\mu^{4-n}i\int \frac{d^n k}{(2\pi)^n}\frac{k^2}{k^2-M^2+i0^+}=
\mu^{4-n}i\int \frac{d^n k}{(2\pi)^n}\frac{k^2-M^2+M^2}{k^2-M^2+i0^+},
\end{displaymath}
   where we have added $0=-M^2+M^2$ in the numerator.
We make use of
$$\mu^{4-n}i\int \frac{d^n k}{(2\pi)^n}=0$$
in dimensional regularization which is ``shown'' as follows.
   Consider the (more general) integral
\begin{equation}
\label{4:9:dnkk2p}
\int d^n k (k^2)^p,
\end{equation}
substitute $k=\lambda k'\ (\lambda > 0)$, and relabel $k'=k$
\begin{displaymath}
=\lambda^{n+2p}\int d^n k (k^2)^p.
\end{displaymath}
   Since $\lambda>0$ is arbitrary and, for fixed $p$, the result
is to hold for arbitrary $n$, Eq.\ (\ref{4:9:dnkk2p})
is set to zero in dimensional regularization.
   We emphasize that the vanishing of Eq.\ (\ref{4:9:dnkk2p}) has
the character of a prescription.
   The integral does not depend on any scale and its analytic continuation
is ill defined in the sense that there is no dimension $n$ where it is
meaningful.
  It is ultraviolet divergent for $n+2p\geq 0$ and infrared divergent
for $n+2p\leq 0$.

   We then obtain
$$
\mu^{4-n}i\int \frac{d^n k}{(2\pi)^n}\frac{k^2}{k^2-M^2+i0^+}=
M^2 I(M^2,\mu^2,n),
$$
with $I(M^2,\mu^2,n)$ of Eq.\ (\ref{4:9:im22}).
   The pion-loop contribution to the $\pi^0$ self energy is thus
$$\frac{i}{6 F^2_0}(-4p^2+M^2_{\pi,2})I(M^2_{\pi,2},\mu^2,n),$$
which is indeed of the type discussed in Eq.\ (\ref{4:9:sigmaphi})
and diverges as $n\to 4$.

   After analyzing all loop contributions and combining them with the contact
contributions of Eqs.\ (\ref{4:9:ab}), the constants
$A_\phi$ and $B_\phi$ of Eq.\ (\ref{4:9:sigmaphi}) are given by
\begin{eqnarray}
\label{4:9:AB}
A_\pi&=&\frac{M^2_\pi}{F^2_0}\Bigg\{
\underbrace{-\frac{1}{6}I(M^2_\pi)
-\frac{1}{6}I(M^2_\eta)-\frac{1}{3}I(M^2_K)}_{\mbox{one-loop contribution}}
\nonumber\\
&&\underbrace{+32[(2m+m_s)B_0L_6+mB_0L_8]}_{\mbox{contact contribution}}
\Bigg\},\nonumber\\
B_\pi&=&\frac{2}{3}\frac{I(M^2_\pi)}{F^2_0}+\frac{1}{3}
\frac{I(M^2_K)}{F^2_0}-\frac{16B_0}{F^2_0}\left[
(2m+m_s)L_4+mL_5\right],\nonumber\\
A_K&=&\frac{M^2_K}{F^2_0}\Bigg\{\frac{1}{12}I(M^2_\eta)
-\frac{1}{4}I(M^2_\pi)-\frac{1}{2}I(M^2_K)\nonumber\\
&&+32\left[(2m+m_s)B_0L_6+\frac{1}{2}(m+m_s)B_0L_8\right]\Bigg\},\nonumber\\
B_K&=&\frac{1}{4}\frac{I(M^2_\eta)}{F^2_0}
+\frac{1}{4}\frac{I(M^2_\pi)}{F^2_0}
+\frac{1}{2}\frac{I(M^2_K)}{F^2_0}\nonumber\\
&&-16 \frac{B_0}{F^2_0}\left[(2m+m_s)L_4+\frac{1}{2}(m+m_s)L_5\right],
\nonumber\\
A_\eta&=&\frac{M^2_\eta}{F^2_0}\left[-\frac{2}{3}I(M^2_\eta)\right]
+\frac{M^2_\pi}{F^2_0}\left[\frac{1}{6}I(M^2_\eta)-\frac{1}{2}I(M^2_\pi)
+\frac{1}{3}I(M^2_K)\right]\nonumber\\
&&+\frac{M^2_\eta}{F^2_0}[16M^2_\eta L_8+32(2m+m_s)B_0L_6]\nonumber\\
&&+\frac{128}{9}\frac{B^2_0(m-m_s)^2}{F^2_0}(3L_7+L_8),\nonumber\\
B_\eta&=&\frac{I(M^2_K)}{F^2_0}-\frac{16}{F^2_0}(2m+m_s)B_0L_4
-8\frac{M^2_\eta}{F^2_0}L_5,
\end{eqnarray}
   where, for simplicity, we have suppressed the dependence on the
scale $\mu$ and the number of dimensions $n$ in
the integrals $I(M^2,\mu^2,n)$ [see Eq.\ (\ref{4:9:im22})].
   Furthermore, the squared masses appearing in the loop integrals of
Eq.\ (\ref{4:9:AB}) are given by the predictions of lowest order,
Eqs.\ (\ref{4:3:mpi2}) - (\ref{4:3:meta2}).
   Finally, the integrals $I$ as well as the bare coefficients $L_i$
(with the exception of $L_7$) have $1/(n-4)$ poles and finite pieces.
   In particular, the coefficients $A_\phi$ and $B_\phi$ are {\em not}
finite as $n\to 4$.

   The masses at ${\cal O}(p^4)$ are determined by solving the
general equation
\begin{equation}
\label{4:9:mse}
M^2=M_0^2+\Sigma(M^2)
\end{equation}
with the predictions of Eq.\ (\ref{4:9:sigmaphi}) for the self energies,
$$
M^2=M_0^2+A+BM^2,
$$
where the lowest-order terms, $M^2_0$, are given in Eqs.\
(\ref{4:3:mpi2}) - (\ref{4:3:meta2}).
   We then obtain
\begin{displaymath}
M^2=\frac{M_0^2+A}{1-B}=M_0^2(1+B)+A+
{\cal O}(p^6),
\end{displaymath}
   because $A={\cal O}(p^4)$ and $\{B, M_0^2\}={\cal O}(p^2)$.
   Expressing the bare coefficients $L_i$ in Eq.\ (\ref{4:9:AB}) in terms of
the renormalized coefficients by using Eq.\ (\ref{4:8:li}),
the results for the masses of the Goldstone bosons at ${\cal O}(p^4)$
read
\begin{eqnarray}
\label{4:9:mpi24}
M^2_{\pi,4}&=&M^2_{\pi,2}\Bigg\{1+\frac{M^2_{\pi,2}}{32\pi^2F^2_0}
\ln\left(\frac{M^2_{\pi,2}}{\mu^2}\right)-\frac{M^2_{\eta,2}}{96\pi^2F^2_0}
\ln\left(\frac{M^2_{\eta,2}}{\mu^2}\right)\nonumber\\
&&+\frac{16}{F^2_0}\left[(2m+m_s)B_0(2L^r_6-L^r_4)
+mB_0(2L^r_8-L^r_5)\right]\Bigg\},\\
\label{4:9:mk24}
M^2_{K,4}&=&M^2_{K,2}\Bigg\{1+\frac{M^2_{\eta,2}}{48\pi^2F^2_0}
\ln\left(\frac{M^2_{\eta,2}}{\mu^2}\right)\nonumber\\
&&+\frac{16}{F^2_0}\left[(2m+m_s)B_0(2L^r_6-L^r_4)
+\frac{1}{2}(m+m_s)B_0(2L^r_8-L^r_5)\right]\Bigg\},\nonumber\\
&&\\
\label{4:9:meta24}
M^2_{\eta,4}&=&M^2_{\eta,2}\left[1+\frac{M^2_{K,2}}{16\pi^2F^2_0}
\ln\left(\frac{M^2_{K,2}}{\mu^2}\right)
-\frac{M^2_{\eta,2}}{24\pi^2F^2_0}\ln\left(\frac{M^2_{\eta,2}}{\mu^2}\right)
\right.\nonumber\\
&&\left.+\frac{16}{F^2_0}(2m+m_s)B_0(2L^r_6-L^r_4)
+8\frac{M^2_{\eta,2}}{F^2_0}(2L^r_8-L^r_5)\right]\nonumber\\
&&+M^2_{\pi,2}\left[\frac{M^2_{\eta,2}}{96\pi^2F^2_0}
\ln\left(\frac{M^2_{\eta,2}}{\mu^2}\right)
-\frac{M^2_{\pi,2}}{32\pi^2F^2_0}
\ln\left(\frac{M^2_{\pi,2}}{\mu^2}\right)\right.\nonumber\\
&&\left.
+\frac{M^2_{K,2}}{48\pi^2F^2_0}
\ln\left(\frac{M^2_{K,2}}{\mu^2}\right)\right]\nonumber\\
&&+\frac{128}{9}\frac{B^2_0(m-m_s)^2}{F^2_0}
(3L^r_7+L^r_8).
\end{eqnarray}
   In Eqs.\ (\ref{4:9:mpi24}) - (\ref{4:9:meta24}) we have included the
subscripts 2 and 4 in order to indicate from which chiral order the predictions
result.
   First of all, we note that the expressions for the masses are finite.
   The infinite parts of the coefficients $L_i$ of the Lagrangian
of Gasser and Leutwyler exactly cancel the divergent terms
resulting from the integrals.
   This is the reason why the bare coefficients $L_i$ must be infinite.
   Furthermore, at ${\cal O}(p^4)$ the masses of the Goldstone bosons
vanish, if the quark masses are sent to zero.
   This is, of course, what we had expected from QCD in the chiral limit
but it is comforting to see that the self interaction in ${\cal L}_2$
(in the absence of quark masses) does not generate Goldstone
boson masses at higher order.
   The quark masses appear in combination with $B_0$ and therefore
Eqs.\ (\ref{4:9:mpi24}) - (\ref{4:9:meta24}) (and their generalization for
$m_u\neq m_d$) are used to extract quark mass ratios.
   At ${\cal O}(p^4)$, the squared Goldstone boson masses contain terms
which are analytic in the quark masses, namely, of the form $m^2_q$
multiplied by the renormalized low-energy coupling constants $L_i^r$.
   However, there are also non-analytic terms  of the
type $m^2_q \ln(m_q)$---so-called  chiral logarithms---which do not involve
new parameters.
   Such a behavior is an illustration of the mechanism found by Li and
Pagels \cite{Li:1971vr:11}, who noticed that a perturbation theory
around a symmetry which is realized in the Nambu-Goldstone mode
results in both analytic as well as non-analytic expressions in
the perturbation.
   Finally, the scale dependence of the renormalized coefficients
$L_i^r$ of Eq.\ (\ref{4:8:limu1mu2}) is by construction such that it cancels
the scale dependence of the chiral logarithms.
   Thus, physical observables do not depend on the scale $\mu$.
   Let us verify this statement by differentiating Eqs.\ (\ref{4:9:mpi24}) -
(\ref{4:9:meta24}) with respect to $\mu$.
   Using Eq.\ (\ref{4:8:limu1mu2}),
$$L_i^r(\mu)=L_i^r(\mu')+\frac{\Gamma_i}{16\pi^2}\ln\left(\frac{\mu'}{\mu}
\right),
$$
we obtain
$$
\frac{d L_i^r(\mu)}{d\mu}=-\frac{\Gamma_i}{16\pi^2\mu}
$$
and, analogously, for the chiral logarithms
$$\frac{d}{d\mu}\ln\left(\frac{M^2}{\mu^2}\right)=
2\frac{d}{d\mu}\left[\ln(M)-\ln(\mu)\right]=-\frac{2}{\mu}.
$$
   As a specific example, let us differentiate the expression for the
pion mass
\begin{eqnarray*}
\frac{d M^2_{\pi,4}}{d\mu}&=&\frac{M^2_{\pi,2}}{16\pi^2\mu F_0^2}\Bigg\{
\frac{M^2_{\pi,2}}{2}(-2)-\frac{M^2_{\eta,2}}{6}(-2)\\
&&+16[(2m+m_s)B_0(-2\Gamma_6+\Gamma_4)+mB_0(-2\Gamma_8+\Gamma_5)]\Bigg\}\\
&=&\frac{M^2_{\pi,2}}{16\pi^2\mu F_0^2}\Bigg\{-2B_0m +\frac{2}{9}(m+2m_s)B_0\\
&&+16\Bigg[(2m+m_s)B_0
\underbrace{\left(-2\frac{11}{144}+\frac{1}{8}\right)}_{\mbox{$-\frac{1}{36}$}}
+mB_0\underbrace{\left(-2\frac{5}{48}+\frac{3}{8}\right)
}_{\mbox{$\frac{1}{6}$}}
\Bigg]\Bigg\}\\
&=&\frac{M^2_{\pi,2}}{16\pi^2\mu F_0^2}\Bigg\{
B_0m
\left(-2+\frac{2}{9}-\frac{8}{9}+\frac{8}{3}\right)
+B_0m_s\left(\frac{4}{9}-\frac{16}{36}\right)\Bigg\}\\
&=&0,
\end{eqnarray*}
where we made use of the $\Gamma_i$ of Table \ref{4:8:tableli}.

\begin{exercise}
\label{exercise_L42pi}
\rm
  For the SU(2) calculation of the Goldstone boson self energies at
${\cal O}(p^4)$ we need the interaction Lagrangian
\begin{displaymath}
{\cal L}_{\rm int}={\cal L}_2^{4\phi}+{\cal L}_4^{2\phi}.
\end{displaymath}
   Consider the Lagrangians of Gasser and Leutwyler and of
Gasser, Sainio, and \v{S}varc, respectively:
\begin{eqnarray*}
{\cal L}^{\rm GL}_4 &=&
\frac{l_1}{4} \left\{\mbox{Tr}[D_{\mu}U (D^{\mu}U)^{\dagger}] \right\}^2
+\frac{l_2}{4}\mbox{Tr}[D_{\mu}U (D_{\nu}U)^{\dagger}]
\mbox{Tr}[D^{\mu}U (D^{\nu}U)^{\dagger}]
\nonumber \\
&&+\frac{l_3}{16}\left[\mbox{Tr}(\chi U^\dagger+ U\chi^\dagger)\right]^2
+\frac{l_4}{4}\mbox{Tr}[D_\mu U(D^\mu\chi)^\dagger
+D_\mu\chi(D^\mu U)^\dagger]\nonumber\\
&&+l_5\left[\mbox{Tr}(f^R_{\mu\nu}U f^{\mu\nu}_LU^\dagger)
-\frac{1}{2}\mbox{Tr}(f_{\mu\nu}^L f^{\mu\nu}_L
+f_{\mu\nu}^R f^{\mu\nu}_R)\right]\nonumber\\
&&+i\frac{l_6}{2}\mbox{Tr}[ f^R_{\mu\nu} D^{\mu} U (D^{\nu} U)^{\dagger}
+ f^L_{\mu\nu} (D^{\mu} U)^{\dagger} D^{\nu} U]\nonumber\\
&&-\frac{l_7}{16}\left[\mbox{Tr}(\chi U^\dagger-U\chi^\dagger)\right]^2
\nonumber\\
&&+\frac{h_1+h_3}{4}\mbox{Tr}(\chi\chi^\dagger)
+\frac{h_1-h_3}{16}\left\{
\left[\mbox{Tr}(\chi U^\dagger + U\chi^\dagger)\right]^2\right.
\nonumber\\
&&\left.
+\left[\mbox{Tr}(\chi U^\dagger-U\chi^\dagger)\right]^2
-2\mbox{Tr}(\chi U^\dagger\chi U^\dagger + U\chi^\dagger U\chi^\dagger)
\right\}\nonumber\\
&&-2h_2 \mbox{Tr}(f_{\mu\nu}^L f^{\mu\nu}_L
+f_{\mu\nu}^R f^{\mu\nu}_R).
\end{eqnarray*}
\begin{eqnarray*}
{\cal L}^{\rm GSS}_4& =&
\frac{l_1}{4} \left\{\mbox{Tr}[D_{\mu}U (D^{\mu}U)^{\dagger}] \right\}^2
+\frac{l_2}{4}\mbox{Tr}[D_{\mu}U (D_{\nu}U)^{\dagger}]
\mbox{Tr}[D^{\mu}U (D^{\nu}U)^{\dagger}]
\nonumber \\
&&+\frac{l_3+l_4}{16}\left[\mbox{Tr}(\chi U^\dagger+ U\chi^\dagger)\right]^2
+\frac{l_4}{8}\mbox{Tr}[D_\mu U(D^\mu U)^\dagger]\mbox{Tr}(\chi U^\dagger
+U\chi^\dagger)\nonumber\\
&&+l_5\mbox{Tr}(f^R_{\mu\nu}U f^{\mu\nu}_LU^\dagger)
+i\frac{l_6}{2}\mbox{Tr}[ f^R_{\mu\nu} D^{\mu} U (D^{\nu} U)^{\dagger}
+ f^L_{\mu\nu} (D^{\mu} U)^{\dagger} D^{\nu} U]\nonumber\\
&&-\frac{l_7}{16}\left[\mbox{Tr}(\chi U^\dagger-U\chi^\dagger)\right]^2
+\frac{h_1+h_3-l_4}{4}\mbox{Tr}(\chi\chi^\dagger)\nonumber\\
&&+\frac{h_1-h_3-l_4}{16}\left\{
\left[\mbox{Tr}(\chi U^\dagger + U\chi^\dagger)\right]^2
+\left[\mbox{Tr}(\chi U^\dagger-U\chi^\dagger)\right]^2\right.\nonumber\\
&&\left.
-2\mbox{Tr}(\chi U^\dagger\chi U^\dagger + U\chi^\dagger U\chi^\dagger)
\right\}
-\frac{4h_2+l_5}{2}\mbox{Tr}(f_{\mu\nu}^L f^{\mu\nu}_L
+f_{\mu\nu}^R f^{\mu\nu}_R).
\end{eqnarray*}
   Setting the external fields to zero and inserting $\chi=2 B m$, derive
the terms involving two pion fields.

Remark: The bare and the renormalized low-energy constants $l_i$
and $l_i^r$ are related by
\begin{displaymath}
l_i=l_i^r+\gamma_i\frac{R}{32\pi^2},
\end{displaymath}
where $R=2/(n-4)-[\ln(4\pi)+\Gamma'(1)+1]$ and
\begin{displaymath}
\gamma_1=\frac{1}{3},\quad \gamma_2=\frac{2}{3},\quad
\gamma_3=-\frac{1}{2},\quad
\gamma_4=2,\quad
\gamma_5=-\frac{1}{6},\quad
\gamma_6=-\frac{1}{3},\quad
\gamma_7=0.
\end{displaymath}
   In the SU(2) sector one often uses the scale-independent
parameters $\bar{l}_i$ which are defined by
\begin{displaymath}
l_i^r=\frac{\gamma_i}{32\pi^2}\left[\bar{l}_i+\ln\left(\frac{M^2}{\mu^2}\right)
\right],\quad i=1,\cdots,6,
\end{displaymath}
where $M^2=B(m_u+m_d)$.
   Since $\ln(1)=0$, the $\bar{l}_i$ are proportional to the renormalized
coupling constant at the scale $\mu=M$.
\end{exercise}

\begin{exercise}
\label{exercise_self_energy_coeffiecients}
\rm
  Using isospin symmetry, at ${\cal O}(p^4)$ the pion
self energy is of the form
\begin{displaymath}
\Sigma_{ba}(p^2)=\delta_{ab}(A+B p^2).
\end{displaymath}
   The constants $A$ and $B$ receive a tree-level contribution from
${\cal L}_4$ and a one-loop contribution from ${\cal L}_2$ (see
Figure \ref{figure:pion:selfenergy}).
   Using the results of exercises \ref{exercise_pion_pion_scattering},
\ref{exercise_standard_integral}, and \ref{exercise_L42pi},
derive the expressions of Table \ref{app:dp:tab:abz} for the
self-energy coefficients.

\begin{figure}
\begin{center}
\epsfig{file=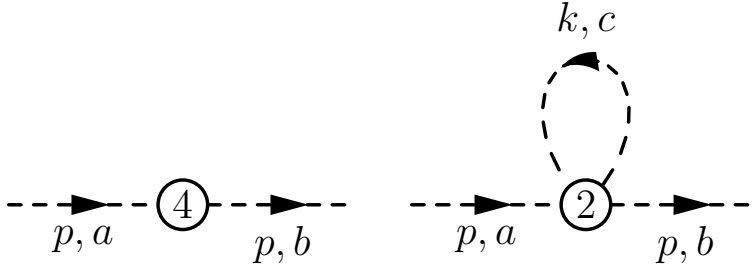,width=10cm}
\caption{\label{figure:pion:selfenergy}
Self-energy diagrams at ${\cal O}(p^4)$.
   Vertices derived from ${\cal L}_{2n}$ are denoted by $2n$ in the
interaction blobs.}
\end{center}
\end{figure}

\begin{table}[htb]
\begin{center}
\label{app:dp:tab:abz}
\begin{tabular}
{|c|c|c|} \hline
&$A$&$B$\\
\hline
&&\\
GL, \mbox{exponential}&
$-\frac{1}{6}\frac{M^2}{F^2} I+2 l_3 \frac{M^4}{F^2}$&
$\frac{2}{3}\frac{I}{F^2}$\\
&&\\
\hline
&&\\
GL, \mbox{square root}&
$\frac{3}{2}\frac{M^2}{F^2} I+2 l_3 \frac{M^4}{F^2}$&
$-\frac{I}{F^2}$\\
&&\\
\hline
&&\\
GSS, \mbox{exponential}&
$-\frac{1}{6}\frac{M^2}{F^2} I+2 (l_3+l_4)\frac{M^4}{F^2}$&
$\frac{2}{3}\frac{I}{F^2}-2l_4\frac{M^2}{F^2}$\\
&&\\
\hline
&&\\
GSS, \mbox{square root}& $\frac{3}{2}\frac{M^2}{F^2} I+2
(l_3+l_4)\frac{M^4}{F^2}$& $-\frac{I}{F^2}-2l_4\frac{M^2}{F^2}$
\\
&&\\
\hline
\end{tabular}
\end{center}
\caption{Self-energy coefficients and wave function
renormalization constants.
   $I$ denotes the dimensionally regularized integral
$I=I(M^2,\mu^2,n)=\frac{M^2}{16\pi^2}\left[
R+\ln\left(\frac{M^2}{\mu^2}\right)\right]+O(n-4)$,
$R=\frac{2}{n-4}-\left[\ln(4\pi)+\Gamma'(1)+1\right]$, $M^2=2B m$.
}
\end{table}
Using
\begin{displaymath}
M^2_{\pi,4}=\frac{M^2_{\pi,2}+A}{1-B}=M^2_{\pi,2}(1+B)
+A+{\cal O}(p^6),
\end{displaymath}
derive the squared pion mass at ${\cal O}(p^4)$:
\begin{displaymath}
M_{\pi,4}^2=M^2-\frac{\bar{l}_3}{32\pi^2 F^2}M^4 + {\cal O}(M^6),
\end{displaymath}
where $M^2= 2B m$.

\end{exercise}

\begin{exercise}
\label{exercise_masses_SU(3)}
\rm

You may repeat the {\em full} calculation in
SU(3) to obtain the masses of the Goldstone boson octet.

\noindent Remark: Conceptionally the calculation is completetly analogous
to the SU(2) calculation. Due to the SU(3) algebra and the fact that
the loop integrals contain different mass scales it is now
considerably more work.

\end{exercise}

\chapter{Chiral Perturbation Theory for Baryons}
\label{chap_cptb}

   So far we have considered the purely mesonic sector involving
the interaction of Goldstone bosons with each other and with
the external fields.
   However, ChPT can be extended to also describe the dynamics of baryons
at low energies.
   Here we will concentrate on matrix elements with a single baryon in the
initial and final states.
   With such matrix elements we can, e.g.,  describe static properties such as
masses or magnetic moments, form factors, or, finally, more complicated
processes, such as pion-nucleon scattering, Compton scattering, pion
photoproduction etc.
   Technically speaking, we are interested in the baryon-to-baryon
transition amplitude in the presence of external fields (as
opposed to the vacuum-to-vacuum transition amplitude of Section
\ref{subsec_qcdpefgf}),
$$
{\cal F}(\vec{p}\,',\vec{p};v,a,s,p)=\langle {\vec{p}\,'};{\rm out}|
{\vec{p}\,};{\rm in}\rangle^{\rm c}_{v,a,s,p},
\quad
\vec{p}\neq\vec{p}\,',
$$
determined by the Lagrangian of Eq.\ (\ref{2:6:lqcds}),
$$
{\cal L}={\cal L}^0_{\rm QCD}+{\cal L}_{\rm ext} ={\cal L}^0_{\rm
QCD}+\bar{q}\gamma_\mu (v^\mu +\frac{1}{3}v^\mu_{(s)} +\gamma_5
a^\mu )q -\bar{q}(s-i\gamma_5 p)q.
$$
   In the above equation
$|\vec{p};{\rm in}\rangle$ and
$|\vec{p}\,';{\rm out}\rangle$ denote asymptotic one-baryon in- and
out-states, i.e., states which in the remote past and distant future behave as
free one-particle states of momentum $\vec{p}$ and $\vec{p}\,'$, respectively.
   The functional ${\cal F}$ consists of connected diagrams only (superscript
c).
   For example, the matrix elements of the vector and axial-vector currents
between one-baryon states are given by
\begin{eqnarray*}
\label{5:vc}
\langle \vec{p}\,'|V^{\mu,a}(x)|\vec{p}\,\rangle
&=&\left.
\frac{\delta}{i\delta v^a_\mu(x)}
{\cal F}(\vec{p}\,',\vec{p};v,a,s,p)\right|_{v=0,a=0,s=M,p=0},\\
\label{5:avc}
\langle \vec{p}\,'|A^{\mu,a}(x)|\vec{p}\,\rangle
&=&\left.
\frac{\delta}{i\delta a^a_\mu(x)}
{\cal F}(\vec{p}\,',\vec{p};v,a,s,p)\right|_{v=0,a=0,s=M,p=0},
\end{eqnarray*}
where $M=\mbox{diag}(m_u,m_d,m_s)$ denotes the quark-mass matrix
and
\begin{displaymath}
V^{\mu,a}(x)=\bar{q}(x)\gamma^\mu\frac{\lambda^a}{2} q(x),\quad
A^{\mu,a}(x)=\bar{q}(x)\gamma^\mu \gamma_5
\frac{\lambda^a}{2} q(x).
\end{displaymath}
   As in the mesonic sector the method of calculating the Green functions
associated with the above functional consists of an
effective-Lagrangian approach in combination with an appropriate
power counting.
   Specific matrix elements will be calculated applying the Gell-Mann
and Low formula of perturbation theory.

\section{Transformation Properties of the Fields}
\label{sec_tpf}
   The group-theoretical foundations of constructing phenomenological
Lagrangians in the presence of spontaneous symmetry breaking have been
developed in Refs.\ \cite{Weinberg:de:1,Coleman:sm:1,Callan:sn:1}.
   The fields entering the Lagrangian are assumed to transform under
irreducible representations of the subgroup $H$ which leaves the
vacuum invariant whereas the symmetry group $G$ of the Hamiltonian
or Lagrangian is nonlinearly realized (for the transformation
behavior of the Goldstone bosons, see Section \ref{sec_tpgb}).

   Our aim is a description of the interaction of baryons with the Goldstone
bosons as well as the external fields at low energies.
   To that end we need to specify the transformation properties of the
dynamical fields entering the Lagrangian.
   Our discussion follows Refs.\ \cite{Georgi:1,Gasser:1987rb:1}.

   To be specific, we consider the octet of the $\frac{1}{2}^+$ baryons (see
Figure \ref{sec:tf:tab:octetbaryons}).
   With each member of the octet we associate a complex, four-component Dirac
field which we arrange in a traceless $3\times 3$ matrix $B$,
\begin{figure}[t]

\vspace{2em}
\unitlength1cm
\begin{center}
\begin{picture}(10,6)
\thicklines
\put(0,0){\vector(1,0){10}}
\put(10,-1){$I_3$}
\put(0,0){\vector(0,1){6}}
\put(-1,6){$S$}
\put(3,1){\circle*{0.2}}
\put(2.0,0.5){$n(940)(udd)$}
\put(7,1){\circle*{0.2}}
\put(6.0,0.5){$p(938)(uud)$}
\put(1,3){\circle*{0.2}}
\put(0,3.5){$\Sigma^-(1197)(dds)$}
\put(5,3){\circle*{0.2}}
\put(4.0,3.5){$\Sigma^0(1193)(uds)$}
\put(4.0,2.5){$\Lambda(1116)(uds)$}
\put(9,3){\circle*{0.2}}
\put(8.0,3.5){$\Sigma^+(1189)(uus)$}
\put(3,5){\circle*{0.2}}
\put(2.0,4.5){$\Xi^-(1321)(dss)$}
\put(7,5){\circle*{0.2}}
\put(6.0,4.5){$\Xi^0(1315)(uss)$}
\put(0,1){\circle*{0.1}}
\put(-0.5,1){0}
\put(0,3){\circle*{0.1}}
\put(-0.5,3){-1}
\put(0,5){\circle*{0.1}}
\put(-0.5,5){-2}
\put(1,0){\circle*{0.1}}
\put(1,-0.5){-1}
\put(3,0){\circle*{0.1}}
\put(3,-0.5){-1/2}
\put(5,0){\circle*{0.1}}
\put(5,-0.5){0}
\put(7,0){\circle*{0.1}}
\put(7,-0.5){1/2}
\put(9,0){\circle*{0.1}}
\put(9,-0.5){1}
\end{picture}
\vspace{2em}
\end{center}
\caption{\label{sec:tf:tab:octetbaryons} The baryon octet in an
$(I_3,S)$ diagram. We have included the masses in MeV as well as
the quark content.}
\end{figure}
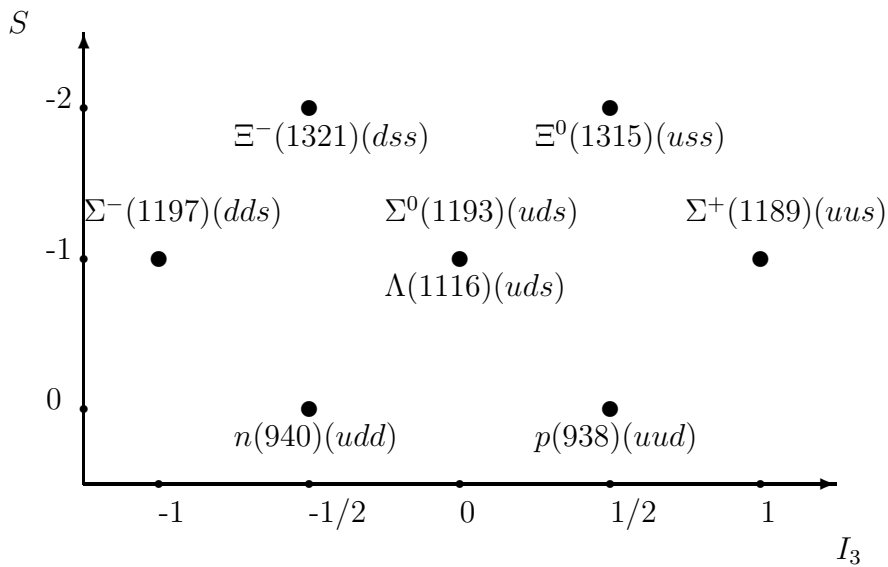
\begin{equation}
\label{5:1:su3oktett}
B=\sum_{a=1}^8 \frac{\lambda_a B_a}{\sqrt{2}}=
\left(\begin{array}{ccc}
\frac{1}{\sqrt{2}}\Sigma^0+\frac{1}{\sqrt{6}}\Lambda&\Sigma^+&p\\
\Sigma^-&-\frac{1}{\sqrt{2}}\Sigma^0+\frac{1}{\sqrt{6}}\Lambda&n\\
\Xi^-&\Xi^0&-\frac{2}{\sqrt{6}}\Lambda
\end{array}\right),
\end{equation}
   where we have suppressed the dependence on $x$.
   For later use, we have to keep in mind that each entry of
Eq.\ (\ref{5:1:su3oktett}) is a Dirac field, but for the purpose of
discussing the transformation properties under global flavor SU(3)
this can be ignored, because these transformations
act on each of the four components in the same way.
   In contrast to the mesonic case of Eq.\ (\ref{4:3:upar}),
where we collected the fields of the Goldstone octet in a Hermitian
traceless matrix $\phi$, the $B_a$ of the spin-$1/2$ case are not
real (Hermitian), i.e., $B\neq B^\dagger$.
\begin{exercise}
\label{exercise_baryon_octet}
\rm
   Using Eq.\ (\ref{5:1:su3oktett}),
express the physical fields in terms of Cartesian fields.
\end{exercise}

   Now let us define the set
\begin{equation}
\label{5:1:setm}
M\equiv\{B(x)|B(x)\, \mbox{complex, traceless $3\times 3$ matrix}\}
\end{equation}
which under the addition of matrices is a complex vector space.
   The following homomorphism is a representation of the abstract
group $H=\mbox{SU}(3)_V$ on the vector space $M$
[see also Eq.\ (\ref{4:2:uhtrafo})]:
\begin{eqnarray}
\label{5:1:su3hom}
&&\varphi: H\to \varphi(H),\quad V\mapsto \varphi(V)\quad \mbox{where}\quad
\varphi(V): M\to M,\nonumber\\
&&B(x)\mapsto B'(x)=\varphi(V)B(x)\equiv V B(x) V^\dagger.
\end{eqnarray}
   First of all, $B'(x)$ is again an element of $M$, because $\mbox{Tr}[B'(x)]
=\mbox{Tr}[VB(x)V^\dagger]\newline=\mbox{Tr}[B(x)]=0$.
   Equation (\ref{5:1:su3hom}) satisfies the homomorphism property
\begin{eqnarray*}
\varphi(V_1)\varphi(V_2) B(x)&=&\varphi(V_1)V_2B(x) V_2^\dagger
=V_1 V_2 B(x) V_2^\dagger V_1^\dagger=(V_1 V_2)B(x) (V_1 V_2)^\dagger\\
&=&\varphi(V_1V_2) B(x)
\end{eqnarray*}
and is indeed a {\em representation} of SU(3), because
\begin{eqnarray*}
\varphi(V)[\lambda_1B_1(x)+\lambda_2 B_2(x)]&=&
V[\lambda_1B_1(x)+\lambda_2 B_2(x)]V^\dagger\\
&=&\lambda_1 VB_1(x)V^\dagger+\lambda_2VB_2(x)V^\dagger\\
&=&\lambda_1\varphi(V)B_1(x)
+\lambda_2\varphi(V)B_2(x).
\end{eqnarray*}
   Equation (\ref{5:1:su3hom}) is just the familiar statement that $B$
transforms as an octet under (the adjoint representation of)
SU(3)$_V$.\footnote{Technically speaking the adjoint representation is faithful
(one-to-one) modulo the center $Z$ of SU(3) which is defined as the set of all
elements commuting with all elements of SU(3) and is given by $Z=\{1_{3\times 3},
\exp(2\pi i/3)1_{3\times 3}, \exp (4\pi i/3) 1_{3\times 3} \}$.}

   Let us now turn to various representations and realizations of the
group $\mbox{SU(3)}_L\times\mbox{SU(3)}_R$.
   We consider two explicit examples and refer the interested reader
to the textbook by Georgi \cite{Georgi:1} for more details.
   In analogy to the discussion of the quark fields in QCD, we may introduce
left- and right-handed components of the baryon fields
[see Eq.\ (\ref{2:4:qlr})]:
\begin{equation}
\label{5:1:blr}
B_1=P_L B_1+P_R B_1=B_L + B_R.
\end{equation}
   We define the set $M_1\equiv\{(B_L(x),B_R(x))\}$ which under the
addition of matrices is a complex vector space.
   The following homomorphism is a representation of the abstract group
$G=\mbox{SU}(3)_L\times \mbox{SU(3)}_R$ on $M_1$:
\begin{eqnarray}
\label{5:1:su3lrhom} (B_L,B_R)\mapsto (B'_L,B'_R) \equiv (L B_L L^\dagger,R B_R
R^\dagger),
\end{eqnarray}
   where we have suppressed the $x$ dependence.
   The proof proceeds in complete analogy to that of Eq.\ (\ref{5:1:su3hom}).

   As a second example, consider the set $M_2\equiv\{B_2(x)\}$ with the
homomorphism
\begin{equation}
\label{5:1:su3lrhomb2}
B_2\mapsto B_2'\equiv L B_2 L^\dagger,
\end{equation}
i.e.\ the transformation behavior is independent of $R$.
   The mapping defines a representation of the group
$\mbox{SU(3)}_L\times\mbox{SU(3)}_R$, although the transformation
behavior is drastically different from the first example.
   However, the important feature which both mappings have in common
is that under the subgroup $H=\{(V,V)|V\in \mbox{SU($3$)}\}$ of $G$
both fields $B_i$ transform as an octet:
\begin{eqnarray*}
B_1=B_L+B_R&\stackrel{H}{\mapsto}&VB_LV^\dagger+VB_R V^\dagger=VB_1V^\dagger,\\
B_2&\stackrel{H}{\mapsto}&VB_2V^\dagger.
\end{eqnarray*}

   We will now show how in a theory also containing Goldstone bosons
the various realizations may be connected to each other using field
redefinitions.
   Here we consider the second example, with the fields $B_2$ of
Eq.\ (\ref{5:1:su3lrhomb2}) and $U$ of
Eq.\ (\ref{4:3:upar}) transforming as
\begin{displaymath}
B_2\mapsto LB_2L^\dagger,\quad U\mapsto RUL^\dagger,
\end{displaymath}
and define new baryon fields by
\begin{displaymath}
\tilde{B}\equiv UB_2,
\end{displaymath}
   so that the new pair $(\tilde B,U)$ transforms as
$$\tilde{B}\mapsto RUL^\dagger L B_2 L^\dagger=R\tilde{B}L^\dagger,\quad
U\mapsto RU L^\dagger.$$
   Note in particular that $\tilde{B}$ still transforms as an octet under
the subgroup $H=\mbox{SU(3)}_V$.

   Given that physical observables are invariant under field transformations
we may choose a description of baryons that is maximally convenient for the
construction of the effective Lagrangian \cite{Georgi:1} and which is commonly
used in chiral perturbation theory.
   We start with $G=\mbox{SU(2)}_L\times\mbox{SU(2)}_R$ and consider
the case of $G=\mbox{SU(3)}_L\times\mbox{SU(3)}_R$ later.
   Let
\begin{equation}
\label{5:1:Psi}
\Psi=\left(\begin{array}{c}p\\n\end{array}\right)
\end{equation}
denote the nucleon field with two four-component Dirac fields for the proton
and the neutron and $U$ the SU(2) matrix containing the pion fields.
   We have already seen in Section \ref{subsec_aqcd} that the mapping
$U\mapsto RUL^\dagger$ defines a nonlinear realization of $G$.
   We denote the square root of $U$ by $u$, $u^2(x)=U(x)$, and define
the SU(2)-valued function $K(L,R,U)$ by
\begin{equation}
\label{5:1:kdef}
u(x)\mapsto u'(x)=\sqrt{RUL^\dagger}\equiv RuK^{-1}(L,R,U),
\end{equation}
i.e.
\begin{displaymath}
K(L,R,U)=u'^{-1}Ru=\sqrt{RUL^\dagger}^{-1}R \sqrt{U}.
\end{displaymath}
   The following homomorphism defines an operation of $G$ on the set
$\{(U,\Psi)\}$ in terms of a nonlinear realization:
\begin{equation}
\label{5:1:su2real}
\varphi(g):\left(\begin{array}{c}U\\ \Psi\end{array}\right)\mapsto
\left(\begin{array}{c}U'\\ \Psi'\end{array}\right)
=\left(\begin{array}{c}RUL^\dagger\\K(L,R,U)\Psi\end{array}\right),
\end{equation}
because the identity leaves $(U,\Psi)$ invariant and
\begin{eqnarray*}
\varphi(g_1)\varphi(g_2)
\left(\begin{array}{c}U\\ \Psi\end{array}\right)&=&\varphi(g_1)
\left(\begin{array}{c}R_2UL_2^\dagger\\K(L_2,R_2,U)\Psi\end{array}\right)\\
&=&\left(\begin{array}{c}R_1R_2UL_2^\dagger L_1^\dagger\\
K(L_1,R_1,R_2UL_2^\dagger)K(L_2,R_2,U)\Psi
\end{array}\right)\\
&=&\left(\begin{array}{c}R_1 R_2U(L_1L_2)^\dagger\\
K(L_1L_2,R_1R_2,U)\Psi\end{array}\right)\\
&=&\varphi(g_1g_2)\left(\begin{array}{c}U\\ \Psi\end{array}\right).
\end{eqnarray*}

\begin{exercise}
\label{exercise_K_homomorphism}
\rm
 Consider the SU(3)-valued function
\begin{displaymath}
K(L,R,U)=\sqrt{RUL^\dagger}^{-1}R \sqrt{U}.
\end{displaymath}
   Verify the homomorphism property
\begin{displaymath}
K(L_1, R_1, R_2 U L_2^\dagger)
K (L_2,R_2,U)=K((L_1 L_2), (R_1 R_2),U).
\end{displaymath}
\end{exercise}
   Note that for a general group element $g=(L,R)$ the transformation behavior
of $\Psi$ depends on $U$.
   For the special case of an isospin transformation, $R=L=V$, one obtains
$u'=VuV^\dagger$, because
$$U'=u'^2=VuV^\dagger VuV^\dagger=Vu^2V^\dagger=VUV^\dagger.$$
   Comparing with Eq.\ (\ref{5:1:kdef}) yields
$K^{-1}(V,V,U)=V^\dagger$ or $K(V,V,U)=V$,
i.e., $\Psi$ transforms linearly as an isospin doublet
under the isospin subgroup $\mbox{SU(2)}_V$ of
$\mbox{SU(2)}_L\times\mbox{SU(2)}_R$.
   A general feature here is that the transformation behavior under
the subgroup which leaves the ground state invariant is independent of $U$.
   Moreover, as already discussed in Section \ref{subsec_aqcd}, the Goldstone
bosons $\phi$ transform according to the adjoint representation of
SU(2)$_V$, i.e., as an isospin triplet.

   For the case $G=\mbox{SU(3)}_L\times \mbox{SU(3)}_R$ one uses the
nonlinear realization
\begin{equation}
\label{5:1:su3real}
\varphi(g):\left(\begin{array}{c}U\\ B\end{array}\right)\mapsto
\left(\begin{array}{c}U'\\ B'\end{array}\right)
=\left(\begin{array}{c}RUL^\dagger\\K(L,R,U)B K^\dagger(L,R,U)
\end{array}\right),
\end{equation}
where $K$ is defined completely analogously to Eq.\ (\ref{5:1:kdef}) after
inserting the corresponding SU(3) matrices.

\section{Baryonic Effective Lagrangian at Lowest Order}
\label{sec_loebl}

   Given the dynamical fields of Eqs.\ (\ref{5:1:su2real}) and
(\ref{5:1:su3real}) and their transformation properties, we will now discuss
the most general effective baryonic Lagrangian at lowest order.
   As in the vacuum sector, chiral symmetry provides constraints among the
single-baryon Green functions.
   Analogous to the mesonic sector, these Ward identities will be satisfied
if the Green functions are calculated from the most general effective
Lagrangian coupled to external fields with a {\em local} invariance under the
chiral group (see Section \ref{sec_gfwi}).

   Let us start with the construction of the $\pi N$ effective Lagrangian
${\cal L}^{(1)}_{\pi N}$ which we demand to have a {\em local}
$\mbox{SU}(2)_L\times\mbox{SU(2)}_R\times\mbox{U(1)}_V$ symmetry.
   The transformation behavior of the external fields is given in
Eq.\ (\ref{2:6:sg}), whereas the nucleon doublet and $U$ transform as
\begin{equation}
\label{5:2:psitrans}
\left(\begin{array}{c}U(x)\\ \Psi(x)\end{array}\right)\mapsto
\left(\begin{array}{c}V_R(x)U(x)V_L^\dagger(x)\\
\exp[-i\Theta(x)]K[V_L(x),V_R(x),U(x)]\Psi(x)\end{array}\right).
\end{equation}
   The local character of the transformation implies that we need to
introduce a covariant derivative $D_\mu \Psi$ with the usual property that
it transforms in the same way as $\Psi$:
\begin{equation}
\label{5:2:kovder}
D_\mu \Psi(x)\mapsto [D_\mu \Psi(x)]'\stackrel{!}{=}
\exp[-i\Theta(x)]K[V_L(x),V_R(x),U(x)]D_\mu\Psi(x).
\end{equation}
   Since $K$ not only depends on $V_L$ and $V_R$ but also on $U$,
we may expect the covariant derivative to contain $u$ and $u^\dagger$
and their derivatives.

   The so-called connection
(recall $\partial_\mu u u^\dagger=-u\partial_\mu u^\dagger$),
\begin{equation}
\label{5:2:gamma}
\Gamma_\mu=\frac{1}{2}\left[u^\dagger(\partial_\mu-ir_\mu)u
+u(\partial_\mu-il_\mu)u^\dagger\right],
\end{equation}
is an integral part of the covariant derivative of the nucleon doublet:
\begin{equation}
\label{5:2:kovderpsi}
D_\mu\Psi=(\partial_\mu+\Gamma_\mu-iv_\mu^{(s)})\Psi.
\end{equation}
   What needs to be shown is
\begin{equation}
\label{5:2tbs}
D'_\mu\Psi'=[\partial_\mu+\Gamma_\mu'-i(v_\mu^{(s)}-\partial_\mu\Theta)]
\exp(-i\Theta)K\Psi
=\exp(-i\Theta)K(\partial_\mu+\Gamma_\mu-iv_\mu^{(s)})\Psi.
\end{equation}
To that end, we make use of the product rule,
$$
\partial_\mu[\exp(-i\Theta)K\Psi]=-i\partial_\mu\Theta \exp(-i\Theta)K\Psi
+\exp(-i\Theta)\partial_\mu K\Psi
+\exp(-i\Theta) K \partial_\mu \Psi,
$$
in Eq.\ (\ref{5:2tbs}) and multiply by $\exp(i\Theta)$,
reducing it to
$$
\partial_\mu K=K\Gamma_\mu-\Gamma_\mu'K.
$$
   Using Eq.\ (\ref{5:1:kdef}),
\begin{eqnarray*}
K&=&u'^\dagger V_R u=\underbrace{u'u'^\dagger}_{\mbox{1}}u'^\dagger V_R u
=u' U'^\dagger V_R u= u' V_L
\underbrace{U^\dagger}_{\mbox{$u^\dagger u^\dagger$}}
\underbrace{V_R^\dagger V_R}_{\mbox{1}} u
=u'V_Lu^\dagger,
\end{eqnarray*}
we find
\begin{eqnarray*}
2(K\Gamma_\mu-\Gamma_\mu'K)&=&
K\left[u^\dagger(\partial_\mu-ir_\mu)u\right]
-\left[u'^\dagger(\partial_\mu-iV_R r_\mu V_R^\dagger
+V_R\partial_\mu V_R^\dagger)u'\right]K\\
&&+(R\to L,r_\mu\to l_\mu,u\leftrightarrow u^\dagger,
u'\leftrightarrow u'^\dagger)\\
&=&u'^\dagger V_R(\partial_\mu u-ir_\mu u)
-u'^\dagger\partial_\mu u'
\underbrace{K}_{\mbox{$u'^\dagger V_R u$}}\\
&&
+iu'^\dagger V_R r_\mu\underbrace{V_R^\dagger u'K}_{\mbox{$u$}}
-u'^\dagger V_R\partial_\mu V^\dagger_R\underbrace{u'K}_{\mbox{$V_R u$}}\\
&&+(R\to L,r_\mu\to l_\mu,
u\leftrightarrow u^\dagger,u'\leftrightarrow u'^\dagger)\\
&=&u'^\dagger V_R\partial_\mu u
-iu'^\dagger V_R r_\mu u
-\underbrace{u'^\dagger\partial_\mu u'u'^\dagger}_{
\mbox{$-\partial_\mu u'^\dagger$}} V_R u\\
&&+iu'^\dagger V_R r_\mu u
-u'^\dagger\underbrace{V_R\partial_\mu V_R^\dagger
V_R}_{\mbox{$-\partial_\mu V_R$}}u\\
&&+(R\to L,r_\mu\to l_\mu,
u\leftrightarrow u^\dagger,u'\leftrightarrow u'^\dagger)\\
&=&u'^\dagger V_R\partial_\mu u +\partial_\mu u'^\dagger V_R u +
u'^\dagger \partial_\mu V_R u\\
&&+(R\to L,u\leftrightarrow u^\dagger,u'\leftrightarrow u'^\dagger)\\
&=&\partial_\mu(u'^\dagger V_Ru+u'V_Lu^\dagger)=2\partial_\mu K,
\end{eqnarray*}
i.e., the covariant derivative defined in Eq.\ (\ref{5:2:kovderpsi}) indeed
satisfies the condition of Eq.\ (\ref{5:2:kovder}).
   At ${\cal O}(p)$ there exists another Hermitian building block,
the so-called vielbein,
\begin{equation}
\label{5:2:chvi}
u_\mu\equiv i\left[u^\dagger(\partial_\mu-i r_\mu)u-u(\partial_\mu-i
l_\mu)u^\dagger\right],
\end{equation}
which under parity transforms as an axial vector:
\begin{displaymath}
u_\mu\stackrel{P}{\mapsto} i\left[u(\partial^\mu-il^\mu)u^\dagger
-u^\dagger(\partial^\mu-ir^\mu)u\right]=-u^\mu.
\end{displaymath}
\begin{exercise}
\label{exercise_umu_trans}
\rm
Using
\begin{displaymath}
u'=V_R u K^\dagger=K u V_L^\dagger
\end{displaymath}
show that, under $\mbox{SU(2)}_L\times\mbox{SU(2)}_R\times \mbox{U}(1)_V$,
$u_\mu$ transforms as
\begin{displaymath}
u_\mu \mapsto Ku_\mu K^\dagger.
\end{displaymath}
\end{exercise}

   The most general effective $\pi N$ Lagrangian describing processes with a
single nucleon in the initial and final states is then of the type
$\bar{\Psi} \widehat{O} \Psi$, where $\widehat{O}$ is an operator acting
in Dirac and flavor space, transforming under
$\mbox{SU(2)}_L\times\mbox{SU(2)}_R\times \mbox{U}(1)_V$
as $K\widehat{O}K^\dagger$.
   As in the mesonic sector, the Lagrangian must be a Hermitian Lorentz scalar
which is even under the discrete symmetries $C$, $P$, and $T$.

   The most general such Lagrangian with the smallest number of derivatives
is given by \cite{Gasser:1987rb:2} \footnote{The power counting will be discussed
below.}
\begin{equation}
\label{5:2:l1pin}
{\cal L}^{(1)}_{\pi N}= \bar{\Psi}\left(iD\hspace{-.6em}/
-\stackrel{\circ}{m}_N
+\frac{\stackrel{\circ}{g}_A}{2}\gamma^\mu \gamma_5 u_\mu\right)\Psi.
\end{equation}
   It contains two parameters not determined by chiral symmetry:
the nucleon mass $\stackrel{\circ}{m}_N$ and the axial-vector coupling
constant $\stackrel{\circ}{g}_A$, both taken in the chiral limit
(denoted by $\circ$). [Physical nucleon mass: $m_N=939$ MeV. Theoretical
analysis: $\stackrel{\circ}{m}_N\approx 883$ MeV (at fixed $m_s\neq 0$).
Physical axial-vector coupling constant from neutron beta decay:
$g_A=1.267$.]
   The overall normalization of the Lagrangian is chosen such that in the
case of no external fields and no pion fields it reduces to that
of a free nucleon of mass $\stackrel{\circ}{m}_N$.

\begin{exercise}
\label{exercise_pion_nucleon_interaction}
\rm
Consider the lowest-order $\pi N$ Lagrangian of Eq.\ (\ref{5:2:l1pin}).
   Assume that there are no external fields, $l_\mu=r_\mu=v_\mu^{(s)}=0$,
so that
\begin{eqnarray*}
\Gamma_\mu&=&\frac{1}{2}(u^\dagger\partial_\mu u+u\partial_\mu u^\dagger),\\
u_\mu&=&i(u^\dagger\partial_\mu u-u\partial_\mu u^\dagger).
\end{eqnarray*}
By expanding
\begin{displaymath}
u=\exp\left(i\frac{\vec{\tau}\cdot\vec{\phi}}{2 F}\right)=1
+i\frac{\vec{\tau}\cdot\vec{\phi}}{2 F}-\frac{\vec{\phi}\,^2}{8F^2}+\cdots,
\end{displaymath}
derive the interaction Lagrangians containing one and two pion fields,
respectively.
\end{exercise}

\begin{exercise}
\label{exercise_interactions_L}
\rm
\renewcommand{\labelenumi}{(\alph{enumi})}
Consider the two-flavor Lagrangian
\begin{displaymath}
{\cal L}_{\rm eff}={\cal L}^{(1)}_{\pi N}+{\cal L}_2^{\pi},
\end{displaymath}
where
\begin{eqnarray*}
{\cal L}^{(1)}_{\pi N}&=& \bar{\Psi}\left(iD\hspace{-.6em}/
-\stackrel{\circ}{m}_N
+\frac{\stackrel{\circ}{g}_A}{2}\gamma^\mu \gamma_5 u_\mu\right)\Psi,\\
{\cal L}_2^\pi&=&
\frac{F^2}{4}\mbox{Tr}[D_\mu U (D^\mu U)^\dagger]
+\frac{F^2}{4}\mbox{Tr}(\chi U^\dagger + U\chi^\dagger).
\end{eqnarray*}
\begin{enumerate}
\item We would like to study this Lagrangian
in the presence of an electromagnetic field ${\cal A}_\mu$.
For that purpose we need to insert for the external fields
\begin{displaymath}
r_\mu=l_\mu=-e\frac{\tau_3}{2}{\cal A}_\mu,\quad
v_\mu^{(s)}=-\frac{e}{2}{\cal A}_\mu.
\end{displaymath}
Derive the interaction Lagrangians ${\cal L}_{\gamma NN}$, ${\cal L}_{\pi NN}$,
${\cal L}_{\gamma\pi NN}$, and ${\cal L}_{\gamma\pi\pi}$.
   Here, the nomenclature is such that ${\cal L}_{\gamma NN}$ denotes the
interaction Lagrangian describing the interaction of an external
electromagnetic field with a single nucleon in the initial and final
states, respectively.
   For example, ${\cal L}_{\gamma\pi NN}$ must be symbolically of the type
$\bar{\Psi}\phi {\cal A} \Psi$.
   Using Feynman rules, these four interaction Lagrangians would be sufficient
to describe pion photoproduction of the nucleon, $\gamma N\to \pi N$,
at lowest order in ChPT.
\item Now we would like to describe the interaction with
a massive charged weak boson
${\cal W}^\pm_\mu=({\cal W}_{1\mu}\mp i {\cal W}_{2\mu})/\sqrt{2}$,
\begin{displaymath}
r_\mu=0,\quad l_\mu=-\frac{g}{\sqrt{2}} ({\cal W}^+_\mu T_+ + \mbox{H.c.}),
\end{displaymath}
where H.c.~refers to the Hermitian conjugate and
$$
T_+=\left(\begin{array}{rr}0&V_{ud}\\0&0\end{array}\right).
$$
   Here, $V_{ud}$ denotes an element of the
Cabibbo-Kobayashi-Maskawa quark-mixing matrix,
$$|V_{ud}|=0.9735\pm 0.0008.$$
   At lowest order in perturbation theory, the Fermi constant is related
to the gauge coupling $g$ and the $W$ mass as
$$
G_F=\sqrt{2} \frac{g^2}{8 M^2_W}=1.16639(1)\times 10^{-5}\,\mbox{GeV}^{-2}.
$$
Derive the interaction Lagrangians ${\cal L}_{WNN}$ and ${\cal L}_{W\pi}$.
\item  Finally, we consider the neutral weak interaction
\begin{eqnarray*}
r_\mu&=&e \tan(\theta_W) \frac{\tau_3}{2} {\cal Z}_\mu,\nonumber\\
l_\mu&=&-\frac{g}{\cos(\theta_W)}\frac{\tau_3}{2} {\cal Z}_\mu+
e \tan(\theta_W) \frac{\tau_3}{2} {\cal Z}_\mu,
\nonumber\\
v_\mu^{(s)}&=&\frac{e\tan(\theta_W)}{2}{\cal Z}_\mu,
\end{eqnarray*}
where  $\theta_W$ is the weak angle, $e=g\sin(\theta_W)$.
Derive the interaction Lagrangians ${\cal L}_{ZNN}$ and ${\cal L}_{Z\pi}$.
\end{enumerate}
\end{exercise}

   Since the nucleon mass $m_N$ does not vanish in the chiral limit, the
zeroth component $\partial^0$ of the partial derivative acting on the nucleon
field does not produce a ``small'' quantity.
   We thus have to address the new features of chiral power counting in the
baryonic sector.
   The counting of the external fields as well as of covariant derivatives
acting on the mesonic fields remains the same as in mesonic chiral
perturbation theory [see Eq.\ (\ref{4:5:powercounting}) of Section
\ref{sec_cel}].
   On the other hand, the counting of bilinears $\bar{\Psi}\Gamma\Psi$ is
probably easiest understood by investigating the matrix elements of
positive-energy plane-wave solutions to the free Dirac equation in the
Dirac representation:
\begin{equation}
\psi^{(+)}(\vec{x},t)=\exp(-ip_N\cdot x) \sqrt{E_N+m_N}\left(
\begin{array}{c}
\chi\\
\frac{\vec{\sigma}\cdot\vec{p}_N}{E_N+m_N}\chi
\end{array}
\right),
\end{equation}
where $\chi$ denotes a two-component Pauli
spinor and $p_N^\mu=(E_N,\vec{p}_N)$ with $E_N=\sqrt{\vec{p}\,_N^2+m_N^2}$.
   In the low-energy limit, i.e.~for nonrelativistic kinematics, the lower
(small) component is suppressed as $|\vec{p}_N|/m_N$ in comparison with the
upper (large) component.
   For the analysis of the bilinears it is convenient
to divide the 16 Dirac matrices into even and odd ones, ${\cal E}=
\{1, \gamma_0,\gamma_5 \gamma_i,\sigma_{ij}\}$ and ${\cal
O}=\{\gamma_5,\gamma_5 \gamma_0,\gamma_i,\sigma_{i0}\}$
\cite{Foldy:1949wa:2,Fearing:ii:2}, respectively, where odd
matrices couple large and small components but not large with
large, whereas even matrices do the opposite.
   Finally, $i\partial^\mu$ acting on the nucleon solution produces $p_N^\mu$
which we write symbolically as $p_N^\mu=(m_N,\vec{0})+(E_N-m_N,\vec{p}_N\,)$
where we count the second term as ${\cal O}(p)$, i.e., as a small quantity.
   We are now in the position to summarize the chiral counting scheme for
the (new) elements of baryon chiral perturbation theory \cite{Krause:xc:2}:
\begin{eqnarray}
\label{5:2:powercounting}
&&\Psi,\bar{\Psi} =  {\cal O}(p^0),\, D_{\mu} \Psi = {\cal  O}(p^0),\,
(iD\hspace{-.6em}/ -\stackrel{\circ}{m}_N)\Psi={\cal O}(p),\nonumber\\
&&1,\gamma_\mu,\gamma_5\gamma_\mu,\sigma_{\mu\nu}={\cal O}(p^0),\,
\gamma_5 ={\cal O}(p),
\end{eqnarray}
   where the order given is the minimal one.
   For example, $\gamma_\mu$ has both an ${\cal O}(p^0)$ piece, $\gamma_0$,
as well as an ${\cal O}(p)$ piece, $\gamma_i$.
   A rigorous nonrelativistic reduction may be achieved in the framework
of the Foldy-Wouthuysen method \cite{Foldy:1949wa:2,Fearing:ii:2} or the
heavy-baryon approach \cite{Jenkins:1990jv:2,Bernard:1992qa:2}.

   The construction of the $\mbox{SU(3)}_L\times\mbox{SU(3)}_R$ Lagrangian
proceeds similarly except for the fact that the baryon fields are contained
in the $3\times 3$ matrix of Eq.\ (\ref{5:1:su3oktett})
transforming as $K B K^\dagger$.
   As in the mesonic sector, the building blocks are written as products
transforming as $K\cdots K^\dagger$ with a trace taken at the end.
   The lowest-order Lagrangian reads \cite{Krause:xc:2,Georgi:2}
\begin{equation}
\label{5:2:l1su3}
{\cal L}^{(1)}_{MB}=\mbox{Tr}\left[\bar{B}\left(iD\hspace{-.7em}/\hspace{.2em}
-M_0\right)B\right]
-\frac{D}{2}\mbox{Tr}\left(\bar{B}\gamma^\mu\gamma_5\{u_\mu,B\}\right)
-\frac{F}{2}\mbox{Tr}\left(\bar{B}\gamma^\mu\gamma_5[u_\mu,B]\right),
\end{equation}
where $M_0$ denotes the mass of the baryon octet in the chiral limit.
  The covariant derivative of $B$ is defined as
\begin{equation}
\label{5:2:kovderb}
D_\mu B=\partial_\mu B +[\Gamma_\mu,B],
\end{equation}
with $\Gamma_\mu$ of Eq.\ (\ref{5:2:gamma}) [for $\mbox{SU(3)}_L\times\mbox{
SU(3)}_R$].
   The constants $D$ and $F$ may be determined by fitting the semi-leptonic
decays $B\to B'+e^-+\bar{\nu}_e$ at tree level \cite{Borasoy:1998pe:2}:
\begin{equation}
\label{5:2:df}
D=0.80,\quad
F=0.50.
\end{equation}
Other ``popular'' values are: ($D=0.75$, $F=0.5$), ($D=0.804$, $F=0.463$).

\begin{exercise}
\label{exercise_three_flavor_lagrangian} \rm Consider the three-flavor Lagrangian
of Eq.\ (\ref{5:2:l1su3}) in the absence of external fields:
\begin{eqnarray*}
D_\mu B&=&\partial_\mu B+\frac{1}{2}[u^\dagger\partial_\mu u+
u\partial_\mu u^\dagger, B],\\
u_\mu&=&i(u^\dagger\partial_\mu u-u\partial_\mu u^\dagger).
\end{eqnarray*}
Using
\begin{displaymath}
B=\frac{B_a\lambda_a}{\sqrt{2}},\quad
\bar{B}=\frac{\bar{B}_b\lambda_b}{\sqrt{2}},
\end{displaymath}
show that the interaction Lagrangians with one and two mesons can be
written as
\begin{eqnarray*}
{\cal L}^{(1)}_{\phi BB}&=&\frac{1}{F_0}(d_{abc}D+if_{abc} F)\bar{B}_b
\gamma^\mu\gamma_5 B_a \partial_\mu \phi_c,\\
{\cal L}^{(1)}_{\phi\phi BB}&=&-\frac{i}{2 F_0^2}f_{abe} f_{cde}
\bar{B}_b\gamma^\mu B_a \phi_c\partial_\mu \phi_d.
\end{eqnarray*}
Hint: $u^\dagger\partial_\mu u+u\partial_\mu u^\dagger=u^\dagger\partial_\mu
u-\partial_\mu u u^\dagger=[u^\dagger,\partial_\mu u]$.
\end{exercise}

\section{Application at Lowest Order:
Goldberger-Treiman Relation and the Axial-Vec\-tor Current Matrix Element}
\label{subsec_gtravcme}

   We have seen in Section \ref{subsec_csbdqm} that the quark masses in
QCD give rise to a non-vanishing divergence of the axial-vector current
operator [see Eq.\ (\ref{2:4:dsva})].
   Here we will discuss the implications for the matrix elements of the
pseudoscalar density and of the axial-vector current evaluated between
single-nucleon states in terms of the lowest-order
Lagrangians of Eqs.\ (\ref{4:5:l2}) and (\ref{5:2:l1pin}).
   In particular, we will see that the Ward identity
\begin{equation}
\label{5:3:axwi}
\langle N(p')|\partial_\mu A^\mu_i(0)|N(p)\rangle=
\langle N(p')|m_q P_i(0)|N(p)\rangle,
\end{equation}
where $m_q=m_u=m_d$, is satisfied.

   The nucleon matrix element of the pseudoscalar density
can be parameterized as
\begin{equation}
\label{5:3:def_gt}
   m_q\langle N(p')| P_i (0) | N(p) \rangle =
         \frac{M_\pi^2 F_\pi}{M_\pi^2 - t}
         G_{\pi N}(t)i\bar{u}(p') \gamma_5 \tau_i u(p),
\end{equation}
   where $t=(p'-p)^2$.
   Equation (\ref{5:3:def_gt}) {\em defines} the form factor $G_{\pi N}(t)$ in
terms of the QCD operator $m_q P_i(x)$.
   The operator $m_q P_i(x)/(M_\pi^2 F_\pi)$ serves as
an interpolating pion field and thus $G_{\pi N}(t)$ is also referred to as the
pion-nucleon form factor (for this specific choice of the interpolating pion
field).
   The pion-nucleon coupling constant $g_{\pi N}$ is defined through
$G_{\pi N}(t)$ evaluated at $t=M_\pi^2$.

\begin{figure}[t]
\begin{center}
\epsfig{file=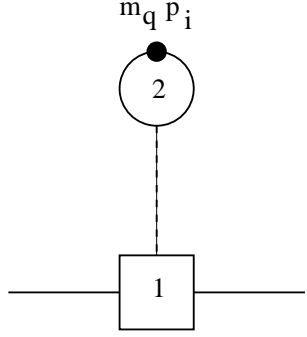,width=4cm}
\caption{\label{5:3:fig:pionnucleonformfactor} Lowest-order
contribution to the single-nucleon matrix element of the
pseudoscalar density. Mesonic and baryonic vertices are denoted by
a circle and a box, respectively, with the numbers 2 and 1
referring to the chiral order of ${\cal L}_2$ and ${\cal
L}^{(1)}_{\pi N}$. }
\end{center}
\end{figure}

   The Lagrangian ${\cal L}_{\pi N}^{(1)}$ of Eq.\ (\ref{5:2:l1pin})
does not generate a direct coupling of an external pseudoscalar field
$p_i(x)$ to the nucleon, i.e.,
it does not contain any terms involving $\chi$ or $\chi^\dagger$.
   At lowest order in the chiral expansion, the matrix element of the
pseudoscalar density is therefore given in terms of the diagram of
Figure \ref{5:3:fig:pionnucleonformfactor}, i.e., the pseudoscalar
source produces a pion which propagates and is then absorbed by
the nucleon.
   The coupling of a pseudoscalar field to the pion in the
framework of ${\cal L}_2$ is given by
\begin{equation}
\label{5:3:l2ext2} {\cal L}_{\rm ext} =i\frac{F^2
B}{2}\mbox{Tr}(pU^\dagger-Up)=2B F p_i\phi_i+\cdots.
\end{equation}
    When working with the nonlinear realization of Eq.\ (\ref{5:1:su2real})
it is convenient to use the so-called exponential parameterization
\begin{displaymath}
U(x)=\exp\left[i\frac{\vec{\tau}\cdot\vec{\phi}(x)}{F}\right],
\end{displaymath}
   because in that case the square root is simply given by
\begin{displaymath}
u(x)=\exp\left[i\frac{\vec{\tau}\cdot\vec{\phi}(x)}{2F}\right].
\end{displaymath}
   According to Figure \ref{5:3:fig:pionnucleonformfactor}, we need to
identify the interaction term of a nucleon with a single pion.
   In the absence of external fields the vielbein of Eq.\ (\ref{5:2:chvi}) is
odd in the pion fields,
\begin{equation}
\label{5:3:umupin}
u_\mu=i\left[u^\dagger\partial_\mu u-u\partial_\mu u^\dagger
\right]\stackrel{\phi^a\mapsto -\phi^a}{\mapsto}
i\left[u\partial_\mu u^\dagger -u^\dagger \partial_\mu u\right]
=-u_\mu.
\end{equation}
   Expanding $u$ and $u^\dagger$ as
\begin{equation}
\label{5:3:uentwi} u=1+i\frac{\vec{\tau}\cdot\vec{\phi}}{2 F}+{\cal
O}(\phi^2),\quad u^\dagger=1-i\frac{\vec{\tau}\cdot\vec{\phi}}{2 F} +{\cal
O}(\phi^2),
\end{equation}
we obtain
\begin{eqnarray}
\label{5:3:umuentwi} u_\mu
&=&-\frac{\vec{\tau}\cdot\partial_\mu\vec{\phi}}{F}+{\cal O}(\phi^3),
\end{eqnarray}
which, when inserted into ${\cal L}^{(1)}_{\pi N}$ of Eq.\ (\ref{5:2:l1pin}),
generates the following interaction Lagrangian (see Exercise
\ref{exercise_pion_nucleon_interaction}):
\begin{equation}
\label{5:3:lpin1} {\cal L}_{\rm int}=-\frac{1}{2}\frac{\stackrel{\circ}{g}_A}{F}
\bar{\Psi}\gamma^\mu\gamma_5
\underbrace{\vec{\tau}\cdot\partial_\mu\vec{\phi}}_{\mbox{$
\tau^b\partial_\mu\phi^b$}}\Psi.
\end{equation}
   (Note that the sign is opposite to the conventionally used pseudovector
pion-nucleon coupling.\footnote{In fact, also the definition of the
pion-nucleon form factor of Eq.\ (\ref{5:3:def_gt}) contains a sign
opposite to the standard convention so that, in the end, the
Goldberger-Treiman relation emerges with the conventional sign.})
   The Feynman rule for the vertex of an incoming pion with four-momentum
$q$ and Cartesian isospin index $a$ is given by
\begin{equation}
\label{5:3:pionnucleonvertex}
i\left(-\frac{1}{2}\frac{\stackrel{\circ}{g}_A}{F}\right)
\gamma^\mu\gamma_5\tau^b \delta^{ba}(-i q_\mu)=
-\frac{1}{2}\frac{\stackrel{\circ}{g}_A}{F} q\hspace{-.45em}/ \gamma_5 \tau^a.
\end{equation}
   On the other hand, the connection of Eq.\ (\ref{5:2:gamma}) with the
external fields set to zero is even in the pion fields,
\begin{equation}
\label{5:3:gammapin}
\Gamma_\mu=\frac{1}{2}\left[u^\dagger\partial_\mu u+u\partial_\mu u^\dagger
\right]\stackrel{\phi^a\mapsto -\phi^a}{\mapsto}
\frac{1}{2}\left[u\partial_\mu u^\dagger +u^\dagger \partial_\mu u\right]
=\Gamma_\mu,
\end{equation}
i.e., it does not contribute to the single-pion vertex.

   We now put the individual pieces together and obtain for
the diagram of Figure \ref{5:3:fig:pionnucleonformfactor}
\begin{eqnarray*}
\lefteqn{m_q 2 B F \frac{i}{t-M_\pi^2}\bar{u}(p')\left(
-\frac{1}{2}\frac{\stackrel{\circ}{g}_A}{F} q\hspace{-.45em}/
\gamma_5 \tau_i\right)u(p)}\\
&=&M_\pi^2 F  \frac{\stackrel{\circ}{m}_N \stackrel{\circ}{g}_A}{F}
\frac{1}{M_\pi^2-t}\bar{u}(p')\gamma_5 i \tau_i u(p),
\end{eqnarray*}
where we used $M_\pi^2=2B m_q$, and the Dirac equation to show
$\bar{u}q\hspace{-.5em}/ \gamma_5 u=2 \stackrel{\circ}{m}_N\bar{u}\gamma_5 u$.
   At ${\cal O}(p^2)$ $F_\pi=F$ so that, by comparison with
Eq.\ (\ref{5:3:def_gt}), we can read off the lowest-order result
\begin{equation}
\label{5:3:gt1} G_{\pi N}(t)= \frac{\stackrel{\circ}{m}_N}{F}
{\stackrel{\circ}{g}}_A,
\end{equation}
   i.e., at this order the form factor does not depend on $t$.
   In general, the pion-nucleon coupling constant is defined at $t=M_\pi^2$
which, in the present case, simply yields
\begin{equation}
\label{5:3:gpinn} g_{\pi N}=G_{\pi N}(M_\pi^2)= \frac{\stackrel{\circ}{m}_N}{F}
{\stackrel{\circ}{g}}_A.
\end{equation}
   Equation (\ref{5:3:gpinn}) represents the famous Goldberger-Treiman
relation \cite{Goldberger:1958tr:3:1,Nambu:xd:3:1} which establishes a connection
between quantities entering weak processes, $F_\pi$ and $g_A$ (to be discussed
below), and a typical strong-interaction quantity, namely the pion-nucleon
coupling constant $g_{\pi N}$.
   The numerical violation of the Goldberger-Treiman relation, as expressed
in the so-called Goldberger-Treiman discrepancy
\begin{equation}
\label{5:3:gtd}
\Delta_{\pi N}\equiv1-\frac{g_A m_N}{g_{\pi N}F_\pi},
\end{equation}
is at the percent level,\footnote{ Using $m_N=938.3$ MeV, $g_A=1.267$,
$F_\pi=92.4$ MeV, and $g_{\pi N}=13.21$,  \cite{Schroder:rc:3:1}, one obtains
$\Delta_{\pi N}=2.6$ \%.} although one has to keep in mind that {\em all four}
physical quantities move from their chiral-limit values $\stackrel{\circ}{g}_A$
etc.\ to the empirical ones $g_A$ etc.

   Using Lorentz covariance and isospin symmetry, the matrix element of the
axial-vector current between initial and final nucleon states---excluding
second-class currents \cite{Weinberg:1958ut:3:1}--- can be parameterized
as\footnote{The terminology ``first and second classes'' refers to the
transformation property of strangeness-conserving semi-leptonic weak interactions
under ${\cal G}$ conjugation \cite{Weinberg:1958ut:3:1} which is the product of
charge symmetry and charge conjugation ${\cal G}={\cal C}\exp(i\pi I_2)$.
   A second-class contribution would show up in terms
of a third form factor $G_T$ contributing as
\begin{displaymath}
G_T(t) \bar{u}(p') i\frac{\sigma^{\mu\nu} q_\nu}{2 m_N} \gamma_5
\frac{\tau_i}{2} u(p).
\end{displaymath}
   Assuming a perfect ${\cal G}$-conjugation symmetry,
the form factor $G_T$ vanishes.}
\begin{equation}
\label{5:3:axial_current}
   \left<N(p')| A_i^\mu (0) | N(p) \right> =
          \bar{u}(p') \left[ \gamma^\mu G_A(t) + \frac{(p'-p)^\mu}{2m_N}
          G_P(t) \right] \gamma_5 \frac{\tau_i}{2} u(p),
\end{equation}
where $t = (p'-p)^2$, and $G_A(t)$ and $G_P(t)$ are the
axial and induced pseudoscalar form factors, respectively.

   At lowest order, an external axial-vector field $a_\mu^i$ couples directly
to the nucleon as
\begin{equation}
\label{5:3:lnucleonavfc}
{\cal L}_{\rm ext}={\stackrel{\circ}{g}}_A\bar{\Psi}\gamma^\mu \gamma_5
\frac{\tau_i}{2}\Psi a_\mu^i+\cdots,
\end{equation}
which is obtained from ${\cal L}^{(1)}_{\pi N}$ through
$u_\mu=(r_\mu-l_\mu)+\cdots = 2 a_\mu+\cdots$.
   The coupling to the pions is obtained from ${\cal L}_2$ with
$r_\mu=-l_\mu=a_\mu$,
\begin{equation}
\label{5:3:lpionavfc} {\cal L}_{\rm ext}=-F \partial^\mu \phi_i a_\mu^i+\cdots,
\end{equation}
which gives rise to a diagram similar to Figure
\ref{5:3:fig:pionnucleonformfactor}, with  $m_q p_i$ replaced by
$a^\mu_i$.

   The matrix element is thus given by
\begin{eqnarray*}
\bar{u}(p')\left\{{\stackrel{\circ}{g}}_A \gamma^\mu\gamma_5 \frac{\tau_i}{2}
+\left[-\frac{1}{2}\frac{{\stackrel{\circ}{g}}_A}{F}
(p'\hspace{-.70em}/\hspace{.45em}-p\hspace{-.45em}/)\gamma_5 \tau_i\right]
\frac{i}{q^2-M^2_\pi} (-iF q^\mu)\right\}u(p),
\end{eqnarray*}
   from which we obtain, by applying the Dirac equation,
\begin{eqnarray}
\label{5:3:ga}
   G_A(t) &=& {\stackrel{\circ}{g}}_A,\\
\label{5:3:gp}
   G_P(t) &=& - \frac{ 4 \stackrel{\circ}{m}^2_N {\stackrel{\circ}{g}}_A}{
              t-M_\pi^2}.
\end{eqnarray}
   At this order the axial form factor does not yet show a $t$ dependence.
   The axial-vector coupling constant is defined as $G_A(0)$ which is
simply given by ${\stackrel{\circ}{g}}_A$.
   We have thus identified the second new parameter of ${\cal L}^{(1)}_{\pi N}$
besides the nucleon mass $\stackrel{\circ}{m}_N$.
   The induced pseudoscalar form factor is determined by the pion exchange
which is the simplest version of the so-called pion-pole dominance.
   The $1/(t-M_\pi^2)$ behavior of $G_P$ is not in conflict with
the book-keeping of a calculation
at chiral order ${\cal O}(p)$, because, according to Eq.\
(\ref{4:5:powercounting}), the external axial-vector field $a_\mu$ counts
as ${\cal O}(p)$, and the definition of the matrix element contains a
momentum $(p'-p)^\mu$ and the Dirac matrix $\gamma_5$ [see Eq.\
(\ref{5:2:powercounting})] so that the combined order of all elements
is indeed ${\cal O}(p)$.

   It is straightforward to verify that the form factors of
Eqs.\ (\ref{5:3:gt1}), (\ref{5:3:ga}), and (\ref{5:3:gp}) satisfy the relation
\begin{equation}
\label{ff_relation}
   2m_N G_A(t) + \frac{t}{2m_N} G_P(t) =
       2\frac{M_\pi^2 F_\pi}{M_\pi^2 - t} G_{\pi N}(t),
\end{equation}
which is required by the Ward identity of Eq.\ (\ref{5:3:axwi}) with
the parameterizations of Eqs.\ (\ref{5:3:def_gt}) and
(\ref{5:3:axial_current}) for the matrix elements.
   In other words, only two of the three form factors $G_A$, $G_P$, and
$G_{\pi N}$ are independent.
   Note that this relation is not restricted
to small values of $t$ but holds for any $t$.

\begin{exercise}
\label{exercise_gt}
\rm
\renewcommand{\labelenumi}{(\alph{enumi})}
According to Eq.\ (\ref{2:4:dsva}), the divergence of the axial-vector
current in the SU(2) sector is given by
\begin{displaymath}
\partial_\mu A^\mu_i(x)=m_q P_i(x), \quad i=1,2,3,
\end{displaymath}
where we have assumed $m_q=m_u=m_d$.
   Let $|A\rangle$ and $|B\rangle$ denote some (arbitrary) hadronic states
which are eigenstates of the four-momentum operator $P^\mu$ with eigenvalues
$p_A^\mu$ and $p_B^\mu$, respectively.
   Evaluating the above operator equation between
$|A\rangle$ and $\langle B|$ and using translational invariance, one obtains
\begin{eqnarray*}
\langle B|\partial_\mu A^\mu_i(x)|A\rangle&=&
\partial_\mu\langle B| A^\mu_i(x)|A\rangle
=\partial_\mu (\langle B|e^{iP\cdot x}A^\mu_i(0)e^{-iP\cdot x}|A\rangle)\\
&=&\partial_\mu(e^{i(p_B-p_A)\cdot x}
\langle B|A^\mu_i(0)|A\rangle)=iq_\mu e^{iq\cdot x}
\langle B|A^\mu_i(0)|A\rangle\\
&\stackrel{!}{=}&e^{iq\cdot x} m_q \langle B|P_i(0)|A\rangle,
\end{eqnarray*}
where we introduced $q=p_B-p_A$.
   Dividing both sides by $e^{iq\cdot x}\neq 0$, we obtain
\begin{displaymath}
i q_\mu \langle B|A^\mu_i(0)|A\rangle
= m_q \langle B|P_i(0)|A\rangle.
\end{displaymath}

\begin{enumerate}
\item Make use of the parameterizations of
Eqs.\ (\ref{5:3:def_gt}) and (\ref{5:3:axial_current})
for the nucleon matrix elements and derive Eq.\ (\ref{ff_relation}).

\noindent Hint: Make use of the Dirac equation.
\item Verify that the lowest-order predictions
\begin{displaymath}
G_A(t)={\stackrel{\circ}{g}}_A,\quad
G_P(t)= - \frac{ 4 \stackrel{\circ}{m}^2_N {\stackrel{\circ}{g}}_A}{
              t-M_\pi^2},\quad
G_{\pi N}(t)= \frac{\stackrel{\circ}{m}_N}{F} {\stackrel{\circ}{g}}_A,
\end{displaymath}
indeed satisfy this constraint.
\end{enumerate}
\end{exercise}

\section{Application at Lowest Order:
Pion-Nu\-cle\-on Scattering} \label{subsec_apnstl}

   As another example, we will consider pion-nucleon scattering and show
how the effective Lagrangian of Eq.\ (\ref{5:2:l1pin}) reproduces the
Weinberg-Tomozawa predictions for the $s$-wave scattering lengths
\cite{Weinberg:1966kf:3:2,Tomozawa:3:2}.
   We will contrast the results with those of a tree-level calculation
within pseudoscalar (PS) and pseudovector (PV) pion-nucleon couplings.

    Before calculating the $\pi N$ scattering amplitude within ChPT we introduce
a general parameterization of the invariant amplitude ${\cal M}=iT$ for the
process $\pi^a(q)+N(p)\to\pi^b(q')+N(p')$:\footnote{One also finds the
parameterization
\begin{displaymath}
T=\bar{u}(p')\left(D-\frac{1}{4m_N}
[q\,'\hspace{-.9em}/\hspace{.2em},q\hspace{-.45em}/\hspace{.1em}]B\right)u(p)
\end{displaymath}
with $D=A+\nu B$,
where, for simplicity, we have omitted the isospin indices.}
\begin{eqnarray}
\label{5:3:mpinpariso}
T^{ab}(p,q;p',q')&=&\frac{1}{2}\{\tau^b,\tau^a\}T^{+}(p,q;p',q') +
\frac{1}{2}[\tau^b,\tau^a]T^{-}(p,q;p',q')\nonumber\\
&=&\delta^{ab}T^{+}(p,q;p',q')-i\epsilon_{abc}\tau^c T^{-}(p,q;p',q'),
\end{eqnarray}
where
\begin{equation}
\label{5:3:AB} T^{\pm}(p,q;p',q')=\bar{u}(p')\left[A^\pm(\nu,\nu_B)
+\frac{1}{2}(q\hspace{-.45em}/+q'\hspace{-.7em}/\hspace{.2em})
B^\pm(\nu,\nu_B)\right] u(p).
\end{equation}
   The amplitudes $A^\pm$ and $B^\pm$ are functions of
two independent scalar kinematical variables
\begin{eqnarray} \label{5:3:nu}
\nu&=&\frac{s-u}{4m_N}=\frac{(p+p')\cdot q}{2m_N}
=\frac{(p+p')\cdot q'}{2m_N},\\
\label{5:3:nub}
\nu_B&=&-\frac{q\cdot q'}{2 m_N}=\frac{t-2M_\pi^2}{4 m_N},
\end{eqnarray}
where $s=(p+q)^2$, $t=(p'-p)^2$, and $u=(p'-q)^2$ are the usual
Mandelstam variables satisfying $s+t+u=2 m_N^2+ 2M_\pi^2$.
   From pion-crossing symmetry $T^{ab}(p,q;p',q')=
T^{ba}(p,-q';p',-q)$ we obtain for the crossing behavior of
the amplitudes
\begin{eqnarray}
\label{5:3:crossingab}
&&A^+(-\nu,\nu_B)=A^+(\nu,\nu_B),\quad
A^-(-\nu,\nu_B)=-A^-(\nu,\nu_B),\nonumber\\
&&B^+(-\nu,\nu_B)=-B^+(\nu,\nu_B),\quad
B^-(-\nu,\nu_B)=B^-(\nu,\nu_B).
\end{eqnarray}
   As in $\pi\pi$ scattering one often also finds the isospin
decomposition as in Exercise \ref{exercise_pion_pion_scattering},
\begin{displaymath}
\langle I',{I'}\hspace{-.3em}_3|T|I,I_3\rangle=T^I \delta_{II'} \delta_{I_3
{I'}\hspace{-.1em}_3}.
\end{displaymath}
   In this context we would like to point out that our convention for
the physical pion fields (and states) (see Exercise
\ref{exercise_physical_fields}) differs by a minus for the $\pi^+$ from the
spherical convention which is commonly used in the context of applying the
Wigner-Eckart theorem.
   Taking for each $\pi^+$ in the initial and final states a factor of $-1$ into
account, the relation between the two sets is given by
\begin{eqnarray}
\label{5:3:trel}
T^{\frac{1}{2}}&=&T^++2 T^-,\nonumber\\
T^{\frac{3}{2}}&=&T^+- T^-.
\end{eqnarray}
   To verify Eqs.\ (\ref{5:3:trel}), we consider
\begin{eqnarray*}
T^{\pi^+\pi^+}&=&\frac{1}{2}(T^{11}-iT^{12}+iT^{21}+T^{22})=T^+-\tau_3 T^-,\\
T^{\pi^+\pi^0}&=&\frac{1}{\sqrt{2}}(T^{13}+iT^{23})=\tau_+ T^-,
\end{eqnarray*}
and evaluate the matrix elements
\begin{eqnarray*}
\langle p \pi^+|T|p \pi^+\rangle&=& T^+-T^-,\nonumber\\
\langle p \pi^0|T|n\pi^+\rangle&=&\sqrt{2} T^-.
\end{eqnarray*}
   A comparison with the results of Exercise
\ref{exercise_pion_nucleon_scattering} below,
\begin{eqnarray*}
{}_{\rm sph.}\langle p\pi^+|T|p \pi^+\rangle_{\rm sph.}&=&T^\frac{3}{2} =(-1)^2
\langle p \pi^+|T|p \pi^+\rangle=T^+-T^-,\\
{}_{\rm sph.}\langle p\pi^0|T|n \pi^+\rangle_{\rm sph.}&=&
\frac{\sqrt{2}}{3}(T^\frac{3}{2}-T^\frac{1}{2})=(-1)\langle
p\pi^0|T|n\pi^+\rangle=-\sqrt{2}T^-,
\end{eqnarray*}
results in Eqs.\ (\ref{5:3:trel}). (The subscript sph.~serves to distinguish the
spherical convention from our convention.)

\begin{exercise}
\label{exercise_pion_nucleon_scattering}
\renewcommand{\labelenumi}{(\alph{enumi})}
\rm Consider the general parameterization of the invariant amplitude ${\cal
M}=iT$ for the process $\pi^a(q)+N(p)\to\pi^b(q')+N(p')$ of Eqs.\
(\ref{5:3:mpinpariso}) and (\ref{5:3:AB})
with the kinematical variables of Eqs.\
(\ref{5:3:nu}) and (\ref{5:3:nub}).
\begin{enumerate}
\item
Show that
$$s-m_N^2=2m_N(\nu-\nu_B),\quad u-m_N^2=-2 m_N(\nu+\nu_B).
$$
Hint: Make use of four-momentum conservation, $p+q=p'+q'$, and of
the mass-shell conditions, $p^2=p'^2=m_N^2$, $q^2=q'^2=M_\pi^2$.

   Derive the threshold values
$$\nu|_{\rm thr}=M_\pi,\quad
\nu_B|_{\rm thr}=-\frac{M_\pi^2}{2m_N}.
$$
\item
Show that from pion-crossing symmetry
$$T^{ab}(p,q;p',q')=
T^{ba}(p,-q';p',-q)$$ we obtain the crossing behavior of Eq.\
(\ref{5:3:crossingab}).
\item The physical $\pi N$ channels may be expressed in terms of
the isospin eigenstates as  (a spherical convention is understood)
\begin{eqnarray*}
|p\pi^+\rangle&=&|\frac{3}{2},\frac{3}{2}\rangle,\\
|p\pi^0\rangle&=&\sqrt{\frac{2}{3}}|\frac{3}{2},\frac{1}{2}\rangle
+\frac{1}{\sqrt{3}}|\frac{1}{2},\frac{1}{2}\rangle,\\
|n\pi^+\rangle&=&\frac{1}{\sqrt{3}}|\frac{3}{2},\frac{1}{2}\rangle
-\sqrt{\frac{2}{3}}|\frac{1}{2},\frac{1}{2}\rangle,\\
|p\pi^-\rangle&=&\frac{1}{\sqrt{3}}|\frac{3}{2},-\frac{1}{2}\rangle
+\sqrt{\frac{2}{3}}|\frac{1}{2},-\frac{1}{2}\rangle,\\
|n\pi^0\rangle&=&\sqrt{\frac{2}{3}}|\frac{3}{2},-\frac{1}{2}\rangle
-\frac{1}{\sqrt{3}}|\frac{1}{2},-\frac{1}{2}\rangle,\\
|n\pi^-\rangle&=&|\frac{3}{2},-\frac{3}{2}\rangle.
\end{eqnarray*}
Using
\begin{displaymath}
\langle I',{I'}\hspace{-.3em}_3|T|I,I_3\rangle=T^I \delta_{II'}
\delta_{I_3 {I'}\hspace{-.1em}_3},
\end{displaymath}
derive the expressions for $\langle p\pi^0|T|n\pi^+\rangle$,
$\langle p\pi^0|T|p\pi^0\rangle$, and $\langle n\pi^+|T|n\pi^+\rangle$.
   Verify that
$$
\langle p\pi^0|T|p\pi^0\rangle-\langle n\pi^+|T|n\pi^+\rangle
=\frac{1}{\sqrt{2}}\langle p\pi^0|T|n\pi^+\rangle.
$$
\end{enumerate}
\end{exercise}

\begin{exercise}
\label{exercise_ps_coupling} \rm
 Consider the so-called pseudoscalar pion-nucleon
interaction
\begin{displaymath}
{\cal L}_{\pi NN}^{\rm PS}=-ig_{\pi N}
\bar{\Psi}\gamma_5\vec{\tau}\cdot\vec{\phi}\Psi.
\end{displaymath}
   The Feynman rule for both the absorption and the emission of a pion
with Cartesian isospin index $a$ is given by
\begin{displaymath}
g_{\pi N}\gamma_5\tau_a.
\end{displaymath}
   Derive the $s$- and $u$-channel contributions
to the invariant amplitude of pion-nucleon scattering.
\end{exercise}

   Let us turn to the tree-level approximation to the $\pi N$ scattering
amplitude as obtained from ${\cal L}^{(1)}_{\pi N}$ of Eq.\ (\ref{5:2:l1pin}).
   In order to derive  the relevant interaction Lagrangians from
Eq.\ (\ref{5:2:l1pin}), we reconsider the connection of Eq.\
(\ref{5:2:gamma}) with the
external fields set to zero and obtain
\begin{eqnarray}
\label{5:3:gammaentwi} \Gamma_\mu&=&
\frac{i}{4F^2}\vec{\tau}\cdot\vec{\phi}\times\partial_\mu\vec{\phi}
+ {\cal O}(\phi^4).
\end{eqnarray}
   The linear pion-nucleon interaction term was already derived in Eq.\
(\ref{5:3:lpin1}) so that we end up with the following
interaction Lagrangian:
\begin{equation}
\label{5:3:lpin} {\cal L}_{\rm int}=-\frac{1}{2}\frac{\stackrel{\circ}{g}_A}{F}
\bar{\Psi}\gamma^\mu\gamma_5 \tau^b\partial_\mu\phi^b\Psi
-\frac{1}{4F^2}\bar{\Psi}\gamma^\mu\underbrace{\vec{\tau}\cdot\vec{\phi}
\times\partial_\mu\vec{\phi}}_{\mbox{$\epsilon_{cde}\tau^c
\phi^d\partial_\mu\phi^e$}}\Psi.
\end{equation}
   The first term is the pseudovector pion-nucleon coupling and the second
the contact interaction with two factors of the pion field interacting
with the nucleon at a single point.
   The Feynman rules for the vertices derived from  Eq.\ (\ref{5:3:lpin}) read
\begin{itemize}
\item for an incoming pion with four-momentum $q$ and Cartesian isospin
index $a$:
\begin{equation}
\label{5:3fr1} -\frac{1}{2}\frac{\stackrel{\circ}{g}_A}{F} q\hspace{-.45em}/
\gamma_5 \tau^a,
\end{equation}
\item for an incoming pion with $q,a$ and an outgoing pion with $q',b$:
\begin{equation}
\label{5:3:fr2} i\left(-\frac{1}{4F^2}\right)\gamma^\mu\epsilon_{cde}\tau^c\left(
\delta^{da}\delta^{eb}iq'_\mu+\delta^{db}\delta^{ea}(-iq)_\mu\right)
=\frac{q\hspace{-.45em}/ +q'\hspace{-.7em}/}{4 F^2}\epsilon_{abc}\tau^c.
\end{equation}
\end{itemize}
\begin{figure}[t]
\begin{center}
\epsfig{file=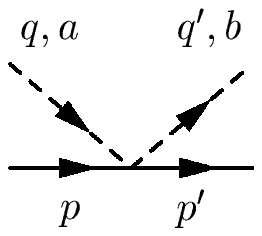,width=0.3\textwidth}
\caption{\label{5:3:fig:piNcontact} Contact contribution to the
pion-nucleon scattering amplitude.}\end{center}
\end{figure}
   The latter gives the contact contribution to ${\cal M}$
(see Figure \ref{5:3:fig:piNcontact}),
\begin{equation}
\label{5:3:cont} {\cal M}_{\rm cont}=\bar{u}(p')
\frac{q\hspace{-.45em}/+q'\hspace{-.7em}/\hspace{.2em}}{4 F^2}
\underbrace{\epsilon_{abc}\tau^c}_{\mbox{$i\frac{1}{2}[\tau^b,\tau^a]$}} u(p) =i
\frac{1}{2 F^2}\bar{u}(p')\frac{1}{2}[\tau^b,\tau^a]
\frac{1}{2}(q\hspace{-.45em}/+q'\hspace{-.7em}/\hspace{.2em})u(p).
\end{equation}
   We emphasize that such a term is not present in a conventional calculation
with either a pseudoscalar or a pseudovector pion-nucleon interaction.

\begin{figure}[t]
\begin{center}
\epsfig{file=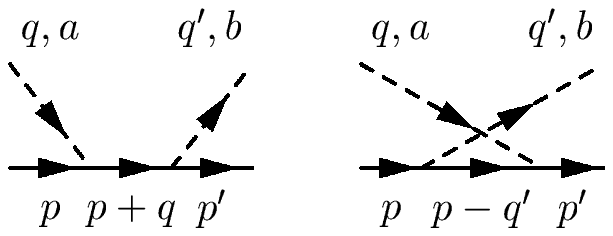,width=0.6\textwidth}
\caption{\label{5:3:fig:piNsu} $s$- and $u$-channel pole
contributions to the pion-nucleon scattering amplitude.}
\end{center}
\end{figure}

   For the $s$- and $u$-channel nucleon-pole diagrams the pseudovector
vertex appears twice (see Figure \ref{5:3:fig:piNsu}) and we
obtain
\begin{eqnarray}
\label{5:3:sukanal} {\cal M}_{s+u}&=&i\frac{\stackrel{\circ}{g}_A^2}{4
F^2}\bar{u}(p') \tau^b\tau^a(-q'\hspace{-.7em}/\hspace{.2em})\gamma_5 \frac{1}{
p'\hspace{-.7em}/\hspace{.2em}+q'\hspace{-.7em}/\hspace{.2em}
-\stackrel{\circ}{m}_N}
q\hspace{-.45em}/\gamma_5u(p)\nonumber\\
&&+i\frac{\stackrel{\circ}{g}_A^2}{4 F^2}\bar{u}(p') \tau^a\tau^b
q\hspace{-.45em}/\gamma_5
\frac{1}{p'\hspace{-.7em}/\hspace{.2em}-q\hspace{-.45em}/\hspace{.2em}
-\stackrel{\circ}{m}_N} (-q'\hspace{-.7em}/\hspace{.2em})\gamma_5 u(p).
\end{eqnarray}
   The $s$- and $u$-channel contributions are related to each other
through pion crossing $a\leftrightarrow b$ and $q\leftrightarrow -q'$.
   In what follows we explicitly calculate only the $s$ channel and
make use of pion-crossing symmetry at the end to obtain the $u$-channel
result.
   Moreover, we perform the manipulations such that the result
of pseudoscalar coupling may also be read off.
   Using the Dirac equation, we rewrite
$$q\hspace{-.45em}/\gamma_5 u(p)=(p'\hspace{-.7em}/\hspace{.2em}
+q'\hspace{-.7em}/\hspace{.2em}
-\stackrel{\circ}{m}_N
+\stackrel{\circ}{m}_N\!-p\hspace{-.45em}/)\gamma_5
u(p)=
(p'\hspace{-.7em}/\hspace{.2em}+q'\hspace{-.7em}/\hspace{.2em}
-\stackrel{\circ}{m}_N)\gamma_5
u(p)
+2\stackrel{\circ}{m}_N\gamma_5 u(p)
$$
and obtain
\begin{eqnarray*}
{\cal M}_s&=&i\frac{\stackrel{\circ}{g}_A^2}{4 F^2}\bar{u}(p')\tau^b\tau^a
(-q'\hspace{-.7em}/\hspace{.2em})\gamma_5 \frac{1}{p'\hspace{-.7em}/\hspace{.2em}
+q'\hspace{-.7em}/\hspace{.2em}-\stackrel{\circ}{m}_N}
\left[(p'\hspace{-.7em}/\hspace{.2em}+q'\hspace{-.7em}/\hspace{.2em}
-\stackrel{\circ}{m}_N)
+2\stackrel{\circ}{m}_N\right]\gamma_5 u(p)\nonumber\\
&\stackrel{\gamma_5^2=1}{=}& i\frac{\stackrel{\circ}{g}_A^2}{4
F^2}\bar{u}(p')\tau^b\tau^a \left[(-q'\hspace{-.7em}/\hspace{.2em})
+(-q'\hspace{-.7em}/\hspace{.2em})\gamma_5
\frac{1}{p'\hspace{-.7em}/\hspace{.2em}+q'\hspace{-.7em}/\hspace{.2em}
-\stackrel{\circ}{m}_N} 2\stackrel{\circ}{m}_N\gamma_5 \right]u(p).
\end{eqnarray*}
   We repeat the above procedure
$$\bar{u}(p')q'\hspace{-.7em}/\hspace{.2em}\gamma_5
=\bar{u}(p')[-2
\stackrel{\circ}{m}_N\!\gamma_5 -\gamma_5(p\hspace{-.45em}/+q\hspace{-.45em}/
-\stackrel{\circ}{m}_N)],
$$
yielding
\begin{equation}
\label{5:2ms1} {\cal M}_s=i\frac{\stackrel{\circ}{g}_A^2}{4
F^2}\bar{u}(p')\tau^b\tau^a [(-q'\hspace{-.7em}/\hspace{.2em})
+\underbrace{4m_N^2\gamma_5
\frac{1}{p'\hspace{-.7em}/\hspace{.2em}+q'\hspace{-.7em}/\hspace{.2em}
-\stackrel{\circ}{m}_N} \gamma_5}_{\mbox{PS coupling}}
+2\stackrel{\circ}{m}_N]u(p),
\end{equation}
   where, for the identification of the PS-coupling result, one has to
make use of the Goldberger-Treiman relation (see Section
\ref{subsec_gtravcme})
$$\frac{\stackrel{\circ}{g}_A}{F}=\frac{\stackrel{\circ}{g}_{\pi N}
}{\stackrel{\circ}{m}_N},$$
where $\stackrel{\circ}{g}_{\pi N}$ denotes the pion-nucleon coupling
constant in the chiral limit.
   Using
$$s-m_N^2=2m_N(\nu-\nu_B),
$$
we find
\begin{eqnarray*}
\bar{u}(p')\gamma_5
\frac{1}{p'\hspace{-.7em}/\hspace{.2em}+q'\hspace{-.7em}/\hspace{.2em}-
\stackrel{\circ}{m}_N}
\gamma_5 u(p)&=&
\bar{u}(p')\gamma_5
\frac{p'\hspace{-.7em}/\hspace{.2em}+q'\hspace{-.7em}/\hspace{.2em}+
\stackrel{\circ}{m}_N}{
(p'+q')^2-\stackrel{\circ}{m}_N^2}\gamma_5 u(p)\nonumber\\
&=&\frac{1}{2\stackrel{\circ}{m}_N(\nu-\nu_B)}\left[-\frac{1}{2}\bar{u}(p')
(q\hspace{-.45em}/+q'\hspace{-.7em}/\hspace{.2em})u(p)\right],
\end{eqnarray*}
where we again made use of the Dirac equation.
   We finally obtain for the $s$-channel contribution
\begin{equation}
\label{5:3:msf} {\cal M}_{s}=i\frac{\stackrel{\circ}{g}_A^2}{4 F^2}
\bar{u}(p')\tau^b\tau^a\left[ 2\stackrel{\circ}{m}_N
+\frac{1}{2}(q\hspace{-.45em}/+q'\hspace{-.7em}/\hspace{.2em})\left(
-1-\frac{2\stackrel{\circ}{m}_N}{\nu-\nu_B}\right)\right]u(p).
\end{equation}
   As noted above, the expression for the $u$ channel results from
the substitution $a\leftrightarrow b$ and $q\leftrightarrow -q'$
\begin{equation}
\label{5:3:ukanal} {\cal M}_{u}=i\frac{\stackrel{\circ}{g}_A^2}{4 F^2}
\bar{u}(p')\tau^a\tau^b\left[ 2\stackrel{\circ}{m}_N
+\frac{1}{2}(q\hspace{-.45em}/+q'\hspace{-.7em}/\hspace{.2em})\left(
1-\frac{2\stackrel{\circ}{m}_N}{\nu+\nu_B}\right)\right]u(p).
\end{equation}
   We combine the $s$- and $u$-channel contributions using
$$\tau^b\tau^a=\frac{1}{2}\{\tau^b,\tau^a\}+\frac{1}{2}[\tau^b,\tau^a],
\quad
\tau^a\tau^b=\frac{1}{2}\{\tau^b,\tau^a\}-\frac{1}{2}[\tau^b,\tau^a],
$$
and
$$
\frac{1}{\nu-\nu_B}\pm\frac{1}{\nu+\nu_B}=
\frac{\left\{\begin{array}{c}2\nu\\ 2\nu_B\end{array}\right\}}{
\nu^2-\nu_B^2}
$$
and summarize the contributions to the functions $A^\pm$ and $B^\pm$ of Eq.\
(\ref{5:3:AB}) in Table \ref{5:3:tableresults}.

\begin{table}[htb]
\begin{center}
\begin{tabular}{|l|c|c|c|c|}
\hline
amplitude$\backslash$origin&PS&$\Delta$PV&contact&sum\\
\hline $A^+$&0&$\frac{\stackrel{\circ}{g}_A^2 \stackrel{\circ}{m}_N}{F^2}$&0&
$\frac{\stackrel{\circ}{g}_A^2 \stackrel{\circ}{m}_N}{F^2}$\\
\hline
$A^-$&0&0&0&0\\
\hline $B^+$&$-\frac{\stackrel{\circ}{g}_A^2}{F^2}
\frac{\stackrel{\circ}{m}_N\nu}{\nu^2-\nu_B^2}$&0&0&
$-\frac{\stackrel{\circ}{g}_A^2}{F^2}
\frac{\stackrel{\circ}{m}_N\nu}{\nu^2-\nu_B^2}$\\
\hline $B^-$& $-\frac{\stackrel{\circ}{g}_A^2}{F^2}
\frac{\stackrel{\circ}{m}_N\nu_B}{\nu^2-\nu_B^2}$&
$-\frac{\stackrel{\circ}{g}_A^2}{2F^2}$& $\frac{1}{2F^2}$&
$\frac{1-\stackrel{\circ}{g}_A^2}{2F^2} -\frac{\stackrel{\circ}{g}_A^2}{F^2}
\frac{\stackrel{\circ}{m}_N\nu_B}{\nu^2-\nu_B^2}$
\\
\hline
\end{tabular}
\caption{\label{5:3:tableresults} Tree-level contributions to the
functions $A^\pm$ and $B^\pm$ of Eq.\ (\ref{5:3:AB}).
   The second column (PS) denotes the result using pseudoscalar pion-nucleon
coupling (using the Goldberger-Treiman relation).
   The sum of the second and third column (PS+$\Delta$PV) represents the
result of  pseudovector pion-nucleon coupling.
   The contact term is specific to the chiral approach.
The last column, the sum of the second, third, and fourth columns,
is the lowest-order ChPT result.}
\end{center}
\end{table}

   In order to extract the scattering lengths, let us consider threshold
kinematics
\begin{equation}
\label{5:3:schkin}
p^\mu=p'^\mu=(m_N,0),\quad
q^\mu=q'^\mu=(M_\pi,0),\quad
\nu|_{\rm thr}=M_\pi,\quad
\nu_B|_{\rm thr}=-\frac{M_\pi^2}{2m_N}.
\end{equation}
   Since we only work at lowest-order tree level, we replace
$\stackrel{\circ}{m}_N\to m_N$, etc.
   Together with\footnote{Recall that we use the normalization
$\bar{u}u=2 m_N$.}
$$u(p)\to \sqrt{2 m_N}\left(\begin{array}{c}\chi\\ 0\end{array}\right),\quad
\bar{u}(p')\to \sqrt{2 m_N}\left(\chi'^\dagger\,\, 0\right)
$$
we find for the threshold matrix element
\begin{equation}
\label{5:3:mthr}
T|_{\rm thr}=2 m_N
\chi'^\dagger\left[\delta^{ab}\left(A^++M_\pi B^+
\right)-i\epsilon_{abc}\tau^c\left(A^-+M_\pi B^-\right)\right]_{\rm
thr}
\chi.
\end{equation}
   Using
$$\left[\nu^2-\nu_B^2\right]_{\rm thr}
=M^2_\pi\left(1-\frac{\mu^2}{4}\right),\quad
\mu=\frac{M_\pi}{m_N}\approx \frac{1}{7},
$$
we obtain
\begin{eqnarray}
\label{5:3:mthrres}
T|_{\rm thr}&=& 2m_N\chi'^\dagger\Bigg[\delta^{ab}
\underbrace{\Bigg(
\frac{g^2_A m_N}{F_\pi^2}+\underbrace{
M_\pi\left(-\frac{g^2_A}{F_\pi^2}\right)
\frac{m_N}{M_\pi}\frac{1}{1-\frac{\mu^2}{4}}}_{\mbox{PS}}\Bigg)
 }_{\mbox{ChPT = PV}}
\nonumber\\
&&-i\epsilon_{abc}\tau^c M_\pi\underbrace{\Bigg(
\frac{1}{2F_\pi^2}
\underbrace{-\frac{g^2_A}{2F_\pi^2}
\underbrace{-\frac{g^2_A}{F_\pi^2}\left(-\frac{1}{2}\right)
\frac{1}{1-\frac{\mu^2}{4}}}_{\mbox{PS}}}_{\mbox{PV}}
\Bigg)}_{\mbox{ChPT}}\Bigg]\chi,
\end{eqnarray}
  where we have indicated the results for the various coupling schemes.

   Let us discuss the $s$-wave scattering lengths resulting from
Eq.\ (\ref{5:3:mthrres}).
   Using the above normalization for the Dirac spinors, the
differential cross section in the center-of-mass frame is given by
\begin{equation}
\label{5:3:dscm}
\frac{d\sigma}{d\Omega}=\frac{|\vec{q}\,'|}{|\vec{q}\,|}
\left(\frac{1}{8\pi \sqrt{s}}\right)^2 |T|^2,
\end{equation}
which, at threshold, reduces to
\begin{equation}
\label{5:3:dsthr}
\left.\frac{d\sigma}{d\Omega}\right|_{\rm thr}
=\left(\frac{1}{8\pi(m_N+M_\pi)}\right)^2
|T|^2_{\rm thr}\stackrel{!}{=}|a|^2.
\end{equation}
   The $s$-wave scattering lengths are defined as\footnote{
The threshold parameters are defined in terms of a multipole expansion of the
$\pi N$ scattering amplitude.
   The sign convention for the $s$-wave scattering parameters $a_{0+}^{(\pm)}$
is opposite to the convention of the effective range expansion.}
\begin{equation}
\label{5:3:apm}
a^\pm_{0+}=
\frac{1}{8\pi(m_N+M_\pi)}T^{\pm}|_{\rm thr}
=\frac{1}{4\pi(1+\mu)}\left[A^\pm+M_\pi B^\pm\right]_{\rm thr}.
\end{equation}
   The subscript $0+$ refers to the fact that the $\pi N$ system is in
an orbital $s$ wave ($l=0$) with total angular momentum $1/2=0+1/2$.
   Inserting the results of Table \ref{5:3:tableresults} we obtain\footnote{We
do not expand the fraction $1/(1+\mu)$, because the $\mu$ dependence is not of
dynamical origin.}
\begin{eqnarray}
\label{5:3:aminus}
a^-_{0+}&=&\frac{M_\pi}{8\pi(1+\mu)F_\pi^2}\left(1+\frac{g_A^2\mu^2}{4}
\frac{1}{1-\frac{\mu^2}{4}}\right)
=\frac{M_\pi}{8\pi(1+\mu)F_\pi^2}[1+{\cal O}(p^2)],\nonumber\\
&&\\
\label{5:3:aplus}
a^+_{0+}&=&
-\frac{g_A^2 M_\pi}{16\pi(1+\mu)F_\pi^2}\frac{\mu}{1-\frac{\mu^2}{4}}
= {\cal O}(p^2),
\end{eqnarray}
   where we have also indicated the chiral order.
   Taking the linear combinations $a^\frac{1}{2}=a^+_{0+}+2 a^-_{0+}$
and $a^\frac{3}{2}=a^+_{0+}-a^-_{0+}$ [see Eq.\ (\ref{5:3:trel})], we see that
the results of Eqs.\ (\ref{5:3:aminus}) and (\ref{5:3:aplus}) indeed satisfy the
Weinberg-Tomozawa relation \cite{Weinberg:1966kf:3:2,Tomozawa:3:2}:\footnote{The
result, in principle, holds for a general target of isospin $T$ (except for the
pion) after replacing 3/4 by $T(T+1)$ and $\mu$ by $M_\pi/M_T$.}
\begin{equation}
\label{5:3:weinbergtomozawa}
a^I=-\frac{M_\pi}{8\pi (1+\mu)F_\pi^2}[I(I+1)-\frac{3}{4}-2].
\end{equation}
   As in $\pi\pi$ scattering, the scattering lengths vanish in the chiral
limit reflecting the fact that the interaction of Goldstone bosons
vanishes in the zero-energy limit.
   The pseudoscalar pion-nucleon interaction produces a scattering length
$a^+_{0+}$ proportional to $m_N$ instead of $\mu M_\pi$ and is clearly in
conflict with the requirements of chiral symmetry.
   Moreover, the scattering length $a^-_{0+}$ of the pseudoscalar coupling
is too large by a factor $g_A^2$ in comparison
with the two-pion contact term of Eq.\ (\ref{5:3:cont}) (sometimes also
referred to as the Weinberg-Tomozawa term) induced by the
nonlinear realization of chiral symmetry.
   On the other hand, the pseudovector pion-nucleon interaction
gives a totally wrong result for $a^-_{0+}$, because it misses
the two-pion contact term of Eq.\ (\ref{5:3:cont}).

   Using the values
\begin{eqnarray}
\label{5:3:par}
&&g_A=1.267,\quad F_\pi=92.4\,\mbox{MeV},\nonumber\\
&&
m_N=m_p=938.3\,\mbox{MeV},\quad
M_\pi=M_{\pi^+}=139.6\,\mbox{MeV},
\end{eqnarray}
   the numerical results for the scattering lengths are given in
Table \ref{5:3:tecomp}.
   We have included the full results of Eqs.\ (\ref{5:3:aminus})
and (\ref{5:3:aplus}) and the consistent corresponding prediction
at ${\cal O}(p)$.
    The empirical results quoted have been taken from
low-energy partial-wave analyses \cite{Koch:bn:3:2,Matsinos:1997pb:3:2}
and recent precision X-ray
experiments on pionic hydrogen and deuterium \cite{Schroder:rc:3:2}.

\begin{table}[htb]
\begin{center}
\begin{tabular}{|l|c|c|}
\hline
Scattering length &$a^+_{0+}$ [MeV$^{-1}$]&$a^-_{0+}$ [MeV$^{-1}$]\\
\hline
Tree-level result &$-6.80\times 10^{-5}$ & $+5.71\times 10^{-4}$ \\
\hline
ChPT ${\cal O}(p)$ & $0$ & $+5.66\times 10^{-4}$ \\
\hline
HBChPT ${\cal O}(p^2)$ \cite{Mojzis:1997tu:3:2} & $-1.3\times 10^{-4}$
&$+5.5\times 10^{-4}$ \\
\hline
HBChPT ${\cal O}(p^3)$ \cite{Mojzis:1997tu:3:2}& $(-7\pm 9)\times 10^{-5}$ &
$(+6.7\pm1.0)\times 10^{-4}$\\
\hline
HBChPT ${\cal O}(p^4)$  [I] \cite{Fettes:2000xg:3:2} & $-6.9\times 10^{-5}$ &
$+6.47\times 10^{-4}$\\
\hline
HBChPT ${\cal O}(p^4)$ [II] \cite{Fettes:2000xg:3:2} & $+3.2\times 10^{-5}$ &
$+5.52\times 10^{-4}$\\
\hline
HBChPT ${\cal O}(p^4)$ [III] \cite{Fettes:2000xg:3:2} & $+1.9\times 10^{-5}$ &
$+6.21\times 10^{-4}$\\
\hline
RChPT ${\cal O}(p^4)$ (a) \cite{Becher:2001hv:3:2}
&$-6.0\times 10^{-5}$ & $+6.55 \times 10^{-4}$\\
\hline
RChPT ${\cal O}(p^4)$ (b) \cite{Becher:2001hv:3:2}
&$-9.4\times 10^{-5}$ & $+6.55 \times 10^{-4}$\\
\hline
PS & $-1.23 \times 10^{-2}$ & $+9.14 \times 10^{-4}$ \\
\hline
PV &$-6.80\times 10^{-5}$ & $+5.06 \times 10^{-6}$\\
\hline
\hline
Empirical values \cite{Koch:bn:3:2} & $ (-7\pm 1)\times 10^{-5}$ &
$(6.6 \pm 0.1) \times
10^{-4}$\\
\hline
Empirical values \cite{Matsinos:1997pb:3:2}
& $(2.04\pm 1.17) \times 10^{-5}$ &
$(5.71\pm 0.12) \times 10^{-4}$\\
&& $(5.92\pm 0.11)  \times 10^{-4}$\\
\hline
Experiment \cite{Schroder:rc:3:2} & $(-2.7\pm 3.6)\times 10^{-5}$  &
$(+6.59\pm0.30)\times 10^{-4}$
 \\
\hline
\end{tabular}
\caption{\label{5:3:tecomp} $s$-wave scattering lengths
$a_{0+}^\pm$.}
\end{center}
\end{table}

\section{The Next-To-Leading-Order Lagrangian}
\label{sec_ntlol}
   The next-to-leading-order pion-nucleon Lagrangian contains seven
low-energy constants $c_i$ \cite{Gasser:1987rb:4,Fettes:2000gb:4},
\begin{eqnarray}
\label{nucl2}
{\cal L}_{\pi N}^{(2)} &=&
c_1 \mbox{Tr}(\chi_{+})\bar\Psi\Psi
- \frac{c_2}{4m^2}\mbox{Tr}(u_\mu u_\nu)
(\bar{\Psi}D^\mu D^\nu \Psi + \mbox{H.c.})
\nonumber\\
&&+\frac{c_3}{2}\mbox{Tr}(u^\mu u_\mu)\bar\Psi\Psi
-\frac{c_4}{4}\bar\Psi\gamma^\mu\gamma^\nu [u_\mu ,u_\nu ]\Psi
+c_5\bar{\Psi}\left[\chi_+ -\frac{1}{2}\mbox{Tr}(\chi_+)\right]\Psi\nonumber\\
&&+\bar{\Psi}\left[\frac{c_6}{2}f_{\mu\nu}^+
    + \frac{c_7}{2} v_{\mu\nu}^{(s)}\right] \sigma^{\mu\nu} \Psi,
\nonumber\\
\end{eqnarray}
where H.c.~refers to the Hermitian conjugate and
\begin{eqnarray*}
\chi_\pm& = &u^\dagger \chi u^\dagger\pm u\chi^\dagger u,\nonumber\\
v_{\mu\nu}^{(s)} &=& \partial_\mu v_\nu^{(s)} - \partial_\nu v_\mu^{(s)},
\nonumber\\
 f_{\mu\nu}^\pm &=& u f_{\mu\nu}^L u^\dagger \pm u^\dagger f_{\mu\nu}^Ru,
\nonumber\\
 f_{\mu\nu}^L &=& \partial_{\mu}l_{\nu} - \partial_{\nu}l_{\mu} -
                    i\left[l_{\mu},l_{\nu}\right],
\nonumber\\
 f_{\mu\nu}^R &=& \partial_{\mu}r_{\nu} - \partial_{\nu}r_{\mu} -
                    i\left[r_{\mu},r_{\nu}\right].
\end{eqnarray*}
   The low-energy constants $c_1,\cdots,c_4$ may be estimated
from a (tree-level) fit \cite{Becher:2001hv:4} to the $\pi N$ threshold
parameters of Koch \cite{Koch:4}:
\begin{equation}
\label{5:4:parametersci}
c_1=-0.9\,m_N^{-1},\quad
c_2=2.5\, m_N^{-1},\quad
c_3=-4.2\, m_N^{-1},\quad
c_4=2.3\, m_N^{-1}.
\end{equation}
Note that other determinations of these parameters exist in the
literature.
   The constant $c_5$ is related to the strong contribution to the
neutron-proton mass difference.

   Finally, the constants $c_6$ and $c_7$ are related to the isovector
and isoscalar magnetic moments of the nucleon in the chiral limit.
   This is seen by considering the coupling to an external electromagnetic
field:
\begin{displaymath}
r_\mu=l_\mu=-e\frac{\tau_3}{2}{\cal A}_\mu,
\quad
v_\mu^{(s)}=-e\frac{1}{2}{\cal A}_\mu.
\end{displaymath}
We then obtain
\begin{displaymath}
v^{(s)}_{\mu\nu}=-e\frac{1}{2}{\cal F}_{\mu\nu},\quad
{\cal F}_{\mu\nu}=\partial_\mu{\cal A}_\nu-\partial_\nu{\cal A}_\mu,
\end{displaymath}
\begin{displaymath}
f_{\mu\nu}^L=\partial_\mu l_\nu-\partial_\nu l_\mu
-i\underbrace{[l_\mu,l_\nu]}_{0}=-e\frac{\tau_3}{2}{\cal F}_{\mu\nu}=
f_{\mu\nu}^R,
\end{displaymath}
and thus
\begin{displaymath}
f_{\mu\nu}^+=uf_{\mu\nu}^L u^\dagger+u^\dagger f^R_{\mu\nu} u
= f_{\mu\nu}^L+f_{\mu\nu}^R+\cdots
=-e\tau_3 {\cal F}_{\mu\nu}+\cdots.
\end{displaymath}
   We thus obtain for the terms without pion fields
\begin{displaymath}
-\frac{e}{2}\bar{\Psi}\left(c_6\tau_3+\frac{1}{2}c_7\right)\sigma^{\mu\nu}\Psi
{\cal F}_{\mu\nu}.
\end{displaymath}
   Comparing with the interaction Lagrangian of
a magnetic field with the {\em anomalous} magnetic moment of the nucleon,
\begin{displaymath}
-\frac{e}{4m_N}\bar\Psi\frac{1}{2}(\kappa^{(s)}+\tau_3\kappa^{(v)})
\sigma^{\mu\nu}\Psi{\cal F}_{\mu\nu},
\end{displaymath}
we obtain
\begin{displaymath}
c_7=\frac{\stackrel{\circ}{\kappa}^{(s)}}{2\stackrel{\circ}{m}_N},\quad
c_6=\frac{\stackrel{\circ}{\kappa}^{(v)}}{4\stackrel{\circ}{m}_N},
\end{displaymath}
where $\circ$ denotes the chiral limit.
   The physical values read
\begin{displaymath}
\kappa_p=\frac{1}{2}(\kappa^{(s)}+\kappa^{(v)})=1.793,\quad
\kappa_n=\frac{1}{2}(\kappa^{(s)}-\kappa^{(v)})=-1.913,
\end{displaymath}
and thus $\kappa^{(s)}=-0.120$ and $\kappa^{(v)}=3.706$.

\section{Example for a Loop Diagram}
\label{sec_eld}
   In Section \ref{sec_elwpcs} we saw that, in the purely mesonic sector,
contributions of $n$-loop diagrams are at least of order ${\cal O}(p^{2n+2})$,
i.e., they are suppressed by $p^{2n}$ in comparison with tree-level diagrams.
   An important ingredient in deriving this result was the fact that we
treated the squared pion mass as a small quantity of order $p^2$.
   Such an approach is motivated by the observation that the masses of the
Goldstone bosons must vanish in the chiral limit.
   In the framework of ordinary chiral perturbation theory $M_\pi^2\sim m_q$
[see Eq.\ (\ref{4:3:mpi2})]
which translates into a momentum expansion of observables
at fixed ratio $m_q/p^2$.
   On the other hand, there is no reason to believe that the masses of
hadrons other than the Goldstone bosons should vanish or become small in the
chiral limit.
   In other words, the nucleon mass entering the pion-nucleon Lagrangian
of Eq.\ (\ref{5:2:l1pin}) should---as already anticipated in the discussion
following Eq.\ (\ref{5:2:l1pin})---not be treated as a small quantity of,
say, order ${\cal O}(p)$.

   Naturally the question arises how all this affects the calculation of
loop diagrams and the setup of a consistent power counting scheme.
   Our goal is to propose a renormalization procedure generating a power counting for
tree-level and loop diagrams of the (relativistic) EFT for baryons
which is analogous to that given in Section \ref{sec_elwpcs} for mesons.
   Choosing a suitable renormalization condition will
allow us to apply the following power counting:
a loop integration in $n$ dimensions counts as $p^n$,
pion and fermion propagators count as $p^{-2}$ and
$p^{-1}$, respectively, vertices derived from ${\cal L}_{2k}$ and
${\cal L}_{\pi N}^{(k)}$ count as $p^{2k}$ and $p^k$, respectively.
   Here, $p$ generically denotes a small expansion parameter such as,
e.g., the pion mass.
   In total this yields for the power $D$ of a diagram  the standard formula
\begin{equation}
\label{5:5:dimension1}
D=n N_L - 2 I_\pi - I_N +\sum_{k=1}^\infty 2k N^\pi_{2k}
+\sum_{k=1}^\infty k N_k^N,
\end{equation}
where $N_L$, $I_\pi$, $I_N$, $N^\pi_{2k}$, and $N^N_k$ denote the number
of independent loop momenta, internal pion lines, internal nucleon
lines, vertices originating from ${\cal L}_{2k}$, and vertices originating
from ${\cal L}_{\pi N}^{(k)}$, respectively.
   We make use of the relation\footnote{This relation can be understood as
follows: For each internal line we have a propagator in combination with an
integration with measure $d^4 k/(2\pi)^4$. So we end up with $I_\pi+I_N$
integrations. However, at each vertex we have a four-momentum conserving delta
function, reducing the number of integrations by $N_\pi+N_N-1$, where the $-1$ is
related to the overall four-momentum conserving delta function
$\delta^4(P_f-P_i)$.}
\begin{displaymath}
N_L=I_\pi+I_N-N_\pi-N_N+1
\end{displaymath}
with $N_\pi$ and $N_N$ the total number of pionic and baryonic vertices,
respectively, to eliminate $I_\pi$:
\begin{displaymath}
D=(n-2)N_L+I_N+2+\sum_{k=1}^\infty 2(k-1)N^\pi_{2k}
+\sum_{k=1}^\infty(k-2)N^N_k.
\end{displaymath}
   Finally, for processes containing exactly one nucleon in the initial
and final states we have\footnote{In the low-energy effective field
theory one has no closed fermion loops. In other words, in the
single-nucleon sector exactly one fermion line runs through the diagram
connecting the initial and final states.}
$N_N=I_N+1$ and we thus obtain
\begin{eqnarray}
\label{5:5:dimension2}
D&=&1+(n-2)N_L+\sum_{k=1}^\infty 2(k-1) N^\pi_{2k}
+\sum_{k=1}^\infty (k-1) N_k^N\nonumber\\ &&\\
&\geq&\mbox{1 in 4 dimensions}.\nonumber
\end{eqnarray}
   According to Eq.\ (\ref{5:5:dimension2}), one-loop calculations in the
single-nucleon sector should
start contributing at ${\cal O}(p^{n-1})$.
   For example, let us consider the one-loop contribution of the first
diagram of Figure \ref{SEDia} to the nucleon self-energy.
   According to Eq.\ (\ref{5:5:dimension1}), the renormalized result should
be of order
\begin{equation}
\label{5:5:dexample}
D=n\cdot 1-2\cdot 1-1\cdot 1+1\cdot 2=n-1.
\end{equation}
   We will see below that the corresponding renormalization scheme is
more complicated than in the mesonic sector.

\subsection{One-Loop Correction to the Nucleon Mass}
\label{subsec_feolcnm}

\begin{exercise}
\label{exercise_nucleon_mass}
\rm
\renewcommand{\labelenumi}{(\alph{enumi})}
In the following we will calculate the mass $m_N$ of the nucleon up to and
including order ${\cal O}(p^3)$. As in the case of pions, the physical mass is
defined through the pole of the full propagator (at $\slashed{p}=m_N$ for the
nucleon). The propagator is given by
\begin{equation}\label{NProp}
    S_0(p)=\frac{1}{\slashed{p}-m_0-\Sigma_0(\slashed{p})}\equiv
    \frac{1}{\slashed{p}-\stackrel{\circ}{m}_N-\Sigma(\slashed{p})},
\end{equation}
where $m_0$ refers to the bare mass, $\stackrel{\circ}{m}_N$ is
the nucleon mass in the chiral limit and $\Sigma_0(\slashed{p})$
denotes the nucleon self energy. To determine the mass, the
equation
\begin{equation}\label{MassDef}
    m_N-m_0-\Sigma_0(m_N)=m_N-\stackrel{\circ}{m}_N-\Sigma(m_N)=0
\end{equation}
has to be solved, so the task is to calculate the nucleon self
energy $\Sigma(\slashed{p})$.

\begin{enumerate}
\item The $\pi N$ Lagrangian at order ${\cal O}(p^2)$ is given by
\begin{eqnarray}\label{LpN2}
 \mathcal{L}_{\pi N}^{(2)}&=& c_1\,\mbox{Tr}(\chi_+)\bar{\Psi}\Psi
-\frac{c_2}{4m^2}\left[\bar{\Psi}\,
\mbox{Tr}(u_{\mu}u_{\nu})D^{\mu}D^{\nu}\Psi+\mbox{H.c.}\right]
\nonumber\\
 &+& \bar{\Psi}\left[\frac{c_3}{2}\,\mbox{Tr}(u_\mu u^\mu)
+i\frac{c_4}{4}[u_{\mu},u_{\nu}]+c_5\left[\chi_+-\frac{1}{2}\mbox{Tr}(\chi_+)
\right]\right.\nonumber\\
&&\left.
+\frac{c_6}{2}f_{\mu\nu}^+
+\frac{c_7}{2}v_{\mu\nu}^{(s)}\right]\sigma^{\mu\nu}\Psi.\nonumber\\
\end{eqnarray}
Which of these terms contain only the nucleon fields and therefore
give a contact contribution to the self energy? Determine
$-i\Sigma^{\rm contact}(\slashed{p})$ from $i \langle \bar{\Psi}|
    \mathcal{L}_{{\pi}N}^{(2)}|\Psi\rangle.$

\noindent Remark: There are no contact contributions from the
Lagrangian $\mathcal{L}_{\pi N}^{(3)}$.

\item By using the expansion of $\mathcal{L}_{\pi N}^{(1)}$ up to
two pion fields from Exercise~\ref{exercise_pion_nucleon_interaction} verify the
following Feynman rules:\footnote{Here, the subscripts 0 denote bare quantities.
The generation of counterterms is discussed in Section \ref{genct}.}

\vspace{2em}

\begin{tabular}{cc}
\parbox{4cm}{\epsfig{file=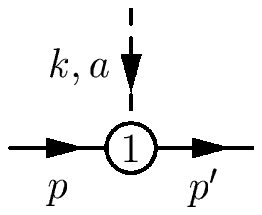,width=2.5cm}}&
\parbox{8cm}{\begin{displaymath}
-\frac{\stackrel{\circ}{g}_{A0}}{2F_0}\,\slashed{k}\gamma_5\tau_a\end{displaymath}}\\
\\
\parbox{4cm}{\epsfig{file=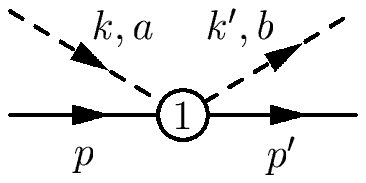,width=3cm}}&
\parbox{8cm}{\begin{displaymath}
\frac{1}{4F_0^2}\,(\slashed{k}+\slashed{k}')\epsilon_{abi}\tau_i
\end{displaymath}}\\
\\
\end{tabular}

There are two types of loop contributions at order ${\cal
O}(p^3)$, shown in Figure \ref{SEDia}.

\begin{figure}[t]\begin{center}
\epsfig{file=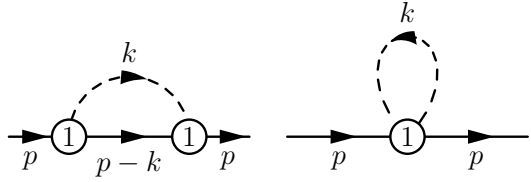,width=7cm}\caption{\label{SEDia}One-loop
contributions to the nucleon self-energy}
\end{center}\end{figure}

\item Use the Feynman rules to show that the second diagram in
Figure \ref{SEDia} does not contribute to the self energy.

\item Use the Feynman rules and the expressions for the propagators,
\begin{eqnarray*}
  i\Delta_\pi(p) &=& \frac{i}{p^2-M^2+i0^+}, \\
  i S_N(p) &=& i\,\frac{\slashed{p}+\stackrel{\circ}{m}_N-i0^+}
  {p^2-\stackrel{\circ}{m}_N^2+i0^+},
\end{eqnarray*}
to verify that in dimensional regularization the first diagram in
Figure \ref{SEDia} gives the contribution
\begin{equation}\label{Loop1}
    -i\Sigma^{\rm loop}(\slashed{p})
=-i\frac{3\stackrel{\circ}{g}_{A0}^2}{4F_0^2}\,
    i\mu^{4-n}\int\frac{d^nk}{(2\pi)^n}\frac{\slashed{k}
    (\slashed{p}-\stackrel{\circ}{m}_N-\slashed{k})\slashed{k}}
    {[(p-k)^2-\stackrel{\circ}{m}_N^2+i0^+][k^2-M^2+i0^+]}.
\end{equation}

\item Show that the numerator can be simplified to
\begin{equation}\label{Num}
    -(\slashed{p}+\stackrel{\circ}{m}_N)k^2+(p^2-\stackrel{\circ}{m}_N^2)\slashed{k}
    -\left[(p-k)^2-\stackrel{\circ}{m}_N^2\right]\slashed{k},
\end{equation}
which, when inserted in Eq.~(\ref{Loop1}), gives
\begin{eqnarray}\label{LoopRes}
\lefteqn{\Sigma^{\rm loop}(\slashed{p}) =
  \frac{3\stackrel{\circ}{g}_{A0}^2}{4F_0^2}\,
  \left\{-(\slashed{p}+\stackrel{\circ}{m}_N)\mu^{4-n}i\int\frac{d^nk}{(2\pi)^n}
  \frac{1}{[(p-k)^2-\stackrel{\circ}{m}_N^2+i0^+]}\right.}\nonumber\\
&&-(\slashed{p}+\stackrel{\circ}{m}_N)M^2\mu^{4-n}i\int\frac{d^nk}{(2\pi)^n}
  \frac{1}{[(p-k)^2-\stackrel{\circ}{m}_N^2+i0^+][k^2-M^2+i0^+]}\nonumber\\
&&
+(p^2-\stackrel{\circ}{m}_N^2)\mu^{4-n}i\int\frac{d^nk}{(2\pi)^n}
  \frac{\slashed{k}}{[(p-k)^2-\stackrel{\circ}{m}_N^2+i0^+][k^2-M^2+i0^+]}\nonumber\\
&& \left.-\mu^{4-n}i\int\frac{d^nk}{(2\pi)^n}
\frac{\slashed{k}}{[k^2-M^2+i0^+]}\right\}.
\end{eqnarray}

Hint: $ \{\gamma_\mu, \gamma_\nu\}=2g_{\mu\nu},\quad \{\gamma_\mu,
\gamma_5\}=0,\quad \gamma_5 \gamma_5=1,\quad k^2=k^2-M^2+M^2.$

\item The last term in Eq.~(\ref{LoopRes}) vanishes since the
integrand is odd in $k$. We use the following convention for
scalar loop integrals
\begin{eqnarray*}\label{scalarInt}
    \lefteqn{I_{N\cdots\pi\cdots}(p_1,\cdots,q_1,\cdots)}\\
    &&=\mu^{4-n}i\int\frac{d^nk}{(2\pi)^n}
    \frac{1}{[(k+p_1)^2-\stackrel{\circ}{m}_N^2+i0^+]\cdots[(k+q_1)^2-M^2+i0^+]\cdots}\,.
\end{eqnarray*}

To determine the vector integral use the ansatz
\begin{equation}\label{vecInt}
    \mu^{4-n}i\int\frac{d^nk}{(2\pi)^n}
  \frac{k_\mu}{[(p-k)^2-\stackrel{\circ}{m}_N^2+i0^+][k^2-M^2+i0^+]}=
  p_\mu \, C.
\end{equation}
Multiply Eq.~(\ref{vecInt}) by $p^\mu$ to show that $C$ is given
by
\begin{equation}\label{C}
    C=\frac{1}{2p^2}\left[I_N-I_\pi+(p^2-\stackrel{\circ}{m}_N^2+M^2)I_{N\pi}(-p)\right].
\end{equation}

Using the above convention the loop contribution to the nucleon
self energy reads
\begin{eqnarray}\label{LoopResInt}
\Sigma^{\rm loop}(\slashed{p})
&=&-\frac{3\stackrel{\circ}{g}_{A0}^2}{4F_0^2}\,\Bigg\{
(\slashed{p}+\stackrel{\circ}{m}_N)I_N+(\slashed{p}+\stackrel{\circ}{m}_N)M^2
I_{N\pi}(-p,0)\nonumber\\
&&-(p^2-\stackrel{\circ}{m}_N^2)\frac{\slashed{p}}{2p^2}
\left[I_N-I_\pi+(p^2-\stackrel{\circ}{m}_N^2+M^2)I_{N\pi}(-p)\right]\Bigg\}.\nonumber\\
\end{eqnarray}

The explicit expressions for the integrals are given by
\begin{eqnarray}\label{Int}
  I_{\pi}&=&\frac{M^2}{16\pi^2}\left[R+\ln{\left(\frac{M^2}{\mu^2}\right)}\right],\nonumber\\
  I_{N}&=&\frac{\stackrel{\circ}{m}_N^2}{16\pi^2}
\left[R+\ln{\left(\frac{\stackrel{\circ}{m}_N^2}{\mu^2}\right)}\right],\nonumber\\
  I_{N\pi}(p,0)&=&\frac{1}{16\pi^2}
\left[R+\ln{\left(\frac{\stackrel{\circ}{m}_N^2}{\mu^2}\right)}-1
\right.\nonumber\\
&&\left.
+
\frac{p^2-\stackrel{\circ}{m}_N^2-M^2}{p^2}\ln{\left(\frac{M}{\stackrel{\circ}{m}_N}\right)}
+\frac{2\stackrel{\circ}{m}_N M}{p^2}F(\Omega)\right],\nonumber\\
\end{eqnarray}
where
\begin{eqnarray*}
R&=&\frac{2}{n-4}-[\ln(4\pi)+{\Gamma}'(1)+1],\\
\Omega&=&\frac{p^2-\stackrel{\circ}{m}_N^2-M^2}{2\stackrel{\circ}{m}_NM},
\end{eqnarray*}
and
\begin{displaymath}
F(\Omega)=\left\{\begin{array}{l@{\quad}l}
\sqrt{\Omega^2-1}\ln{\left(-\Omega-\sqrt{\Omega^2-1}\right)}, &
\Omega\le -1,\\
\sqrt{1-\Omega^2}\;\arccos(-\Omega),&-1 \le\Omega\le 1,\\
\sqrt{\Omega^2-1}\ln{\left(\Omega+\sqrt{\Omega^2-1}\right)}
-i\pi\sqrt{\Omega^2-1}, & 1 \le \Omega\,.
\end{array}\right.
\end{displaymath}

\item The result for the self energy contains divergences as
$n\rightarrow 4$ (the terms proportional to $R$), so it has to be
renormalized. For convenience, choose the renormalization
parameter $\mu=\stackrel{\circ}{m}_N$. The $\widetilde{\rm MS}$
renormalization can be performed by simply dropping the terms
proportional to $R$ and by replacing all bare coupling constants
($c_{1}, \stackrel{\circ}{g}_{A0}, F_0$) with the renormalized
ones, now indicated by a subscript $r$. The $\widetilde{\rm MS}$
renormalized self energy contribution then reads
\begin{eqnarray}\label{SErenorm}
\Sigma^{\rm
loop}_r(\slashed{p})&=&-\frac{3\stackrel{\circ}{g}_{Ar}^2}{4F_r^2}\,\Bigg\{
(\slashed{p}+\stackrel{\circ}{m}_N)M^2
I_{N\pi}^r(-p,0)\nonumber\\
&&-(p^2-\stackrel{\circ}{m}_N^2)\frac{\slashed{p}}{2p^2}
\left[(p^2-\stackrel{\circ}{m}_N^2+M^2)I_{N\pi}^r(-p)-I_\pi^r\right]\Bigg\},\nonumber\\
\end{eqnarray}
where the superscript $r$ on the integrals means that the terms
proportional to $R$ have been dropped. Using the definition of the
integrals, show that Eq.~(\ref{SErenorm}) contains a term of order
${\cal O}(p^2)$. What does the presence of this term tell you
about the applicability of the $\widetilde{\rm MS}$ scheme in baryon
ChPT?

Hint: What chiral order did the power counting assign to the
diagram from which we calculated $\Sigma^{\rm loop}$?

\item  We can now solve Eq.~(\ref{MassDef}) for the nucleon mass,
\begin{eqnarray}
    m_N&=&\stackrel{\circ}{m}_N
         +\Sigma^{\rm contact}_r(m_N)
         +\Sigma^{\rm loop}_r(m_N)\nonumber\\
       &=&\stackrel{\circ}{m}_N-4c_{1r}M^2+\Sigma^{\rm loop}_r(m_N).
\end{eqnarray}
We have $m_N-\stackrel{\circ}{m}_N={\cal O}(p^2)$. Since our
calculation is only valid up to order ${\cal O}(p^3)$, determine
$\Sigma^{\rm loop}_r(m_N)$ to that order. Check that you only need an
expansion of $I_{N\pi}^r$, which, using
$$\arccos{(-\Omega)}=\frac{\pi}{2}+\cdots,$$
verify to be
\begin{equation}\label{INpExp}
    I_{N\pi}^r=\frac{1}{16\pi^2}\left(-1+\frac{\pi
    M}{\stackrel{\circ}{m}_N}\cdots\right).
\end{equation}
Show that this yields
\begin{equation}\label{Mass}
    m_N=\stackrel{\circ}{m}_N-4c_{1r}M^2+
    \frac{3\stackrel{\circ}{g}_{Ar}^2M^2}{32\pi^2F_r^2}\stackrel{\circ}{m}_N
    -\frac{3\stackrel{\circ}{g}_{Ar}^2M^3}{32\pi^2F_r^2}.
\end{equation}

\item The solution to the power counting problem is the observation
that the term violating the power counting (the third on the right
of Eq.~(\ref{Mass})) is \emph{analytic} in small quantities and
can thus be absorbed in counter terms. In addition to the
$\widetilde{\rm MS}$ scheme we have to perform an additional
\emph{finite} renormalization. Rewrite
\begin{equation}\label{cRenorm}
    c_{1r}=c_1+\delta c_1
\end{equation}
in Eq.~(\ref{Mass}) and determine $\delta c_1$ so that the term
violating the power counting is absorbed, which then gives the
final result for the nucleon mass at order ${\cal O}(p^3)$
\begin{equation}\label{MassFinal}
    m_N=\stackrel{\circ}{m}_N-4c_{1}M^2
    -\frac{3\stackrel{\circ}{g}_{Ar}^2M^3}{32\pi^2F_r^2}.
\end{equation}

\end{enumerate}

\end{exercise}

   We saw in Exercise \ref{exercise_nucleon_mass} that, for the case
of the nucleon self energy, the expression for loop diagrams
renormalized by applying dimensional regularization in combination
with the $\widetilde{\rm MS}$ scheme as in the mesonic sector
contained terms not consistent with the power counting.
   The appearance of terms violating the power counting when using
the $\widetilde{\rm MS}$ scheme is a general feature of loop calculations in
baryonic chiral perturbation theory \cite{Gasser:1987rb:5}.
   However, in the example above these terms were analytic in small parameters and could be absorbed
by an additional finite renormalization.
   The question arises if this can be done in general.
   Indeed, there are several renormalization schemes that yield a consistent power
counting for the baryonic sector of chiral perturbation theory.
   They make use of the observation that the terms violating
the power counting are analytic in small parameters and can thus be absorbed in
the available parameters.
   We briefly mention two methods without going into any
details.

   In the infrared regularization of Becher and Leutwyler \cite{Becher:1999he:5},
one-loop integrals are split into two pieces,
\begin{eqnarray*}
    H=\int_0^1 dx \cdots &=& \int_0^\infty dx \cdots - \int_1^\infty dx \cdots\\
    &=&I+R,
\end{eqnarray*}
where $I$ satisfies the power counting, whereas $R$ violates it and is absorbed
in counterterms.

   In the extended-on-mass-shell (EOMS) scheme
\cite{Fuchs:2003qc:5}, the integrand of loop integrals is expanded in small
parameters, and the (integrated) terms violating the power counting are then
subtracted.
   The advantage of the EOMS scheme is that it can also be applied
in the case of diagrams containing resonances \cite{Fuchs:2003sh:5} as well as
multi-loop diagrams \cite{Schindler:2003je:5}.

   Applying similar techniques as in Exercise \ref{exercise_nucleon_mass},
the result for the mass of the nucleon at ${\cal O}(p^4)$ in the EOMS scheme is
given by \cite{Fuchs:2003qc:5,Fuchs:2003kq:5:1}
\begin{equation}
\label{5:5:mnoq4}
m_N=\stackrel{\circ}{m}_N
+k_1 M^2+k_2 M^3+k_3 M^4\ln\left(\frac{M}{\stackrel{\circ}{m}_N}\right)
+k_4 M^4
+{\cal O}(M^5),
\end{equation}
where the coefficients $k_i$ are given by
\begin{eqnarray}
\label{5:5:parki}
&&k_1=-4 c_1,\quad
k_2=-\frac{3 {\stackrel{\circ}{g_A}}^2}{32\pi F^2},\quad
k_3=\frac{3}{32\pi^2 F^2}\left(8c_1-c_2-4 c_3
-\frac{{\stackrel{\circ}{g_A}}^2}{\stackrel{\circ}{m}_N}\right),
\nonumber\\
&&k_4=\frac{3 {\stackrel{\circ}{g_A}}^2}{32 \pi^2 F^2 \stackrel{\circ}{m}_N}
(1+4 c_1 \stackrel{\circ}{m}_N)
+\frac{3}{128\pi^2F^2}c_2+\frac{1}{2}\alpha.
\end{eqnarray}
Here, $\alpha=-4(8 e_{38}+e_{115}+e_{116})$ is a linear combination
of ${\cal O}(p^4)$ coefficients.
  In order to obtain an estimate for the various contributions
of Eq.\ (\ref{5:5:mnoq4}) to the nucleon mass, we make use of
the set of parameters $c_i$ of Eq.\ (\ref{5:4:parametersci}).
   Using the numerical values
\begin{displaymath}
g_A=1.267,\quad F_\pi=92.4\,\mbox{MeV},\quad
m_N=938.3\,\mbox{MeV},\quad
M_\pi=139.6\,\mbox{MeV},
\end{displaymath}
we obtain for the mass of nucleon in the chiral limit (at
fixed $m_s\neq 0$):
\begin{displaymath}
\stackrel{\circ}{m}_N=m_N-\Delta m=[938.3-74.8+15.3+4.7+1.6-2.3]\,\mbox{MeV}
=882.8\, \mbox{MeV}
\end{displaymath}
with $\Delta m=55.5\,\mbox{MeV}$.
   Here, we have made use of an estimate for $\alpha$ obtained from
the $\sigma$ term (see Ref.\ \cite{Fuchs:2003kq:5:1} for more details).
   The chiral expansion reveals a good convergence and it will be interesting
to further study the convergence at the two-loop level.

\begin{table}
\begin{tabular}[t]{|c|c|c|c|c|}
\hline
& Chiral limit: $M_0$ & $M_{(2)}$ & $M_{(3)}$ & Sum at ${\cal O}(p^3)$: $M_3$\\
\hline
$M_N$&1039&240&$-339$&940\\
$M_\Sigma$&1039&849&$-696$&1192\\
$M_\Lambda$&1039&811&$-737$&1113\\
$M_\Xi$&1039&1400&$-1120$&1319\\
$\sigma_{\pi N}$&---&85&$-40$&45\\
\hline
\end{tabular}
\caption{\label{table:masses:su3}
Baryon masses in MeV and their individual contributions. Note that the free
parameters have been fit so that the masses at ${\cal O}(p^3)$ essentially agree
with the physical masses.}
\end{table}

\begin{figure}
\begin{center}
\epsfbox{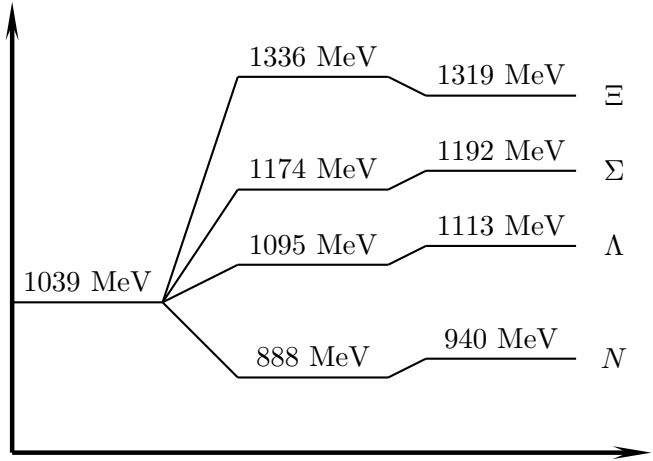}
\end{center}
\caption{\label{massen:fig} Mass level diagram depending on the various
symmetries. Left panel: $\mbox{SU(3)}_L\times\mbox{SU(3)}_R$ symmetry; middle
panel: $\mbox{SU(2)}_L\times\mbox{SU(2)}_R$ symmetry; right panel:
$\mbox{SU(2)}_V$ symmetry.}
\end{figure}

   Finally, it is straightforward but more tedious to apply the same techniques
to an SU(3) calculation of the masses of the baryon octet
\cite{Lehnhart:2004vi:5:1}.
   The results and their individual contributions are shown in Table
\ref{table:masses:su3}.
   Large cancellations between the contributions $M_{(2)}$ and $M_{(3)}$ at ${\cal
O}(p^2)$ and ${\cal O}(p^3)$, respectively, are a well-known feature.
   In Figure \ref{massen:fig} we show how ``switching on'' the quark masses affects
the masses of the baryon octet.
   In the chiral limit all masses reduce to $M_0=1039$ MeV.
   Keeping the up and down quarks massless, we still have an
$\mbox{SU(2)}_L\times\mbox{SU(2)}_R$ symmetry resulting in
\begin{equation}
M_{\pi,2}^2 = 0,~~~ M_{K,2}^2 ~=~ B_0 m_s,~~~ M_{\eta,2}^2 ~=~ \frac{4}{3} B_0
m_s ~=~ \frac{4}{3} M_{K,2}^2.
\end{equation}
   The corresponding values of the mass spectrum are shown in the middle panel of
Figure \ref{massen:fig} ($M_\pi=0$, $M_K=486$ MeV and $M_\eta=562$ MeV), while
the final results, exhibiting only an $\mbox{SU(2)}_V$ symmetry, are shown in the
right panel.

\subsection{\label{genct}The Generation of Counterterms *}
   The renormalization of the effective field theory (of pions
and nucleons) is performed by expressing all the bare parameters and bare fields
of the effective Lagrangian in terms of renormalized quantities (see Ref.\
\cite{Collins:xc:5:2} for details).
   In this process, one generates counterterms which are responsible for the
absorption of all the divergences occurring in the calculation of loop
diagrams.
   In order to illustrate the procedure let us discuss
${\cal L}_{\pi N}^{(1)}$
and consider the free part in combination with the $\pi N$
interaction term with the smallest number of pion fields,
\begin{equation}
{\cal L}_{\pi N}^{(1)}=\bar \Psi_0 \left( i\gamma_\mu
\partial^\mu -m_0 -\frac{1}{2}\frac{{\stackrel{\circ}{g}_{A}}_0}{F_0}
\gamma_\mu
\gamma_5 \tau^a \partial^\mu \pi^a_0\right) \Psi_0 +\cdots,
\label{pieceoflolagr}
\end{equation}
given in terms of bare fields and parameters denoted by subscripts 0.
   Introducing renormalized fields (we work in the isospin-symmetric
limit) through
\begin{equation}
\label{renf}
\Psi=\frac{\Psi_0}{\sqrt{Z_\Psi}},\quad
\pi^a=\frac{\pi^a_0}{\sqrt{Z_\pi}},
\end{equation}
we express the field redefinition constants $\sqrt{Z_\Psi}$ and
$\sqrt{Z_\pi}$
and the bare quantities in terms of renormalized parameters:
\begin{eqnarray}
Z_\Psi&=& 1+\delta Z_\Psi\left(\stackrel{\circ}{m}_N,
\stackrel{\circ}{g}_A, g_i, \nu \right),
\nonumber\\
Z_{\pi}&=&1+\delta Z_\pi\left(\stackrel{\circ}{m}_N,
\stackrel{\circ}{g}_A, g_i, \nu \right),
\nonumber \\
m_0 &=& \stackrel{\circ}{m}_N(\nu )+\delta m\left(\stackrel{\circ}{m}_N,
\stackrel{\circ}{g}_{A}, g_i,
\nu \right), \nonumber\\
{\stackrel{\circ}{g}_{A}}_0&=&\stackrel{\circ}{g}_{A}(\nu)+\delta g_A \left(
\stackrel{\circ}{m}_N, \stackrel{\circ}{g}_{A}, g_i, \nu \right),
\label{bare}
\end{eqnarray}
where $g_i$, $i=1,\cdots, \infty$, collectively denote all the
renormalized parameters which correspond to bare parameters
${g_i}_0$ of the full effective Lagrangian.
   The parameter $\nu$
indicates the dependence on the choice of the
renormalization prescription.\footnote{Note that our choice
$\stackrel{\circ}{m}_N(\nu
)=\,\stackrel{\circ}{m}_N$, where $\stackrel{\circ}{m}_N$
is the nucleon pole mass in the chiral limit, is
only one among an infinite number of possibilities.}
   Substituting Eqs.\ (\ref{renf}) and (\ref{bare}) into
Eq.\ (\ref{pieceoflolagr}), we obtain
\begin{equation}
\label{bct}
{\cal L}_{\pi N}^{(1)}={\cal L}_{\rm basic}+{\cal L}_{\rm ct}+\cdots
\end{equation}
with the so-called basic and counterterm Lagrangians,
respectively,\footnote{Ref.\ \cite{Collins:xc:5:2} uses a slightly different
convention which is obtained through the replacement $(\delta Z_\Psi
m+Z_\Psi\delta m)\to\delta m$.}
\begin{eqnarray}
\label{lbasic}
{\cal L}_{\rm basic}&=&\bar \Psi \left( i\gamma_\mu
\partial^\mu -\stackrel{\circ}{m}_N
-\frac{1}{2} \frac{\stackrel{\circ}{g}_{A}}{F}\gamma_\mu
\gamma_5 \tau^a \partial^\mu \pi^a\right) \Psi,\\
\label{lcounterterm}
{\cal L}_{\rm ct}&=&\delta Z_\Psi \bar \Psi i\gamma_\mu\partial^\mu
\Psi
-\delta\{m\}\bar{\Psi}\Psi
-\frac{1}{2}\delta\left\{\frac{\stackrel{\circ}{g}_{A}}{F}\right\}
\bar{\Psi}\gamma_\mu \gamma_5 \tau^a \partial^\mu \pi^a \Psi, \nonumber\\
\end{eqnarray}
where we introduced the abbreviations
\begin{eqnarray*}
\delta\{m\}&\equiv&\delta Z_\Psi m+Z_\Psi\delta m,\\
\delta\left\{\frac{\stackrel{\circ}{g}_{A}}{F}\right\}&\equiv&
\delta Z_\Psi \frac{\stackrel{\circ}{g}_{A}}{F}\sqrt{Z_\pi}
+Z_\Psi\left(\frac{{\stackrel{\circ}{g}_{A}}_0}{F_0}-
\frac{\stackrel{\circ}{g}_{A}}{F}
\right)\sqrt{Z_\pi}
+\frac{\stackrel{\circ}{g}_{A}}{F}(\sqrt{Z_\pi}-1).
\end{eqnarray*}
   In Eq.\ (\ref{lbasic}), $\stackrel{\circ}{m}_N$, $\stackrel{\circ}{g}_A$,
and $F$ denote
the chiral limit of the physical nucleon mass, the axial-vector coupling
constant, and the pion-decay constant, respectively.
   Expanding the counterterm Lagrangian of Eq.\ (\ref{lcounterterm})
in powers of the renormalized coupling constants generates an infinite
series, the individual terms of which are responsible for the subtraction
of loop diagrams.

\end{document}